\documentclass[prb,twocolumn,notitlepage,superscriptaddress]{revtex4-2}
\usepackage{amsmath,amssymb}
\usepackage[hidelinks,colorlinks,linkcolor=blue,
citecolor=blue,urlcolor=blue]{hyperref}
\usepackage{graphicx,siunitx}
\usepackage[dvipsnames]{xcolor}

\usepackage{pifont}

\usepackage[normalem]{ulem}

\usepackage{xcolor}
\DeclareUnicodeCharacter{3000}{\textcolor{red}{BAD!!}}

\newcommand{\be}{\begin{equation}}
\newcommand{\ee}{\end{equation}}
\newcommand{\bea}{\begin{eqnarray}}
\newcommand{\eea}{\end{eqnarray}}
\newcommand{\bi}{\begin{itemize}}
\newcommand{\ei}{\end{itemize}}

\renewcommand{\be}{\beta}

\newcommand{\bpm}{\begin{pmatrix}}
\newcommand{\epm}{\end{pmatrix}}

\usepackage{amsmath,amsfonts,amssymb,amsthm,epsfig,array}
\usepackage{dsfont}
\usepackage{slashed}
\usepackage{graphics}
\usepackage{float}
\usepackage{verbatim}
\usepackage{color}
\usepackage{tabularx}
\usepackage[mathscr]{euscript}
\usepackage{mathtools}

\newcommand{\bsl}[1]{\boldsymbol{#1}}

\newcommand{\ket}[1]{|#1 \rangle}

\newcommand{\ii}{\mathrm{i}}

\newcommand{\dsZ}{\mathbb{Z}}

\newcommand{\dsR}{\mathbb{R}}

\newcommand{\eqnref}[1]{Eq.\,\eqref{#1}}
\newcommand{\figref}[1]{Fig.\,\ref{#1}}

\newcommand{\refcite}[1]{Ref.\,\cite{#1}}

\newcommand{\mat}[1]{\left(\begin{matrix}#1\end{matrix}\right)}

\newcommand{\eq}[1]{\begin{equation} #1 \end{equation}}

\newcommand{\eqa}[1]{\begin{align}\begin{split} #1 \end{split}\end{align}}

\usepackage{environ}
\NewEnviron{eqs}{%
\begin{equation}\begin{split}
    \BODY
\end{split}\end{equation}
}

\let\oldAA\AA
\renewcommand{\AA}{\text{\normalfont\oldAA}}

\newcommand{\ie}{{\emph{i.e.}}}

\newcommand{\cc}{\mathcal{K}}

\newcommand{\tmt}{\text{$t$MoTe$_2$}}

\newtheorem{proposition}{Proposition}
\newcommand{\propref}[1]{Prop.\,\ref{#1}}

\usepackage{cleveref}
\crefname{appendix}{App.}{Apps.}
\crefname{equation}{Eq.}{Eqs.}
\crefname{figure}{Fig.}{Figs.}
\crefname{table}{Tab.}{Tabs.}
\crefname{section}{Sec.}{Secs.}
\creflabelformat{appendix}{[#2#1#3]}

\begin{document}

\title{Large scale theoretical investigation of the phase diagram of twisted bilayer MoTe$_2$ at fractional fillings: agreements and contradictions with current experiments}

\author{Heqiu Li}
\affiliation{Donostia International Physics Center, P. Manuel de Lardizabal 4, 20018 Donostia-San Sebastian, Spain}

\author{Jiabin Yu}
\affiliation{Department of Physics and Quantum Theory Project, University of Florida, Gainesville, FL, USA}
\affiliation{Department of Physics, Princeton University, Princeton, New Jersey 08544, USA}

\author{Xiaodong Xu}
\affiliation{Geballe Laboratory for Advanced Materials, Departments of Applied Physics, Stanford University, Stanford, California 94305, USA}

\author{B. Andrei Bernevig}
\affiliation{Department of Physics, Princeton University, Princeton, New Jersey 08544, USA}
\affiliation{Donostia International Physics Center, P. Manuel de Lardizabal 4, 20018 Donostia-San Sebastian, Spain}
\affiliation{IKERBASQUE, Basque Foundation for Science, Bilbao, Spain}

\author{N. Regnault}
\affiliation{Center for Computational Quantum Physics, Flatiron Institute, 162 5th Avenue, New York, NY 10010, USA}
\affiliation{Laboratoire de Physique de l'Ecole normale sup\'{e}rieure, ENS, Universit\'{e} PSL, CNRS, Sorbonne Universit\'{e}, Universit\'{e} Paris-Diderot, Sorbonne Paris Cit\'{e}, 75005 Paris, France}
\affiliation{Department of Physics, Princeton University, Princeton, New Jersey 08544, USA}

\begin{abstract}

We present a comprehensive exact-diagonalization study of interaction-driven phases in twisted bilayer MoTe$_2$ across experimentally relevant twist angles ($2.13^\circ$--$4^\circ$) and hole fillings. Using continuum-model moir\'e bands, we compare the one-band-per-valley (1BPV) projection with a two-band-per-valley (2BPV) calculation that includes interaction-driven band mixing, and we benchmark both the widely used first-harmonic continuum model and a parameter-free DFT ``fitting-free'' model. At odd-denominator fillings, the 2BPV calculation reproduces the experimentally observed hierarchy of fractional Chern insulators (FCIs) around $\theta\approx 3.7^\circ$, including robust incompressible states at $\nu=-2/3$, $-3/5$, and $-4/7$ while correctly finding the absence of an FCI at $\nu=-3/7$, and it favors a charge density wave ground state at $\nu=-1/3$ over the FCI. At half filling $\nu=-1/2$, the 1BPV calculation exhibits clear composite Fermi liquid (CFL) signatures, whereas band mixing in 2BPV calculations destabilizes the CFL ground state. Finally, motivated by the Landau-level analogy at $\theta\approx 2.13^\circ$, we test the proposed non-Abelian Pfaffian state at $\nu=-3/2$ in the fully spin-polarized sector but find no evidence for this state within the models and parameters studied. Our results establish a unified numerical benchmark for correlated and topological phases in twisted bilayer MoTe$_2$ and clarify where multi-band physics is essential for a quantitative comparison with experiments.

\end{abstract}

\maketitle

\section{Introduction}

Fractional Chern insulators (FCIs) are lattice realizations of fractional quantum Hall (FQH) physics without an external magnetic field~\cite{Neupert2011Quantum,Sheng2011quantum,RegnaultBernevig2011Chern,Tang2011High-Temperature,Sun2011Flatbands,BergholtzLiu2013Flat,Parameswaran2013quantum}. Moir\'e superlattices~\cite{Cao2018ainsulator,Cao2018bsuperconductivity} provide a practical route to realize this phase, motivating extensive theoretical studies of FCIs in moir\'e systems~\cite{Abouelkomsan2020Duality,Ledwith2020Chern,RepellinSenthil2020Chern,Parker2021_arXiv2112tuned,Wilhelm2021fractional,ShefferStern2021magic-angle,Li2021fractional,Devakul2021NatCommu,Yu2020GiantBerryHomobilayer,Pan2020WSe2BandTopology,Zhang2021ChargeOrderTMD,DongWangFu2022electron}. On the experimental side, early studies devoted to reducing the magnetic field needed to stabilize lattice FCIs~\cite{Spanton2018fractional,Xie2021Chern}. In 2023, the experimental observation of FCIs was reported in twisted bilayer MoTe$_2$ ($t$MoTe$_2$) at multiple rational fillings~\cite{cai2023signatures,zeng2023integer,park2023observation,Xu2023FCItMoTe2}. To date, experiments have established robust FCIs with spontaneous magnetism at various fractional hole fillings including $\nu=-2/3,-3/5,-4/7,-5/9$ in \tmt ~\cite{cai2023signatures,zeng2023integer,park2023observation,Xu2023FCItMoTe2,Ji2024LocalProbetMoTe2,Young2024MagtMoTe2,Xu2024tMoTe2_3.15,Kang2024_tMoTe2_2.13,Park2024tMoTe2_2.6_3.8,park2025obsfci,xu_txl2025FCI,sun2026twistangleevolutionvalleypolarizedfractional,Chang2026ecgs,LiXiaodong2026sfc,WangHidden2025}, and FCI platforms have also been extended to include the superlattice of hBN-aligned rhombohedral multilayer graphene~\cite{Lu2024PGexp,Aronson2025DisplacementFCI,Xiaobohexa2025,Choi_tetralayer2024,Waters2025RPG,LuLong2025,Huo2025MoireMatter,Xie2025OrbitalMagnetism,Li2025StackingOrientation}. These experimental breakthroughs have in turn stimulated theoretical studies of many different aspects of \tmt, including band geometry, multi-band effects and competing phases~\cite{CrepelFu2023Hall,wang2023fractional,Qiu2023Topological,MoralesDuran2023bAngles,ReddyFu2023_arXiv2308_10406global,Abouelkomsan2023metric,Xu2024maximallfim,Li2024contrasting,MoralesDuran2023enhanced,SongZhangSenthil2023transitions,WangDevakulZaletelFu2023magnetic,Wu2023TRInvariantm,kwan2026FTI,liu2025orbitalmagnetizationcorrelatedstates,goncalves2025spinful,LuWuSantos2025ele,Zaklama2025StructureFactor,QiuWu2025MagnonsDomainWalls,Liu2025FCIQuasiparticles,Wu2024MetricInversionFCI,Luo2025solvingfra,Hart2026representability,CHEN20261034,Wang2026fti,Yu2024mote,Fu2023BandMixingFCItMoTe2,He2025fractionalcherninsulatorscompeting,hou2025stabilizingfractionalchernstates,Wang2023HigherIntegerMoTe2,kwan2024abelianfractionaltopologicalinsulators,Sheng2024QAHCrystal,Tuo2025fqah,Shen2024ExchangeFCI,SongSenthil2024AnyonHalo,LiWu2025VariationalMapping,Shen2026Magnetorotons,CrepelMillis2024TMDTightBinding,Zeng2024SublatticeTopology,Li2025DeepLearningTopologicalInsulators}, magnetism and field-driven transitions~\cite{LiuWangZhangCaoXiao2023antiferromagnetic,Fengcheng2023tMoTe2HFnum1,Sharma_2024qp,WangVafek2024MagneticFieldMoTe2,ShiLiu2026}, composite Fermi liquid (CFL) physics~\cite{Dong2023CFLtMoTe2,Goldman2023Composite}, and possible non-Abelian states~\cite{Wang2025higherLL,Xu2025nonabelian,Ahn2024nonabelian,Chen2025nonabelian,Reddy2024NonAbelianMinibands,Reddy2026nonabelian,LiWuAbelian2026}.

In this work, we focus on $t$MoTe$_2$, and provide a comprehensive theoretical study of the various experimental observations related to FCIs and magnetism.
The key experimental observations that we will study for $t$MoTe$_2$ are summarized as follows:
\begin{itemize}
    \item Robust magnetization has been observed for hole fillings $\nu$ ranging from $\sim -0.35$ to $\sim -1.1$ at twist angles $\theta=3.4^\circ$~\cite{zeng2023integer}, $\theta=3.7^\circ$~\cite{cai2023signatures,park2023observation}, and $\theta=3.9^\circ$~\cite{Young2024MagtMoTe2}.
    
    \item Experiments near $\theta \approx 3.7^\circ$ report Chern insulators (CIs) at $\nu=-1$~\cite{cai2023signatures,zeng2023integer,park2023observation,Xu2023FCItMoTe2,park2025obsfci,xu_txl2025FCI} and FCIs at $\nu=-2/3$~\cite{cai2023signatures,zeng2023integer,park2023observation,Xu2023FCItMoTe2,park2025obsfci,xu_txl2025FCI} and $\nu=-3/5$~\cite{cai2023signatures,park2023observation,park2025obsfci,xu_txl2025FCI}. 
    More recently, FCIs at $\nu=-4/7,\ -5/9$ have been observed in high-quality samples~\cite{park2025obsfci,xu_txl2025FCI}. At $\nu=-1/3$, the results differ between groups. A charge-density-wave (CDW) state has been reported in experiments~\cite{Xiaodongnew}, whereas recent experiments reported an FCI with ferromagnetism from optical measurements~\cite{Pan1v3FCI2026}. Such an FCI had not been observed in previous experiments~\cite{cai2023signatures,zeng2023integer,park2023observation,Xu2023FCItMoTe2,park2025obsfci,xu_txl2025FCI}. Overall, most experiments near $3.7^\circ$ have observed robust FCIs at $\nu=-2/3,-3/5,-4/7,-5/9$ but not at $\nu=-3/7,-2/5$.
    
    \item For other twist angles, recent experiments~\cite{Xiaodongnew} reported FCI from $2.5^\circ$ to $4^\circ$ for $\nu=-2/3$; from $2.6^\circ$ to $4^\circ$ for $\nu=-3/5$; and from 2.6$^\circ$ to $3^\circ$ for $\nu=-2/5$ and $\nu=-1/3$. CDW has also been reported for $\nu=-1/3$ at twist angles $3^\circ - 4^\circ$~\cite{Xiaodongnew}. At $2.1^\circ$, \refcite{Kang2024_tMoTe2_2.13} observed a CI at $\nu=-1$, and Ref.~\cite{KangTRBFQSH2025} reported spontaneous time-reversal-symmetry breaking over a broad range of fillings from $\nu<1$ to $\nu>6$ except for even integer fillings. \refcite{Park2024tMoTe2_2.6_3.8} and \refcite{Xu2024tMoTe2_3.15} observed incipient FCIs at $\nu=-2/3$ for $\theta=2.6^\circ$ and $\theta=3.15^\circ$, respectively, though in both cases the longitudinal resistance $R_{xx}$ remains large.

    \item Around $\nu=-1/2$, transport measurements in a $\theta=3.7^\circ$ device~\cite{park2023observation} show that the Hall resistance $R_{xy}$ varies linearly with $\nu$, while $R_{xx}$ remains small without a clear dip. Under a magnetic field, $R_{xy}$ shows hysteresis at $\pm 2 h/e^2$, with similar behavior also reported in a $\theta=3.9^\circ$ device. These features are consistent with a CFL or an anomalous Hall metal~\cite{CrepelFu2023Hall}. Optical trion-sensing measurements in $\theta=3.4^\circ$ and $3.6^\circ$ devices~\cite{Anderson2024TrionCFL} reported signatures of a zero-field CFL near $\nu=-1/2$. In contrast, microwave impedance microscopy in a $\theta\approx3.2^\circ$ device~\cite{Ji2024LocalProbetMoTe2} reveals a dip at $\nu=-1/2$, suggesting an insulating state, similar to the clearer dips observed at $\nu=-1$, $-2/3$, and $-3/5$.

\end{itemize}

\begin{figure}
    \centering
    \includegraphics[width=1.0\columnwidth]{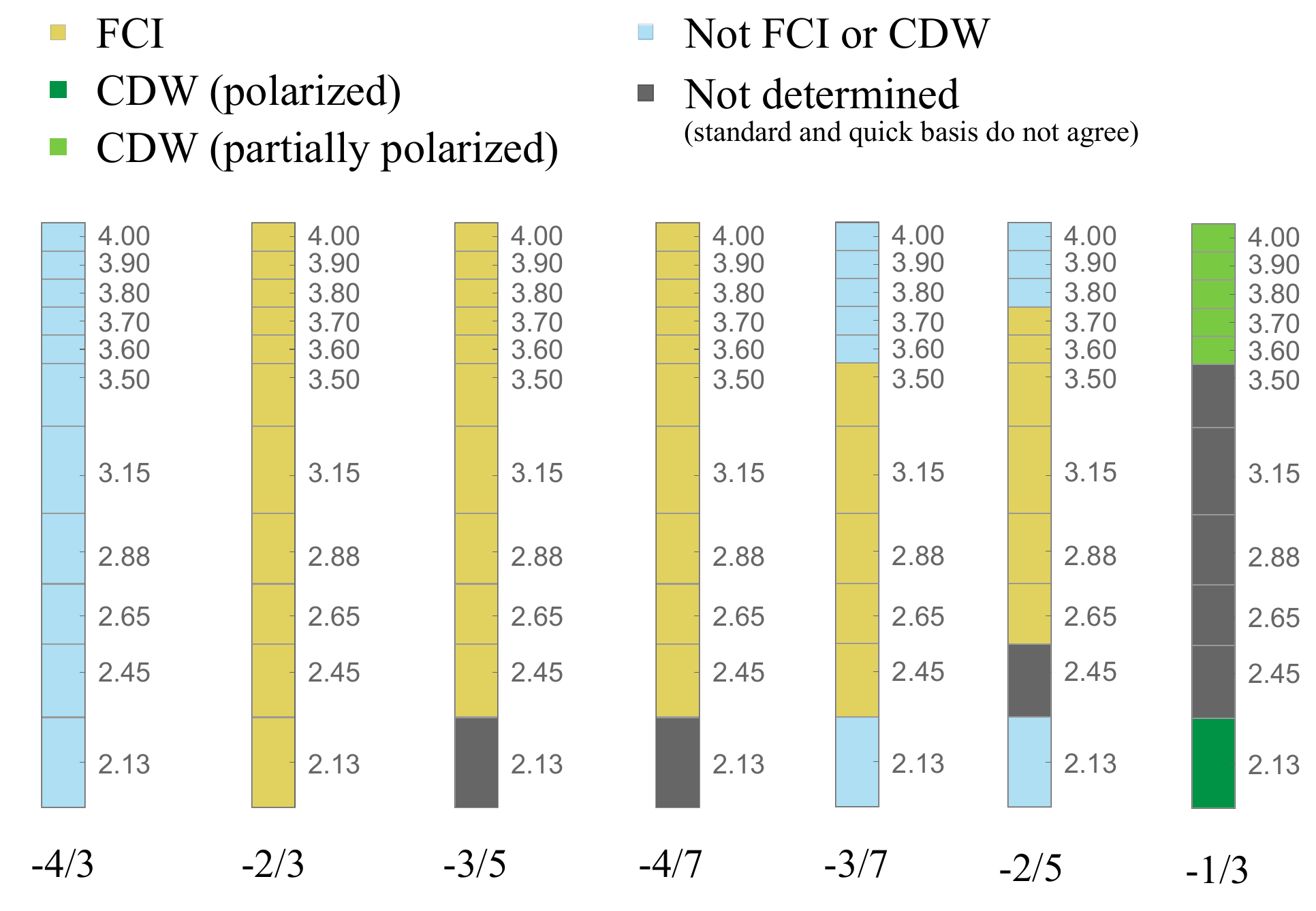}
    \caption{ Phase diagrams of ground states for various twist angles and filling fractions at interaction strength $10/\epsilon=0.9$. The ground states are obtained from 2BPV calculations in 21-site system for $\nu=-2/3,-4/7,-3/7,-1/3$ and 20-site system for $\nu=-3/5,-2/5$. The fillings $\nu=-2/3,-3/5,-4/7,-3/7$ are calculated in the sector with full spin polarization, and the calculations in $\nu=-2/5,-1/3$ include sectors with and without full spin polarization (more details can be found in Sec.~\ref{Sec_largephasediagram}). The $\nu=-4/3$ filling is obtained from 2BPV calculation with all spin sectors in $3\times 3$ system. The light green region with partially polarized CDW refers to the case where the ground state of polarized sector is CDW but there exist states in the other spin sectors with energy in-between the CDW ground states. 
    }
    \label{fig:phaselargesize}
\end{figure}

Capturing these experimental results requires a reliable and accurate single-particle description. Continuum-model descriptions of
twisted TMD homobilayers has been studied extensively~\cite{Wu2019TIintTMD,MFCII,wang2023fractional,Mao2024translearn,Xu2024maximallfim,Zhang2024pol,DFTnofitting2024,Xu2024maximallfim,Zhangtwisttransferable2025}. In particular, Ref.~\cite{DFTnofitting2024} proposed a DFT ``no-fitting'' method that can build a \text{moir\'e} model from DFT without continuous parameter fitting or Wannierizing the bands. The TMD materials have strong spin–valley locking, therefore the top \text{moir\'e} valence band in each valley is spin-polarized and carries Chern number $\pm 1$. At fractional hole fillings, Coulomb interactions can drive valley and spin polarization, allowing FCIs to emerge within this framework. Most many-body studies based on continuum models employ the one-band-per-valley (1BPV) projection onto the top moir\'e Chern band. As in fractional quantum Hall systems, where Landau-level mixing can modify the effective interaction and correlated ground states~\cite{PhysRevB.95.195105,PhysRevLett.130.186302}, such a single-band projection can overestimate spin polarization and FCI stability. Going beyond this approximation, previous works have shown that coupling to remote moir\'e bands can substantially reshape the phase diagram by suppressing spin polarization and modifying the stability of FCIs and competing phases~\cite{Yu2024mote,Fu2023BandMixingFCItMoTe2,kwan2026FTI,He2025fractionalcherninsulatorscompeting,hou2025stabilizingfractionalchernstates}. In particular, Ref.~\cite{Yu2024mote} demonstrated that two-band-per-valley (2BPV) calculations can improve the consistency with experiment at $\nu=-1/3$, $-2/3$, and $-4/3$. This observation was then confirmed by subsequent works~\cite{kwan2026FTI,Fu2023BandMixingFCItMoTe2,hou2025stabilizingfractionalchernstates}.

In this article, we provide a unified exact-diagonalization benchmark against experiments across twist angles from $2.13^\circ$ to $4^\circ$ and several experimentally relevant fillings, using both the first-harmonic (FH) continuum model and the DFT no-fitting model. At odd-denominator fillings with $\nu=-1/3, -2/3, -4/3, -2/5, -3/5, -3/7,-4/7$, we examine signatures of FCIs and their competing phases. At even-denominator fillings, we restrict the analysis to the fully spin-polarized sector, where we study CFL at $\nu=-1/2$ and search for a possible non-Abelian Pfaffian state at $\nu=-3/2$.
Our calculations include the top two moir\'e valence bands per valley (two-band-per-valley, or 2BPV).
We aim to capture key experimental phenomena across the relevant filling fractions, highlight the role of band mixing, and examine the scaling of FCI gaps.

Our findings for the FCIs at odd-denominator fillings are summarized in the schematic phase diagram in Fig.~\ref{fig:phaselargesize}, while the even-denominator fillings $\nu=-1/2$ and $\nu=-3/2$ are discussed separately in Secs.~\ref{sec:HalfFilling} and \ref{Sec_3v2}, where we restrict to the fully spin-polarized sector. Several aspects of this phase diagram are consistent with previous theoretical studies. In particular, the robust FCI at $\nu=-2/3$ agrees with earlier ED calculations that identified this filling as a stable FCI regime in $t$MoTe$_2$~\cite{wang2023fractional,ReddyFu2023_arXiv2308_10406global,Yu2024mote}, and the suppression of spurious spin-polarized FCIs at $\nu=-1/3$ and $-4/3$ by multi-band effects is closely related to the remote-band analysis of Ref.~\cite{Yu2024mote,kwan2026FTI,He2025fractionalcherninsulatorscompeting,hou2025stabilizingfractionalchernstates}. We find that robust FCI emerges at fillings $\nu=-2/3,\ -3/5,\ -4/7$ for a wide range of twist angles from 2.45$^\circ$ to 4$^\circ$. At $\nu=-3/7$ the FCI emerges only for twist angles smaller than 3.5$^\circ$. This is consistent with recent experiments~\cite{park2025obsfci,xu_txl2025FCI} around 3.7$^\circ$ that observed FCI at $\nu=-2/3,\ -3/5,\ -4/7$ but not at $\nu=-3/7,-2/5,-1/3$. At $\nu=-1/3$ near 3.7$^\circ$ we found that ground state is a charge density wave with ordering vector at \text{moir\'e} momentum $K_M$ (K-CDW) instead of FCI, consistent with recent experimental reports~\cite{Xiaodongnew}. At even-denominator filling $\nu=-1/2$, the 1BPV model shows clear CFL features with the same ground-state degeneracy as the lowest Landau level at half filling, whereas band mixing in the 2BPV model weakens these CFL signatures. We also considered $\nu=-3/2$ in the fully spin-polarized sector where the second-top moir\'e band is half-filled, but we do not find signatures of a non-Abelian Pfaffian state.

The remainder of this article is organized as follows. In Sec.~\ref{sec:ReviewModel}, we introduce the single-particle models and the interaction Hamiltonian used in the many-body calculations. In Sec.~\ref{Sec:phasediagrammain}, we study the phase diagrams at $\nu=-1/3$, $-2/3$, and $-4/3$, emphasizing the role of band mixing and the comparison between 1BPV and 2BPV results. In Sec.~\ref{Sec:v5v7}, we extend the analysis to the Jain-sequence fillings and discuss the scaling of the FCI gaps. In Sec.~\ref{sec:HalfFilling}, we examine the even-denominator filling $\nu=-1/2$ and the signatures of CFL physics. In Sec.~\ref{Sec_3v2}, we discuss the fully spin-polarized $\nu=-3/2$ sector and search for signatures of non-Abelian phases. We conclude in Sec.~\ref{Sec:conclusion} with a comparison to experiments and a discussion about open questions.

\section{Review of the Models}
\label{sec:ReviewModel}

In this section, we review the different models used to describe \tmt\ in the AA stacking configuration.
We focus on the valence-band manifold in each valley (labeled $+K$ and $-K$), where strong spin-orbit coupling (SOC)
locks spin up (down) to the $+K$ ($-K$) valley.
Recent theoretical studies~\cite{MFCII,Zhang2024pol,reddy2023fractional,Qiu2023Topological,wang2023fractional,LiuWangZhangCaoXiao2023antiferromagnetic,Mao2024translearn,Zhang2024pol,Wang2025higherLL} have shown that the top two valence bands per valley are highly relevant for strong-correlation
physics in $\tmt$. Here, we outline how the single-particle (SP) Hamiltonian is built, and how we project onto either one or two bands per valley for many-body computations.

\subsection{Continuum Hamiltonian}
\label{sec:ContinuumHamiltonian}

Following Refs.~\cite{Wu2019TIintTMD,MFCII}, we define $c_{\eta,l,\mathbf{r}}^\dagger$ as the operator that creates an electron at position $\mathbf{r}$ in layer $l \in \{t,b\}$ (top or bottom) and valley $\eta = \pm 1$ representing $\pm K$. In the AA stacking, the single-particle Hamiltonian can be written as $H_0=\sum_{\eta=\pm 1} H_{\eta,0}$, where
\eq{
H_{\eta,0} = \int d^2 r \left( c^\dagger_{\eta,b,\bsl{r}} , c^\dagger_{\eta,t,\bsl{r}} \right) \mat{ h_{\eta,b}(\bsl{r}) & t_\eta(\bsl{r}) \\ t_\eta^*(\bsl{r}) & h_{\eta,t}(\bsl{r})  } \mat{ c_{\eta,b,\bsl{r}} \\ c_{\eta,t,\bsl{r}} } \ .
}
The quantities $h_{\eta,l}(\bsl r)$ and $t_{\eta}(\bsl r)$ encode the intralayer moir\'e potentials and interlayer tunneling, respectively. This real-space Hamiltonian in valley $\eta$ respects the moir\'e translation and $C_3$, $C_{2y}T$ symmetries, while time-reversal $T$ maps $H_{\eta,0}$ to $H_{-\eta,0}$ in the opposite valley. A complete discussion can be found in Appendix~\ref{App_H0}.

To capture the top two valence bands around the $\pm K$ valleys, we expand the moir\'e potential in the reciprocal moir\'e vectors $\{\mathbf{G}_M\}$. Concretely, keeping only the first harmonic terms, we write
\bea
h_{\eta,l}(\bsl{r}) &=& \frac{\hbar^2 \nabla^2}{2 m^*} + V_{\eta,l}(\bsl{r}) \ ,\label{Hsppara}\\
V_{\eta,l}(\mathbf{r}) &=&
\left(V \, e^{-\,(-1)^l\, i\,\psi} \sum_{i=1}^3 \,e^{\,i\,\mathbf{g}_i\cdot \mathbf{r}}+h.c.\right)
\nonumber\\
t_{\eta}(\mathbf{r}) &=&
w \sum_{i=1}^3 \,e^{-\,\eta \, i\,\mathbf{q}_i\cdot \mathbf{r}}\nonumber
\eea
where $l \in \{t,b\}$ corresponds to top or bottom layer (with $(-1)^t = +1$ and $(-1)^b=-1$). The parameters $(V,\psi,w)$ describe the first-harmonic (FH) moir\'e potentials and interlayer tunneling. Here $\{\mathbf{g}_i\}$ are the three smallest moir\'e reciprocal lattice vectors related by $C_3$. The vectors $\{\mathbf{q}_i\}$ encode the momentum shift between layers. The parameters are obtained from fitting \emph{ab initio} band structure at twist angle $\theta= 3.89^\circ$ in Ref.~\cite{MFCII}, which are given in Table~\ref{tab:parameters_DFT} in App.~\ref{App_H0}. In this FH model, there is an emergent effective intra-valley inversion symmetry $\mathcal{I}$ given by $\mathcal I c^\dagger_{\eta,l,\bsl r}\mathcal{I}^{-1}=c^\dagger_{\eta,\overline{l},-\bsl r}$ which commutes with the Hamiltonian. This makes the single-particle energies at $\bsl k$ and $-\bsl k$ identical. This effective inversion symmetry will be broken if additional terms beyond the FH moir\'e potential are added.

\subsection{No-fitting models}\label{sec:nofittingmodels}

In addition to twist angles close to 3.7$^\circ$~\cite{cai2023signatures,zeng2023integer,park2023observation,Xu2023FCItMoTe2,park2025obsfci,xu_txl2025FCI}, recent experiments~\cite{Kang2024_tMoTe2_2.13,KangTRBFQSH2025,Park2024tMoTe2_2.6_3.8,Xu2024tMoTe2_3.15,Anderson2024TrionCFL,Ji2024LocalProbetMoTe2} approached smaller twist angles in the range $2^\circ-3.4^\circ$.
For these angles, the band structure can no longer be accurately captured by the FH continuum model with parameters determined from fitting density-functional theory (DFT) band structure at $3.89^\circ$ (see Table~\ref{tab:parameters_DFT} in App.~\ref{App_H0}). 
Moreover, as pointed out by ab-initio calculations~\cite{Mao2024translearn,MFCII}, relaxation effects are more important at smaller twist angles, making the FH continuum models less practical. 
Of course, one can include many extra harmonics and still fit the continuum model to the DFT results; however, doing so can easily lead to overfitting, resulting in a severe mismatch in the wavefunctions.

\begin{figure}
\centering
\includegraphics[width=0.9\columnwidth]{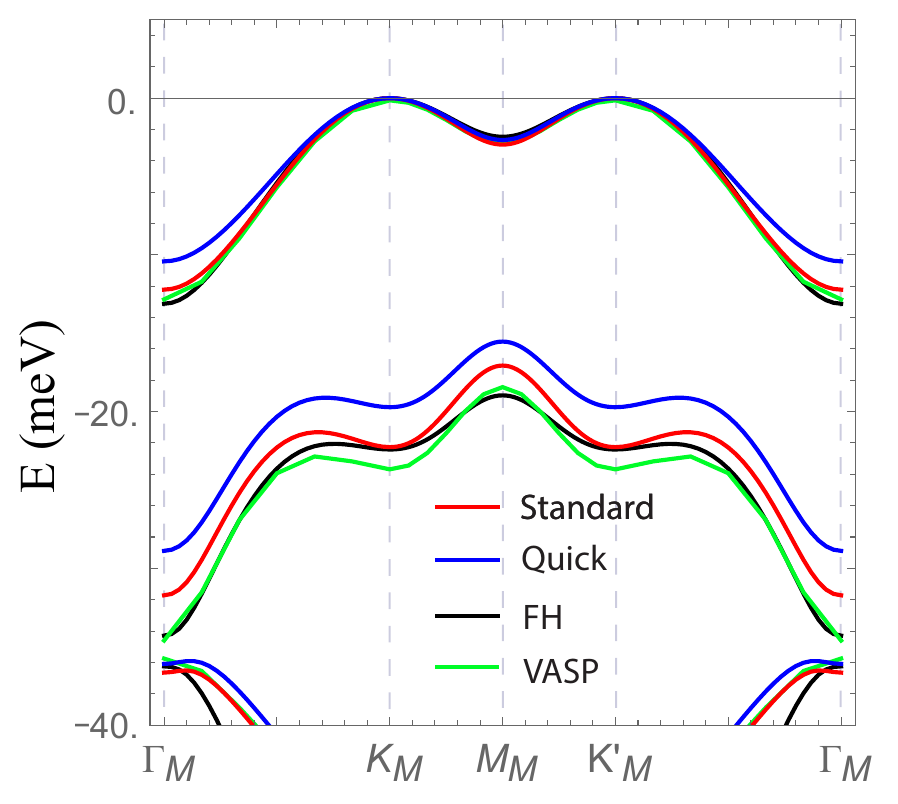}
\caption{ Band structures at 3.89$^\circ$ of VASP (green) and the continuum model with FH parameters (black), the DFT no-fitting model in the standard basis (red) and quick basis (blue).}
\label{fig:BandFHSQ}
\end{figure}

To overcome this challenge, different strategies have been proposed~\cite{DFTnofitting2024,wang2023fractional,Zhang2024pol,Zhangtwisttransferable2025,Shi2024adiabatic,Xu2024maximallfim}. 
Here we adopt the method 
of Ref.~\cite{DFTnofitting2024} to construct a single-particle model of $t$MoTe$_2$ directly from DFT, without any continuous parameter fitting, which we refer to as the no-fitting model. Because each twist angle requires an independent DFT calculation, this approach yields models only at a discrete set of angles where DFT is feasible. This model does not possess the effective inversion symmetry $\mathcal{I}$ that is present in the FH approximation because it includes terms beyond the FH approximation. In our implementation, the no-fitting model is generated from openMX, whereas the FH parameters were fit to DFT bands obtained with VASP at $3.89^{\circ}$. This difference in the underlying DFT leads to a small energy difference between the FH and no-fitting band structures in standard and quick bases in Fig.~\ref{fig:BandFHSQ}.

We construct the no-fitting model based on DFT calculations for seven discrete twist angles: 2.13$^\circ$, 2.45$^\circ$, 2.65$^\circ$, 2.88$^\circ$, 3.15$^\circ$, 3.48$^\circ$ and 3.89$^\circ$. Depending on the number of orbitals selected in the DFT calculation, the single-particle Hamiltonian in the no-fitting model can be slightly different. We consider two choices of DFT orbital basis~\cite{DFTnofitting2024}, which are denoted as the standard basis and the quick basis. The standard basis contains Mo-$s3p2d2$ and Te-$s3p2d2f1$, where Mo-$s3p2d2$ denotes 3 $s$-orbitals, 2 sets of $p$-orbitals, and 2 sets of $d$-orbitals for Mo, and similarly for Te-$s3p2d2f1$. The quick basis involves a different number of orbitals given by Mo-$s3p2d1$ and Te-$s3p2d2$. 

The band structures of VASP and the DFT no-fitting model in the standard basis, quick basis and the continuum model with FH parameters at 3.89$^\circ$ are shown in Fig.~\ref{fig:BandFHSQ}. The top valence bands in these models are qualitatively similar, with slightly different bandwidths. The bands obtained from the standard basis are closer to VASP than those from the quick basis, hence we will focus on the standard basis, and we also perform calculations in the quick basis as a comparison to examine the dependence of ground states on the single-particle model. From a practical perspective, the standard basis and the quick basis both lead to qualitatively similar band structures (see Fig.~\ref{fig:BandFHSQ} and App.~\ref{App_quickbasis}). But the quick basis tends to produce a top valence band with a smaller dispersion ($\simeq 2\ {\rm meV}$) while keeping the indirect gap almost unchanged. As such, the spread over gap ratio is more favorable for the quick basis (typically $0.6$ times smaller). Thus based on the energetics, we expect the quick basis to help stabilize FCIs at lower interaction.

\subsection{Normal-Ordered Interaction}

The \emph{ab initio} band structure is computed at charge neutrality, where all the valence bands are fully filled. Therefore, we choose the following form of interaction that is normal-ordered against charge neutrality:
\eq{
H_{\text{int}} = \frac{1}{2}\sum_{l \eta,l' \eta'} \int d^2 r d^2 r' V(\bsl{r}-\bsl{r}')  \widetilde{c}^\dagger_{\eta, l, \bsl{r}} \widetilde{c}^\dagger_{\eta', l', \bsl{r}'} \widetilde{c}_{\eta', l', \bsl{r}'} \widetilde{c}_{\eta, l, \bsl{r}} \ .
\label{Vint}
}
Here $\tilde c^\dagger_{\eta, l, \bsl{r}}\equiv \cc c_{\eta,l,\bsl{r}} \cc^{-1}$ is the creation operator of a hole in valley $\eta$ and layer $l$, and $\cc$ is the particle-hole operator. Note that in Eq.~\eqref{Vint} we order the hole annihilation operators to the right, which ensures that the interaction can annihilate the state at charge neutrality that has all valence bands fully occupied. We use the double-gated screened Coulomb potential with gate distance $\xi=20$ nm:
\bea
V(\bsl{p}) = \pi \xi^2 V_{\xi} \frac{\tanh(\xi |\bsl{p}|/2)}{\xi |\bsl{p}|/2},
\eea
where $V_{\xi} = \frac{e^2 }{4\pi \epsilon \epsilon_0 \xi}$ and $V(\bsl{r}) =\int_{\dsR^2} \frac{d^2 p }{ (2\pi)^2} V(\bsl{p}) e^{\ii \bsl{p}\cdot\bsl{r}}$. 
Following previous works, we use the dimensionless parameter $10/\epsilon$ as a convenient way to tune the interaction strength relative to the kinetic energy~\cite{Yu2024mote,ReddyFu2023_arXiv2308_10406global,CrepelFu2023Hall}. This parameter should be viewed as an effective screening strength rather than a directly measured dielectric constant, since the experimental screening depends on the hBN environment, gates, and the anisotropic dielectric response of MoTe$_2$. Typical values used in modeling place $\epsilon$ in the range of $10$ to $20$~\cite{laturia_dielectric_2018,kwan2026FTI}, which gives a typical range of $10/\epsilon$ of $0.5$--$1$.

In our ED computation, we always use the hole basis since we study the moir\'e system under hole doping. The single-particle Hamiltonian in the hole basis can be diagonalized as:
\eq{
\label{eq:H_eta_0_hole_eigen}
H_{\eta,0} = \sum_{\bsl{k},\eta,n} \widetilde{\gamma}^\dagger_{\bsl{k},\eta,n} \widetilde{\gamma}_{\bsl{k},\eta,n} (-\epsilon_{\bsl{k},\eta,n})\ , 
}
where $\epsilon_{\bsl{k},\eta,n}$ is the electron band energy and $\tilde \gamma^\dagger_{\bsl{k},\eta,n}$ is the creation operator for the hole eigenstate at $\bsl{k}$ in valley $\eta$ and band $n$. We keep at most the topmost two electron valence bands per valley in the ED computation. Denote the top electron valence band as band 0, and the second-top electron valence band as band 1. Then in the hole basis, band 0 has lower energy and band 1 is about $10-20$ meV above it.

\begin{figure}
(a) $\nu=-1$, $18\times 18$, FH model, 2BPV
\includegraphics[width=\columnwidth]{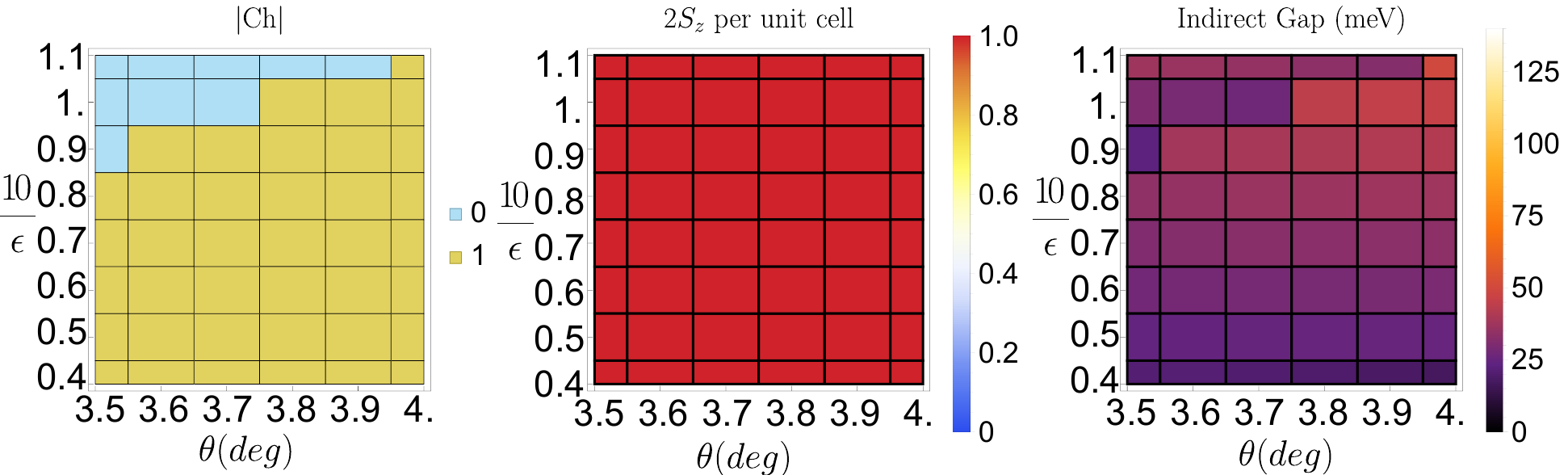}
(b) $\nu=-1$, $18\times 18$, Nofitting model, 2BPV
\includegraphics[width=\columnwidth]{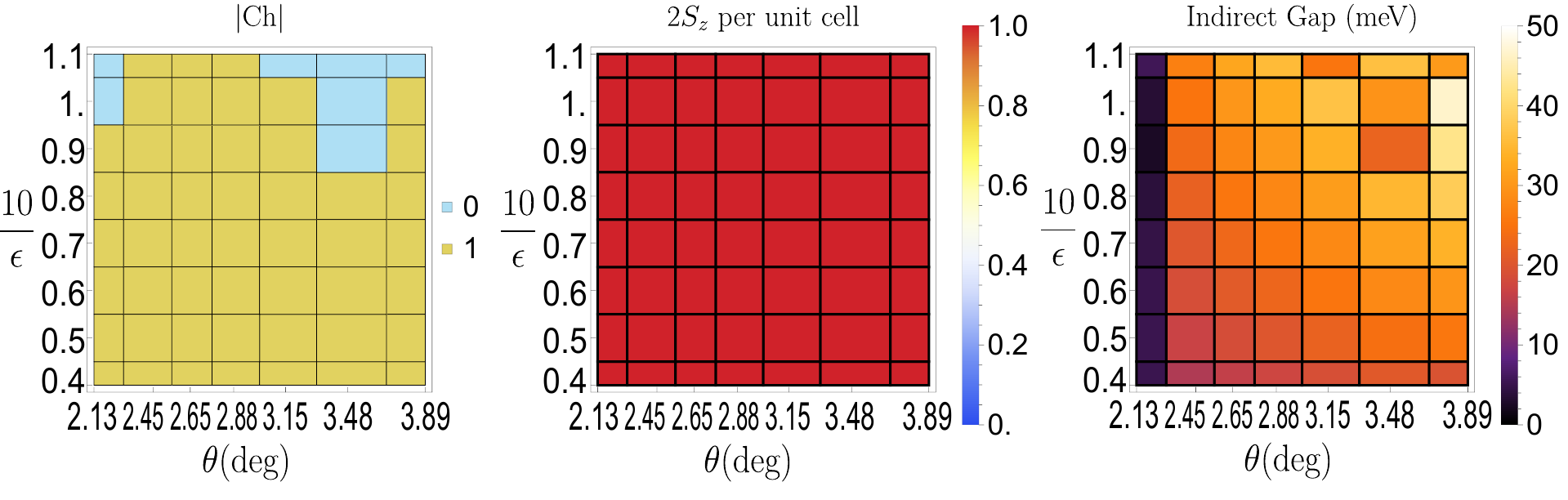}
\caption{ Hartree-Fock results at $\nu=-1$ for (a) FH model and (b) no-fitting model standard basis in $18\times 18$ system.
Here $|\text{Ch}|$ labels the absolute value of the Chern number of the ground states (left panel), $2 S_z$ is twice of the spin in unit of $\hbar$ (middle panel), and the indirect gap refers to the indirect gap of the HF band structure (right panel).
}
\label{fig:HF_results}
\end{figure}

As a first test of the fully interacting model $H_0+H_{\text{int}}$, we performed a Hartree-Fock (HF) calculation at integer filling $\nu=-1$, which is shown in \figref{fig:HF_results}. We found that the ground state in the 2BPV calculation is fully spin-polarized over a wide range of twist angles from 2.13$^\circ$ to 4$^\circ$ and over interaction strength in the range of $10/\epsilon\in [0.4,1.1]$. At small or intermediate interaction with $10/\epsilon<0.9$ the HF ground state has Chern number $C=\pm 1$, whereas at larger interaction with $10/\epsilon\sim 1.1$ the Chern number becomes $C=0$ for $3.15^\circ\le \theta\le4.0^\circ$. This change of Chern number with interaction comes from the presence of the second-top valence band, which indicates the non-negligible effect of band mixing.

In the following, we will perform ED computations for various twist angles and interaction strengths.
{Motivated by previous studies emphasizing the importance of remote-band effects in $t$MoTe$_2$~\cite{Yu2024mote,Fu2023BandMixingFCItMoTe2,kwan2026FTI,He2025fractionalcherninsulatorscompeting,hou2025stabilizingfractionalchernstates}, we consider both 1BPV and 2BPV calculations.} We demonstrate that while 1BPV computations can capture the emergence of valley polarization and FCI at certain fillings (e.g., $-2/3$ filling), it is not consistent with experiments in some other fillings (e.g., $-1/3$ and $-4/3$ fillings). The inconsistency with experiments can be fixed by 2BPV computations by taking the next band into account.

\section{Phase diagrams at filling fractions $\nu=-1/3,-2/3$ and $-4/3$ }
\label{Sec:phasediagrammain}

In this section, we perform ED computations using the FH and no-fitting models and obtain the phase diagrams at filling fractions $\nu=-1/3,-2/3$ and $-4/3$. 

\subsection{Numerical signatures of fractional phase}

Before detailing the results at each filling, we define the criterion used to identify fractional Chern insulators in finite-size ED:

\begin{proposition}
\label{prop:FCI}
     
    Consider a filling $\nu = -p/q$ with $q$ odd. We label the system an FCI if (i) the $q$ lowest states are in the same spin sector labeled by $S_z^{\rm{FCI}}$, (ii) the momenta of the $q$ lowest states match the momenta of an FCI~\cite{1961AnPhy..16..407L,BernevigPhysRevB.85.075128,2014PhRvB..89o5113W}, and (iii) the spread of the $q$ lowest states is smaller than the gap between the $q$th lowest state and $(q+1)$th lowest state in the $S_z^{\rm{FCI}}$ sector.
    
\end{proposition}
Note that the criteria do not fully rule out other competing states. For example, a CDW could appear in the same sectors as an FCI depending on the geometry of the momentum mesh (see App.~\ref{App_largeu}). But we use these conditions to quickly determine potential FCI regions, potentially refining with other probes such as particle entanglement spectrum (PES) \cite{SterdyniakPhysRevLett.106.100405,RegnaultBernevig2011Chern} if needed. We also point out that (iii) is quite restrictive and could be relaxed as in Ref.~\cite{kwan2026FTI}, which only requires the largest separation within the $q$ states rather than the full spread of the $q$ states to be smaller than the gap. In addition to the existence of FCI, we also characterize the magnetic properties of the ground states by the spinful gap. Specifically, given a fixed filling, we can determine the energy of the lowest energy state in each spin sector $S_z$, which is labeled as $E^{S_{z}}$, and we only need to consider $S_z\geq 0$ owing to the TR symmetry. 
Then, the spinful gap is defined as 
\eq{
\text{spinful gap}=\min_{S_z\ne S_z^{\rm{FCI}}}(E^{S_z} - E^{S_z^{\rm{FCI}}})\ ,
\label{Eqspingap}
}
The spinful gap indicates the robustness of the ferromagnetism. We find that when FCI emerges based on \propref{prop:FCI}, $S_z^{\rm{FCI}}$ always equals $S_z^{\rm{max}}$, which is the maximum value of $S_z$ at that filling. In large system sizes the full spinful gap may not be available due to computation limits, we use a spin-1 gap instead which is defined as $E^{S_z^{\rm{max}}-1} - E^{S_z^{\rm{FCI}}}$. As long as the system size allows, we will use spinful gap; otherwise, we use the spin-1 gap.

The ED computation is performed in a momentum mesh in the torus of the moir\'e Brillouin zone (MBZ). The momentum mesh for a $N_x\times N_y$ system is made of momentum points $\bsl k$ of the following form: 
\eq{
\label{eq:momentum_mesh}
\bsl k= \frac{k_x}{N_x} \bsl{f}_1 + \frac{k_y}{N_y} \bsl{f}_2 
}
where $k_i=0,1,...,N_i-1$ for $i=x,y$, and
\eqa{
\label{eq:f_1_f_2}
& \bsl{f}_1 = \widetilde{n}_{11} \bsl{b}_{M,1} + \widetilde{n}_{12} \bsl{b}_{M,2} \\
& \bsl{f}_2 = \widetilde{n}_{21} \bsl{b}_{M,1} + \widetilde{n}_{22} \bsl{b}_{M,2}
}
are two reciprocal lattice vectors (\ie, $\widetilde{n}_{11}$, $\widetilde{n}_{12}$, $\widetilde{n}_{21}$ and $\widetilde{n}_{22}$ are integers). Here $N_x$ and $N_y$ are positive integers, and the total number of momentum points is given by $N_s=N_x N_y$. $\bsl{b}_{M,1}$ and $\bsl{b}_{M,2}$ are shortest nonzero moir\'e reciprocal lattice vectors with $\bsl{b}_{M,2}=R(\frac{2\pi}{6})\bsl{b}_{M,1}$, where $R(\phi)$ is the rotation matrix of vectors by angle $\phi$. These parameters specify the momentum mesh for the ED computation. A more detailed discussion of the properties of the tilted lattice can be found in App.~\ref{App_EDtech}. The 2BPV ED calculation can involve extremely large Hilbert space, which makes the calculation inefficient. In those cases, we can make the approximation to truncate the Hilbert space by only including many-body states with less than $n$ particles in the high energy band, which we refer to as "bandmax $n$"~\cite{MFCIIV,Li2025multibanditer}. Then bandmax 0 is equivalent to the 1BPV calculation. We will take the bandmax approximation when necessary, and if the bandmax of the 2BPV calculation is not mentioned, it is performed without truncation of Hilbert space.

\begin{figure}
    \centering
    \includegraphics[width=\columnwidth]{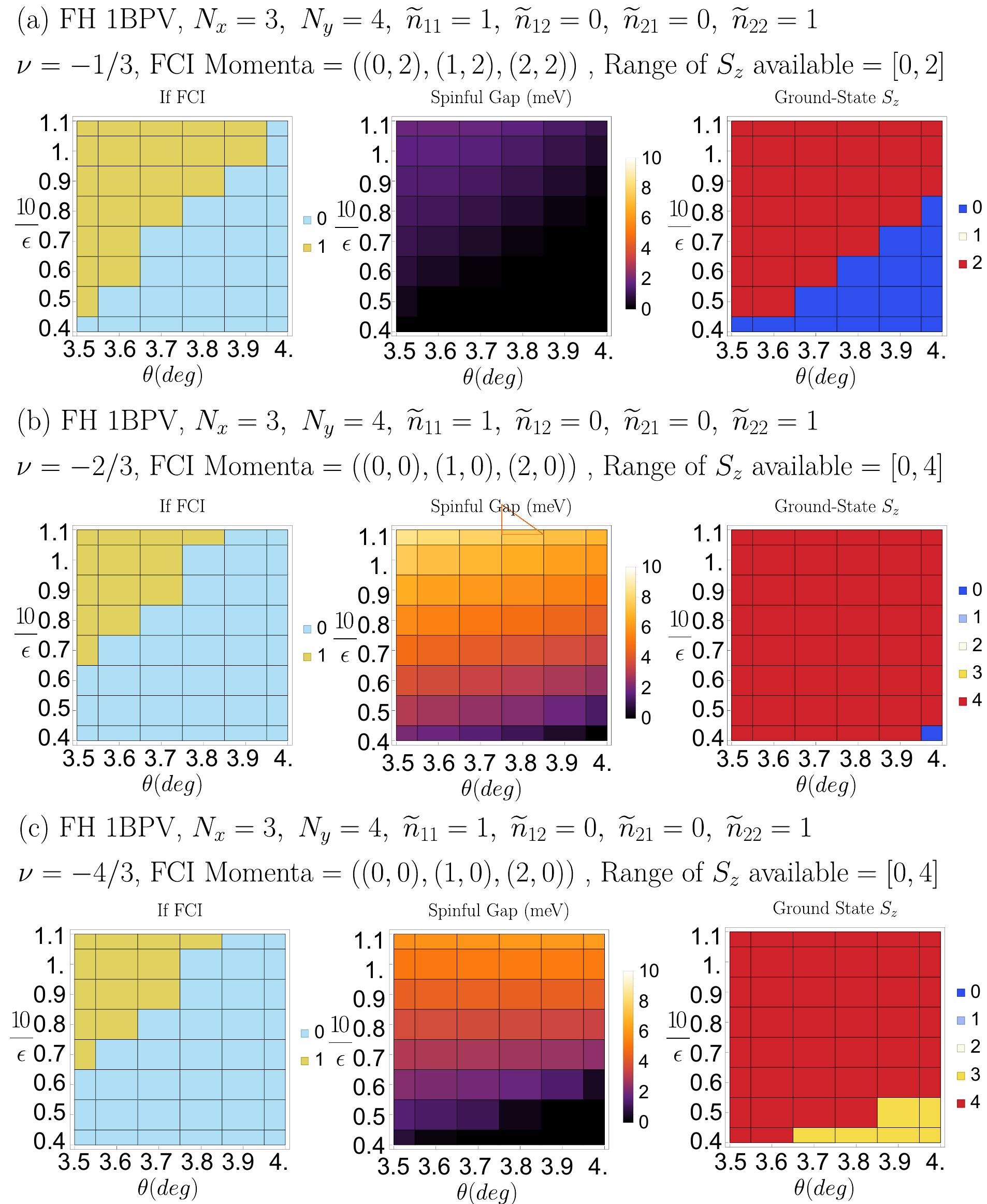}
    \caption{1BPV ED results at $\nu=-1/3,-2/3,-4/3$ for the FH model on a $3 \times 4$ grid. 
    Here the calculations are all done on a momentum mesh of the form in \eqnref{eq:momentum_mesh} with $N_x$, $N_y$, $\widetilde{n}_{11}$, $\widetilde{n}_{12}$, $\widetilde{n}_{21}$ and $\widetilde{n}_{22}$ specified in the figure, and FCI momenta are shown in $(k_x, k_y)$ format. ``Range of $S_z$ available" refers to the range of $S_z$ included in the calculation, and the spin $S_z$ is in the unit of $\hbar$.
    In the left panels, $1$ means the system is in an FCI phase according to \propref{prop:FCI}, while $0$ means otherwise.
    In the middle panels, the spinful gap is shown as zero if its actual value becomes negative.
    }
    \label{fig:FH_1BPV_thirds}
\end{figure}

\begin{figure}
    \centering
    \includegraphics[width=\columnwidth]{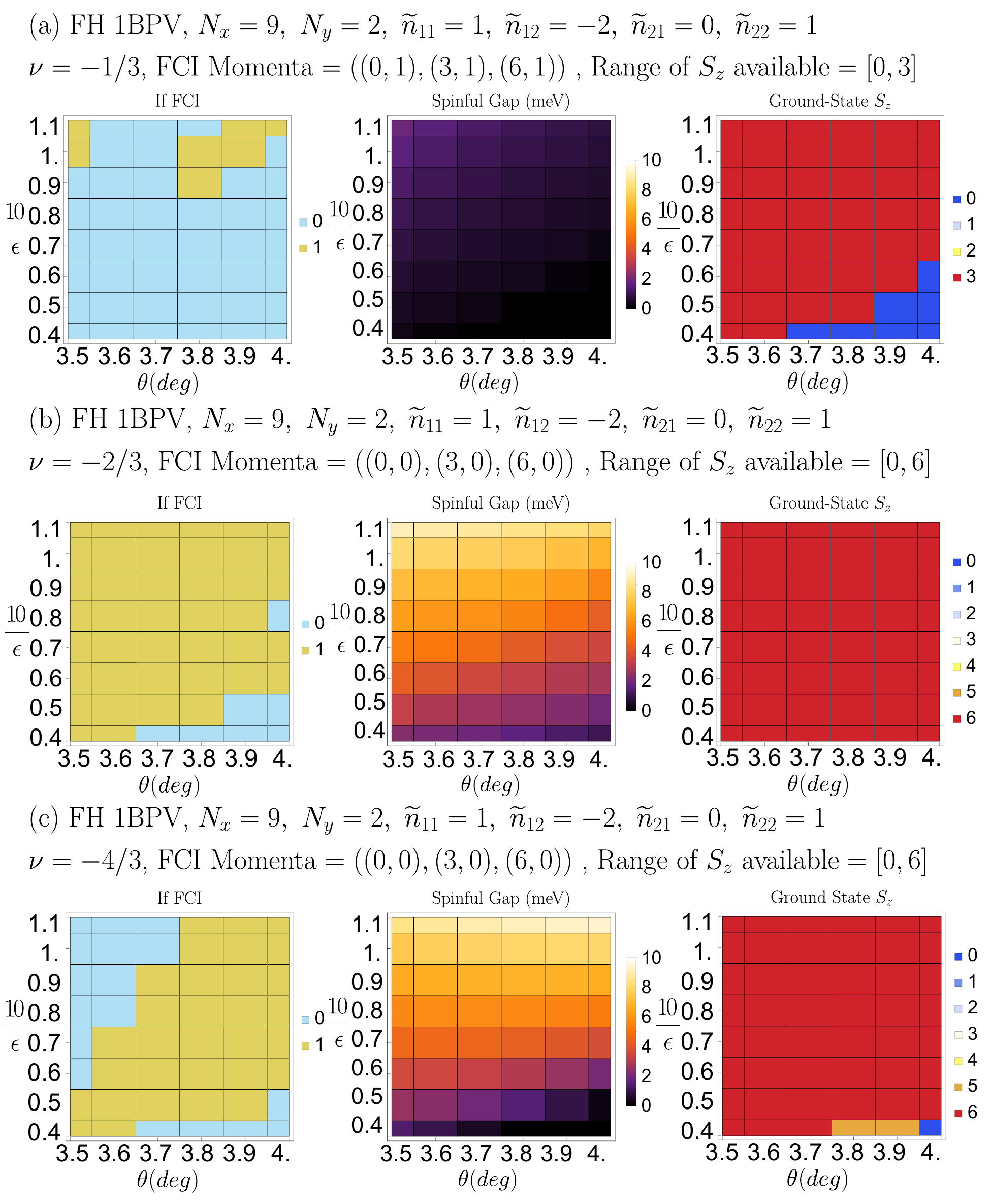}
    \caption{1BPV ED results at $\nu=-1/3,-2/3,-4/3$ for the FH model in a $9\times 2$ system. 
    The meaning of the labels can be found in the caption of \figref{fig:FH_1BPV_thirds}.
    }
    \label{fig:FH_1BPV_thirds_9b2}
\end{figure}

\subsection{Phase diagrams at twist angles $\theta=3.5^\circ-4.0^\circ$}

\subsubsection{1BPV results}
\label{sec:1BPVresults}

In this section, we use~\propref{prop:FCI} to identify fractional Chern insulators at fillings $\nu=-1/3,-2/3$ and $-4/3$ for twist angles $3.5^\circ-4.0^\circ$ in the FH model from 1BPV calculations with all spin sectors taken into account. We present phase diagrams for two different system sizes: $3\times 4$ in Fig.~\ref{fig:FH_1BPV_thirds} and $9\times 2$ in Fig.~\ref{fig:FH_1BPV_thirds_9b2}. The later is the largest size where spinful calculations can be done with a reasonable amount of computing resources for a full phase diagram (the largest Hilbert space being around $2\times 10^7$ while the former will be used as a comparison point for the 2BPV calculations in Sec.~\ref{sec:2BPVresults_thirds}. Additional system sizes are provided in App.~\ref{App:EDresults}.

As shown in Fig.~\ref{fig:FH_1BPV_thirds_9b2}, our 1BPV computations at filling $\nu=-2/3$ exhibit a large parameter region of spin polarization and an FCI according to~\propref{prop:FCI}. The smaller system in Fig.~\ref{fig:FH_1BPV_thirds} exhibits a smaller FCI region. By inspecting the gap above the FCI in the $9 \times 2$ in the fully polarized sector (see App.~\ref{App_FCIgap}) reveals the non-FCI region in the $3\times 4$ is related to a smaller gap in the larger system. As such, it is not surprising that some region of the phase diagram is more sensitive to the finite-size effects. Note that the magnetic properties are less sensitive to such effects.  Overall, these results are consistent with experiments~\cite{cai2023signatures,zeng2023integer,park2023observation,Xu2023FCItMoTe2} that observed ferromagnetism and FCIs at filling $-2/3$.

However, some noticeable discrepancies with experiments remain in the 1BPV phase diagram at fillings $\nu=-1/3$ and $-4/3$. At $\nu=-1/3$, no FCI has been observed experimentally for samples with twist angles near $3.7^\circ$, although those samples can show an FCI at $\nu=-2/3$~\cite{cai2023signatures,zeng2023integer,park2023observation,Xu2023FCItMoTe2}. At this filling factor, a $K$ charge density wave (CDW) has been proposed~\cite{reddy2023fractional,He2025fractionalcherninsulatorscompeting,Yu2024mote,Sharma_2024qp}, in line with the recent experiments~\cite{
Xiaodongnew} reporting a CDW ground state at twist angles $3^\circ -4^\circ$. {The $3\times4$ system in Fig.~\ref{fig:FH_1BPV_thirds} does not contain the moir\'e $K_M$ and $K'_M$ points, which disfavors a $K$-CDW and therefore leads to a larger apparent FCI region. By contrast, the $9\times2$ mesh contains both $K_M$ and $K'_M$, making it compatible with $K$-CDW order. We therefore examine the possible CDW physics at $\nu=-1/3$ more carefully below.}

\begin{figure}
    \centering
    \includegraphics[width=\columnwidth]{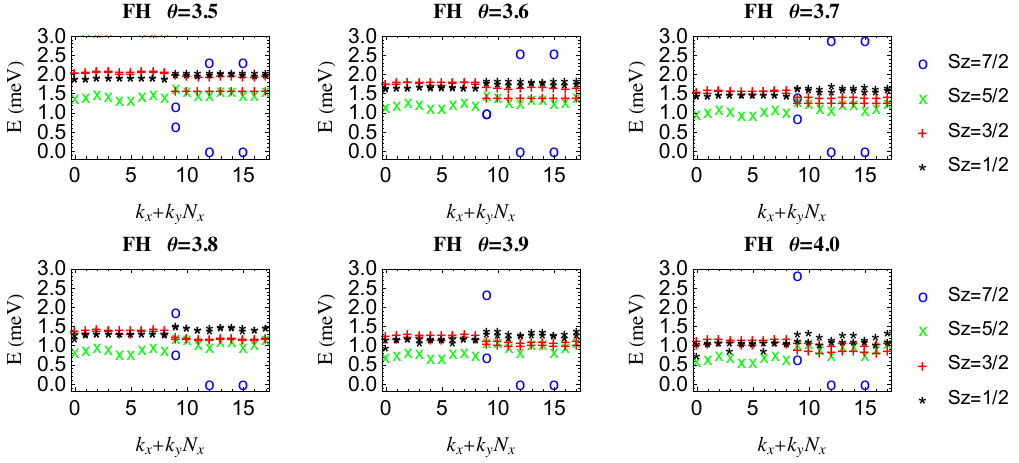}
    \caption{1BPV ED energy spectra in the FH model at $\nu=-1/3$ for $9\times 2$ system with $(N_x,N_y,\widetilde{n}_{11},\widetilde{n}_{12},\widetilde{n}_{21},\widetilde{n}_{22})=(9,2,1,-2,0,1)$ at interaction $10/\epsilon=0.9$. The colors represent different spin sectors. The ground states in the fully-polarized sector has momenta consistent with $K$-CDW and FCI, but there can be low-lying spin excitations with energy similar or smaller than the highest state inside the ground state manifold. Note that the other states in the $S_z=7/2$ sector have energies higher than 3 meV, and we only plot the lowest two energies for each momentum in each spin sector below 3 meV.
    }
    \label{fig:eng_FH_1BPV_thirds_9b2_n6}
\end{figure}

\begin{figure}
    \centering
    \includegraphics[width=\columnwidth]{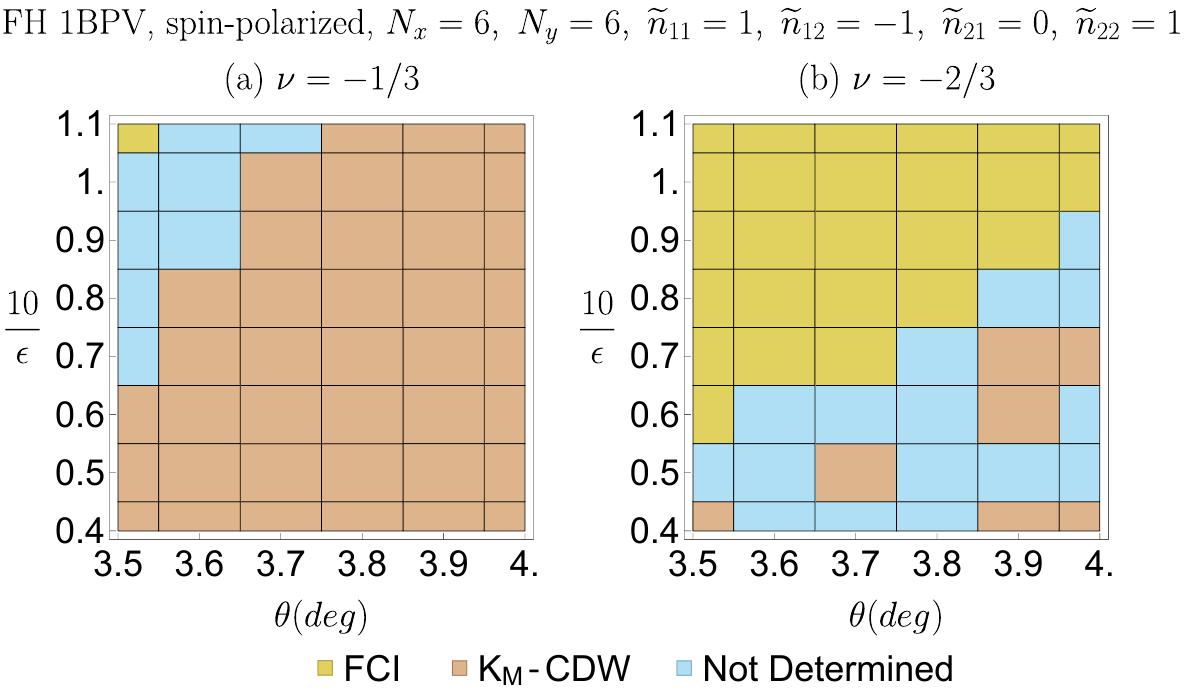}
    \caption{1BPV ED phase diagram in $\nu=-1/3,-2/3$ for the FH model in the fully spin-polarized sector for $6\times 6$ system.
    Yellow means the system is in an FCI phase according to \propref{prop:FCI}, brown means the system is in the $K_M$ charge density wave (CDW), and blue means the ground state cannot be identified as an FCI or CDW by \propref{prop:FCI}
    The other labels have the same meaning as in \figref{fig:FH_1BPV_thirds}.
    }
    \label{fig:FH_1BPV_thirds_6x6}
\end{figure}

The 1BPV spectra at $\nu=-1/3$ in the $9\times2$ system are shown in Fig.~\ref{fig:eng_FH_1BPV_thirds_9b2_n6}. In the fully spin-polarized sector, the three lowest states occur at momenta compatible with a $K$-CDW. However, these three fully polarized states are not cleanly separated from the rest of the spectrum: some lower-spin states have energies below the highest-energy state within this three-state manifold. Moreover, on this particular momentum mesh, the candidate FCI ground states has the same momenta as the $K$-CDW ground states, hence the spectrum in Fig.~\ref{fig:eng_FH_1BPV_thirds_9b2_n6} alone cannot distinguish the two phases. To distinguish the FCI and $K$-CDW more directly, we therefore study a larger $6\times6$ system in Fig.~\ref{fig:FH_1BPV_thirds_6x6}.
There, FCI and CDW ground states appear at different momenta (as opposed to the $9\times 2$ system), making the phase identification easier by simple inspection of the low-energy spectrum. Moreover, the large system minimizes the finite-size effects. Due to the large Hilbert-space dimension of $10^{11}$ for the full calculation in a $6\times 6$ system with all spin sectors, we only perform the calculation in the fully spin-polarized sector, leading to a more reasonable dimension of $3.5 \times 10^7$. The three CDW states are at momenta $\Gamma_M,K_M,K'_M$ whereas the three FCI states are all at $\Gamma_M$.  It confirms that in the FH model the ground state at $3.5^\circ - 3.9^\circ$ is mainly FCI at $\nu=-2/3$ for an interaction strength $10/\epsilon \geq 0.8$, while the phase diagram at $\nu=-1/3$ is dominated by a $K$-CDW in the fully polarized sector. Note that for $3.5^\circ$ and $10/\epsilon \simeq 1.1$, we do obtain a small FCI region that could be compatible with the recent experimental observations~\cite{Pan1v3FCI2026}.

For the 1BPV calculation at $\nu=-4/3$, the system has an almost particle hole symmetry relating its phase diagram to the one at $\nu=-2/3$ \cite{Yu2024mote} as can be seen either in Fig.~\ref{fig:FH_1BPV_thirds} or in Fig.~\ref{fig:FH_1BPV_thirds_9b2}. As such, the system is maximally polarized in contradiction with the experimental observations~\cite{zeng2023integer,cai2023signatures,park2023observation,Young2024MagtMoTe2}. This discrepancy is solved once band mixing is taken into account as we will show in the next subsection.

\subsubsection{2BPV results}
\label{sec:2BPVresults_thirds}

\begin{figure}
    \centering
    \includegraphics[width=\columnwidth]{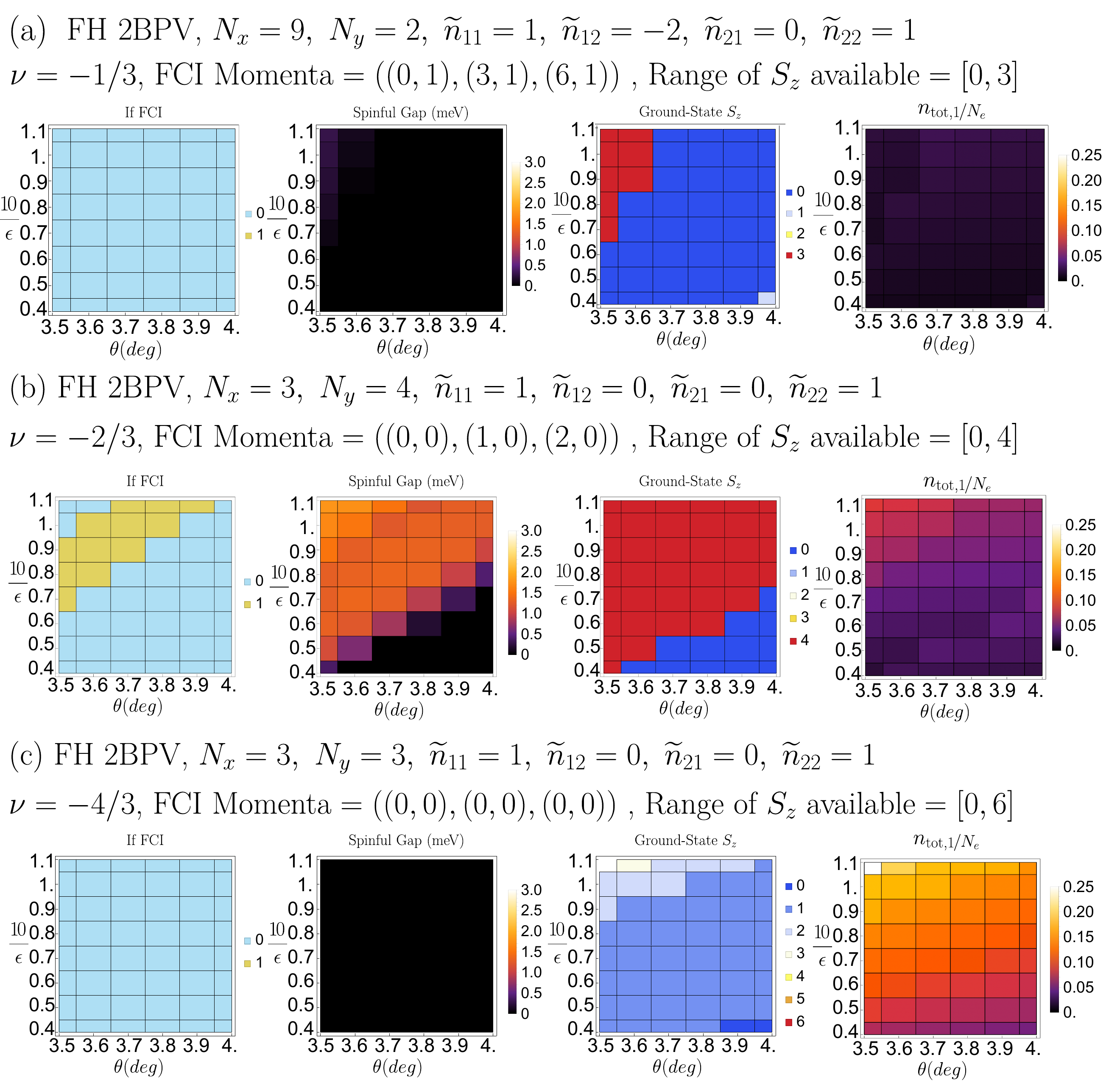}
    \caption{2BPV ED results with $\nu=-1/3,-2/3,-4/3$ for the FH model calculated in $9\times 2,\ 3\times 4$ and $3\times 3$ systems respectively.
    Here $n_{\text{tot},1}$ is the total occupation number in band 1 calculated from the lowest many-body state and $N_e$ is the number of holes. The meaning of the other labels can be found in the caption of \figref{fig:FH_1BPV_thirds}.
    }
    \label{fig:FH_2BPV_thirds_new}
\end{figure}

As was first reported in Ref.~\cite{Yu2024mote} and further emphasized in later works~\cite{kwan2026FTI,Fu2023BandMixingFCItMoTe2,He2025fractionalcherninsulatorscompeting,hou2025stabilizingfractionalchernstates}, band mixing plays a crucial role in $t$MoTe$_2$, especially for the magnetic properties and the stability of FCIs at fillings such as $\nu=-1/3$ and $-4/3$. Here we present the results in the 2BPV approximation at $\nu=-1/3,\ -2/3$ and $-4/3$. Fig.~\ref{fig:FH_2BPV_thirds_new} provides the phase diagram, spin gap and spin polarization for these filling factors for system sizes $9 \times 2$ for $\nu=-1/3$, $3 \times 4$ for $\nu=-2/3$ and $3 \times 3$ for $\nu=-4/3$. Each geometry is chosen to lead to a large but reasonable Hilbert-space dimension where the full 2BPV spinful calculation can be performed.

Starting with $\nu=-1/3$ and $-2/3$, we observe that the ferromagnetic region in 2BPV is smaller than that in 1BPV at both fillings. For $\nu=-2/3$, the system is fully polarized irrespective of the twist angle for $10/\epsilon \geq 0.8$, whereas for $\nu=-1/3$ the ferromagnetic region almost disappears. More precisely for an interaction strength $10/\epsilon \geq 0.8$, it only remains fully polarized for twist angles below $3.7^\circ$. For $\nu=-4/3$, no full spin polarization or FCI was found, and the $S_z$ of the ground states is nearly zero. This is in sharp contrast to the 1BPV results in Sec.~\ref{sec:1BPVresults} where the ground state is spin-polarized in most of the parameter region. The absence of FCI and magnetization at $\nu=-4/3$ obtained by 2BPV calculations aligns well with experiments~\cite{cai2023signatures,zeng2023integer,park2023observation,Xu2023FCItMoTe2}, which highlights the importance of band mixing on the ground states.

Before turning to larger systems, we first choose a representative interaction strength. From the 2BPV phase diagram in Fig.~\ref{fig:FH_2BPV_thirds_new}(b), the $\nu=-2/3$ calculation on a $3\times 4$ momentum mesh exhibits an FCI near $\theta=3.7^\circ$ for $10/\epsilon=0.9$--$1.0$, consistent with experiments at this filling. Based on this observation, we therefore consider this interaction range to faithfully capture the experimental results. Thus, for larger system sizes where a full study of interaction strength is too costly, we will focus on $10/\epsilon=0.9$.

\begin{figure}
    \centering
    \includegraphics[width=\columnwidth]{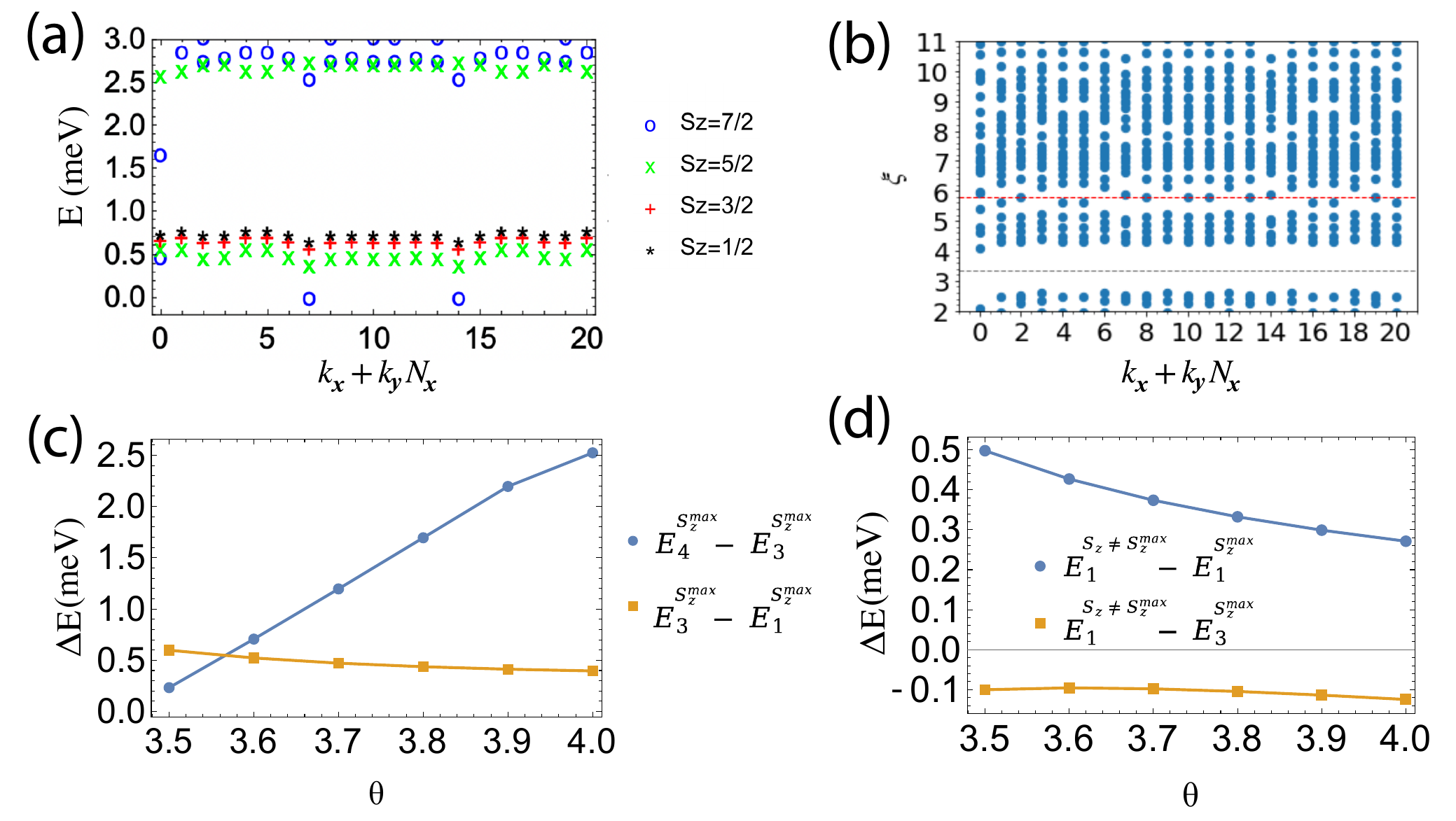}
    \caption{ 2BPV ED results for $\nu=-1/3$ for the FH model in 21-site system with $(N_x,N_y,\widetilde{n}_{11},\widetilde{n}_{12},\widetilde{n}_{21},\widetilde{n}_{22})=(21,1,1,-5,0,1)$ and $10/\epsilon=0.9$. The $\Gamma_M, K_M, K'_M$ points are at $k_x+k_yN_x=0,7,14$ respectively. We use $\Delta E$ to represent energy difference. (a) Energy spectrum with 2BPV at 3.7$^\circ$ with all spin sectors. (b) PES for the lowest three states in the fully polarized sector in (a). The red (black) line represent the FCI (CDW) counting. (c) Energy gap $E_4^{S_z^{\rm{max}}}-E_3^{S_z^{\rm{max}}}$ and spread $E_3^{S_z^{\rm{max}}}-E_1^{S_z^{\rm{max}}}$ at different twist angles in the spin-polarized sector. Here $E_i^{S_z}$ refers to the $i$-th lowest energy across all momenta in the sector with spin $S_z$, with $i=1$ being the lowest state and $S_z^{\rm{max}}$ being the maximum value of $S_z$ (where the $K$-CDW lies). (d) Energy difference $E_1^{S_z\ne S_z^{\rm{max}}}-E_1^{S_z^{\rm{max}}}$ and $E_1^{S_z\ne S_z^{\rm{max}}}-E_3^{S_z^{\rm{max}}}$ between different spin sectors for each twist angle. 
    }
    \label{fig:FH_21b1_1v3}
\end{figure}

At filling $\nu=-1/3$ we can perform calculations at system sizes beyond $9 \times 2$ and look at a 21-site system with $(N_x,N_y,\widetilde{n}_{11},\widetilde{n}_{12},\widetilde{n}_{21},\widetilde{n}_{22})=(21,1,1,-5,0,1)$---which is compatible with $K$-CDW. The energy spectrum at $3.7^\circ$ is shown in Fig.~\ref{fig:FH_21b1_1v3}(a) as an example (the full data is available in App.~\ref{App_EDcombined}). The full spin-polarized sector with $S_z=7/2$ has three lowest states at $\Gamma_M,K_M,K'_M$, which can be either FCI or CDW. To distinguish between them, we calculate the particle entanglement spectrum (PES) for these three states in Fig.~\ref{fig:FH_21b1_1v3}(b). The PES has a gap at CDW counting indicated by the black line, hence these states are CDW. This is consistent with experimental reports of CDW at $\nu=-1/3$~\cite{Xiaodongnew}. We further plot the energy gap above the lowest three states in the polarized sector and the energy spread of these three states in Fig.~\ref{fig:FH_21b1_1v3}(c) for $3.5^\circ-4.0^\circ$, and plot the spin gap in Fig.~\ref{fig:FH_21b1_1v3}(d). The lowest state remains spin-polarized in this range of twist angle, and the momenta of the lowest three states in the fully polarized sector are consistent with CDW. The energy gap above these three states in the fully polarized sector is larger than the spread of them for a large range of twist angle $3.6^\circ-4.0^\circ$. However, there exist states in other spin sectors with energy lying in-between the three CDW states in the fully polarized sector, as seen in Fig.~\ref{fig:FH_21b1_1v3}(a,d). This indicates that the CDW has a soft spin gap and the system is close to the magnetic transition. The closeness to magnetic transition is consistent with experiments that reported a magnetization transition near $\nu=-0.35$ via RMCD measurements~\cite{cai2023signatures}. As pointed out in the previous section, the 1BPV has a potential FCI at $\nu=-1/3$ and $3.5^\circ$ but for large interaction $10 / \epsilon=1.1$. Here in the 2BPV, this FCI phase survives (albeit close to a CDW phase) but at slightly higher interaction $10 / \epsilon \simeq 1.2 - 1.3$ (see App.~\ref{App_largeu}).

\begin{figure}
    \centering
    \includegraphics[width=\columnwidth]{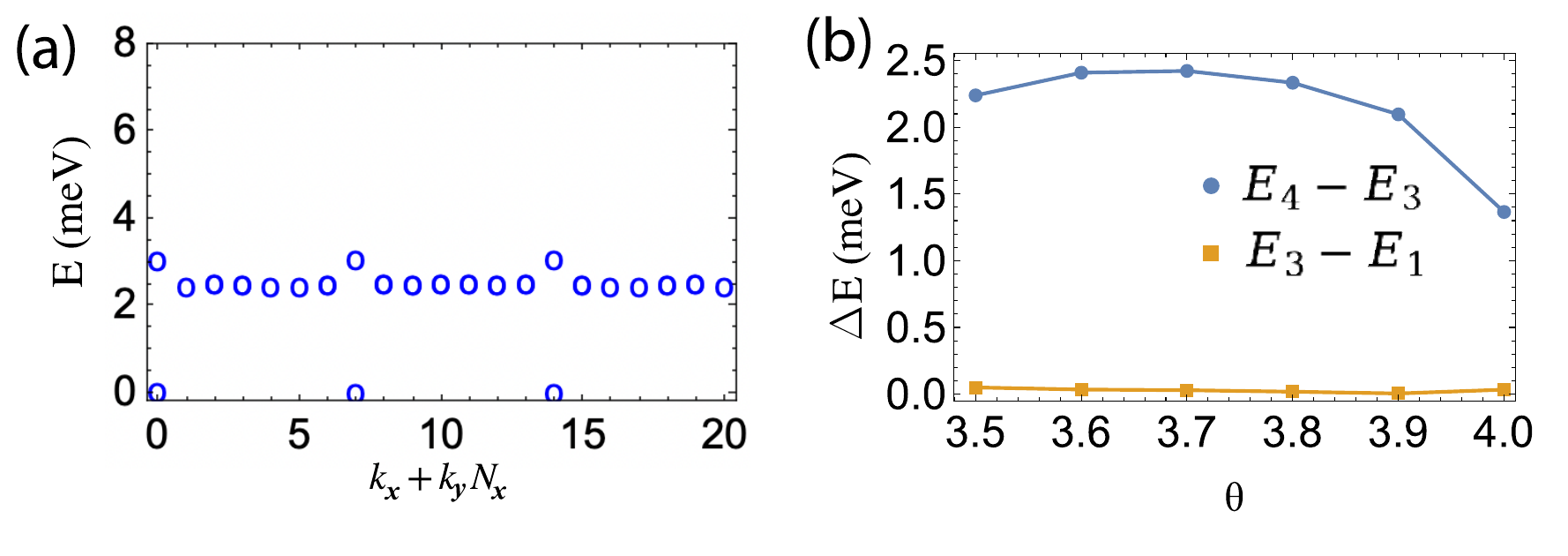}
    \caption{ 2BPV ED results for $\nu=-2/3$ for the FH model in 21-site system with $(N_x,N_y,\widetilde{n}_{11},\widetilde{n}_{12},\widetilde{n}_{21},\widetilde{n}_{22})=(21,1,1,-5,0,1)$ and $10/\epsilon=0.9$. We use $\Delta E$ to represent energy difference. (a) Energy spectrum with 2BPV at 3.7$^\circ$ in the spin-polarized sector. A truncation with no more than 4 particles in the second band is imposed and we show only the lowest energy state per momentum sector with the exception of the GS manifold sectors where we also provide the first excited state. (b) Energy gap $E_4-E_3$ and spread $E_3-E_1$ for $\nu=-2/3$ at different twist angles. Here $E_i$ refers to the $i$-th lowest energy across all momentum sectors, with $E_1$ being the lowest state. 
    }
    \label{fig:FH_21b1_2v3}
\end{figure}

For the sake of comparison, we provide in Fig.~\ref{fig:FH_21b1_2v3} a similar study in a 21-site system for $\nu=-2/3$. Since the Hilbert-space dimension is not accessible for full spinful calculations, we restrict ourselves to the spin-polarized sector and cap the occupation of the first remote band to 4 particles, i.e., bandmax=4 (decreasing the dimension from $2.5 \times 10^9$ to $1.3 \times 10^8$). We found that the spectrum is very similar between bandmax=3 and 4 (see Fig.~\ref{fig:2BPV_finite_size_m2over3} in App.~\ref{App:2bpvscal}), hence convergence to bandmax has been reached and we use result at bandmax=4. There, the system exhibits an FCI phase for all twist angle, even though the gap tends to substantially decrease at $4.0^\circ$. This agrees with the recent experiment~\cite{Xiaodongnew} that indicates FCI at twist angles $2.5^\circ - 4^\circ$. 

To test the robustness of the FCI at $\nu=-2/3$ and reduce finite-size effects, we compute the FCI gap in larger systems with up to $N_s=36$ sites for 1BPV and $N_s=21$ sites for 2BPV in the spin-polarized sector for $10/\epsilon=0.9$ at $3.7^\circ$. We found that the FCI gaps remain finite as system size increases (See Fig.~\ref{fig:finite_size_gap_multi} in Sec.~\ref{Sec_scaling} and App.~\ref{App:EDresults}.)

\subsection{Phase diagrams at smaller angles ($\theta < 3.5^\circ$)}
\label{sec_nofitting}

\begin{figure}
    \centering
\includegraphics[width=\columnwidth]{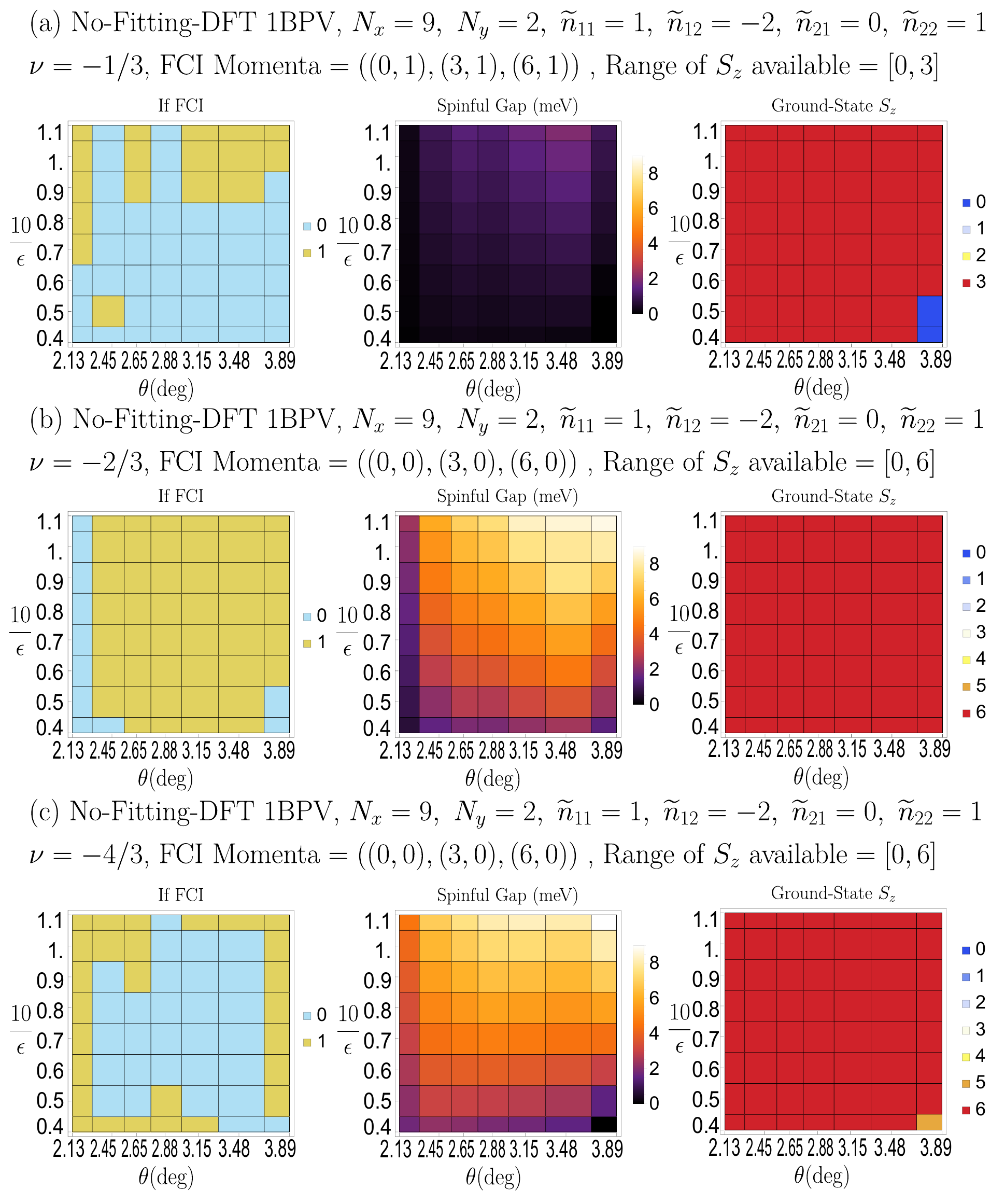}
    \caption{1BPV ED results at $\nu=-1/3,-2/3,-4/3$ for the no-fitting model in the standard basis calculated in $9\times 2$ system with $(N_x,N_y,\widetilde{n}_{11},\widetilde{n}_{12},\widetilde{n}_{21},\widetilde{n}_{22})=(9,2,1,-2,0,1)$.
    The meaning of the labels can be found in the caption of \figref{fig:FH_1BPV_thirds}.
    }
    \label{fig:NoFittingDFT_1BPV_thirds}
\end{figure}

\begin{figure}
    \centering
\includegraphics[width=\columnwidth]{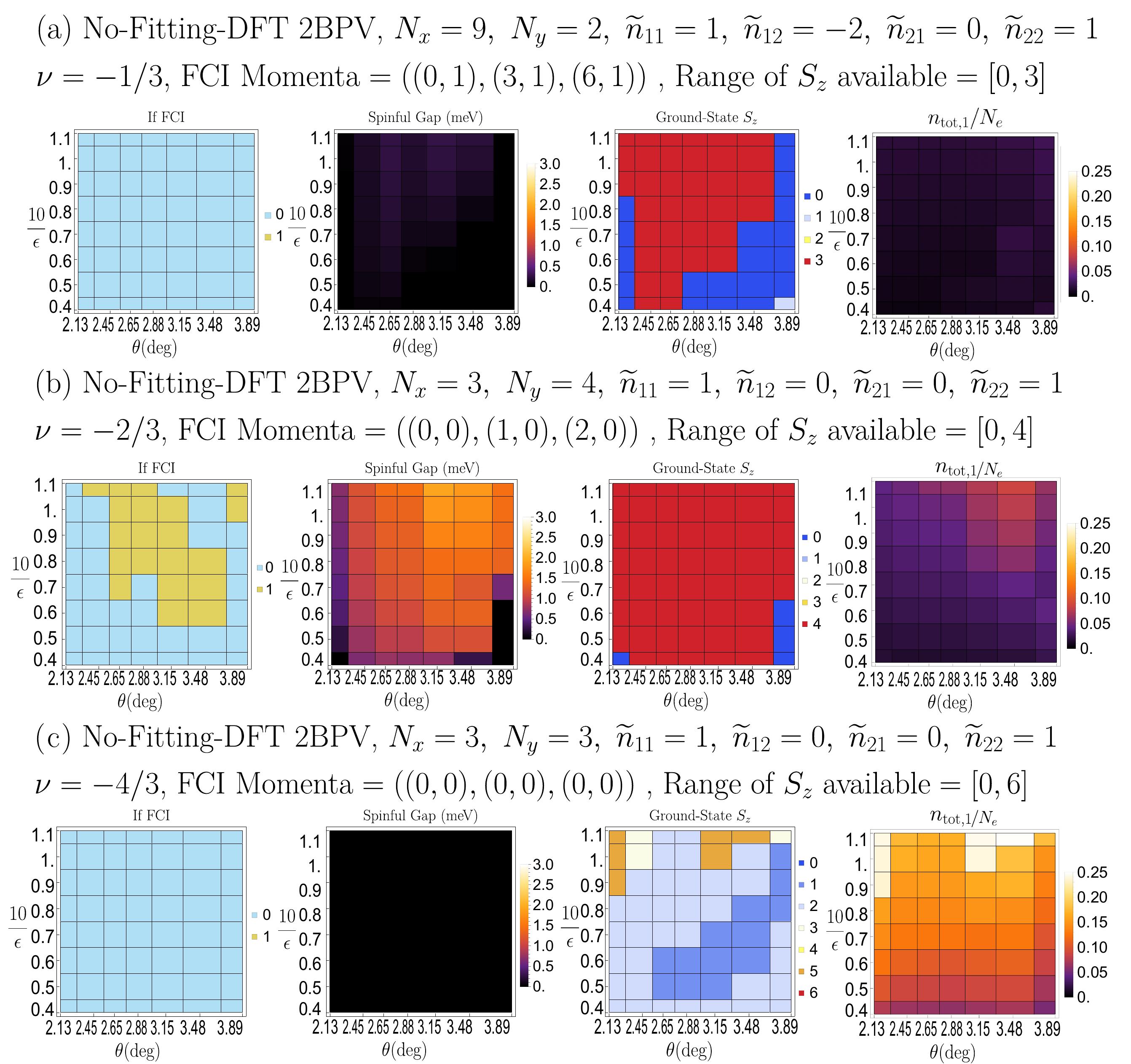}
    \caption{2BPV ED results at $\nu=-1/3,-2/3,-4/3$ for the no-fitting model in the standard basis calculated in $9\times 2,\ 3\times 4$ and $3\times 3$ systems respectively. 
    Here $n_{\text{tot},1}$ is the total occupation number in band 1 calculated from the lowest many-body state and $N_e$ is the number of holes.
    The meaning of the labels can be found in the caption of \figref{fig:FH_1BPV_thirds}.
    }
    \label{fig:NoFittingDFT_2BPV_thirds}
\end{figure}

In this section, we investigate the phase diagrams at smaller twist angles $\theta<3.5^\circ$. For these twist angles, the FH continuum model fitted from DFT band structure at $3.89^\circ$ in the previous section can no longer accurately capture the band structure. Hence, we rely on the DFT no-fitting model with the standard basis presented in Sec.~\ref{sec:nofittingmodels}. The phase diagrams at twist angles from 2.13$^\circ$ to 3.89$^\circ$ in the standard basis from 1BPV and 2BPV calculations are shown in Figs.~\ref{fig:NoFittingDFT_1BPV_thirds} and~\ref{fig:NoFittingDFT_2BPV_thirds} respectively.

Similar to the results in the FH model, when comparing the 1BPV and 2BPV results in Figs.~\ref{fig:NoFittingDFT_1BPV_thirds} and~\ref{fig:NoFittingDFT_2BPV_thirds}, we find that band mixing reduces the stability of spin-polarized FCI phases. At $\nu=-1/3$, the FCI region appearing at large interaction strength $10/\epsilon>0.9$ in the 1BPV calculation is absent in the 2BPV calculation. The FCI region at $\nu=-2/3$ is also reduced once the second moir\'e band is included, although a sizable FCI region remains. At $\nu=-4/3$, the 1BPV ground state is fully spin-polarized, whereas the 2BPV calculation yields a ground state that is generally not fully polarized and is not an FCI. The absence of ferromagnetism and FCI at $\nu=-4/3$ in 2BPV is consistent with experiments~\cite{cai2023signatures,zeng2023integer,park2023observation,Xu2023FCItMoTe2}.

We next compare the no-fitting results with the FH-model results discussed in the previous sections. The twist angles $3.48^\circ$ and $3.89^\circ$ are available in the no-fitting model and lie close to the FH-model calculations, so they provide useful comparison points between the two models. For completeness, the corresponding results obtained with the no-fitting model quick basis are shown in App.~\ref{App_quickbasis}. Overall, the 2BPV no-fitting phase diagram in Fig.~\ref{fig:NoFittingDFT_2BPV_thirds} is in good qualitative agreement with the 2BPV FH-model phase diagram in Fig.~\ref{fig:FH_2BPV_thirds_new}. At $3.89^\circ$, the FCI region in the no-fitting model is very similar to the FCI region calculated from the FH model near $3.9^\circ$, as expected since the FH parameters were fitted to DFT at $3.89^\circ$. At $3.48^\circ$, the range of $10/\epsilon$ where the FCI appears differs only slightly between the two models. Beyond $3.48^\circ$ and $3.89^\circ$, the no-fitting model shows an extended spin-polarized FCI region at $\nu=-2/3$, including intermediate twist angles from $2.65^\circ$ to $3.48^\circ$, consistent with the dip in $R_{xx}$ observed at $\nu=-2/3$ in Ref.~\cite{Xu2024tMoTe2_3.15} at $3.15^\circ$. At other fillings at twist angles from 2.13$^\circ$ to 3.89$^\circ$, the spin gap at $\nu=-1/3$ is nearly vanishing, and the $\nu=-4/3$ ground state is generally not fully spin-polarized, suggesting that these features persist over a broad range of twist angles.

\begin{figure}
    \centering
    \includegraphics[width=\columnwidth]{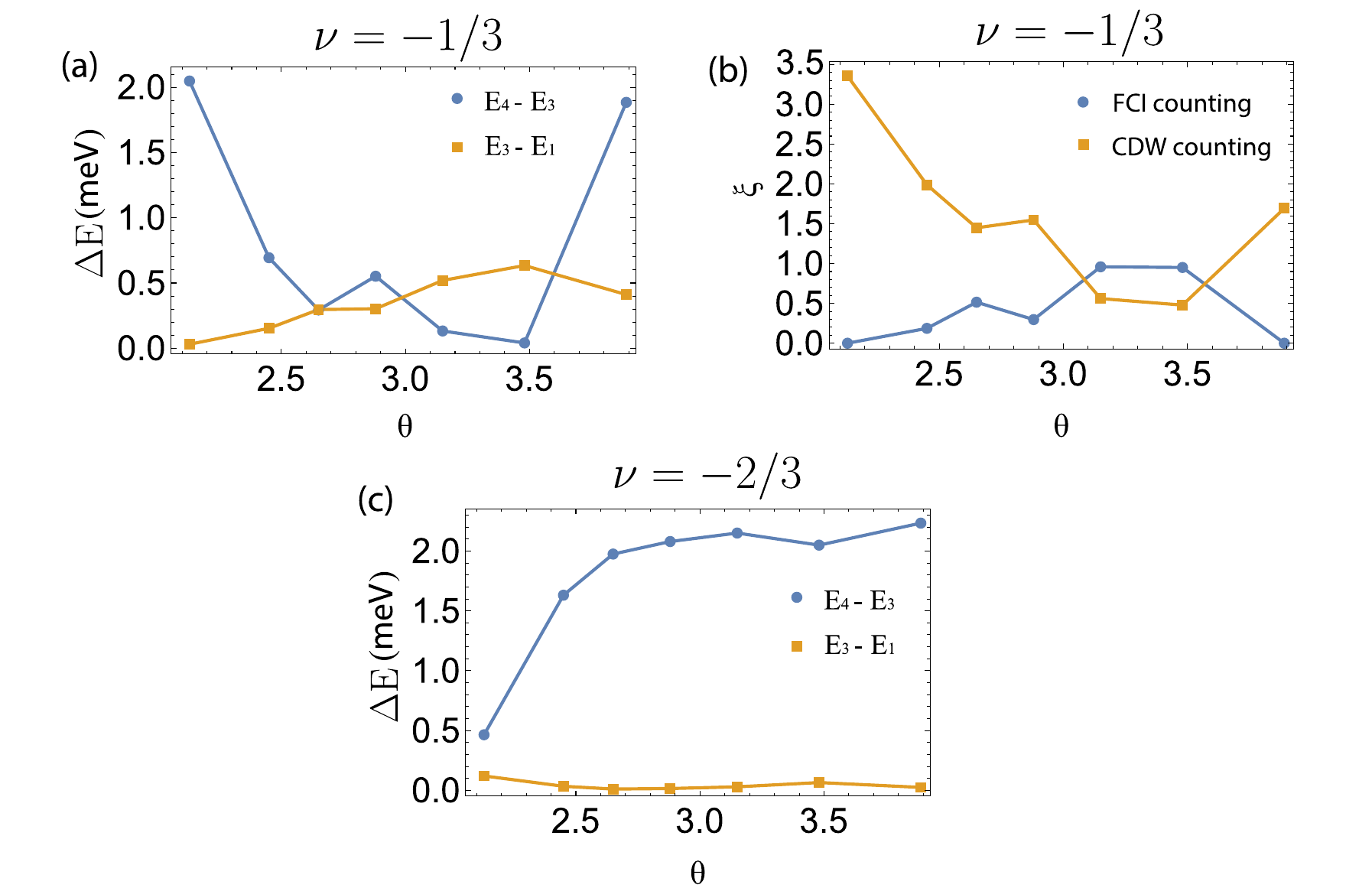}
    \caption{ 2BPV ED results in the spin-polarized sector for $\nu=-1/3,-2/3$ for the no-fitting model in 21-site system with $(N_x,N_y,\widetilde{n}_{11},\widetilde{n}_{12},\widetilde{n}_{21},\widetilde{n}_{22})=(21,1,1,-5,0,1)$ and $10/\epsilon=0.9$. We use $\Delta E$ and $\Delta \xi$ to represent energy difference and the gap in PES respectively. (a) Energy gap $E_4-E_3$ and spread $E_3-E_1$ for the no-fitting in the standard basis for $\nu=-1/3$. (b) The entanglement gap $\Delta \xi$ in PES at the FCI counting and CDW counting for $\nu=-1/3$. The PES is obtained by computing the reduced density matrix for two particles. The phase with the largest entanglement gap should be understood as the most relevant one at a given twist angle. (c) Simiar to (a) but for $\nu=-2/3$ with bandmax 4 (see below Eq.\eqref{eq:f_1_f_2} for the definition of bandmax truncation). 
    }
    \label{fig:nofit_21b1_2v3}
\end{figure}

We also calculate the 2BPV spectrum of the no-fitting model in a larger 21-site system to evaluate finite-size effects. The results are summarized in Fig.~\ref{fig:nofit_21b1_2v3}, and the detailed energy spectrum can be found in App.~\ref{App_EDcombined}. As shown in Fig.~\ref{fig:nofit_21b1_2v3}, at $\nu=-2/3$ and $10/\epsilon=0.9$ the energy gap above the three FCI ground states remains {nonzero} from $2.13^\circ$ to $3.89^\circ$. This is consistent with Ref.~\cite{Xiaodongnew} which reported FCI at $\nu=-2/3$ from $2.13^\circ$ to $4^\circ$. For $\nu=-1/3$, on the contrary, the energy gap no longer remains finite across different twist angles, and phase transition between FCI and CDW occurs. As shown in Fig.~\ref{fig:nofit_21b1_2v3}, at 3.89$^\circ$ the energy gap is much larger than the spread, and the PES shows the ground states at these two angles are CDW. This is consistent with experiment~\cite{Xiaodongnew} that reported CDW at $3^\circ-4^\circ$ at $\nu=-1/3$. The PES gap at the CDW counting is larger than that at FCI counting for most of the twist angles except for $3.15^\circ$ and $3.48^\circ$, but at these two angles the energy gap $E_4-E_3$ above the three ground states is smaller than the spread $E_3-E_1$, hence these ground states are not labeled as FCI.

\begin{figure}
    \centering
    \includegraphics[width=\columnwidth]{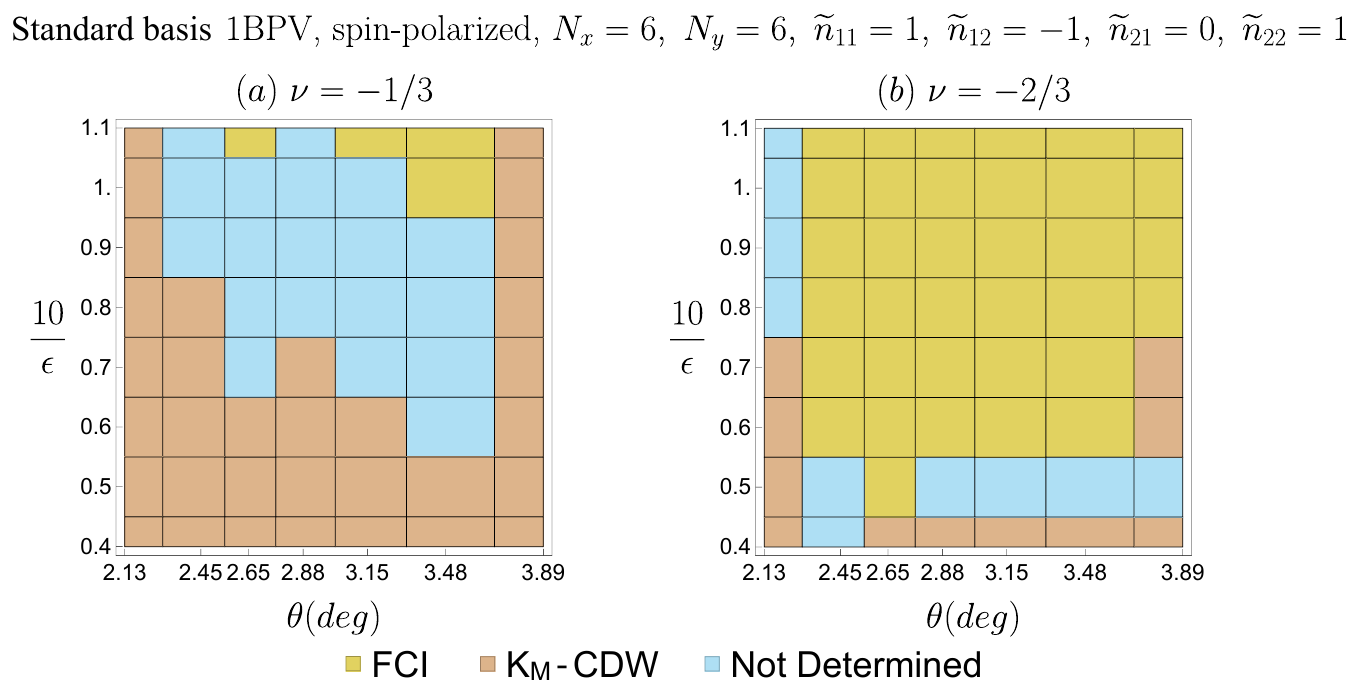}
    \caption{1BPV ED results at $\nu=-1/3,-2/3$ for the DFT no-fitting model in the standard basis in the fully spin-polarized sector for $6\times 6$ system.
    Yellow means the system is in an FCI phase according to \propref{prop:FCI}, brown means the system is in the $K_M$ charge density wave (CDW), and blue means the ground state cannot be identified as an FCI or CDW by \propref{prop:FCI}.
    The other labels have the same meaning as in \figref{fig:FH_1BPV_thirds}.
    }
    \label{fig:FCICDW_s_6by6_1BPV}
\end{figure}

To further investigate the competition between FCI and CDW, we calculate the phase diagram for a larger $6\times 6$ system as shown in Fig.~\ref{fig:FCICDW_s_6by6_1BPV}, where the CDW and FCI ground states have different momenta and can be distinguished by the energy spectra without referring to the PES. In this large system size we only perform 1BPV calculations in the fully polarized sector. At $\nu=-2/3$ there is a large FCI region except for the small twist angle $2.13^\circ$. At $\nu=-1/3$ the CDW dominates at $3.89^\circ$ and $2.13^\circ$. At the intermediate twist angles $2.45^\circ-3.48^\circ$, there is a competition between CDW at weak interaction and FCI at strong interaction. In the region with intermediate interaction $10/\epsilon=0.7-0.9$ the CDW and FCI ground states have similar energy and the ground state cannot be identified as an FCI or CDW by \propref{prop:FCI}. The competition between FCI and CDW at $\nu=-1/3$ is consistent with the different phases reported in experiments from FCI~\cite{Pan1v3FCI2026} to CDW~\cite{Xiaodongnew}.

\section{Phase diagrams for the Jain sequence}
\label{Sec:v5v7}

The fillings considered in this section belong to the principal Jain sequence beyond the third fractions. They are natural candidates for FCI states in a Chern band~\cite{PhysRevB.87.205136}. Several previous works have studied parts of this sequence in $t$MoTe$_2$. In particular, 1BPV ED calculations found evidence for an FCI at $\nu=-3/5$~\cite{wang2023fractional}, while the composite-fermion analysis of Ref.~\cite{Lu2024FCImoteCFL} related the observed $\nu=-2/3$ and $-3/5$ states to a broader Jain-type sequence. Global phase-diagram studies also found a variety of LLL-like incompressible FQAH states near the magic-angle regime~\cite{ReddyFu2023_arXiv2308_10406global}. Here, we extend these studies by performing a systematic 2BPV exact-diagonalization analysis of $\nu=-2/5$, $-3/5$, $-3/7$, and $-4/7$ over a broad range of twist angles, using both the FH and no-fitting models. This allows us to test which members of the Jain sequence remain robust after including band mixing and which ones are suppressed or replaced by competing states. As discussed in Sec.~\ref{sec:2BPVresults_thirds}, $10/\epsilon=0.9$ is a "sweet spot" for the interaction strength. Therefore we will focus on this interaction strength in the following, relegating additional results including the investigation of different interaction strengths to App.~\ref{App_FCIgap}. Moreover, to account for band mixing, we will use the 2BPV approximation unless specified.

\subsection{Phase diagrams at $\nu=-2/5,-3/5$}

\begin{figure}
    \centering
    \includegraphics[width=\columnwidth]{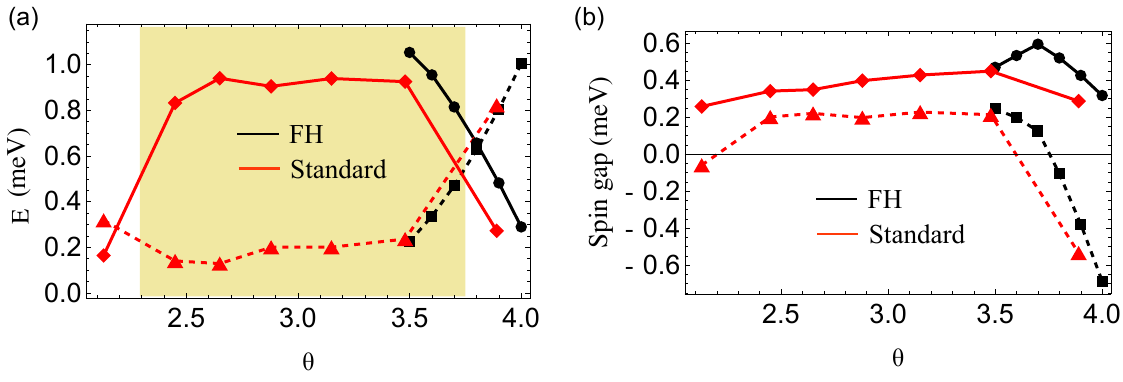}
    \caption{ 2BPV ED results for $\nu=-2/5$ in a 15-site system with $(N_x,N_y,\widetilde{n}_{11},\widetilde{n}_{12},\widetilde{n}_{21},\widetilde{n}_{22})=(15,1,1,-5,0,1)$ and $10/\epsilon=0.9$. The data obtained from the FH model and no-fitting model in the standard basis are shown in black and red colors respectively. (a) The solid lines represent the FCI gap in the polarized sector $E_{q+1}^{S_z^{\rm{max}}}-E_q^{S_z^{\rm{max}}}$, where $q$ is the ground state degeneracy and $q=5$ for $\nu=-2/5$. The dashed lines are the spread of FCI states in the polarized sector $E_{q}^{S_z^{\rm{max}}}-E_1^{S_z^{\rm{max}}}$. The yellow region indicates the ground state is an FCI according to \propref{prop:FCI}. (b) The solid and dashed lines represent the spin gap $E_1^{S_z\ne S_z^{\rm{max}}}-E_1^{S_z^{\rm{max}}}$ as defined in Eq.\eqref{Eqspingap} and $E_1^{S_z\ne S_z^{\rm{max}}}-E_q^{S_z^{\rm{max}}}$ respectively.
    }
    \label{fig:comb2v5}
\end{figure}

\begin{figure}
    \centering
    \includegraphics[width=\columnwidth]{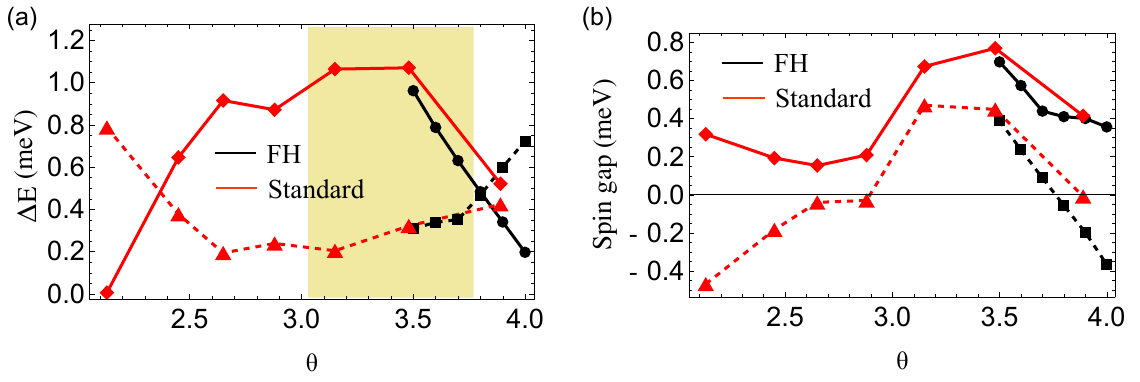}
    \caption{ 2BPV ED results for $\nu=-3/5$ in a 10-site system with $(N_x,N_y,\widetilde{n}_{11},\widetilde{n}_{12},\widetilde{n}_{21},\widetilde{n}_{22})=(5,2,1,1,0,1)$ and $10/\epsilon=0.9$. The meaning of the plots are identical to the caption of Fig.~\ref{fig:comb2v5} with $q=5$.
    }
    \label{fig:comb3v5}
\end{figure}

\begin{figure}
    \centering
    \includegraphics[width=\columnwidth]{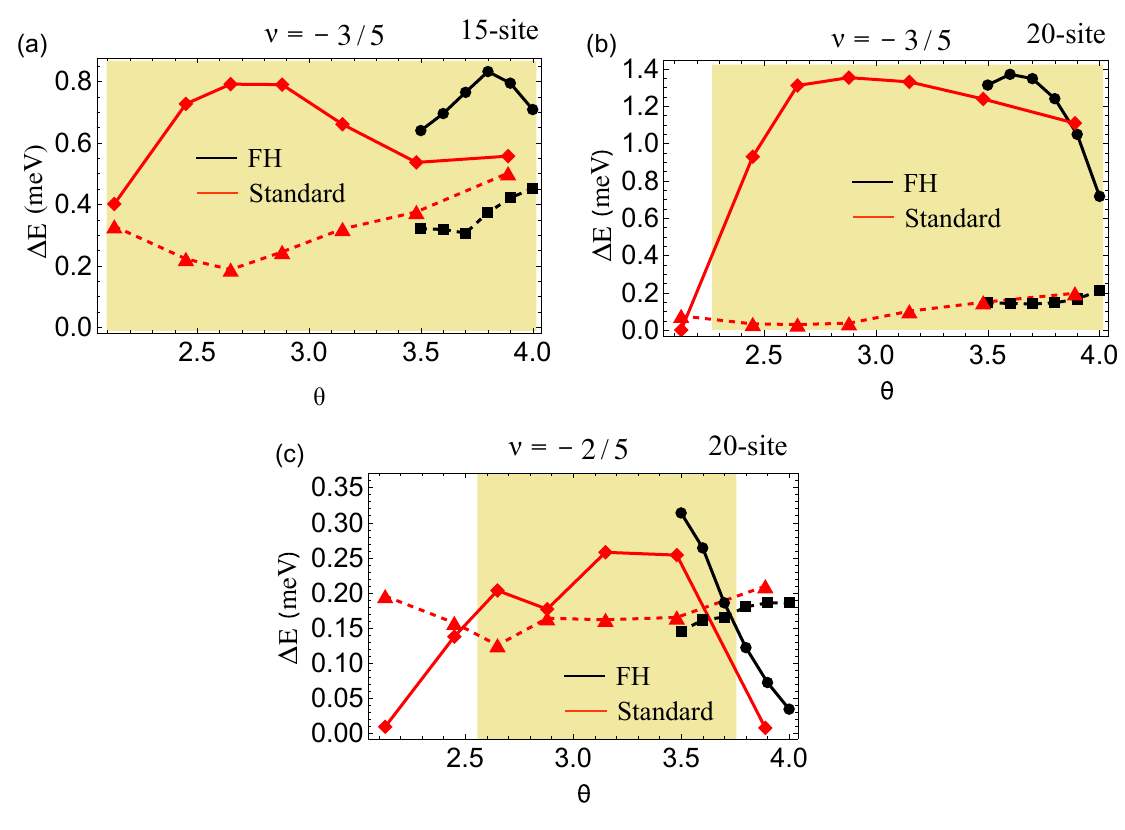}
    \caption{ 2BPV ED results with $10/\epsilon=0.9$ in the fully polarized sector in different systems. (a) 15-site system with $(N_x,N_y,\widetilde{n}_{11},\widetilde{n}_{12},\widetilde{n}_{21},\widetilde{n}_{22})=(15,1,1,-5,0,1)$ at $\nu=-3/5$. (b) 20-site system with $(N_x,N_y,\widetilde{n}_{11},\widetilde{n}_{12},\widetilde{n}_{21},\widetilde{n}_{22})=(5,4,1,0,1,1)$ at $\nu=-3/5$ calculated with bandmax 6. (c) 20-site system with $(N_x,N_y,\widetilde{n}_{11},\widetilde{n}_{12},\widetilde{n}_{21},\widetilde{n}_{22})=(5,4,1,0,1,1)$ at $\nu=-2/5$. The meaning of the plots are identical to the caption of Fig.~\ref{fig:comb2v5} with $q=5$. Note that since the calculation is performed in fully polarized sector (as opposed to Figs.~\ref{fig:comb2v5} and ~\ref{fig:comb3v5}), the identification of FCI in the yellow region does not take into account of partial polarization.
    }
    \label{fig:gap20_fifths}
\end{figure}

We start with the phase diagrams at the two filling factors related by particle-hole symmetry $\nu=-2/5$ and $-3/5$ at different twist angles. We first present the ED results for $\nu=-2/5$ in a 15-site system in Fig.~\ref{fig:comb2v5} and for $\nu=-3/5$ in a 10-site system in Fig.~\ref{fig:comb3v5}. In these small system sizes we are able to perform full 2BPV calculations including all spin sectors. Then we consider larger systems with 15 sites for $\nu=-3/5$ and 20 sites for both $\nu=-2/5$ and $-3/5$ in Fig.~\ref{fig:gap20_fifths} to reduce the finite-size effects, where we only calculate the fully polarized sector. For each twist angle we indicate whether the system is FCI, and show the spin of the lowest-energy state, the FCI gap and the spin gap. 

We first investigate the spin polarization from the calculations that involve all the spin sectors in Figs.~\ref{fig:comb2v5} and \ref{fig:comb3v5} show that the state with lowest energy at both fillings $\nu=-2/5,-3/5$ are spin-polarized in the FH and no-fitting model with standard basis for all the twist angles studied. This is consistent with the experiments that observed spin polarization at $\nu=-0.35\sim -1.1$ at twist angle 3.7$^\circ$~\cite{cai2023signatures}. 

Next we consider the FCI region obtained from these calculations. At $\nu=-3/5$, in the small 10-site system in Fig.~\ref{fig:comb3v5} FCI emerges for twist angles $3.15^\circ-3.7^\circ$. At $\theta>3.7^\circ$ the FCI disappears, but this is a finite-size effects, and in the larger 15-site system and 20-site system in Fig.~\ref{fig:gap20_fifths} the FCI remains robust for $3.7^\circ\le\theta \le 4.0^\circ$. This is consistent with experiments~\cite{cai2023signatures,park2023observation,park2025obsfci,xu_txl2025FCI} that observed robust FCI near 3.7$^\circ$. The broad FCI region across $2.45^\circ-4.0^\circ$ at $\nu=-3/5$ in Fig.~\ref{fig:gap20_fifths} also agrees with experiments~\cite{Xiaodongnew}. At $\nu=-2/5$, there is a FCI region at $\theta=2.65^\circ - 3.7^\circ$ in both the 15-site system in Fig.~\ref{fig:comb2v5} and the 20-site system in Fig.~\ref{fig:gap20_fifths}, showing numerical convergence. Experiments reported FCI at $\nu=-2/5$ for $2.6^\circ -3^\circ$~\cite{Xiaodongnew} but no FCI has been observed near 3.7$^\circ$~\cite{cai2023signatures,park2023observation,park2025obsfci,xu_txl2025FCI}. Therefore, at $\nu=-2/5$ the numerical result overestimates the FCI region compared to experiments, and $3.7^\circ$ is at the boundary between FCI and non-FCI regions.

\subsection{Phase diagrams at $\nu=-3/7,-4/7$}

\begin{figure}
    \centering
    \includegraphics[width=\columnwidth]{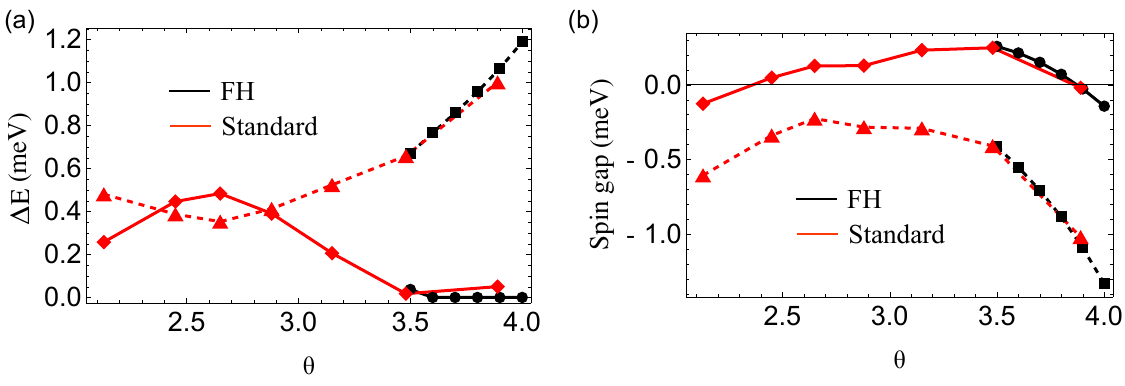}
    \caption{2BPV ED results for $\nu=-3/7$ in a 14-site system with $(N_x,N_y,\widetilde{n}_{11},\widetilde{n}_{12},\widetilde{n}_{21},\widetilde{n}_{22})=(7,2,1,1,0,1)$ and $10/\epsilon=0.9$. The meaning of the plots are identical to the caption of Fig.~\ref{fig:comb2v5} with $q=7$. Note the absence of FCI (which would be labeled by yellow region) in these parameters.
    }
    \label{fig:comb3v7}
\end{figure}

\begin{figure}
    \centering
    \includegraphics[width=\columnwidth]{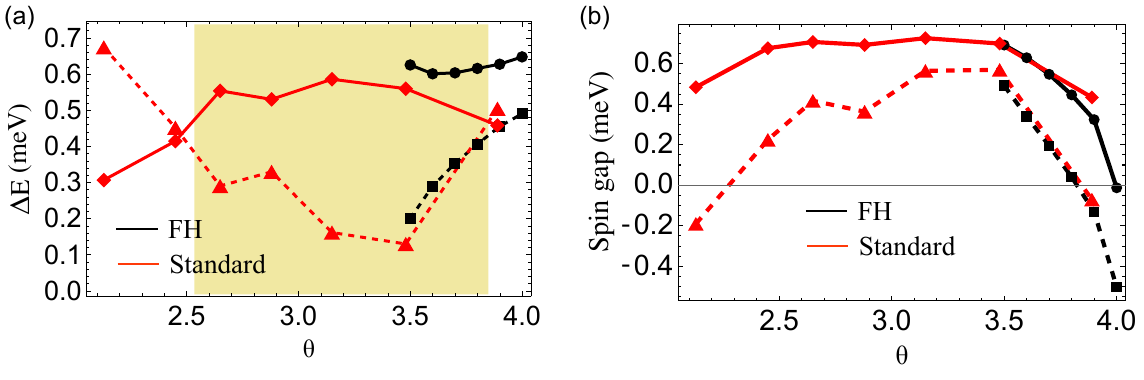}
    \caption{2BPV ED results for $\nu=-4/7$ in a 14-site system with $(N_x,N_y,\widetilde{n}_{11},\widetilde{n}_{12},\widetilde{n}_{21},\widetilde{n}_{22})=(7,2,1,1,0,1)$ and $10/\epsilon=0.9$. The meaning of the plots are identical to the caption of Fig.~\ref{fig:comb2v5} with $q=7$.
    }
    \label{fig:comb4v7}
\end{figure}

\begin{figure}
    \centering
    \includegraphics[width=\columnwidth]{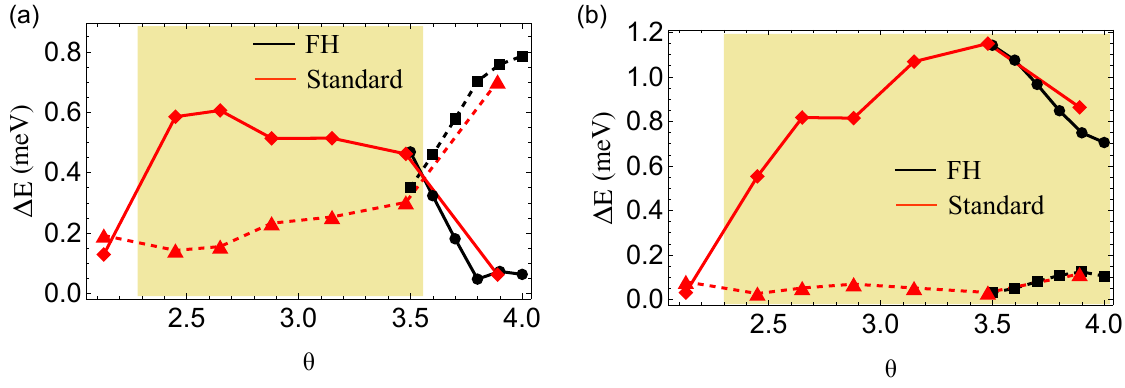}
    \caption{ 2BPV ED results for 21-site system with $(N_x,N_y,\widetilde{n}_{11},\widetilde{n}_{12},\widetilde{n}_{21},\widetilde{n}_{22})=(21,1,1,-5,0,1)$ and $10/\epsilon=0.9$ in the polarized sector. (a) is calculated for $\nu=-3/7$ with bandmax 9 and (b) is calculated for $\nu=-4/7$ with bandmax 5. The meaning of the plots are identical to the caption of Fig.~\ref{fig:comb2v5} with $q=7$ for both $\nu=-3/7$ and $\nu=-4/7$.
    }
    \label{fig:gap21_sevenths}
\end{figure}

In this section, we first show the results at $\nu=-3/7,-4/7$ in a small 14-site system in Figs.~\ref{fig:comb3v7} and \ref{fig:comb4v7} where spinful 2BPV calculations are accessible, and then present the 2BPV results in the fully spin-polarized sector for a larger 21-site system with {twist angle $2.13^\circ - 4.0^\circ$ including the FH model and no-fitting model} in Fig.~\ref{fig:gap21_sevenths}.

In the spinful calculations in Figs.~\ref{fig:comb3v7} and \ref{fig:comb4v7}, there is a large spin-polarized region over $2.45^\circ-3.8^\circ$ at $\nu=-3/7$ and $2.13^\circ-4.0^\circ$ at $\nu=-4/7$. Near $3.7^\circ$, FCI emerges at $\nu=-4/7$, which is characterized by an energy gap much larger than the spread of FCI states in Fig.~\ref{fig:comb4v7}(c), whereas no FCI is found at $\nu=-3/7$. The emergence (absence) of FCI at $\nu=-4/7$ ($\nu=-3/7$) near $3.7^\circ$ also persist in the calculation in the larger 21-site system in Fig.~\ref{fig:gap21_sevenths}. At $\nu=-4/7$, the FCI gap is larger than the spread for all twist angles studied with $\theta>2.13^\circ$. At $\nu=-3/7$, the energy gap above the seven FCI ground states can be larger than the spread of those states only when $\theta<3.5^\circ$ as shown in Fig.~\ref{fig:gap21_sevenths}(a), hence FCI only emerges at twist angle with $\theta<3.5^\circ$. These features are consistent with recent experiments~\cite{park2025obsfci,xu_txl2025FCI,Xiaodongnew} that observed FCI at $\nu=-2/3,\ -3/5,\ -4/7$ but not at $\nu=-3/7$ around 3.7$^\circ$.

\subsection{Combined phase diagrams for different fillings and twist angles}
\label{Sec_largephasediagram}

We present the combined 2BPV phase diagrams that indicate the emergence of FCI/CDW across {twist angles $2.13^\circ,2.45^\circ,2.65^\circ,2.88^\circ,3.15^\circ,3.48^\circ$ for the no-fitting model and  $3.6^\circ-4.0^\circ$ for the FH model} and filling fractions $\nu=-4/3,-2/3,-3/5,-4/7,-3/7,-2/5,-1/3$ from 2BPV calculations discussed in the previous sections, as shown in Fig.~\ref{fig:phaselargesize}. For all fillings and twist angles we set $10/\epsilon=0.9$ as mentioned in Sec.~\ref{sec:2BPVresults_thirds}. The 21-site system is used for $\nu=-2/3,-1/3,-4/7,-3/7$ and the 20-site system is used for $\nu=-3/5,-2/5$. The bandmax truncation (defined below Eq.\eqref{eq:f_1_f_2}) is applied to fillings $\nu=-2/3$ with bandmax 4, $\nu=-3/5$ with bandmax 6 and $\nu=-4/7$ with bandmax 5, and the other fillings have no bandmax truncation applied. The fillings $\nu=-2/3,-3/5,-4/7,-3/7$ are calculated in the spin-polarized sector, $\nu=-2/5$ is calculated in both the spin-polarized sector and the sector with spin-1 excitations, and $\nu=-1/3$ is calculated with all spin sectors. The FCI regions are marked in yellow. The light green region with partially polarized CDW refers to the case where the polarized sector is CDW but there exist states in the other spin sectors with energy in-between the CDW ground states (see Fig.~\ref{fig:FH_21b1_1v3}(a)), otherwise the CDW is marked in deep green. For the no-fitting model we consider both the standard basis and the quick basis. The results in the standard basis have been shown in Sec.~\ref{sec_nofitting} and ~\ref{Sec:v5v7}, and the calculations in the quick basis can be found in App.~\ref{App_EDcombined}. For twist angles where the standard basis and quick basis give different types of ground states, the phase is labeled in gray as undetermined. 

Compared with experiments, this phase diagram captures (i) the FCI at $\nu=-2/3,-3/5$ over a wide range of twist angles~\cite{cai2023signatures,zeng2023integer,park2023observation,Xu2023FCItMoTe2,park2025obsfci,xu_txl2025FCI,Xiaodongnew}, (ii)  the simultaneous emergence of FCI at $\nu=-4/7$ and absence of FCI at $\nu=-3/7$ near 3.7$^\circ$~\cite{park2025obsfci,xu_txl2025FCI,Xiaodongnew}, and (iii)  the CDW at $\nu=-1/3$ near 3.7$^\circ$~\cite{Xiaodongnew}. \emph{Discrepancies with experiments also exist}. At $\nu=-2/5$ the numerical results over-estimate the range of twist angle where FCI emerges to be $2.65^\circ,2.88^\circ,3.15^\circ,3.5^\circ,3.6^\circ,3.7^\circ$, but experiments found no FCI at this filling for $3.0^\circ-3.7^\circ$~\cite{Xiaodongnew}. At $\nu=-1/3$, the comparison with experiment is more delicate. In the no-fitting model standard basis at $10/\epsilon=0.9$, the finite-size spectra do not show a clear FCI in the twist-angle range $2.6^\circ-3.0^\circ$, where an FCI was reported experimentally~\cite{Xiaodongnew}. 
However, the ground state at $\nu=-1/3$ is sensitive to details of the single-particle description and interaction strength. In particular, as shown in Fig.~\ref{fig:ED21_1v3} in App.~\ref{App_EDcombined}, at $2.65^\circ$ and $2.88^\circ$ the quick basis does exhibit FCI signatures, whereas in the standard basis the ground state is a CDW with a soft spin gap. Increasing the interaction strength can separate the spin excitations from the low-energy manifold and lead to a finite spin gap, such that the spectrum calculated in the standard basis at larger interaction becomes consistent with an FCI (see Fig.~\ref{Nofit_1v3_2.88_largeu} in App.~\ref{App_largeu}). 
These results suggest that the $\nu=-1/3$ state in this twist-angle range lies near a phase boundary, so small changes in band dispersion, interaction strength, or sample details can affect the observed ground state.

\subsection{Gap Scaling}
\label{Sec_scaling}

\begin{figure}
    \centering
    \includegraphics[width=\columnwidth]{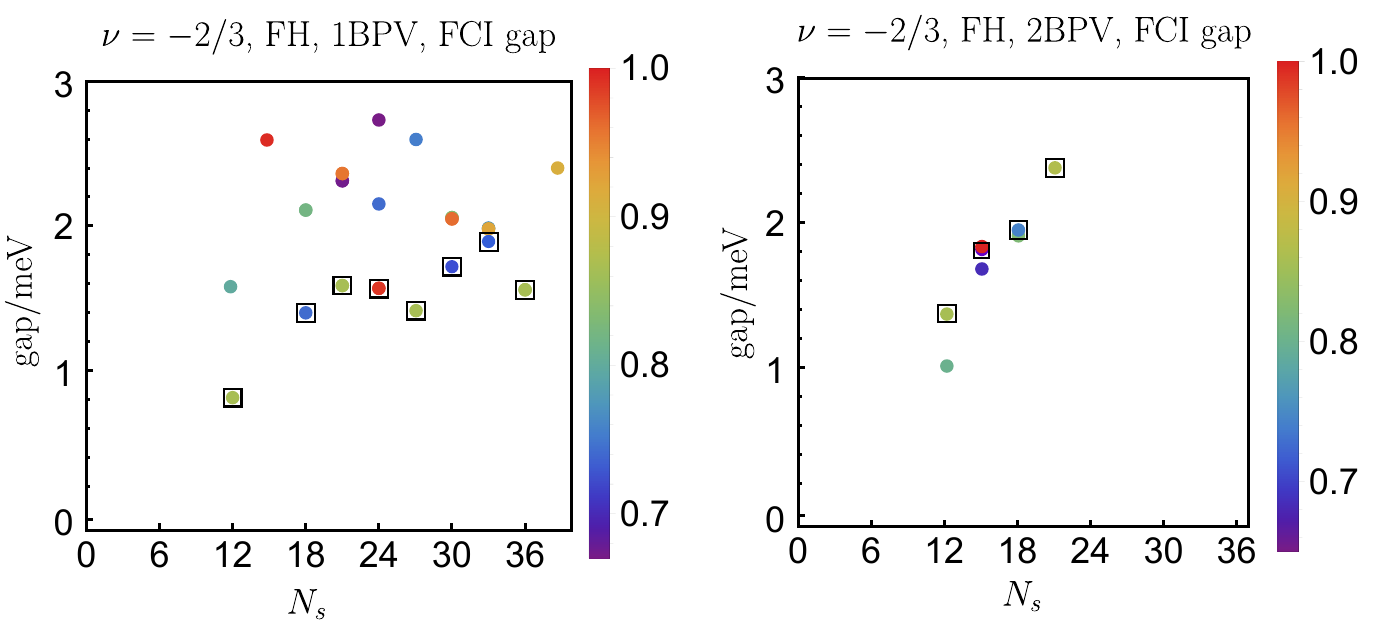}
    \caption{The 1BPV and 2BPV FCI gaps of the spectra in different system sizes at $\nu=-2/3$ for the FH model with $10/\epsilon=0.9$ and $\theta=3.7^\circ$ in the fully polarized sector. The color of the points indicates the aspect ratio of the corresponding lattice. The boxed data points represent momentum meshes that contain the $K_M,K'_M$ points.
    The energy spectrum can be found in App.~\ref{App_scaling}.
    We only plot the gaps of the spectrum that satisfy \propref{prop:FCI}.
    $N_s$ is the number of sites; since there can be more than one momentum mesh with the same $N_s$, we can have more than one values of FCI gap for a single $N_s$.
    }
    \label{fig:finite_size_gap_multi}
\end{figure}

To test the finite-size effects and to try to obtain universal answers, we now examine the scaling of the FCI gap in large systems with up to $N_s=39$ sites. We demonstrate that the FCI gap is stable and converges as system size increases. We choose experimentally relevant parameters $10/\epsilon=0.9$ at $3.7^\circ$ where FCI exists in 2BPV calculations. The FCI gaps in different system sizes at filling $\nu=-2/3$ for 1BPV and 2BPV in the fully polarized sector are summarized in Fig.~\ref{fig:finite_size_gap_multi}, and the corresponding plots for $\nu=-3/5,-4/7$ can be found in App.~\ref{App_scaling}. The color represents the aspect ratio of the momentum mesh, and the black boxes represent momentum meshes that contain the $K_M$ points. 
The FCI gap shows a stronger system-size dependence for momentum meshes that do not contain the $K_M$ and $K'_M$ points, but among the momentum meshes that contain these points, the FCI gap in 1BPV is relatively stable with $N_s=21$ to 36.
The minimum of the top moir\'e band is at $K_M$ points, hence the momentum meshes without $K_M$ can underestimate the bandwidth and lead to unstable FCI gaps. Therefore, we will take the gap in the 36-site system that contains $K_M$ as the FCI gap at $\nu=-2/3$ for 1BPV, and take the 21-site FCI gap for 2BPV.

\begin{figure}
    \centering
    \includegraphics[width=0.49\columnwidth]{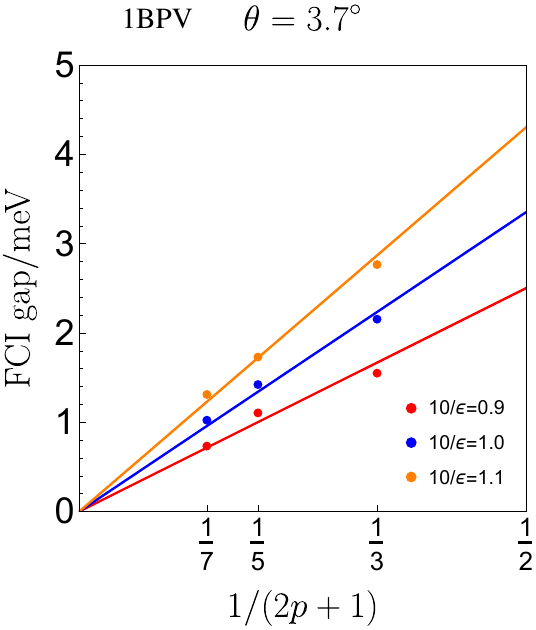}
    \includegraphics[width=0.49\columnwidth]{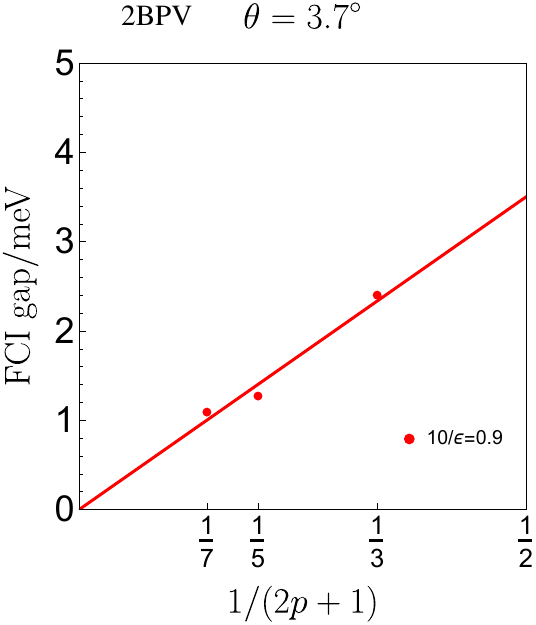}
    \caption{1BPV (left) and 2BPV (right) FCI gaps as a function of $1/(2p+1)$ for the FH model at $\theta=3.7^\circ$. 
    We focus on $p=1,2,3$ and the corresponding filling is $\nu=-(p+1)/(2p+1)$.
    The 1BPV FCI gap is determined by the calculation on $(N_x,N_y,\widetilde{n}_{11},\widetilde{n}_{12},\widetilde{n}_{21},\widetilde{n}_{22})=(35,1,1,-6,0,1)$ for $p=2,3$ and $(6,6,1,-1,0,1)$ for $p=1$.
    The 2BPV FCI gap is determined by $(N_x,N_y,\widetilde{n}_{11},\widetilde{n}_{12},\widetilde{n}_{21},\widetilde{n}_{22})=(21,1,1,-5,0,1)$ for $p=1,3$ and $(5,4,1,0,1,1)$ for $p=2$. The straight lines are guide lines.
    }
    \label{fig:FH_gapscaling_1BPV_linear}
\end{figure}

We can now study the scaling of the FCI gap as a function of the filling. As shown in \figref{fig:FH_gapscaling_1BPV_linear}, the FCI gaps approximately follow a straight line as a function of $1/(2p+1)$, where $\nu=-(p+1)/(2p+1)$ for both 1BPV and 2BPV calculations. Such gap scaling is similar to the FQHE gaps in the lowest Landau level~\cite{Murthy1998scaling,ZhaoJain2022scaling}.

\section{Composite Fermi liquid at filling $\nu=-1/2$}
\label{sec:HalfFilling}

A CFL is a compressible metallic state that can emerge at even-denominator filling in a partially filled Landau level, where electrons bind flux quanta to form composite fermions and develop an emergent Fermi surface~\cite{JainCFL1989,Lopez1991,Halperin1993,Rezayi1994}. In moir\'e Chern bands, previous works proposed a zero-field CFL in projected single-Chern-band models of $t$MoTe$_2$ at different twist angles $\theta\sim 3.6^\circ$~\cite{Dong2023CFLtMoTe2} and $\theta\sim 4.5^\circ$~\cite{Goldman2023Composite}. In this section, we revisit the filling $\nu=-1/2$ by comparing 1BPV and 2BPV calculations in fully spin-polarized sectors, allowing us to test how band mixing affects the CFL signatures.

\begin{figure}
    \centering
    \includegraphics[width=\columnwidth]{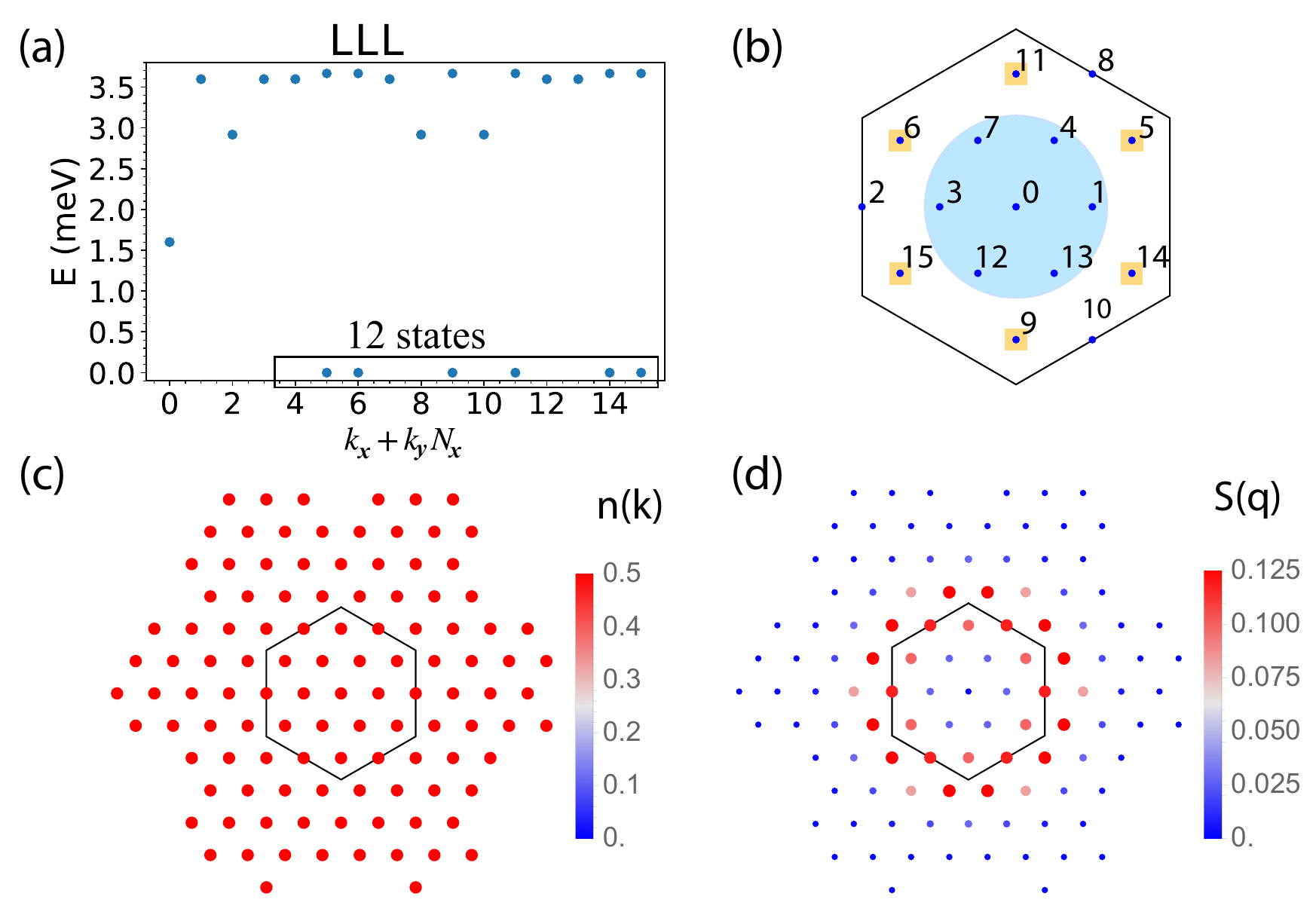}
    \caption{ (a) Energy spectrum of half filled lowest Landau level in the lattice $(N_x,N_y,\widetilde{n}_{11},\widetilde{n}_{12},\widetilde{n}_{21},\widetilde{n}_{22})=(4,4,1,0,0,1)$. The states are twofold degenerate, hence there are 12 states at zero energy inside the box. (b) Momentum points in the Brillouin zone in (a). The blue disk labels momentum $\Gamma_M$ and the six momenta closest to it. The yellow boxes label the six momenta next-closest to $\Gamma_M$, which are the momenta of CFL ground states. (c) Occupation number $n(\bsl k)=\langle \gamma_{\bsl k}^\dagger \gamma_{\bsl k}\rangle$ and (d) structure factor Eq.\eqref{Sqequation} in the lowest Landau level computed from the 12 ground states at zero energy. The black hexagon is the Brillouin zone.
    }
    \label{fig:CFL_LLL}
\end{figure}

Before presenting the calculations in the \tmt\ system, we first discuss the CFL in lowest Landau level (LLL) as a reference. The energy spectrum of the LLL in a non-tilted $4\times 4$ system at half filling is shown in Fig.~\ref{fig:CFL_LLL}. There are 12 states with zero energy at six momentum points $k_x+k_yN_x=5,6,9,11,14,15$ (labels also in Fig.~\ref{fig:CFL_LLL}(b) in the yellow boxes) which we call CFL momenta (each momentum point has two degenerate states). These six momenta can be understood by the filling of 8 particles in a composite Fermi sea. The first 7 composite fermions fill up the state at $\Gamma_M$ and the six momenta $0, 1, 3, 4, 7, 12, 13$ closest to $\Gamma_M$ (the blue disk in Fig.~\ref{fig:CFL_LLL}(b)), and the one remaining composite fermion can occupy any of the six momenta $5, 6, 9, 11, 14, 15$ next-closest to $\Gamma_M$, which is exactly the aforementioned CFL momenta. This leads to the momentum distribution of ground states in Fig.~\ref{fig:CFL_LLL}(a). With the center-of-mass degeneracy at $\nu=-1/2$, there are in total 12 quasi-degenerate ground states. The occupation number in momentum space is written as $n(\bsl k)=\langle \gamma_{\bsl k}^\dagger \gamma_{\bsl k}\rangle$, where $\gamma^\dagger_{\bsl k}$ is the creation operator at momenta $\bsl k$ and the average is taken over all the $N_{\text{GS}}$ ground states $\langle \hat O\rangle\equiv \frac{1}{N_{\text{GS}}}\sum_{a\in\text{GS}}\langle a|\hat O|a\rangle$. The occupation number $n(\bsl k)$ calculated by including all CFL ground states is uniform in momentum space, as shown in Fig.~\ref{fig:CFL_LLL}(c), contrary to a conventional Fermi liquid which has a Fermi surface with discontinuity in occupation number. The CFL at half filling is a compressible phase with an emergent composite-fermion Fermi surface~\cite{Geraedts2016,Dong2023CFLtMoTe2}, which can be diagnosed in finite-size numerics via $2k_F$ features in the static structure factor~\cite{Wilhelm2021fractional}
\bea
S(\bsl q)=\frac{1}{N_0}(\langle \hat \rho_{\bsl q} \hat \rho_{-\bsl q}\rangle-\langle \hat \rho_{\bsl q} \rangle \langle \hat \rho_{-\bsl q} \rangle).
\label{Sqequation}
\eea
Here the average is calculated over all the ground states and $N_0$ is the number of orbitals. The structure factor is shown in Fig.~\ref{fig:CFL_LLL}(d). $S(\bsl q)$ nearly vanishes for large momentum $|\bsl q|>2k_F$, where $k_F$ is the Fermi momentum. 

\begin{figure}
    \centering
    \includegraphics[width=\columnwidth]{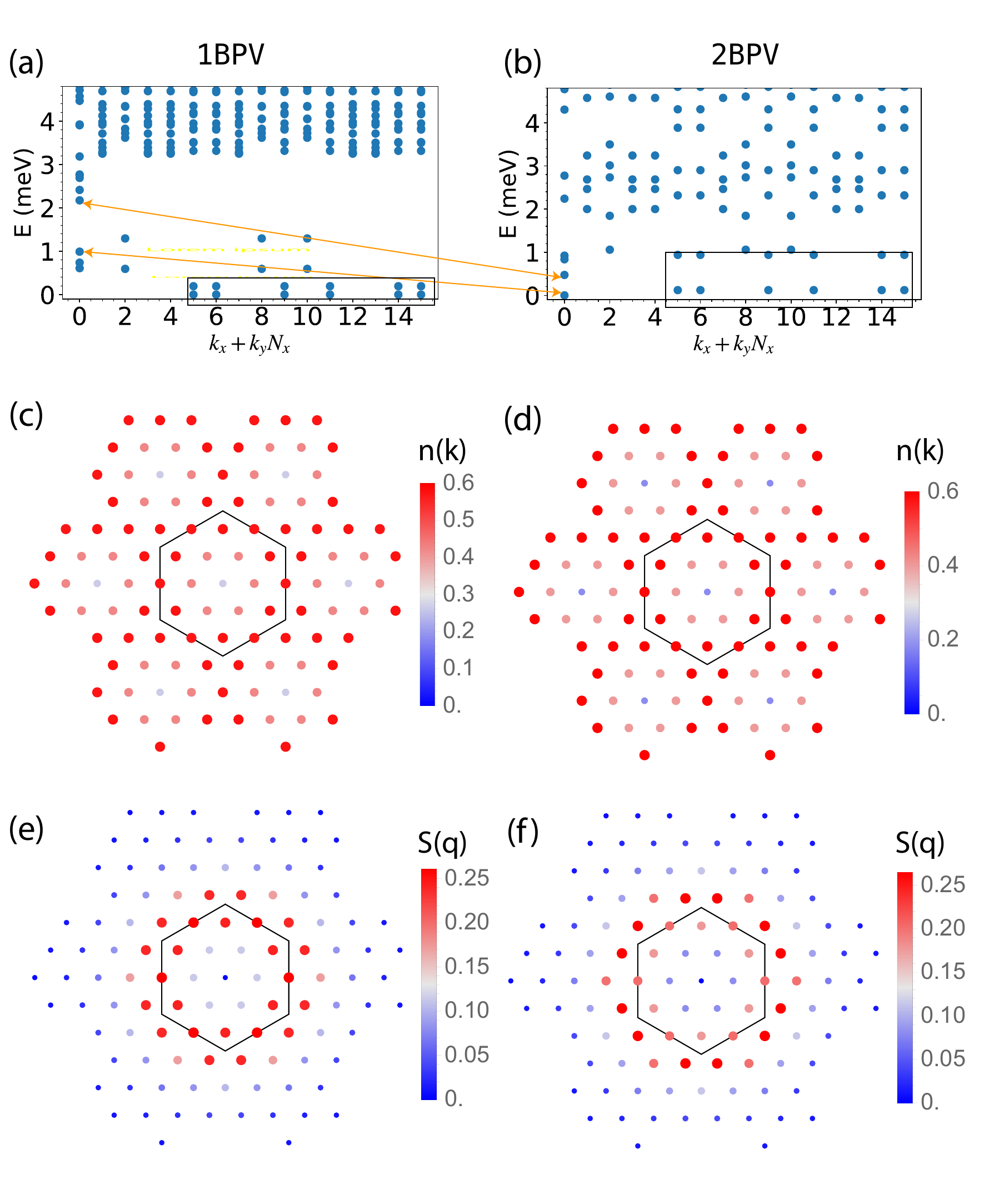}
    \caption{(a,c,e) Energy spectrum, occupation number and structure factor of $t$MoTe$_2$ in the FH model 1BPV calculation in lattice with $(N_x,N_y,\widetilde{n}_{11},\widetilde{n}_{12},\widetilde{n}_{21},\widetilde{n}_{22})=(4,4,1,0,0,1)$ at $3.7^\circ$ and $10/\epsilon=1$ in the fully-polarized sector. The corresponding 2BPV calculations are in (b,d,f). The black hexagon is the Brillouin zone. In (a) the lowest 12 states are at CFL momenta $k_x+k_yN_x=5,6,9,11,14,15$ in the black box. In (b) with the 2BPV calculation, the lowest 12 states are no longer at the CFL momenta, hence the 2BPV ground state is not CFL. The orange arrows track the evolution of the low-energy states at $k_x=k_y=0$ from 1BPV to 2BPV {using wavefunction overlap}. The occupation number in (c,d) and the structure factor Eq.\eqref{Sqequation} in (e,f) are all calculated by averaging over the lowest 12 states in the black box at CFL momenta (two states for each momentum), regardless of whether those 12 states are the absolute lowest energy states.
    }
    \label{fig:CFLcor}
\end{figure}

The above features in the LLL can be utilized to identify CFL in $t$MoTe$_2$. For a CFL to exist in a periodic lattice system, the ground states should lie in the same spin sector and occupy the same momentum sectors as those of the half-filled LLL in the same momentum mesh. Therefore we use the following criteria to identify CFL:
\begin{proposition}
\label{prop:CFL}
Given a momentum mesh defined in \eqnref{eq:momentum_mesh}, we can calculate ED spectrum of half-filled LLL with Coulomb interaction on that mesh, which has $N_{\text{GS}}$ ground states. The system at filling $\nu=-1/2$ with the same momentum mesh is labelled to be in the CFL region if its lowest $N_{\text{GS}}$ ground states (i) have the same spin and (ii) have the same momenta as $N_{\text{GS}}$ ground states of half-filled LLL with Coulomb interaction.
\end{proposition}
In \cref{prop:CFL}, we do not use the spread and gap as criteria since CFL is a gapless phase by definition and any gaps of it shown in the ED spectrum should come from the finite-size effects. We show in Fig.~\ref{fig:CFLcor}(a,b) the energy spectrum of $t$MoTe$_2$ at $3.7^\circ$ in the FH model with $10/\epsilon=1$ in 1BPV and 2BPV calculations for a $4\times 4$ non-tilted lattice at $\nu=-1/2$. We find that the ground state is fully spin-polarized, hence we only show the energy spectrum in the spin-polarized sector. The 1BPV energy spectrum has 12 low-energy states at the same CFL momenta as the LLL, whereas in 2BPV there are additional low-energy many-body states with zero total momentum $k_x=k_y=0$, making the first 12 low-energy states in the 2BPV spectrum at different momenta. This suggests the 2BPV ground state is not consistent with CFL. We have also ruled out the possibility of a competing Pfaffian state, {since it would have a sixfold quasi-degenerate ground states at $k_x=k_y=0$ for this momentum mesh}. {Note that at the $k_x=k_y=0$ sector, there is a switching of low-energy states when comparing the 1BPV and 2BPV spectra, e.g., the lowest two states with $k_x=k_y=0$ in 2BPV were the third- and fourth-lowest energy states in 1BPV, as indicated by the orange arrows.}

We further compute the occupation number and structure factor using the 12 low-energy states \emph{at CFL momenta} in Fig.~\ref{fig:CFLcor}(c-f). The occupation-number distributions in 1BPV and 2BPV are similar, although the fluctuation of occupation number is larger in 2BPV than in 1BPV. The structure factor $S(\bsl q)$ in both 1BPV and 2BPV calculations is still similar to that in the LLL. These results support the identification of the 1BPV ground state as a CFL, while for 2BPV with band mixing the CFL state is still at low-energy, but it is no longer the ground state due to the additional low-energy many-body state with zero total momentum.

\begin{figure}
    \centering
    \includegraphics[width=\columnwidth]{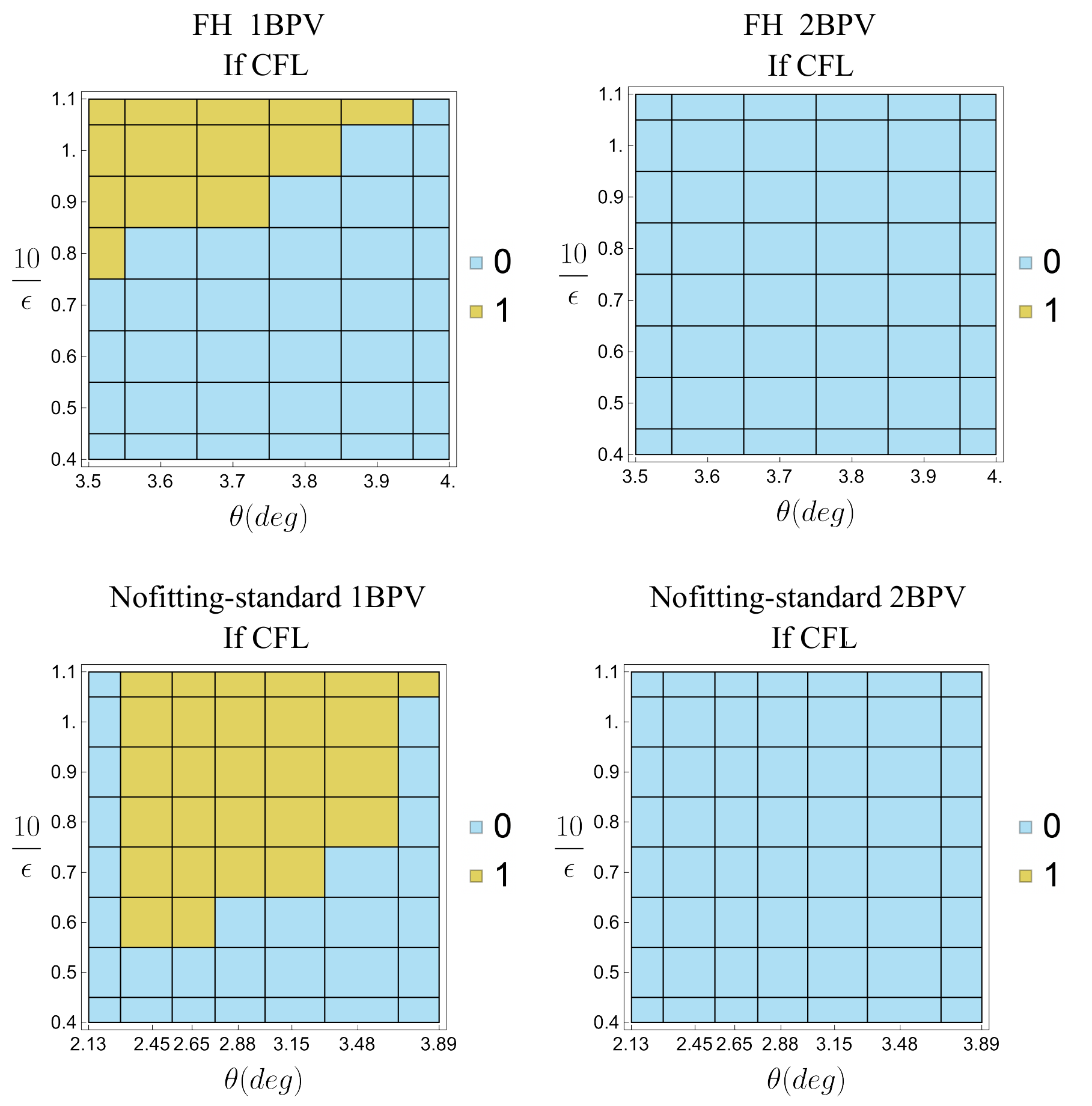}
    \caption{  1BPV and 2BPV phase diagrams in the fully-polarized sector for lattice with $(N_x,N_y,\widetilde{n}_{11},\widetilde{n}_{12},\widetilde{n}_{21},\widetilde{n}_{22})=(4,4,1,0,0,1)$ calculated for the FH model and the no-fitting model in the standard basis. The value 1 represents CFL and 0 represents non-CFL.
    }
    \label{fig:ifCFL4b4}
\end{figure}

Next, we extend the search for CFLs to other parameter regions with different twist angle and interaction strength based on Proposition~\ref{prop:CFL}, and we obtain the phase diagram in Fig.~\ref{fig:ifCFL4b4} for 1BPV and 2BPV calculations in $4\times 4$ non-tilted lattice at filling $\nu=-1/2$. The 1BPV ground state is CFL at large interaction, but the CFL region vanishes in 2BPV when band mixing is taken into account. This absence of CFL in 2BPV calculations does not seem to be related to the system size. As shown in Fig.~\ref{ED20_1v2} in App.~\ref{App_EDcombined}, a similar disappearance of CFL due to band mixing is also observed in a larger $5\times 4$ system, even though there is no clear evidence to identify the nature of competing phases. 

\section{Non-existence of non-Abelian Pfaffian state for $\nu=-3/2$ at $2.13^\circ$}
\label{Sec_3v2}

Previous theoretical works~\cite{Wang2025higherLL,Xu2025nonabelian,Ahn2024nonabelian,Chen2025nonabelian,LiWuAbelian2026} suggest that at small twist angles $\sim 2^\circ$ the non-Abelian Pfaffian state can be realized at $\nu=-3/2$ in the fully-polarized sector where the second-top moir\'e band is half-filled. The polarization of the ground state has been partially probed in numerics using a reduced spinful Hilbert space by either allowing holes in the second-top moir\'e band to flip their valley to occupy the top~\cite{Chen2025nonabelian} or the second-top~\cite{Wang2025higherLL,Ahn2024nonabelian} moir\'e band of the other valley, but no full-fledged 2BPV spinful calculations were performed. Experimentally, $2.1^\circ$ $t$MoTe$_2$ shows signatures of time-reversal-symmetry breaking over a broad range of fillings, including a finite Hall response near $\nu=-3/2$~\cite{Kang2024_tMoTe2_2.13,KangTRBFQSH2025}, which is consistent with a spin-polarized ground state. Motivated by these studies, we investigate in this section whether the non-Abelian Pfaffian state can be realized in our model at $\nu=-3/2$ in the fully-polarized sector. Near $2.13^\circ$, the top two valence bands have the same Chern number $C=1$, in contrast to the situation near $3.7^\circ$, where they have opposite Chern numbers. This makes the top two valence bands analogous to the lowest and first Landau levels. Moreover, as pointed out in Ref.~\cite{Wang2025higherLL}, the quantum geometry of the second moir\'e band at $2.13^\circ$ is close to that of the first Landau level. Therefore, when the top valence band is fully filled by holes and the second valence band is half-filled, a Pfaffian state may be expected by analogy with Landau-level physics. In the following, we compute the ground state at $\theta=2.13^\circ$ and $\nu=-3/2$ under full spin polarization.

\begin{figure*}
    \centering
    \includegraphics[width=\textwidth]{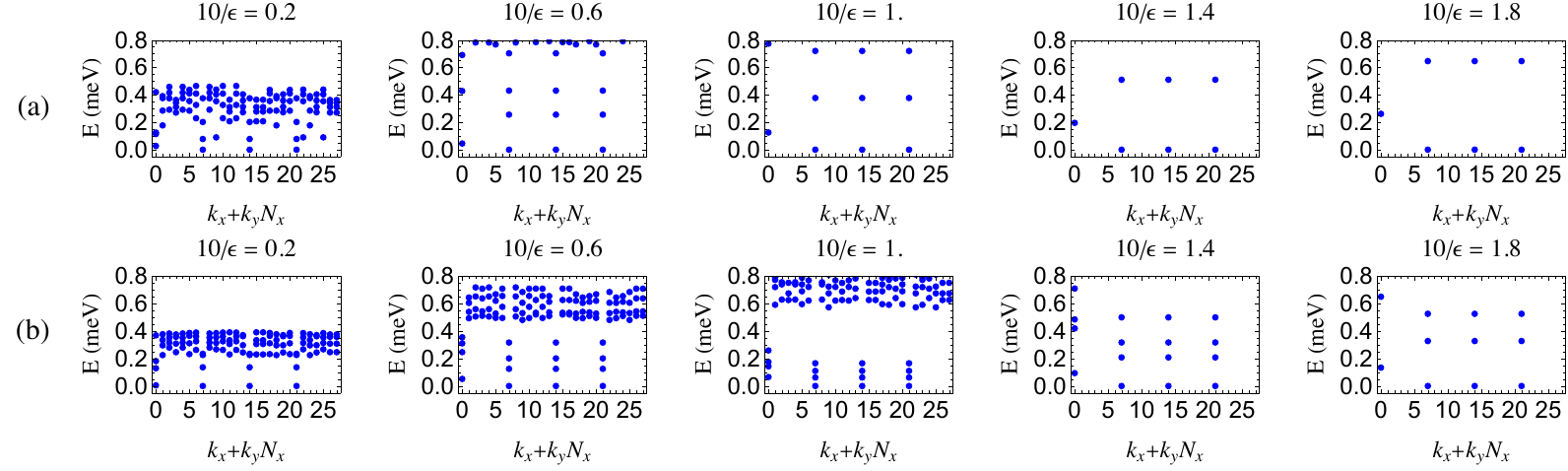}
    
    \caption{Energy spectrum of the DFT no-fitting model in the standard basis at twist angle 2.13$^\circ$ at $-3/2$ filling in the spin-polarized sector with $10/\epsilon=0.2\sim1.8$. The system has 28 sites with $(N_x,N_y,\widetilde{n}_{11},\widetilde{n}_{12},\widetilde{n}_{21},\widetilde{n}_{22})=(14,2,1,2,1,1)$. In (a) the spectrum is computed by assuming the top moir\'e valence band fully filled by holes and the second-top moir\'e valence band is half filled. In (b) the spectrum is computed by considering only the second-top moir\'e valence band and making that band half-filled. 
    }
    \label{fig:nab3v2_14b2}
\end{figure*}

\begin{figure*}
    \centering
    \includegraphics[width=\textwidth]{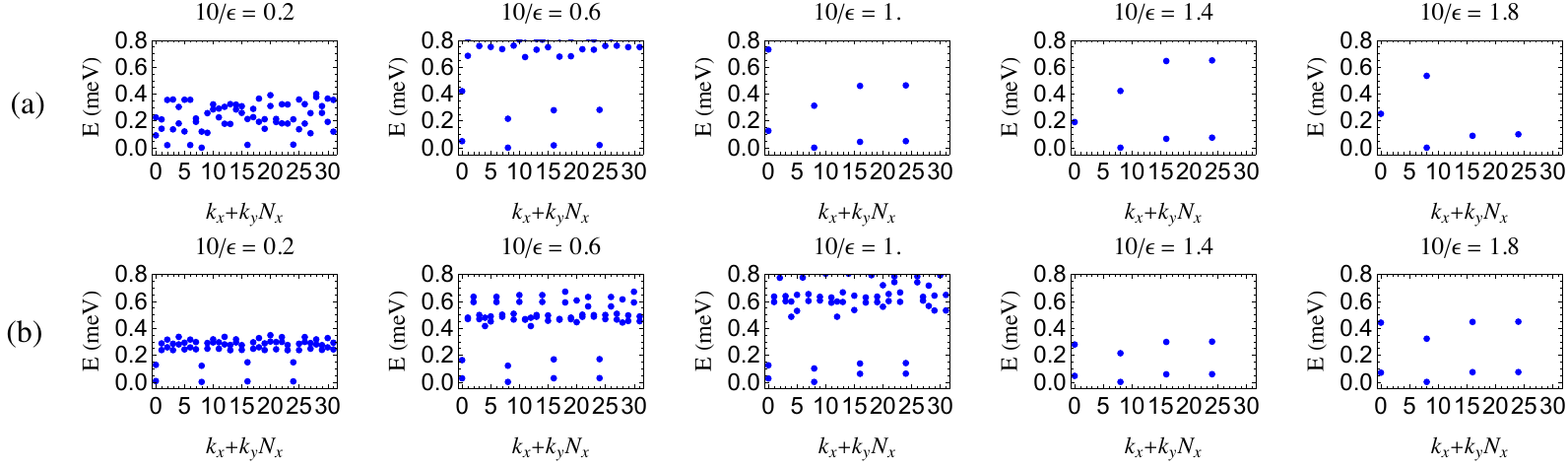}
    
    \caption{Energy spectrum of the DFT no-fitting model in the standard basis at twist angle 2.13$^\circ$ at $-3/2$ filling in the spin-polarized sector with $10/\epsilon=0.2\sim1.8$. The system has 32 sites with $(N_x,N_y,\widetilde{n}_{11},\widetilde{n}_{12},\widetilde{n}_{21},\widetilde{n}_{22})=(16,2,1,-3,0,1)$. In (a) the spectrum is computed by assuming the top moir\'e valence band fully filled by holes and the second-top moir\'e valence band is half filled. In (b) the spectrum is computed by considering only the second-top moir\'e valence band and making that band half-filled.
    }
    \label{fig:nab3v2_16b2}
\end{figure*}

The energy spectrum obtained from the no-fitting model at twist angle 2.13$^\circ$ at $-3/2$ filling for a $14\times 2$ system in the spin-polarized sector is shown in Fig.~\ref{fig:nab3v2_14b2}. Fig.~\ref{fig:nab3v2_14b2}(a) is calculated by including the top two moir\'e bands at $\nu=-3/2$ with the top moir\'e band fully filled by holes and the second-top moir\'e band half-filled. Fig.~\ref{fig:nab3v2_14b2}(b) is calculated by only including the second-top moir\'e band in ED calculation and making that band half-filled. Note that the Hilbert-space dimensions in Fig.~\ref{fig:nab3v2_14b2}(a) and (b) are the same, but Fig.~\ref{fig:nab3v2_14b2}(a) also contains the HF contribution from the fully-filled band. The particle number in this system is an even number, hence the non-Abelian state should have six quasi-degenerate ground states at momenta $k_x+k_y N_x=7,14,21$. This property is not satisfied in Fig.~\ref{fig:nab3v2_14b2} due to the additional low-energy state at $k_x=k_y=0$, therefore the non-Abelian state is not the ground state at $-3/2$ filling in this system. To reduce finite-size effects, we also perform the same calculation for a larger $16\times 2$ system in Fig.~\ref{fig:nab3v2_16b2}, where the six non-Abelian ground states would be at momenta $k_x+k_y N_x=8,16,24$. However, similar to Fig.~\ref{fig:nab3v2_14b2}, the non-Abelian state is still absent. We note that the absence of clear signatures of a non-Abelian state in our calculations differs from the results of Ref.~\cite{Wang2025higherLL,Xu2025nonabelian,Ahn2024nonabelian,Chen2025nonabelian,LiWuAbelian2026}. This difference may arise because those works used energy bands renormalized by a self-consistent HF calculation at $\nu=-2$, which can modify the energetic and geometric properties of bands. This can potentially change the calculation, e.g., in rhombohedral multilayer graphene aligned with hBN the HF renormalized band can bias towards an FCI that is unstable under band mixing, as pointed out in Refs.~\cite{MFCIIV,Li2025multibanditer,MFCI2,MFCI3}.

\section{Conclusion}
\label{Sec:conclusion}
We have presented a systematic study of FCIs and magnetism in \tmt\ over a wide range of twist angles ($2.13^\circ$–$4^\circ$) and filling factors using exact diagonalization that takes the top two moir\'e valence bands into account. 

Many key experimental features are reproduced in our calculations. First, near $\theta\approx 3.7^\circ$ we obtain robust FCIs at $\nu=-2/3$, $-3/5$, and $-4/7$, while no FCI is found at $\nu=-3/7$ in the same regime, consistent with the experimentally established hierarchy around $3.7^\circ$~\cite{cai2023signatures,park2023observation,park2025obsfci,xu_txl2025FCI}.
Second, at $\nu=-1/3$ we find that the ground state near $3.7^\circ$ is a $K$-CDW rather than an FCI, consistent with recent experimental indications of CDW order at $\nu=-1/3$ for $\theta\sim 3^\circ$--$4^\circ$~\cite{Xiaodongnew}.
Third, multi-band effects remove the spurious FCIs that appear in the 1BPV calculation at $\nu=-1/3$ and $-4/3$, improving consistency with the experimentally reported fillings with magnetization~\cite{cai2023signatures}. A similar conclusion was reached in Ref.~\cite{Yu2024mote}, while related works have also highlighted the importance of band mixing and other multi-band effects in $t$MoTe$_2$~\cite{kwan2026FTI,Fu2023BandMixingFCItMoTe2,hou2025stabilizingfractionalchernstates}. In addition, we tested the scaling of the FCI gap in \tmt\ at fillings $\nu=-2/3,-3/5,-4/7$ across the Jain sequence, which recovers a gap scaling similar to the FQHE in lowest Landau level. Combining our numerical results, we assembled a comprehensive phase diagram as a function of twist angle and filling, providing a unified benchmark for the stability of incompressible and competing phases. At $\nu=-3/2$, we tested the possibility for a non-Abelian Pfaffian state at twist angle $2.13^\circ$, motivated by the Landau-level analogy, but we do not find evidence for this state within the models and parameter regimes studied here.

There are also regimes where the comparison with experiments is more subtle. At filling $\nu=-1/3$, the comparison with experiments requires some care because the experimental situation is itself diverse. Ref.~\cite{Xiaodongnew} reported CDW order for twist angles around $3^\circ$--$4^\circ$, whereas Ref.~\cite{Pan1v3FCI2026} observed an FCI at $\theta=3.5^\circ$. Our numerical results indicate that this filling lies close to a competition between CDW and FCI phases. In particular, at $\theta=3.5^\circ$ we find a transition from a CDW ground state to an FCI ground state as the interaction strength is increased. In addition, for twist angles $2.65^\circ$--$3.48^\circ$, the results obtained from the standard and quick bases are not identical: the quick basis gives an FCI ground state with a sizable energy gap larger than the spread of the FCI states, whereas in the standard basis the corresponding gap is smaller than or comparable to the spread of the low-energy states. This difference is consistent with the smaller bandwidth in the quick basis, which makes the FCI phase more favorable. These observations suggest that $\nu=-1/3$ is a particularly delicate regime where small changes in band dispersion, interaction strength, or sample details can shift the balance between CDW and FCI order. At $\nu=-1/2$, the 1BPV calculation exhibits clear CFL-like signatures, consistent with previous single-band theories~\cite{Dong2023CFLtMoTe2,Goldman2023Composite} and experimental indications of zero-field CFL behavior~\cite{park2023observation,Anderson2024TrionCFL}. In the 2BPV calculation, however, the low-energy spectrum becomes more complicated and the CFL signatures can no longer be cleanly observed. This suggests that the half-filled state is more delicate than in the single-band description and that band mixing can play an important role in shaping the low-energy physics at $\nu=-1/2$.

Overall, our comprehensive 2BPV exact-diagonalization results provide a unified numerical benchmark for the experimentally observed hierarchy of FCIs, CDW order, and magnetism in $t$MoTe$_2$, and they identify the fillings where multi-band effects are essential for making reliable contact with experiments.

\acknowledgments

We acknowledge Miguel Gonçalves, Yves Kwan, Jonah Herzog-Arbeitman, D. N. Sheng and Fengcheng Wu for useful discussions. We thank Yan Zhang, Hanqi Pi, Jiaxuan Liu and Quansheng Wu for previous collaborations on related topics. H.L. was supported by the European Research Council (ERC) under the European Union's Horizon 2020 research and innovation program (Grant Agreement No. 101020833). 
J. Y.'s work at Princeton University was supported by the Gordon and Betty Moore Foundation through Grant No. GBMF8685 towards the Princeton theory program. J. Y.'s work at the University of Florida was supported by startup funds at the University of Florida. B.A.B. was supported by the Gordon and Betty Moore Foundation through Grant No.~GBMF8685 towards the Princeton theory program, the Gordon and Betty Moore Foundation’s EPiQS Initiative (Grant No.~GBMF11070), the Global Collaborative Network Grant at Princeton University, the Simons Investigator Grant No.~404513, the NSF-MERSEC (Grant No.~MERSEC DMR 2011750), the Simons Collaboration on New Frontiers in Superconductivity (SFI-MPS-NFS-00006741-01), and the Schmidt Fund at Princeton University and the Princeton Catalysis Initiative. XX acknowledges the support from the DOE BES. The Flatiron Institute is a division of the Simons Foundation.

\onecolumngrid
\newpage

\appendix

\tableofcontents

\section{Hamiltonian}
\label{App_H0}

In this appendix, we summarize the single-particle Hamiltonians used in the main text. We use two descriptions of AA-stacked twisted bilayer MoTe$_2$: the first-harmonic (FH) continuum model and the DFT no-fitting model. The FH model is a compact continuum model with a small number of parameters, while the no-fitting model contains a more general set of continuum-model terms whose coefficients are determined from first-principles calculations~\cite{MFCII,DFTnofitting2024}.

\subsection{First-harmonic continuum model}

We first introduce the first-harmonic (FH) continuum model for AA-stacked twisted bilayer MoTe$_2$, following the convention of Ref.~\cite{MFCII}. The continuum basis is labeled by
$c_{\eta,l,\mathbf{r}}$, where $\eta=\pm$ denotes the $\pm K$ valley, $l=t,b$ denotes the top and bottom layers, and $\mathbf{r}$ is the real-space coordinate. The operator $c_{\eta,l,\mathbf{r}}$ is a slowly varying continuum field obtained by expanding the microscopic electron operator near the rotated monolayer valley momentum $\eta\mathbf{K}_l$ of layer $l$. Schematically,
\begin{equation}
\Psi_l(\mathbf{r})
\simeq
\sum_{\eta=\pm}
e^{i\eta\mathbf{K}_l\cdot\mathbf{r}}
u_{\eta,l}(\mathbf{r})
c_{\eta,l,\mathbf{r}},
\label{eq:microscopic_field_expansion}
\end{equation}
where $u_{\eta,l}(\mathbf{r})$ is the cell-periodic part of the monolayer Bloch wave function. Since the fast factor $e^{i\eta\mathbf{K}_l\cdot\mathbf{r}}$ has been separated out, the continuum field is not strictly periodic under a moir\'e lattice translation. Instead, under a moir\'e lattice vector $\mathbf{R}_M$, it carries a layer- and valley-dependent phase.

For AA-$t$MoTe$_2$, we rotate the top layer by $-\theta/2$ and the bottom layer by $+\theta/2$. The rotated monolayer valley momenta are
\begin{equation}
\mathbf{K}_b
=
\frac{4\pi}{3a_0}
\left(
\cos\frac{\theta}{2},
\sin\frac{\theta}{2}
\right)^T,
\qquad
\mathbf{K}_t
=
\frac{4\pi}{3a_0}
\left(
\cos\frac{\theta}{2},
-\sin\frac{\theta}{2}
\right)^T ,
\label{eq:Kt_Kb_definition}
\end{equation}
where $a_0$ is the monolayer lattice constant. We define
\begin{equation}
\mathbf{q}_1
=
\mathbf{K}_b-\mathbf{K}_t
=
\frac{4\pi}{3a_0}
2\sin\frac{\theta}{2}
\left(
0,
1
\right)^T,
\qquad
\mathbf{q}_2=C_3\mathbf{q}_1,
\qquad
\mathbf{q}_3=C_3^2\mathbf{q}_1 .
\label{eq:q_vectors_definition}
\end{equation}

The symmetry transformations of the continuum creation operators are
\bea
C_3 c^\dagger_{\eta,l,\mathbf{r}} C_3^{-1}
&=&
e^{i\eta\pi/3}
c^\dagger_{\eta,l,C_3\mathbf{r}},\\
(C_{2y}\mathcal{T})
c^\dagger_{\eta,l,\mathbf{r}}
(C_{2y}\mathcal{T})^{-1}
&=&
c^\dagger_{\eta,\bar{l},C_{2y}\mathbf{r}},\\
\mathcal{T}
c^\dagger_{\eta,l,\mathbf{r}}
\mathcal{T}^{-1}
&=&
(-\eta)
c^\dagger_{-\eta,l,\mathbf{r}},\\
T_{\mathbf{R}_M}
c^\dagger_{\eta,l,\mathbf{r}}
T_{\mathbf{R}_M}^{-1}
&=&
e^{-i\eta\mathbf{R}_M\cdot\mathbf{K}_l}
c^\dagger_{\eta,l,\mathbf{r}+\mathbf{R}_M}.
\eea
Here $\bar{l}=b,t$ for $l=t,b$, respectively. The last equation shows explicitly that the continuum field has a quasi-periodic transformation under moir\'e translations. This phase is important for the form of the interlayer tunneling.

The single-particle Hamiltonian is written as
\begin{equation}
H_0
=
\sum_{\eta=\pm} H_{\eta,0}.
\label{eq:H0_valley_sum}
\end{equation}
For a fixed valley $\eta$, the FH Hamiltonian takes the second-quantized form
\begin{equation}
H_{\eta,0}
=
\int d^2\mathbf{r}\,
\begin{pmatrix}
c^\dagger_{\eta,b,\mathbf{r}} &
c^\dagger_{\eta,t,\mathbf{r}}
\end{pmatrix}
\begin{pmatrix}
h_{\eta,b}(\mathbf{r}) & t_{\eta}(\mathbf{r}) \\
t^*_{\eta}(\mathbf{r}) & h_{\eta,t}(\mathbf{r})
\end{pmatrix}
\begin{pmatrix}
c_{\eta,b,\mathbf{r}} \\
c_{\eta,t,\mathbf{r}}
\end{pmatrix}.
\label{eq:FH_realspace_H_eta}
\end{equation}
The layer-diagonal terms are
\begin{equation}
h_{\eta,l}(\mathbf{r})
=
\frac{\hbar^2\nabla^2}{2m^*}
+
V_{\eta,l}(\mathbf{r}),
\label{eq:FH_h_eta_l}
\end{equation}
where $m^*$ is the effective mass of the monolayer valence band. We use the convention appropriate for valence electron bands, so that the kinetic term gives $-\hbar^2|\mathbf{k}|^2/(2m^*)$ in momentum space.

The FH intralayer moir\'e potential is
\begin{equation}
V_{\eta,l}(\mathbf{r})
=
V e^{-(-)^l i\psi}
\sum_{j=1}^{3}
e^{i\mathbf{g}_j\cdot\mathbf{r}}
+
V e^{(-)^l i\psi}
\sum_{j=1}^{3}
e^{-i\mathbf{g}_j\cdot\mathbf{r}},
\label{eq:FH_intralayer_potential}
\end{equation}
where $V$ and $\psi$ are real parameters, $(-)^t=+1$, $(-)^b=-1$, and
\begin{equation}
\mathbf{g}_j
=
C_3^{j-1}\mathbf{b}_{M,1},
\qquad
j=1,2,3 .
\label{eq:g_vectors_definition}
\end{equation}
Here $\mathbf{b}_{M,1}$ is a primitive moir\'e reciprocal lattice vector.

The FH interlayer tunneling is
\begin{equation}
t_{\eta}(\mathbf{r})
=
w
\sum_{j=1}^{3}
e^{-i\eta\mathbf{q}_j\cdot\mathbf{r}},
\label{eq:FH_interlayer_tunneling}
\end{equation}
where $w$ is chosen to be real by fixing the relative phase between the two layers. This form is consistent with the quasi-periodic transformation of $c_{\eta,l,\mathbf{r}}$ under moir\'e translations. In particular, although $t_{\eta}(\mathbf{r})$ is not itself moir\'e periodic, the full interlayer term in Eq.~\eqref{eq:FH_realspace_H_eta} is invariant under moir\'e translations because the layer fields acquire different phase factors. The symmetry constraints on the Hamiltonian terms are
\bea
C_3&:&
\qquad
h_{\eta,l}(C_3\mathbf{r})
=
h_{\eta,l}(\mathbf{r}),
\qquad
t_{\eta}(C_3\mathbf{r})
=
t_{\eta}(\mathbf{r}),\\
C_{2y}\mathcal{T}&:&
\qquad
h_{\eta,\bar{l}}(C_{2y}\mathbf{r})
=
h^*_{\eta,l}(\mathbf{r}),
\qquad
t_{\eta}(C_{2y}\mathbf{r})
=
t_{\eta}(\mathbf{r}),\\
\mathcal{T}&:&
\qquad
h_{-\eta,l}(\mathbf{r})
=
h^*_{\eta,l}(\mathbf{r}),
\qquad
t_{-\eta}(\mathbf{r})
=
t^*_{\eta}(\mathbf{r}),\\
T_{\mathbf{R}_M}&:&
\qquad
h_{\eta,l}(\mathbf{r}+\mathbf{R}_M)
=
h_{\eta,l}(\mathbf{r}),
\qquad
t_{\eta}(\mathbf{r}+\mathbf{R}_M)
=
t_{\eta}(\mathbf{r})
e^{-i\eta\mathbf{q}_1\cdot\mathbf{R}_M}.
\eea
These relations ensure that Eq.~\eqref{eq:FH_realspace_H_eta} is invariant under the symmetries of AA-stacked $t$MoTe$_2$. For numerical calculations, we Fourier transform the continuum fields as
\begin{equation}
c^\dagger_{\eta,l,\mathbf{r}}
=
\frac{1}{\sqrt{A}}
\sum_{\mathbf{k}}
\sum_{\mathbf{Q}\in\mathcal{Q}_{\eta}^{l}}
e^{-i(\mathbf{k}-\mathbf{Q})\cdot\mathbf{r}}
c^\dagger_{\eta,l,\mathbf{k}-\mathbf{Q}},
\label{eq:FH_fourier_transform}
\end{equation}
where $A$ is the system area, $\mathbf{k}$ lies in the first moir\'e Brillouin zone, and
\begin{equation}
\mathcal{Q}_{\eta}^{l}
=
\left\{
\mathbf{G}_M
+
\eta(-)^l\mathbf{q}_1
\right\}.
\label{eq:Q_lattice_definition}
\end{equation}
Here $\mathbf{G}_M$ runs over moir\'e reciprocal lattice vectors. The full momentum set is
\begin{equation}
\mathcal{Q}
=
\mathcal{Q}_{\eta}^{t}
\cup
\mathcal{Q}_{\eta}^{b}.
\label{eq:full_Q_set}
\end{equation}
In this basis, the FH Hamiltonian becomes
\begin{equation}
H_{\eta,0}
=
\sum_{\mathbf{k}}
\sum_{\mathbf{Q},\mathbf{Q}'\in\mathcal{Q}}
c^\dagger_{\eta,\mathbf{k},\mathbf{Q}}
\left[
h_{\eta}(\mathbf{k})
\right]_{\mathbf{Q},\mathbf{Q}'}
c_{\eta,\mathbf{k},\mathbf{Q}'},
\label{eq:FH_momentum_hamiltonian}
\end{equation}
where
$c^\dagger_{\eta,\mathbf{k},\mathbf{Q}}
=
c^\dagger_{\eta,l^\eta_{\mathbf{Q}},\mathbf{k}-\mathbf{Q}}$ and $l^\eta_{\mathbf{Q}}=l$ if $\mathbf{Q}\in\mathcal{Q}_{\eta}^{l}$. The matrix elements are
\begin{equation}
\left[
h_{\eta}(\mathbf{k})
\right]_{\mathbf{Q},\mathbf{Q}'}
=
-\frac{\hbar^2|\mathbf{k}-\mathbf{Q}|^2}{2m^*}
\delta_{\mathbf{Q},\mathbf{Q}'}
+
V
\sum_{j=1}^{3}
\left[
e^{-\eta(-)^{\mathbf{Q}} i\psi}
\delta_{\mathbf{Q}+\mathbf{g}_j,\mathbf{Q}'}
+
e^{\eta(-)^{\mathbf{Q}} i\psi}
\delta_{\mathbf{Q}-\mathbf{g}_j,\mathbf{Q}'}
\right]
+
w
\sum_{j=1}^{3}
\left[
\delta_{\mathbf{Q}+\mathbf{q}_j,\mathbf{Q}'}
+
\delta_{\mathbf{Q},\mathbf{Q}'+\mathbf{q}_j}
\right].
\label{eq:FH_matrix_elements}
\end{equation}
Here $(-)^{\mathbf{Q}}=\pm1$ for $\mathbf{Q}\in\{\mathbf{G}_M\pm\mathbf{q}_1\}$, with $\eta(-)^{\mathbf{Q}}=(-)^{l^\eta_{\mathbf{Q}}}$. The parameters in the FH model obtained in \refcite{MFCII} can be found in Table~\ref{tab:parameters_DFT}, which are similar to the results obtained in Refs~\cite{wang2023fractional,reddy2023fractional}.

\begin{table}[t]
    \centering
    \begin{tabular}{|c|c|c|c|c|c|c|}
    \hline
     Model & $m^*$ ($m_e$) &  $V$   & $\psi (^\circ)$ & $w$  \\ 
    \hline
     \text{FH} &  0.60  & 16.5 & -105.9 & -18.8  \\
    \hline
    \end{tabular}
    \caption{
    The parameter values for the FH continuum model with energy unit in meV, which are obtained in \refcite{MFCII}.
    }
    \label{tab:parameters_DFT}
\end{table}

\subsection{DFT no-fitting model}

We next introduce the DFT no-fitting model used in this work. The no-fitting model provides a more general continuum Hamiltonian than the FH model. It includes a larger set of symmetry-allowed derivative terms and moir\'e harmonics. The general continuum Hamiltonian can be written as
\begin{equation}
H_{{\rm NF},\eta}
=
\sum_{M_x,M_y\ge 0}
\sum_{l,l'=t,b}
\int d^2\mathbf r\,
\left[
i^{M_x+M_y}
\partial_x^{M_x}
\partial_y^{M_y}
c^\dagger_{\eta,l,\mathbf r}
\right]
t^{M_xM_y}_{\eta,ll'}(\mathbf r)
c_{\eta,l',\mathbf r}.
\label{eq:nofit_general_realspace}
\end{equation}
Here $M_x$ and $M_y$ denote the order of derivatives, and $t^{M_xM_y}_{\eta,ll'}(\mathbf r)$ are continuum-model coefficient functions. The terms with $l=l'$ describe layer-diagonal kinetic and intralayer moir\'e-potential contributions, while the terms with $l\neq l'$ describe interlayer tunneling. The FH model corresponds to a small subset of Eq.~\eqref{eq:nofit_general_realspace}: the layer-diagonal quadratic kinetic term, the first-harmonic intralayer potential, and the first-harmonic interlayer tunneling.

Because the continuum fields carry layer-dependent phases under moir\'e translations, the coefficient functions satisfy
\begin{equation}
t^{M_xM_y}_{\eta,ll'}(\mathbf r+\mathbf R_M)
=
e^{-i\eta(\mathbf K_l-\mathbf K_{l'})\cdot\mathbf R_M}
t^{M_xM_y}_{\eta,ll'}(\mathbf r).
\label{eq:nofit_translation_constraint}
\end{equation}
The convenient Fourier expansion is
\begin{equation}
t^{M_xM_y}_{\eta,ll'}(\mathbf r)
=
\sum_{\mathbf G_M}
r^{M_xM_y}_{ll',\mathbf G_M}
e^{-i\eta(\mathbf K_l-\mathbf K_{l'}+\mathbf G_M)\cdot\mathbf r},
\label{eq:nofit_fourier_expansion}
\end{equation}
where $\mathbf G_M$ runs over moir\'e reciprocal lattice vectors. The coefficients $r^{M_xM_y}_{ll',\mathbf G_M}$ specify the continuum Hamiltonian. For example, terms with $l=l'$ and $\mathbf G_M=0$ contribute to the kinetic energy, terms with $l=l'$ and $\mathbf G_M\neq 0$ describe intralayer moir\'e potentials, and terms with $l\neq l'$ describe interlayer tunneling. In Ref.~\cite{DFTnofitting2024}, these coefficients are obtained directly from DFT Hamiltonian matrix elements using the no-fitting construction without the need of parameter-fitting. 

To get the Hamiltonian in momentum space, we write
\begin{equation}
c^\dagger_{\eta,l,\mathbf r}
=
\frac{1}{\sqrt{A}}
\sum_{\mathbf k}
\sum_{\mathbf Q\in\mathcal Q_\eta^l}
e^{-i(\mathbf k-\mathbf Q)\cdot\mathbf r}
c^\dagger_{\eta,l,\mathbf k-\mathbf Q},
\label{eq:nofit_fourier_transform}
\end{equation}
where $\mathbf k$ lies in the first moir\'e Brillouin zone and
\begin{equation}
\mathcal Q_\eta^l
=
\left\{
\mathbf G_M+\eta(-)^l\mathbf q_1
\right\}.
\label{eq:nofit_Q_lattice_definition}
\end{equation}
Here $(-)^t=+1$ and $(-)^b=-1$. The no-fitting Hamiltonian can then be written as
\begin{equation}
H_{{\rm NF},\eta}
=
\sum_{\mathbf k}
\sum_{\mathbf Q,\mathbf Q'\in\mathcal Q}
c^\dagger_{\eta,\mathbf k,\mathbf Q}
\left[
h^{\rm NF}_{\eta}(\mathbf k)
\right]_{\mathbf Q,\mathbf Q'}
c_{\eta,\mathbf k,\mathbf Q'},
\label{eq:nofit_momentum_hamiltonian}
\end{equation}
where $\mathcal Q=\mathcal Q_\eta^t\cup\mathcal Q_\eta^b$ and
\begin{equation}
c^\dagger_{\eta,\mathbf k,\mathbf Q}
=
c^\dagger_{\eta,l^\eta_{\mathbf Q},\mathbf k-\mathbf Q}.
\label{eq:nofit_c_k_Q_definition}
\end{equation}
Here $l^\eta_{\mathbf Q}=l$ if $\mathbf Q\in\mathcal Q_\eta^l$. The matrix elements can be expressed schematically as
\begin{equation}
\left[
h^{\rm NF}_{\eta}(\mathbf k)
\right]_{\mathbf Q,\mathbf Q'}
=
\sum_{M_x,M_y\ge 0}
\sum_{\mathbf G_M}
r^{M_xM_y}_{l^\eta_{\mathbf Q}l^\eta_{\mathbf Q'},\mathbf G_M}
\left(k_x-Q_x\right)^{M_x}
\left(k_y-Q_y\right)^{M_y}
\delta_{\mathbf Q,\,
\mathbf Q'+\eta(\mathbf K_{l^\eta_{\mathbf Q}}-\mathbf K_{l^\eta_{\mathbf Q'}}+\mathbf G_M)} .
\label{eq:nofit_matrix_elements}
\end{equation}
In this work, we use the no-fitting models at $2.13^\circ,\,2.45^\circ,\,2.65^\circ,\,2.88^\circ,\,3.15^\circ,\,3.48^\circ,\,3.89^\circ$. We consider both the standard basis and the quick basis introduced in Ref.~\cite{DFTnofitting2024}. The standard basis contains Mo-s3p2d2 and Te-s3p2d2f1 orbitals, while the quick basis contains Mo-s3p2d1 and Te-s3p2d2 orbitals. The two bases give qualitatively similar low-energy band structures, but they can lead to quantitative differences in bandwidth and band geometry, which in turn affect the stability of the many-body phases. Related continuum-model parameterizations for twisted MoTe$_2$ have also been proposed in Refs.~\cite{wang2023fractional,Zhang2024pol,Zhangtwisttransferable2025,reddy2023fractional}.

\subsection{Normal ordered interaction in the hole basis}

At charge neutrality, all the bands of $H_0$ are fully filled. The interaction $H_{\text{int}}$ is normal ordered with respect to the charge neutrality, leading to the following form 
\eq{
H_{\text{int}} = \frac{1}{2}\sum_{l \eta,l' \eta'} \int d^2 r d^2 r' V(\bsl{r}-\bsl{r}')  \widetilde{c}^\dagger_{\eta, l, \bsl{r}} \widetilde{c}^\dagger_{\eta', l', \bsl{r}'} \widetilde{c}_{\eta', l', \bsl{r}'} \widetilde{c}_{\eta, l, \bsl{r}} \ .
}
Here $\widetilde{c}_{\eta, l,\bsl{r}}^\dagger$ creates a hole at position $\bsl{r}$ in the $l$-th layer in the $\eta$ valley.
$V(\bsl{r})$ is the double-gated screened Coulomb potential with gate distance  $\xi$:
\bea
V(\bsl{r}) =\int_{\dsR^2} \frac{d^2 p }{ (2\pi)^2} V(\bsl{p}) e^{\ii \bsl{p}\cdot\bsl{r}}, \qquad V(\bsl{p}) = \pi \xi^2 V_{\xi} \frac{\tanh(\xi |\bsl{p}|/2)}{\xi |\bsl{p}|/2}, \quad V_{\xi} = \frac{e^2 }{4\pi \epsilon \epsilon_0 \xi} \ ,
\eea
where $\epsilon_0$ is the vacuum permitivity, and $\epsilon$ is the relative dielectric constant.
Throughout of the work, we choose $\xi=20$ nm based on many experimental setups~\cite{cai2023signatures,zeng2023integer,park2023observation,Xu2023FCItMoTe2}.
$\tilde{c}_{\eta,l,\bsl{r}}$ is related to $c_{\eta,l,\bsl{r}}^\dagger$ by 
\eq{
\label{eq:c_ctilde_complexconjugate}
\cc c_{\eta,l,\bsl{r}}^\dagger \cc^{-1} =  \widetilde{c}_{\eta,l,\bsl{r}} \ ,
}
where $\cc$ is the particle-hole operator.
In our ED calculation, we always use hole basis $\tilde{c}_{\eta,l,\bsl{r}}$, and therefore for convenience, we rewrite $H_{0,\eta}$ in the hole basis here
{
\eq{
\label{eq:H_eta_0_hole_basis}
H_{\eta,0} = \int d^2 r \left( \tilde{c}^\dagger_{\eta,b,\bsl{r}} , \tilde{c}^\dagger_{\eta,t,\bsl{r}} \right) (-) \mat{ \frac{\hbar^2 \nabla^2}{2 m^*} + V_{\eta,b}(\bsl{r})  & t_\eta(\bsl{r}) \\ t_\eta^*(\bsl{r}) & \frac{\hbar^2 \nabla^2}{2 m^*} + V_{\eta,t}(\bsl{r})   } \mat{ \tilde{c}_{\eta,b,\bsl{r}} \\ \tilde{c}_{\eta,t,\bsl{r}} } 
}
}
up to a constant.
In the momentum space and diagonal basis, $H_{\eta,0}$ reads
\eq{
\label{eq:H_eta_0_hole_basis_eigen}
H_{\eta,0} = \sum_{\bsl{k},\eta,n} \widetilde{\gamma}^\dagger_{\bsl{k},\eta,n} \widetilde{\gamma}_{\bsl{k},\eta,n} (-\epsilon_{\bsl{k},\eta,n})\ , 
}
where $\epsilon_{\bsl{k},\eta,n}$ is the $n$th top valence electron band in the $\eta$ valley, and $\widetilde{\gamma}^\dagger_{\bsl{k},\eta,n}$ creates a hole at $\bsl{k}$ in valley $\eta$ in band $n$.

\section{Exact-Diagonalization Techniques}
\label{App_EDtech}

In this section, we will introduce two techniques that we used in our ED calculations---tilted lattices and band maximum.

\subsection{Tilted Lattice}

In this section, we elaborate on how we choose the momentum mesh for the ED calculation.
Our discussion will be focused on the moir\'e lattice, but in general it is applicable to all kinds of lattice. 

Our ED calculations are performed in the momentum mesh specified by 
\eq{
\text{momentum mesh}=\left\{ \left. \frac{k_x}{N_x} \bsl{f}_1 + \frac{k_y}{N_y} \bsl{f}_2 \right|  k_x = 0, 1, 2, ..., N_x -1, \ k_y = 0, 1, 2, ..., N_y -1 \right\}\ ,
\label{eq:momentum_mesh_ap}
}
where 
\eqa{
& \bsl{f}_1 = \widetilde{n}_{11} \bsl{b}_{M,1} + \widetilde{n}_{12} \bsl{b}_{M,2},\ \ \  \bsl{f}_2 = \widetilde{n}_{21} \bsl{b}_{M,1} + \widetilde{n}_{22} \bsl{b}_{M,2}
}
are two reciprocal lattice vectors (\ie, $\widetilde{n}_{11}$, $\widetilde{n}_{12}$, $\widetilde{n}_{21}$ and $\widetilde{n}_{22}$ are integers), and $N_x$ and $N_y$ are positive integers with $N_x N_y>1$. $\bsl{b}_{M,1}$ and $\bsl{b}_{M,2}$ are shortest moir\'e reciprocal lattice vectors with $\bsl{b}_{M,2}=R(\frac{2\pi}{6})\bsl{b}_{M,1}$, where $R(\theta)$ is the rotation matrix of vectors by angle $\theta$. $\bsl{a}_{M,1}$ and $\bsl{a}_{M,2}$ are moir\'e lattice vetors in real space that satisfy $\bsl{a}_{M,i}\cdot \bsl{b}_{M,j}=2\pi\delta_{ij}$.

The momentum-space mesh specifies the periodic boundary condition in the real space. To see that, we need to Fourier transform the momentum mesh back to the real space. By defining 
\eqa{
 & \widetilde{\bsl{T}}_1 = \frac{\bsl{f}_1}{N_x}\ ,\ \widetilde{\bsl{T}}_2 = \frac{\bsl{f}_2}{N_y}\ ,
}
their reciprocal lattice vectors $\bsl T_j$ in real space are defined by the requirement $\widetilde{\bsl{T}}_i\cdot \bsl{T}_j=2\pi\delta_{ij}$, which leads to
\eqa{
 & \bsl{T}_1 = 2\pi  \frac{\widetilde{\bsl{T}}_2\times\bsl{e}_z}{\bsl{e}_z\cdot(\widetilde{\bsl{T}}_1\times \widetilde{\bsl{T}}_2)} = 2\pi N_x \frac{ \widetilde{n}_{21} \bsl{b}_{M,1}\times \bsl{e}_z + \widetilde{n}_{22} \bsl{b}_{M,2} \times \bsl{e}_z }{ (\widetilde{n}_{11} \widetilde{n}_{22}- \widetilde{n}_{12} \widetilde{n}_{21}) \bsl{e}_z\cdot(\bsl{b}_{M,1} \times \bsl{b}_{M,2})  } = n_{x1} \bsl{a}_{M,1} + n_{y1} \bsl{a}_{M,2} \\
 & \bsl{T}_2 =-2\pi \frac{\widetilde{\bsl{T}}_1\times\bsl{e}_z}{\bsl{e}_z\cdot(\widetilde{\bsl{T}}_1\times \widetilde{\bsl{T}}_2)} = -2\pi N_y \frac{ \widetilde{n}_{11} \bsl{b}_{M,1}\times \bsl{e}_z + \widetilde{n}_{12} \bsl{b}_{M,2} \times \bsl{e}_z }{ (\widetilde{n}_{11} \widetilde{n}_{22}- \widetilde{n}_{12} \widetilde{n}_{21})\bsl{e}_z\cdot(\bsl{b}_{M,1} \times \bsl{b}_{M,2}) } = n_{x2} \bsl{a}_{M,1} +  n_{y2} \bsl{a}_{M,2}\ ,
}
where
\eqa{
\label{eq:n_for_T}
& n_{x1} = \frac{ N_x \widetilde{n}_{22} }{\widetilde{n}_{11} \widetilde{n}_{22}- \widetilde{n}_{12} \widetilde{n}_{21}},\ \  n_{y1} = \frac{ -N_x \widetilde{n}_{21} }{\widetilde{n}_{11} \widetilde{n}_{22}- \widetilde{n}_{12} \widetilde{n}_{21}} \\
& n_{x2} = \frac{ -N_y \widetilde{n}_{12} }{\widetilde{n}_{11} \widetilde{n}_{22}- \widetilde{n}_{12} \widetilde{n}_{21}}, \ \  n_{y2} = \frac{N_y \widetilde{n}_{11} }{\widetilde{n}_{11} \widetilde{n}_{22}- \widetilde{n}_{12} \widetilde{n}_{21}} \ .
}

There are two conditions for the choice of the momentum mesh.
\eqa{
\label{eq:two_conditions_momentum_mesh}
& \text{1. We must make sure that all points in \eqnref{eq:momentum_mesh_ap} are inequivalent to each other.} \\
& \text{2. We require that $\bsl{T}_{1}$ and $\bsl{T}_{2}$ are moir\'e lattice vectors, \ie, $n_{x1}$, $n_{y1}$, $n_{x2}$ and $n_{y2}$ in \eqnref{eq:n_for_T} are integers.}
}
The two conditions in \eqnref{eq:two_conditions_momentum_mesh} limit the choices of $\bsl{f}_1$, $\bsl{f}_2$, $N_x$ and $N_y$.
$\bsl{T}_1$ and $\bsl{T}_2$ generate a superlattice in the real-space, and owing to the second condition, its supercell should contain an integer number of moir\'e unit cells, which reads 
\eq{
N_s = \frac{\left|\bsl{T}_1\times \bsl{T}_2 \right|}{\left| \bsl{a}_{M,1} \times \bsl{a}_{M,2} \right|} = |n_{x1} n_{y2} - n_{y1} n_{x2}| = \frac{N_x N_y}{\left|\widetilde{n}_{11} \widetilde{n}_{22}- \widetilde{n}_{12} \widetilde{n}_{21}\right|},}
which means $N_x N_y$ is an integer multiple of $\left|\widetilde{n}_{11} \widetilde{n}_{22}- \widetilde{n}_{12} \widetilde{n}_{21}\right|$.
Since $N_s$ equals the number of inequivalent momentum points, we must have 
\eq{
\label{eq:sufficient_for_allowed_mesh}
|\widetilde{n}_{11}\widetilde{n}_{22}  - \widetilde{n}_{12} \widetilde{n}_{21}  | = 1 \ ,
}
owing to condition 1 in \eqnref{eq:two_conditions_momentum_mesh}, leading to 
\eq{
N_s = N_x N_y \ .
}

It turns out that \eqnref{eq:sufficient_for_allowed_mesh} is also a sufficient condition for \eqref{eq:two_conditions_momentum_mesh}, as we will show now.
First note tha \eqnref{eq:sufficient_for_allowed_mesh} naturally achieves the second condition in \eqref{eq:two_conditions_momentum_mesh}, owing to \eqnref{eq:n_for_T}.
For the first condition in \eqref{eq:two_conditions_momentum_mesh}, let us suppose two different points in the mesh are equivalent; the two points can always be expressed as $\frac{k_x}{N_x}\bsl{f}_{1} + \frac{k_y}{N_y}\bsl{f}_{2} $ and $\frac{k_x'}{N_x}\bsl{f}_{1} + \frac{k_y'}{N_y}\bsl{f}_{2} $, where $k_x$, $k_y$, $k_x'$ and $k_y'$ are all integers with $k_x, k_x'\in [0,N_x-1]$ and $k_y, k_y'\in [0,N_y-1]$.
The fact that they are equivalent means that 
\eqa{
& \frac{k_x-k_x'}{N_x}\bsl{f}_{1} + \frac{k_y-k_y'}{N_y}\bsl{f}_{2} = y_1 \bsl{b}_{M,1} + y_2 \bsl{b}_{M,2} \text{ with integers $y_1$ and $y_2$} \\
& \Leftrightarrow \left[ \frac{k_x-k_x'}{N_x} \widetilde{n}_{11} + \frac{k_y-k_y'}{N_y} \widetilde{n}_{21} \right] \bsl{b}_{M,1} + \left[ \frac{k_x-k_x'}{N_x} \widetilde{n}_{12} + \frac{k_y-k_y'}{N_y} \widetilde{n}_{22} \right] \bsl{b}_{M,2}  = y_1 \bsl{b}_{M,1} + y_2 \bsl{b}_{M,2} \\
& \Leftrightarrow \mat{ \frac{k_x-k_x'}{N_x} & \frac{k_y-k_y'}{N_y} } 
\tilde{A}
= \mat{ y_1 & y_2 } \text{ with $\tilde{A}=\mat{ 
\widetilde{n}_{11} & \widetilde{n}_{12} \\
\widetilde{n}_{21} & \widetilde{n}_{22} \\
}$}\ .
}
Since $\bsl{f}_1$ and $\bsl{f}_2$ are linearly independent, we know $\tilde{A}$ is invertiable, which means
\eq{
\label{eq:intermediate_tilted_lattice_1}
\mat{ \frac{k_x-k_x'}{N_x} & \frac{k_y-k_y'}{N_y} }  = \mat{ y_1 & y_2 } \tilde{A}^{-1}  \ .
}
According to 
\eq{
\tilde{A}^{-1} = \frac{1}{\det(\tilde{A})} \text{adj}(\tilde{A})\ ,
}
where $\text{adj}(\tilde{A})$ is the adjoint matrix of $\tilde{A}$.
Since $\tilde{A}$ is an integer matrix, so is $\text{adj}(\tilde{A})$, and thus $\mat{ y_1' & y_2' } =  \mat{ y_1 & y_2 }\text{adj}(\tilde{A})$ is an integer vector.
Then, \eqnref{eq:intermediate_tilted_lattice_1} turns into
\eqa{
& \mat{ \frac{k_x-k_x'}{N_x} & \frac{k_y-k_y'}{N_y} }  = \mat{ y_1' & y_2' } \frac{1}{\det(\tilde{A})} \ .
}
$\mat{ \frac{k_x-k_x'}{N_x} & \frac{k_y-k_y'}{N_y} }$ is nonzero, at least one of $ y_1'$ and $y_2' $ is nonzero, and we assume $y_1'$ is nonzero without loss of generality.
Then, we know 
\eq{
\frac{k_x-k_x'}{N_x} = \frac{y_1'}{\det(\tilde{A})}\ .
}
However, since $\det(\tilde{A}) = \widetilde{n}_{11}\widetilde{n}_{22}  - \widetilde{n}_{12} \widetilde{n}_{21} = \pm 1$, $\frac{k_x-k_x'}{N_x} $ must be a nonzero integer, which is inconsistent with $k_x, k_x'\in [0,N_x-1]$.
Then, we know \eqnref{eq:sufficient_for_allowed_mesh} is a sufficient condition to make sure all points in the momentum mesh (\eqnref{eq:momentum_mesh_ap}) are inequivalent.
Therefore, \eqnref{eq:sufficient_for_allowed_mesh} is a sufficient and necessary condition for \eqnref{eq:two_conditions_momentum_mesh}.
In other words, \eqnref{eq:two_conditions_momentum_mesh} is equivalent to $\tilde A \in GL(2,\dsZ)$.

\begin{figure}
    \centering
    \includegraphics[width=\columnwidth]{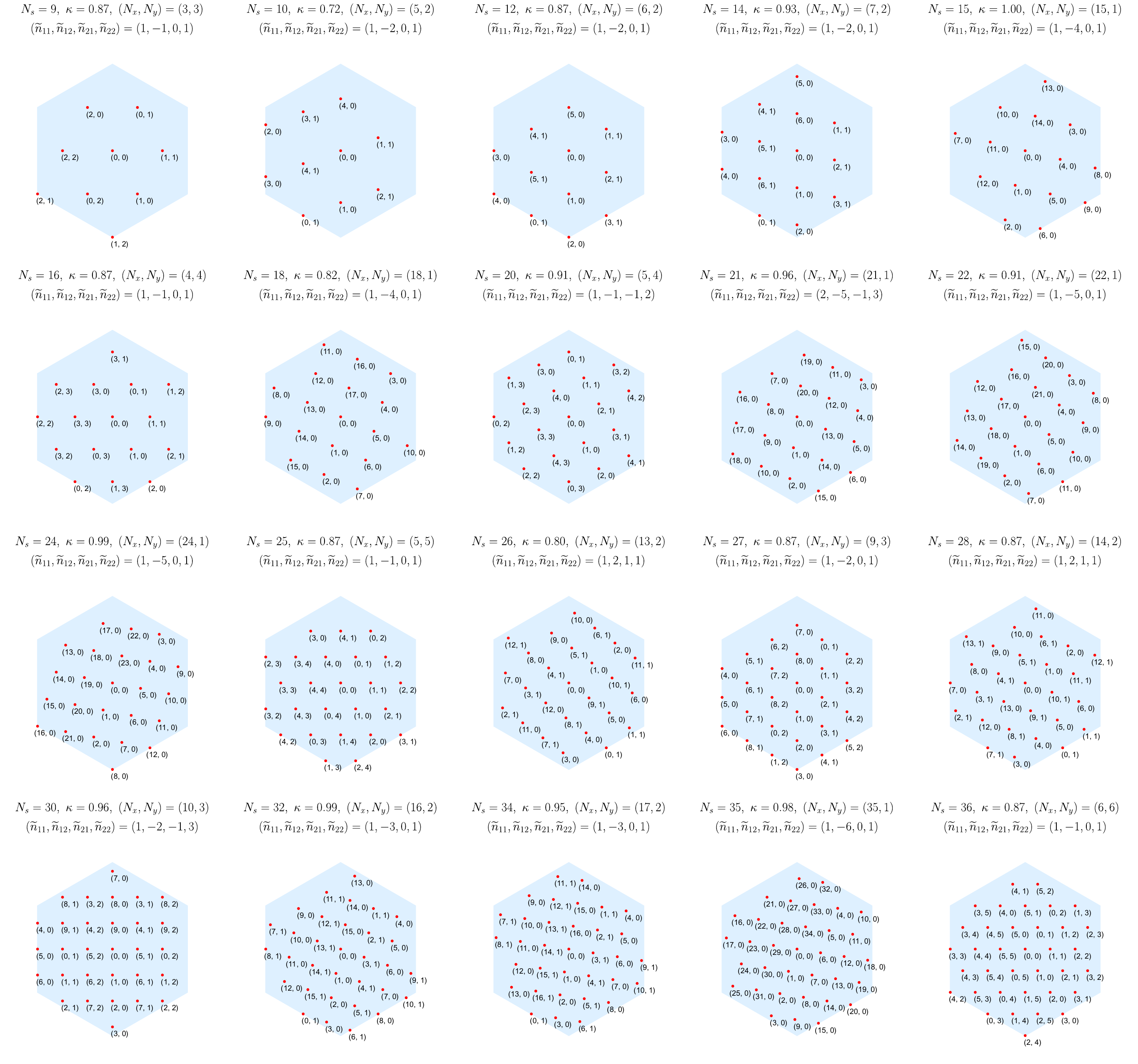}
         \caption{Momentum meshes for the optimal aspect ratio $\kappa$ and for various values of $N_s$ (number of unit cells). 
         $\widetilde{n}_{11}$, $\widetilde{n}_{12}$, $\widetilde{n}_{21}$ and $\widetilde{n}_{22} \bsl{b}_{M,2}$ are defined in \eqnref{eq:f_1_f_2}.
         }
    \label{fig:tilted_lattice}
\end{figure}

For a given $N_s$, we choose the momentum mesh that satisfies the two conditions in \eqref{eq:two_conditions_momentum_mesh} and has the aspect ratio closest to 1.
The aspect ratio $\kappa$ is defined as the following.
We first note that the superlattice generated by $\bsl{T}_1$ and $\bsl{T}_2$, just like any other lattice, has more than one sets of primitive lattice vectors.
We use $\bsl{T}_1'$ and $\bsl{T}_2'$ to label a generic set of  primitive lattice vectors for the superlattice generated by $\bsl{T}_1$ and $\bsl{T}_2$.
For each set $\bsl{T}_1'$ and $\bsl{T}_2'$, we can define an aspect ratio $\kappa'$
\eq{
\kappa' = \frac{\left|\bsl{T}_1'\times \bsl{T}_2' \right|}{ \max(|\bsl{T}_1'|,|\bsl{T}_2'|)^2}\ ,
}
which is essentially the ratio between the short height and the long edge for the parallelogram generated by $\bsl{T}_1'$ and $\bsl{T}_2'$.
Eventually, the aspect ratio $\kappa$ that we choose for the momentum mesh is the largest value of $\kappa'$.
In \figref{fig:tilted_lattice}, we list the momentum meshes with $\kappa$ closest to 1 for various values of $N_s$, \ie,
\eq{
\label{eq:apsect_ratio_kappa}
\kappa = \max_{\bsl{T}_1',\bsl{T}_2'} \kappa'[\bsl{T}_1',\bsl{T}_2']\ .
}

\subsection{Band Maximum}
\label{app:band_max}

The approximation ``band maximum" will only be used for the 2BPV case when the Hilbert space is too large to otherwise perform efficient calculations.
In that case, we will limit the total number of holes in the second-top pair (including both valleys) of valence bands to $N_{\text{band1}}$, where ``1" indicates the second-top band while ``0" labels the top band.
Specifically, with the band maximum limitation, the Slater basis state of the entire many-body Hilbert space in the 2BPV case has the form $\prod_{\bsl{k},n,\eta}(\widetilde{\gamma}_{\bsl{k},n,\eta}^\dagger)^{N_{\bsl{k},n,\eta}}\ket{0}$, where $n=0,1$ is the band index, and $N_{\bsl{k},n,\eta}=0,1$.
The band maximum limitation means that we only consider the subspace generated by $\prod_{\bsl{k},n,\eta}(\widetilde{\gamma}_{\bsl{k},n,\eta}^\dagger)^{N_{\bsl{k},n,\eta}}\ket{0}$ with $\sum_{\bsl{k},\eta}N_{\bsl{k},1,\eta}\leq N_{\text{band1}} $. Reliability of the numerical results with the band maximum truncation is tested by checking the convergence with increasing $N_{\text{band1}}$.

\section{Additional ED Results}
\label{App:EDresults}

\subsection{Energy gap above the FCI ground states}
\label{App_FCIgap}

\begin{figure}
    \centering
    \includegraphics[width=0.9\columnwidth]{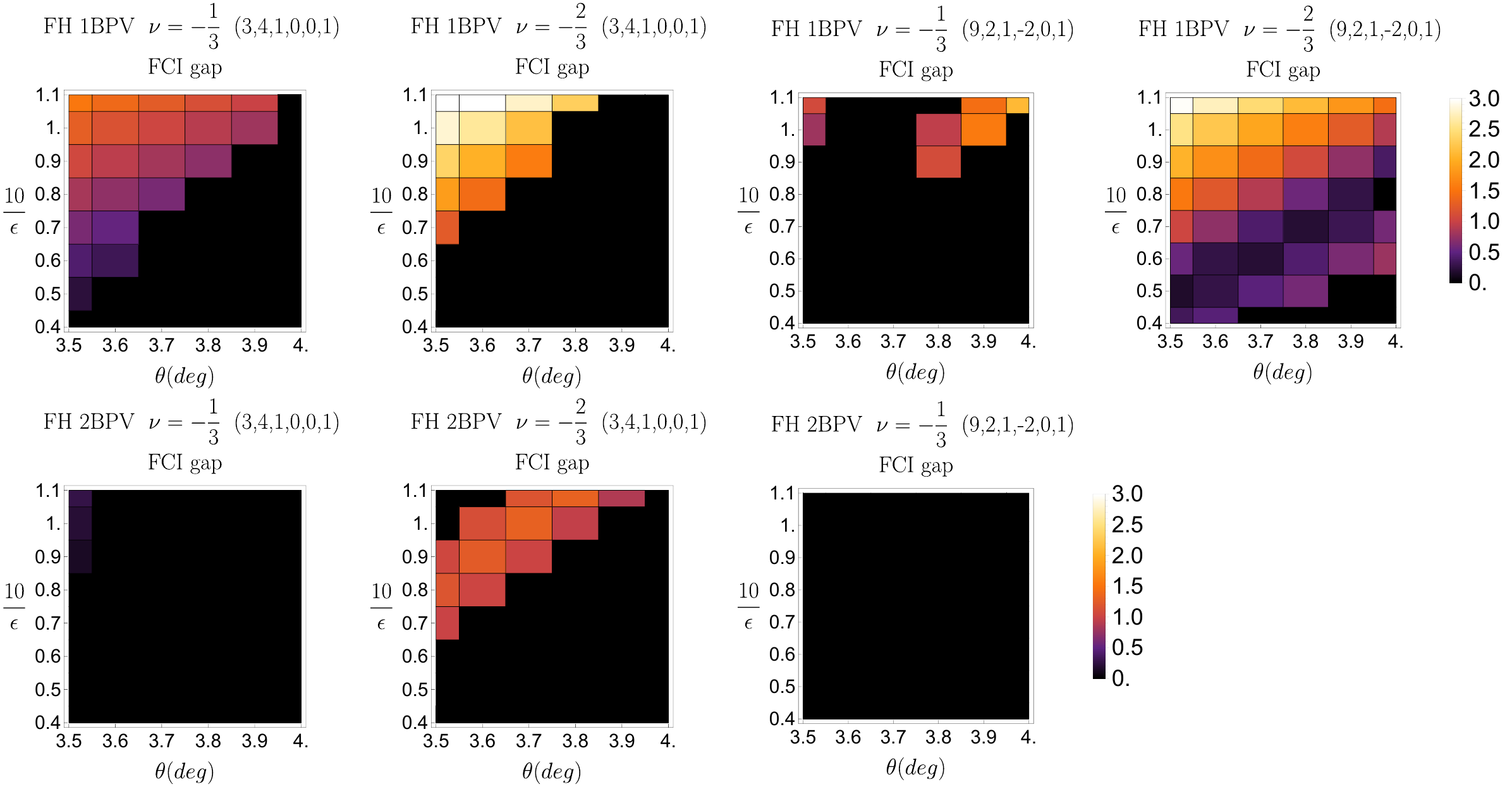}
    \includegraphics[width=0.9\columnwidth]{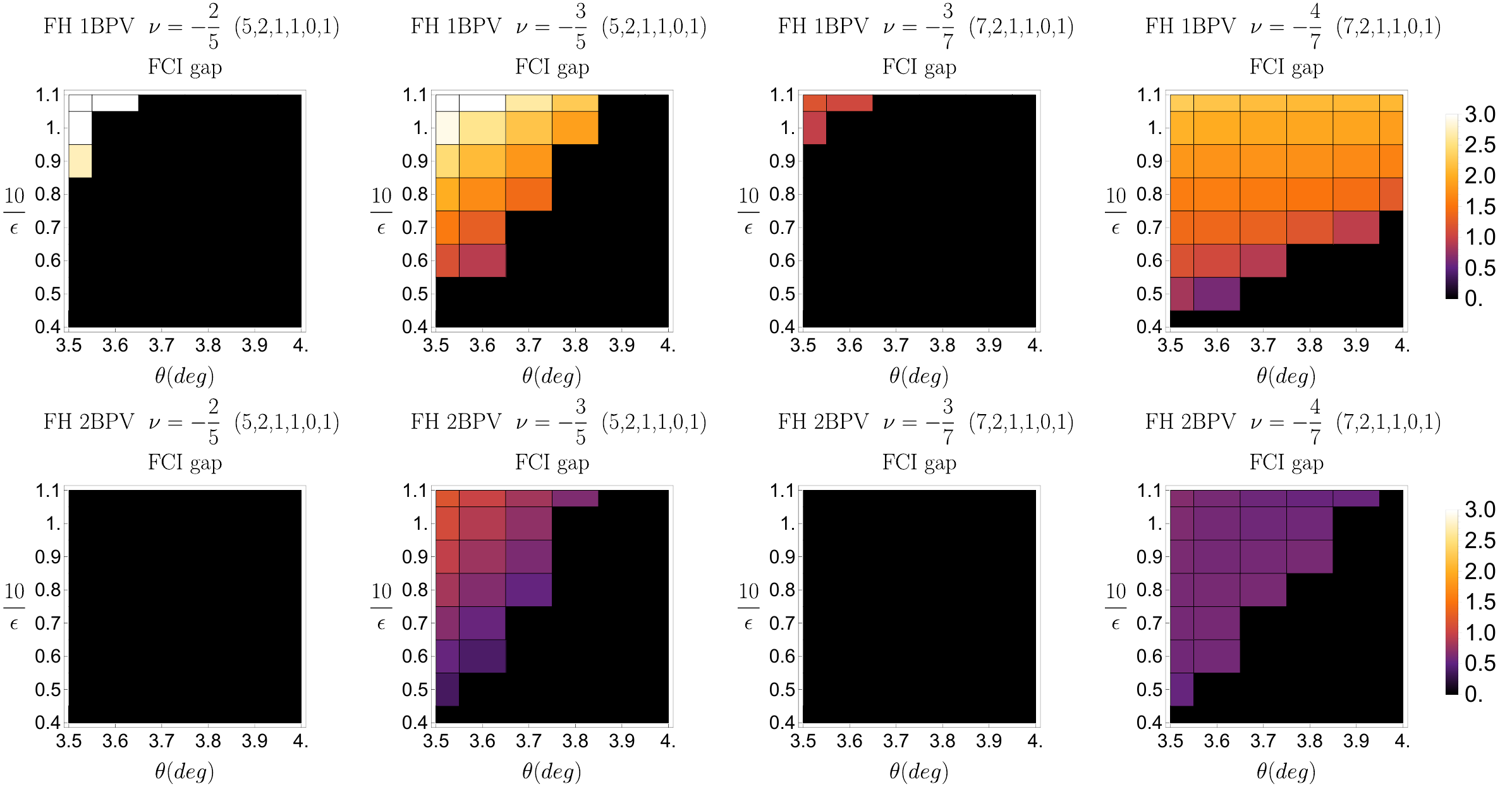}
    \caption{ Energy gap above the FCI ground states identified by \propref{prop:FCI} in the polarized sector for the phase diagrams presented in the main text. The six numbers in parentheses are $(N_x,N_y,\tilde n_{11},\tilde n_{12},\tilde n_{21},\tilde n_{22})$ respectively, and the system size is $N_x\times N_y$. If the ground state is not an FCI, the gap will be set to zero. All the figures share the same color scheme.
    }
    \label{fig:gap_FH}
\end{figure}

\begin{figure}
    \centering
    \includegraphics[width=\textwidth]{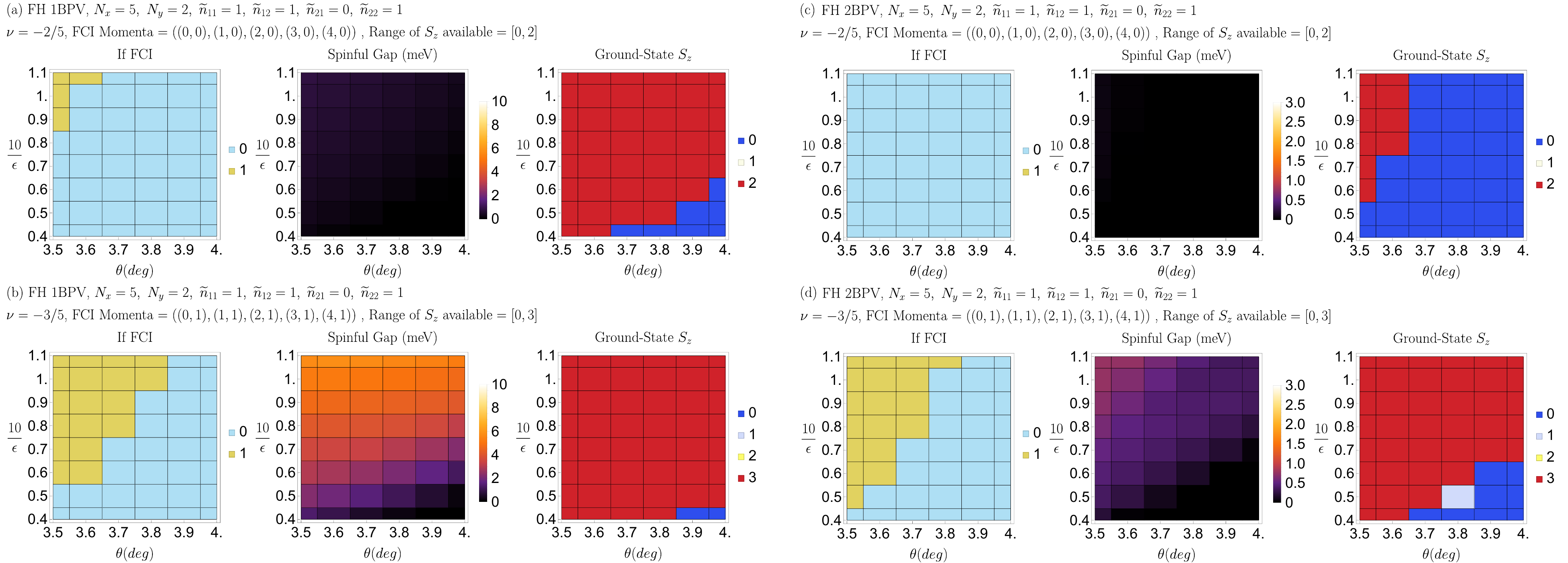}
    \caption{ED results at $\nu=-2/5,-3/5$ for the FH model in a $5\times 2$ system. The 1BPV results are in (a),(b) and the 2BPV results are in (c),(d). 
    The labels have the same meaning as described in the caption of \figref{fig:FH_1BPV_thirds}.
    }
    \label{fig:FH_1B2BPV_fifths}
\end{figure}

\begin{figure}
    \centering
    \includegraphics[width=\textwidth]{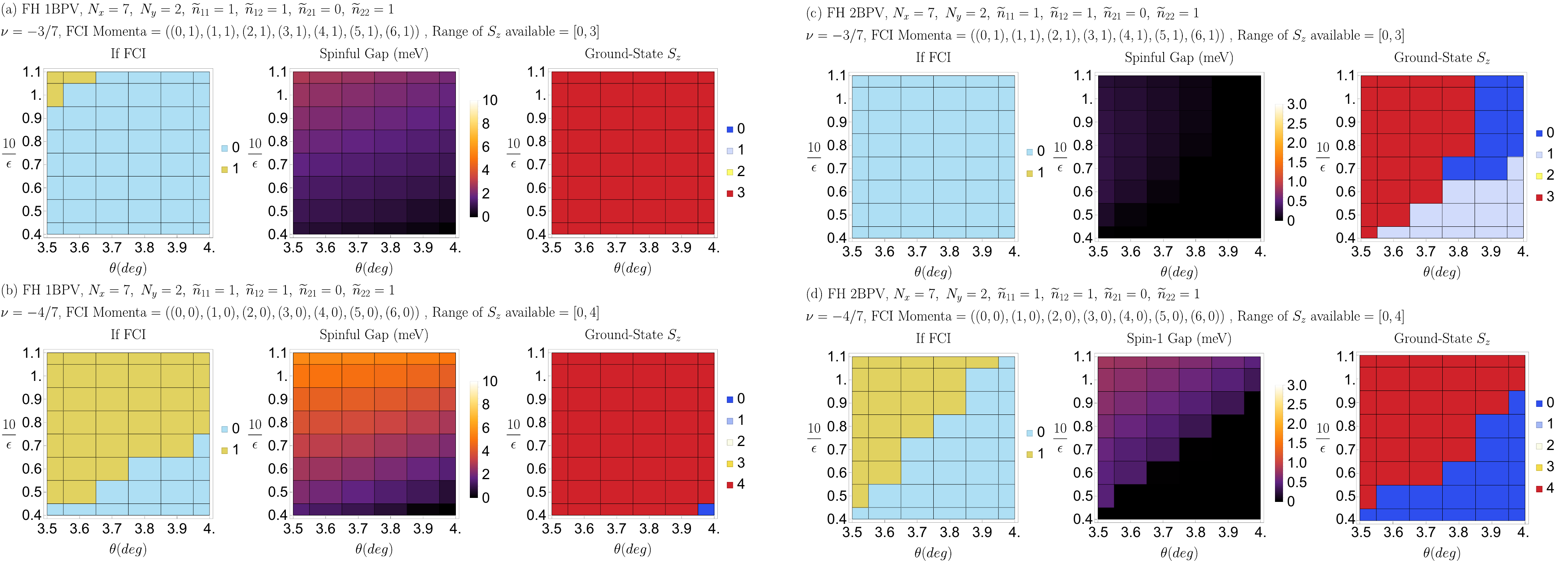}
    \caption{2BPV ED results at $\nu=-3/7,-4/7$ for the FH model in a $7\times 2$ system. The 1BPV results are in (a),(b) and the 2BPV results are in (c),(d). 
    The labels have the same meaning as in \figref{fig:FH_1BPV_thirds}.
    }
    \label{fig:FH_1B2BPV_sevenths}
\end{figure}

In this section we present the dependence of the FCI gap on interaction strength $10/\epsilon=0.4\sim 1.1$ at different fillings, which is not shown in the main text. For each twist angle and interaction strength, if the system is identified as an FCI according to \propref{prop:FCI} with $q$ ground states, the gap is defined by the energy difference between the $(q+1)$th lowest state and the $q$th lowest state $E_{q+1}-E_q$ in the spin sector of FCI; and the gap is set to zero if the system is not an FCI from \propref{prop:FCI}. We show the gap for $\nu=-1/3,-2/5,-3/7,-4/7,-3/5,-2/3$ in both 1BPV and 2BPV in Fig.~\ref{fig:gap_FH}. We also show the dependence of spin gap and polarization of ground state with interaction strength and twist angle for $\nu=-2/5,-3/5,-3/7,-4/7$ in Fig.~\ref{fig:FH_1B2BPV_fifths} and Fig.~\ref{fig:FH_1B2BPV_sevenths} (those for $\nu=-1/3,-2/3$ have been shown in the main text). Comparing the FCI gap in 2BPV and 1BPV calculations at the same setting, it shows in most cases the band mixing reduces the FCI region.

\subsection{Emergence of FCI at $\nu=-1/3$ under large interaction strength}
\label{App_largeu}

At $\nu=-1/3$, the experimental observations are diverse near a twist angle of $3.5^\circ$, with recent reports of FCI~\cite{Pan1v3FCI2026} and CDW~\cite{Xiaodongnew}. Such an FCI had not been observed in previous experiments~\cite{cai2023signatures,zeng2023integer,park2023observation,Xu2023FCItMoTe2}. In this section, we explore the competition between FCI and CDW at $\nu=-1/3$, and we find that a stronger interaction can reinforce ferromagnetism while enabling an FCI ground state.

\begin{figure}
    \centering
    \includegraphics[width=0.8\columnwidth]{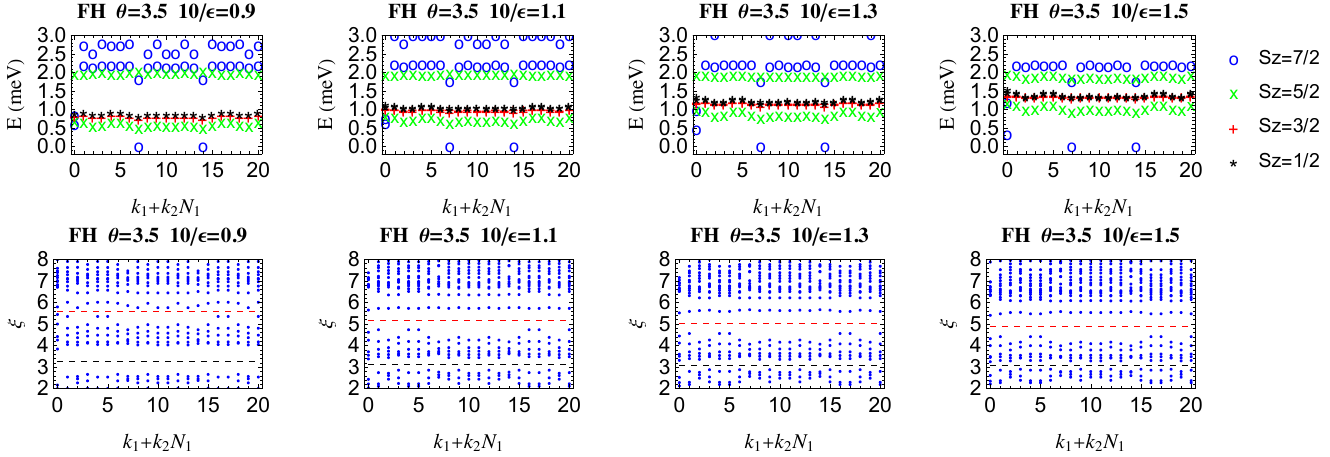}
    \caption{2BPV ED results calculated in the FH model at $3.5^\circ$ with $\nu=-1/3$ in a 21-site system with $(N_x,N_y,\widetilde{n}_{11},\widetilde{n}_{12},\widetilde{n}_{21},\widetilde{n}_{22})=(21,1,1,-5,0,1)$ and $10/\epsilon=0.9-1.5$. The first row is the energy spectrum. The second row is the PES calculated from the lowest three states in the fully polarized sector in the first row. The red and black dashed lines represent the FCI counting and CDW counting respectively.
    }
    \label{FH_21b1_n7_ularge}
\end{figure}

Both the FCI and $K$-CDW phases at $\nu=-1/3$ have three nearly degenerate ground states. In a $K$-CDW phase, the momenta of these three ground states are mutually separated by $\pm K_M$, and in FCI the ground state momenta are determined by the folding rules from FQH to FCI~\cite{RegnaultBernevig2011Chern,BernevigPhysRevB.85.075128}. For 21-site system with $(N_x,N_y,\widetilde{n}_{11},\widetilde{n}_{12},\widetilde{n}_{21},\widetilde{n}_{22})=(21,1,1,-5,0,1)$, both FCI and $K$-CDW phases have three ground states located at $k_x+k_yN_x=0,7,14$ corresponding to $\Gamma_M, K_M, K'_M$ points. We calculate the 2BPV energy spectrum of the FH model at $3.5^\circ$ with $\nu=-1/3$ in this $21$-site system in Fig.~\ref{FH_21b1_n7_ularge} under different interaction strength $10/\epsilon=0.9-1.5$. It shows that as interaction increases, the lowest three states become spin-polarized with momentum $k_x+k_yN_x=0,7,14$ consistent with both FCI and CDW, and they are separated by a large energy gap with the higher states. To determine whether the ground state is FCI or CDW, we further calculate the PES of the lowest three states in the fully polarized sector in Fig.~\ref{FH_21b1_n7_ularge}. The red and black dashed lines represent the FCI counting and CDW counting respectively. It shows that at $10/\epsilon=0.9$ the entanglement gap is dominated by the CDW counting represented by the black line, but as interaction increases, the  dominant entanglement gap becomes the FCI counting at the red line. Therefore, there is a competition between CDW and FCI in this system, and FCI with spin polarization emerges at large interaction. 

\begin{figure}
    \centering
    \includegraphics[width=0.8\columnwidth]{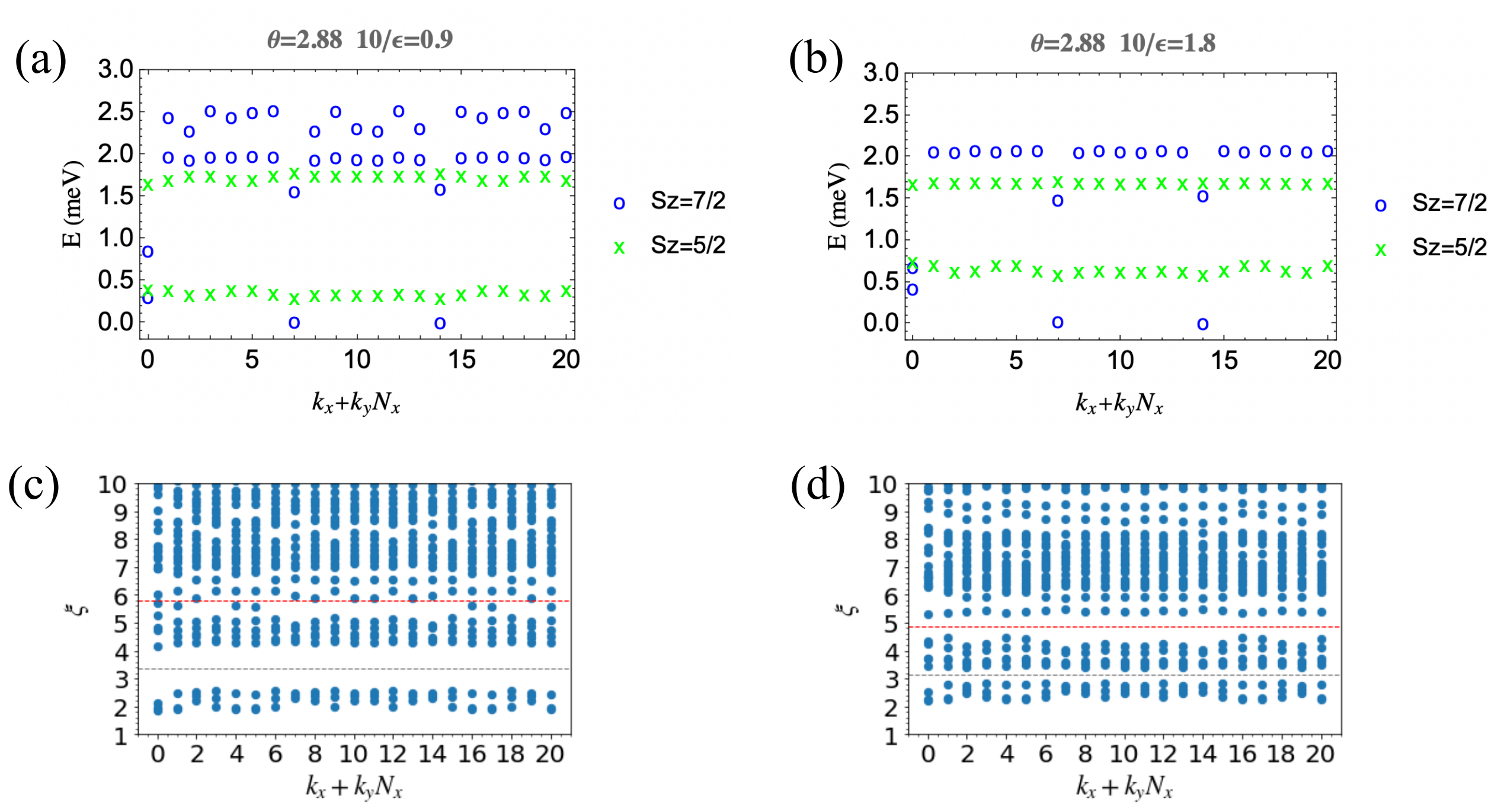}
    \caption{2BPV ED results calculated in the no-fitting model in the standard basis at $2.88^\circ$ with $\nu=-1/3$ in a 21-site system with $(N_x,N_y,\widetilde{n}_{11},\widetilde{n}_{12},\widetilde{n}_{21},\widetilde{n}_{22})=(21,1,1,-5,0,1)$ and $10/\epsilon=0.9,\ 1.8$. The first row is the energy spectrum. The second row is the PES calculated from the lowest three states in the fully polarized sector in the first row. The red and black dashed lines represent the FCI counting and CDW counting respectively.
    }
    \label{Nofit_1v3_2.88_largeu}
\end{figure}

Similar behavior is also observed near twist angle $3^\circ$ where FCI was reported experimentally at $\nu=-1/3$ from $2.6^\circ$ to $3^\circ$~\cite{Xiaodongnew}. As shown in Fig.~\ref{Nofit_1v3_2.88_largeu}, the 2BPV calculations at $2.88^\circ$ in the no-fitting model standard basis show that when interaction increases by a factor of 2 (from $10/\epsilon=0.9$ to 1.8), the energy of the spin excitation states becomes higher than the three ground states in the fully-polarized sector, and the PES of the three ground states opens a gap at the FCI counting.

\subsection{Dependence of FCI gap on system size}
\label{App_scaling}

\begin{figure}
    \centering
    \includegraphics[width=0.24\columnwidth]{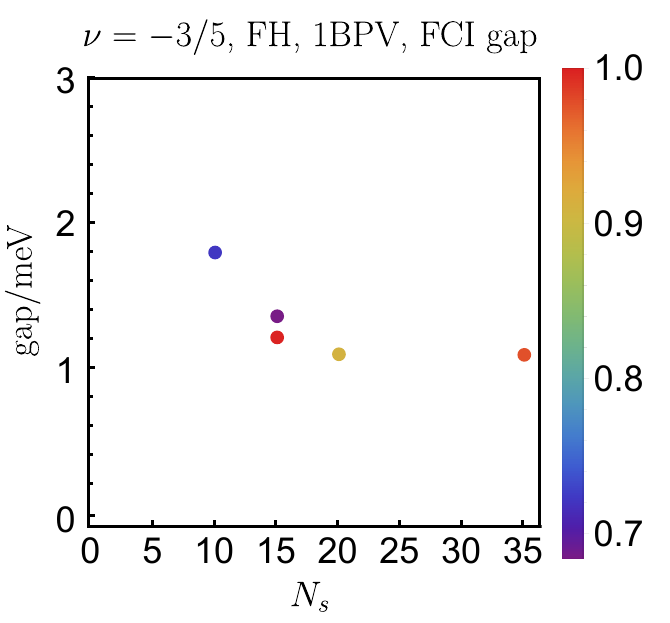}
    \includegraphics[width=0.24\columnwidth]{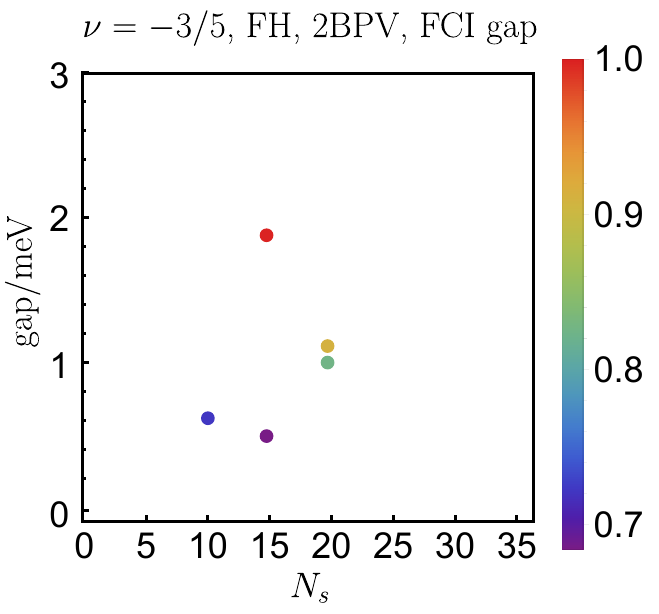}
    \includegraphics[width=0.24\columnwidth]{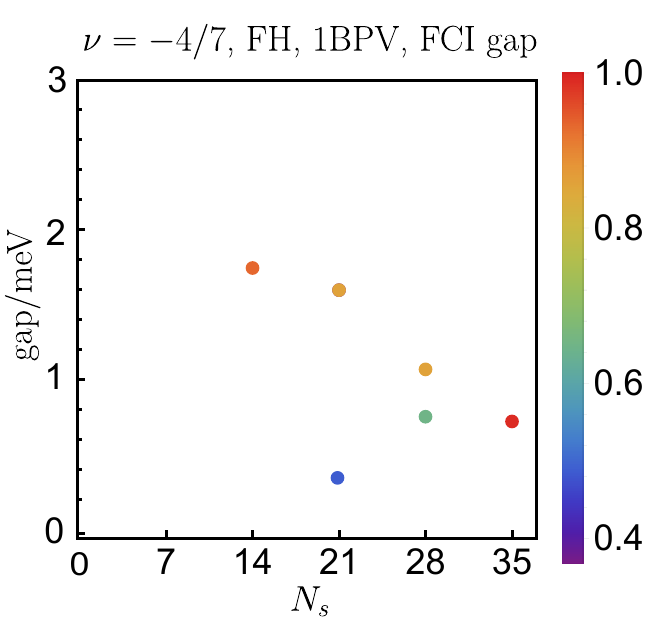}
    \includegraphics[width=0.24\columnwidth]{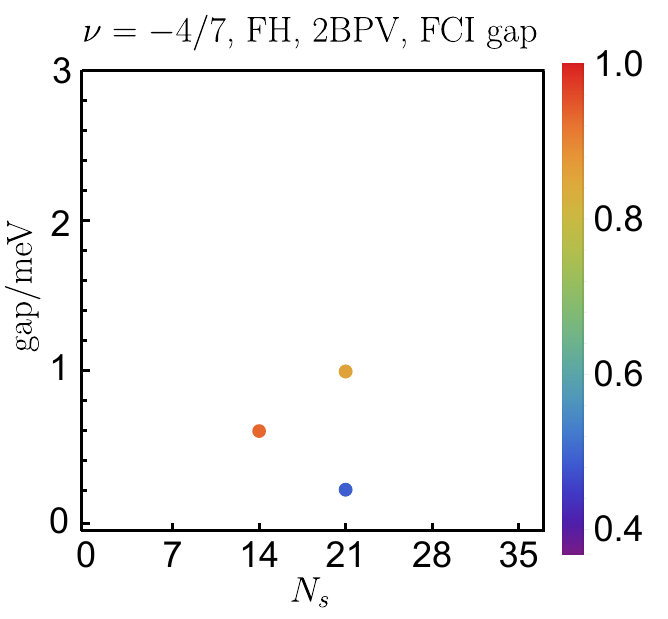}
    \caption{These plots summarize the 1BPV and 2BPV FCI gaps of the spectra in different system sizes at $\nu=-3/5,\ -4/7$ for the FH model with $10/\epsilon=0.9$ and $\theta=3.7^\circ$. $N_s$ is the number of sites.
    The color of the points indicates the aspect ratio of the corresponding lattice. We only include the data where the FCI gap is larger than the spread of ground states. There are no data at $N_s=25,30$ in 1BPV at $\nu=-3/5$ because the spread of the ground states is larger than the gap.
    }
    \label{fig:finite_size_gap_57}
\end{figure}

In addition to the gap scaling of $\nu=-2/3$ in Sec.~\ref{Sec_scaling} of the main text, in this section we present the detailed energy spectrum in different system sizes for ED results on $\nu=-2/3,-3/5,-4/7$. We focus on the parameter at twist angle $3.7^\circ$ and $10/\epsilon=0.9$ in the FH model, where robust FCI and magnetism was found in both 1BPV and 2BPV calculations at fillings $\nu=-2/3,-3/5,-4/7$. The FCI gaps in different momentum meshes at $\nu=-3/5,-4/7$ are plotted in Fig.~\ref{fig:finite_size_gap_57} (those for $\nu=-2/3$ can be found in Sec.~\ref{Sec_scaling} of the main text). We only show the FCI gaps in the cases where the ground state is identified as an FCI by \propref{prop:FCI}. There are no data at $N_s=25,30$ in 1BPV at $\nu=-3/5$ because the spread of the ground states is larger than the gap. The value of FCI gap in the largest system size at $\nu=-3/5,-4/7$ in Fig.~\ref{fig:finite_size_gap_57} is used for the scaling of FCI gap in Fig.~\ref{fig:FH_gapscaling_1BPV_linear} in the main text.

\subsubsection{1BPV energy spectra at $\nu=-2/3,-3/5,-4/7$ in the FH model}

\begin{figure}
    \centering
    \includegraphics[width=\columnwidth]{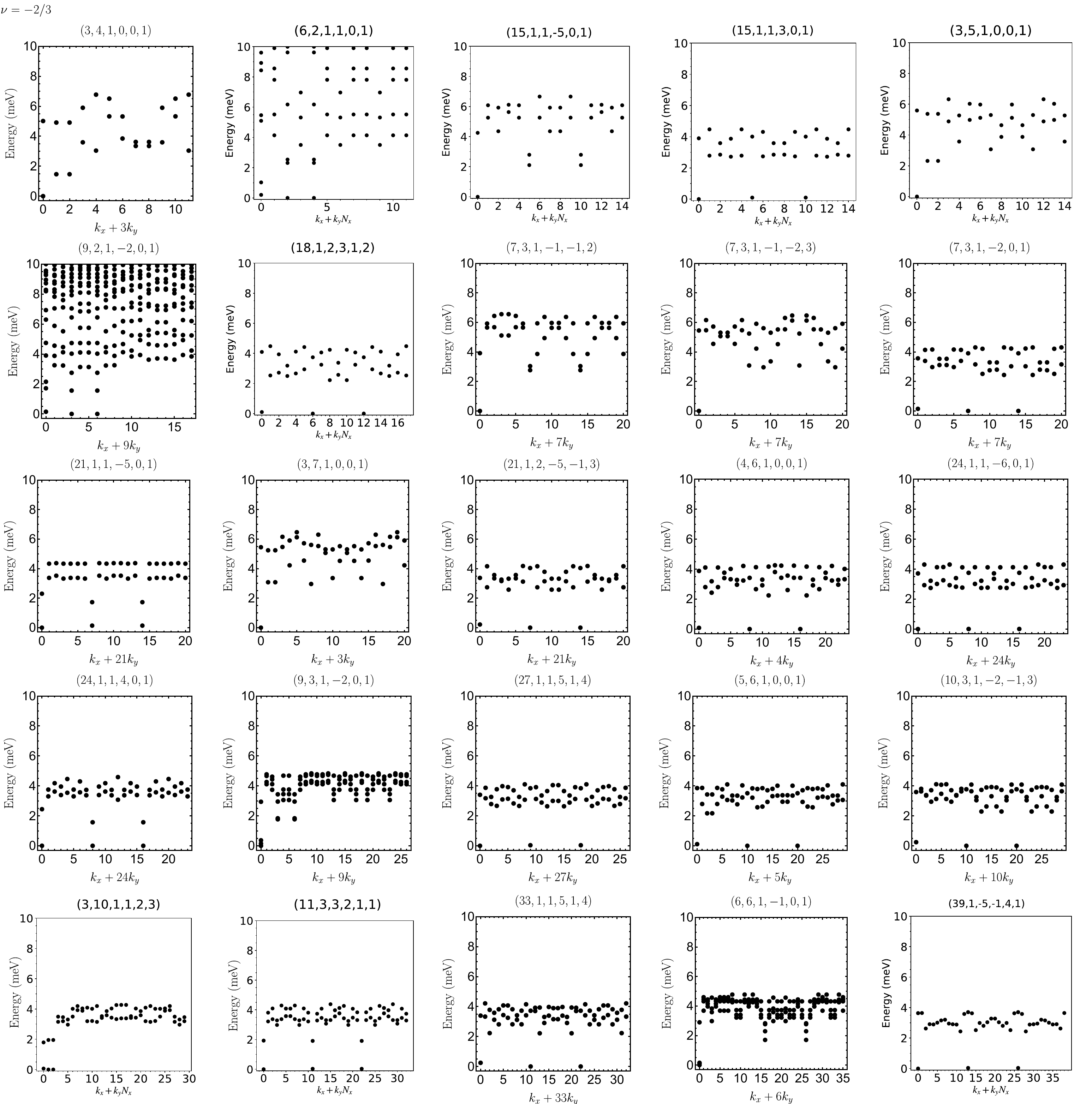}
    \caption{1BPV ED spectra at $\nu=-2/3$ for the FH model with $10/\epsilon=0.9$ and $\theta=3.7^\circ$.
    Since we are interested in the FCI gaps, the plots only contain data in the fully spin-polarized case.
    The six integers in the bracket for each plot represent    $(N_x,N_y,\widetilde{n}_{11},\widetilde{n}_{12},\widetilde{n}_{21},\widetilde{n}_{22})$, which specifies the momentum mesh as defined in \eqnref{eq:momentum_mesh}. In panels where the Hilbert space is large, we show only the lowest few eigenvalues in each momentum sector.
    }
    \label{fig:finite_size_m2over3}
\end{figure}

The 1BPV FCI spectra at $\nu=-2/3$ are shown in \cref{fig:finite_size_m2over3} for various momentum meshes.
As discussed in Sec.~\ref{Sec_scaling} in the main text, the FCI gap has sizable dependence on the system size, but for momentum meshes that contain the $K_M$ points, the FCI gap in 1BPV calculation remains relatively stable from $N_s=21$ to $36$ with deviation within 30\%. Therefore, we will take the gap of 1.5 meV in the 36-site system as the FCI gap at $\nu=-2/3$ for the FCI gap scaling in Fig.~\ref{fig:FH_gapscaling_1BPV_linear}.

\begin{figure}[H]
    \centering
    \includegraphics[width=\columnwidth]{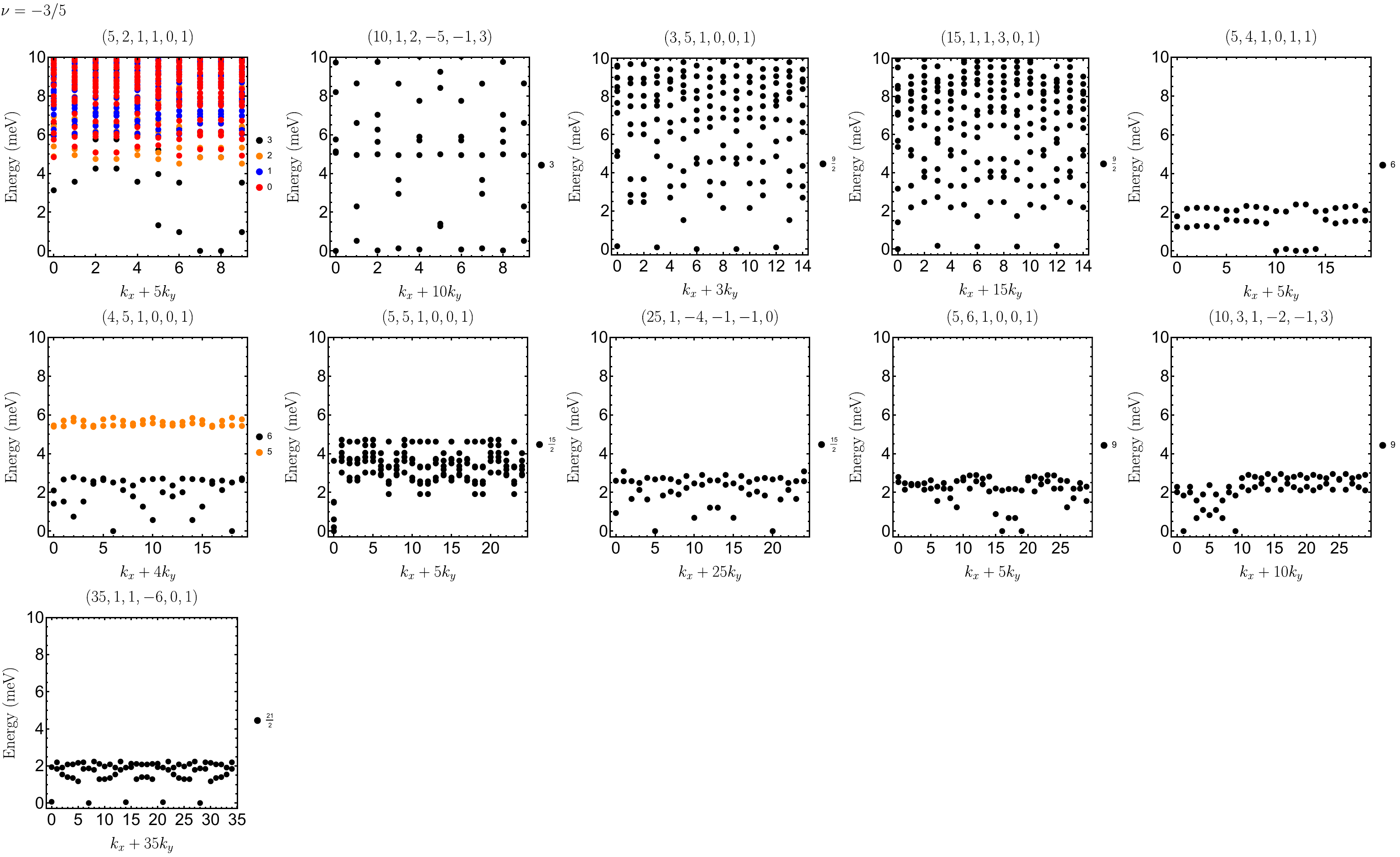}
    \caption{1BPV ED spectra at $\nu=-3/5$ for the FH model with $10/\epsilon=0.9$ and $\theta=3.7^\circ$.
    The six integers in the bracket for each plot represent    $(N_x,N_y,\widetilde{n}_{11},\widetilde{n}_{12},\widetilde{n}_{21},\widetilde{n}_{22})$, which specifies the momentum mesh as defined in \eqnref{eq:momentum_mesh}.
    The different colors represent different spin polarizations. Since we are interested in the FCI gaps, the larger-size plot only contains data in the fully spin-polarized case. In panels where the Hilbert space is large, we show only the lowest few eigenvalues in each momentum sector.
    }
    \label{fig:finite_size_m3over5}
\end{figure}

The 1BPV FCI spectra at $\nu=-3/5$ are shown in \cref{fig:finite_size_m3over5} for various momentum meshes.
Similar to the $-2/3$ filling, we use the FCI gap in the largest system size around 1.1 meV for the 35-site system for the FCI gap scaling in Fig.~\ref{fig:FH_gapscaling_1BPV_linear}.

\begin{figure}[H]
    \centering
    \includegraphics[width=\columnwidth]{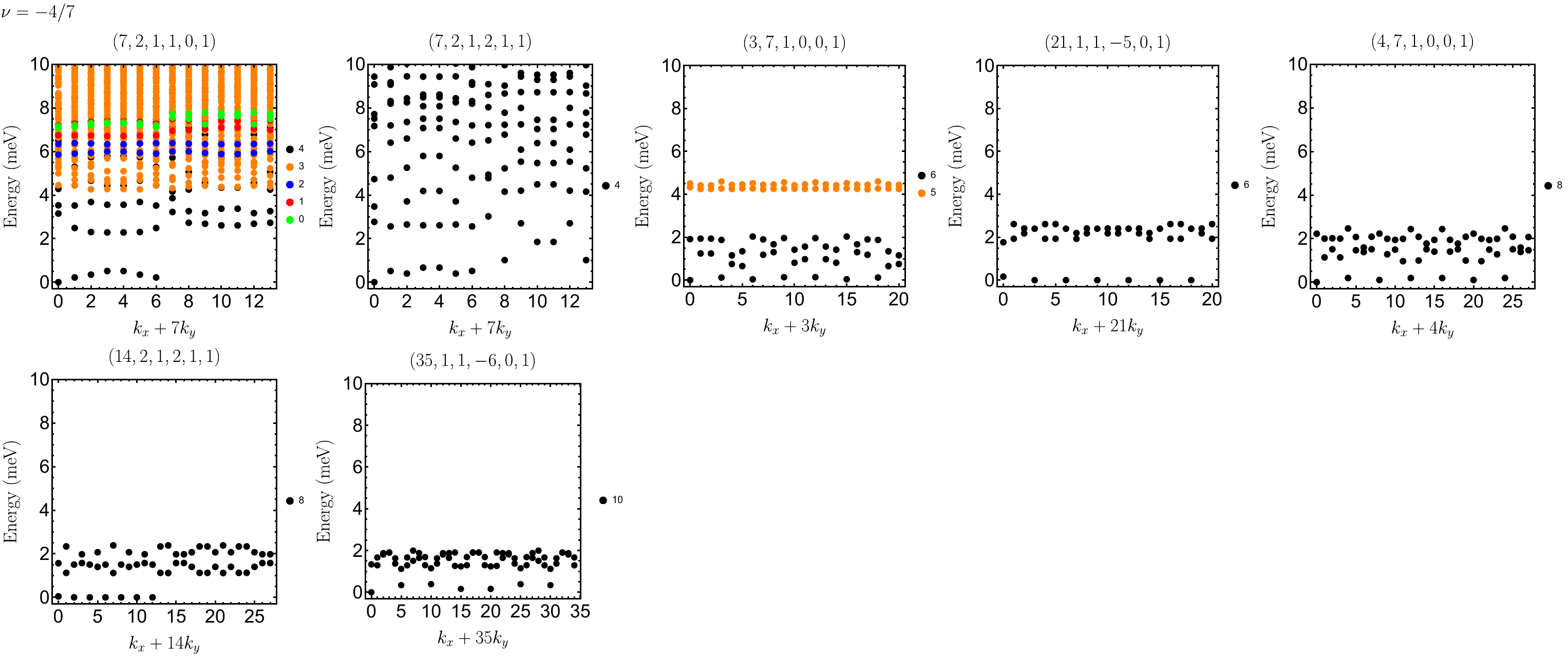}
    \caption{1BPV ED spectra at $\nu=-4/7$ for the FH model with $10/\epsilon=0.9$ and $\theta=3.7^\circ$.
    The six integers in the bracket for each plot represent    $(N_x,N_y,\widetilde{n}_{11},\widetilde{n}_{12},\widetilde{n}_{21},\widetilde{n}_{22})$, which specifies the momentum mesh as defined in \eqnref{eq:momentum_mesh}.
    The different colors represent different spin polarizations. Since we are interested in the FCI gaps, the larger-size plot only contains data in the fully spin-polarized case. In panels where the Hilbert space is large, we show only the lowest few eigenvalues in each momentum sector.
    }
    \label{fig:finite_size_m4over7}
\end{figure}

The 1BPV FCI spectra at $\nu=-4/7$ for various momentum meshes are shown in \cref{fig:finite_size_m4over7}. The gap converges to around 0.7 meV for the 35-site system, which will be used for the FCI gap scaling in Fig.~\ref{fig:FH_gapscaling_1BPV_linear}.

\subsubsection{2BPV energy spectra at $\nu=-2/3,-3/5,-4/7$ in the FH model}
\label{App:2bpvscal}

\begin{figure}[H]
    \centering
    \includegraphics[width=\columnwidth]{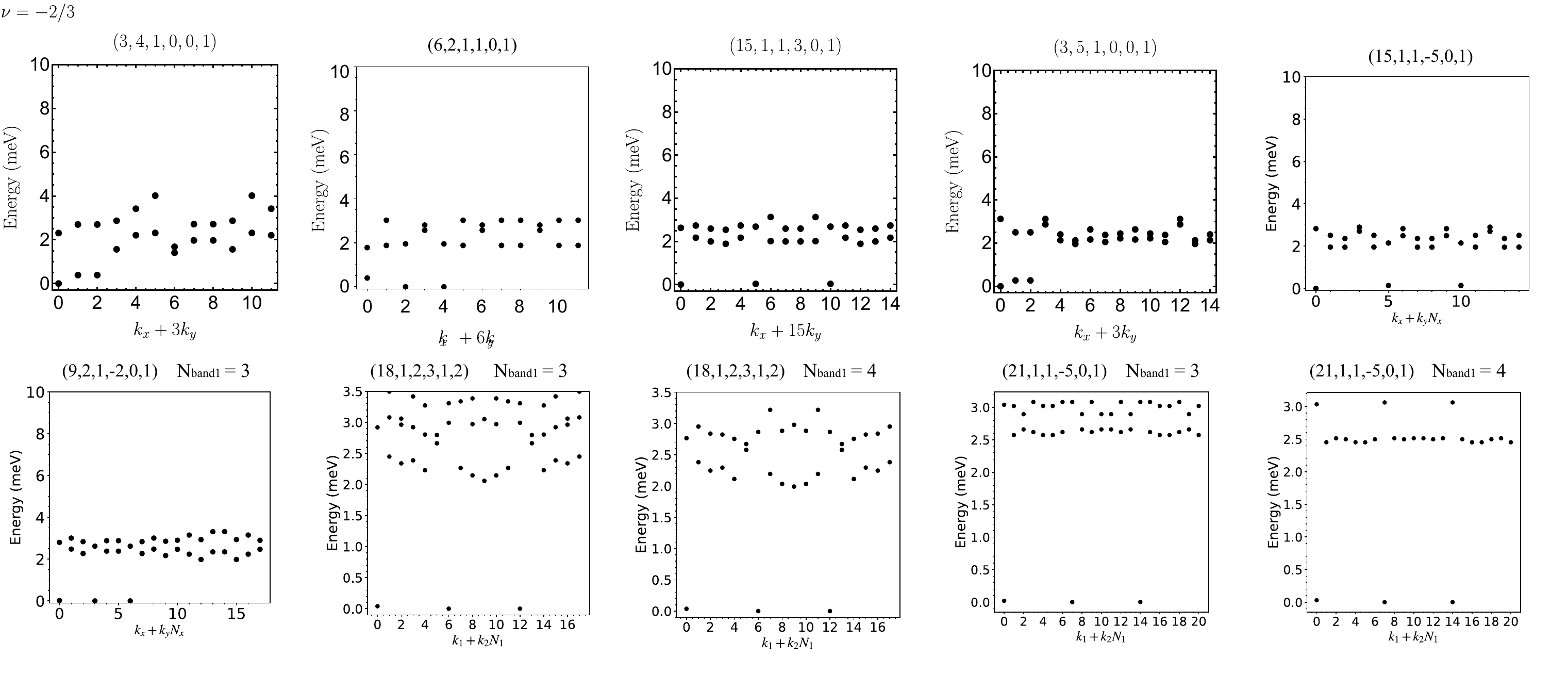}
    \caption{2BPV ED spectra at $\nu=-2/3$ for the FH model with $10/\epsilon=0.9$ and $\theta=3.7^\circ$.
    Since we are interested in the FCI gaps, the plots only contain data in the fully spin-polarized case.
    The six integers in the bracket for each plot represent    $(N_x,N_y,\widetilde{n}_{11},\widetilde{n}_{12},\widetilde{n}_{21},\widetilde{n}_{22})$, which specifies the momentum mesh as defined in \eqnref{eq:momentum_mesh}.
    }
    \label{fig:2BPV_finite_size_m2over3}
\end{figure}

The 2BPV energy spectra at $\nu=-2/3$ for various momentum meshes are shown in \cref{fig:2BPV_finite_size_m2over3}. For large systems with 18 or 21 sites, we restricted the particle number allowed in the high band with $N_{\text{band1}}=3$ or $4$. As shown in the energy spectra, the difference between $N_{\text{band1}}=3$ and $4$ is not significant, hence the spectrum has converged with the truncation of particle number. The gap of the largest system size is 2.4 meV for 21-site system, which will be used for the FCI gap scaling in Fig.~\ref{fig:FH_gapscaling_1BPV_linear} for filling $\nu=-2/3$.

\begin{figure}[H]
    \centering
    \includegraphics[width=\columnwidth]{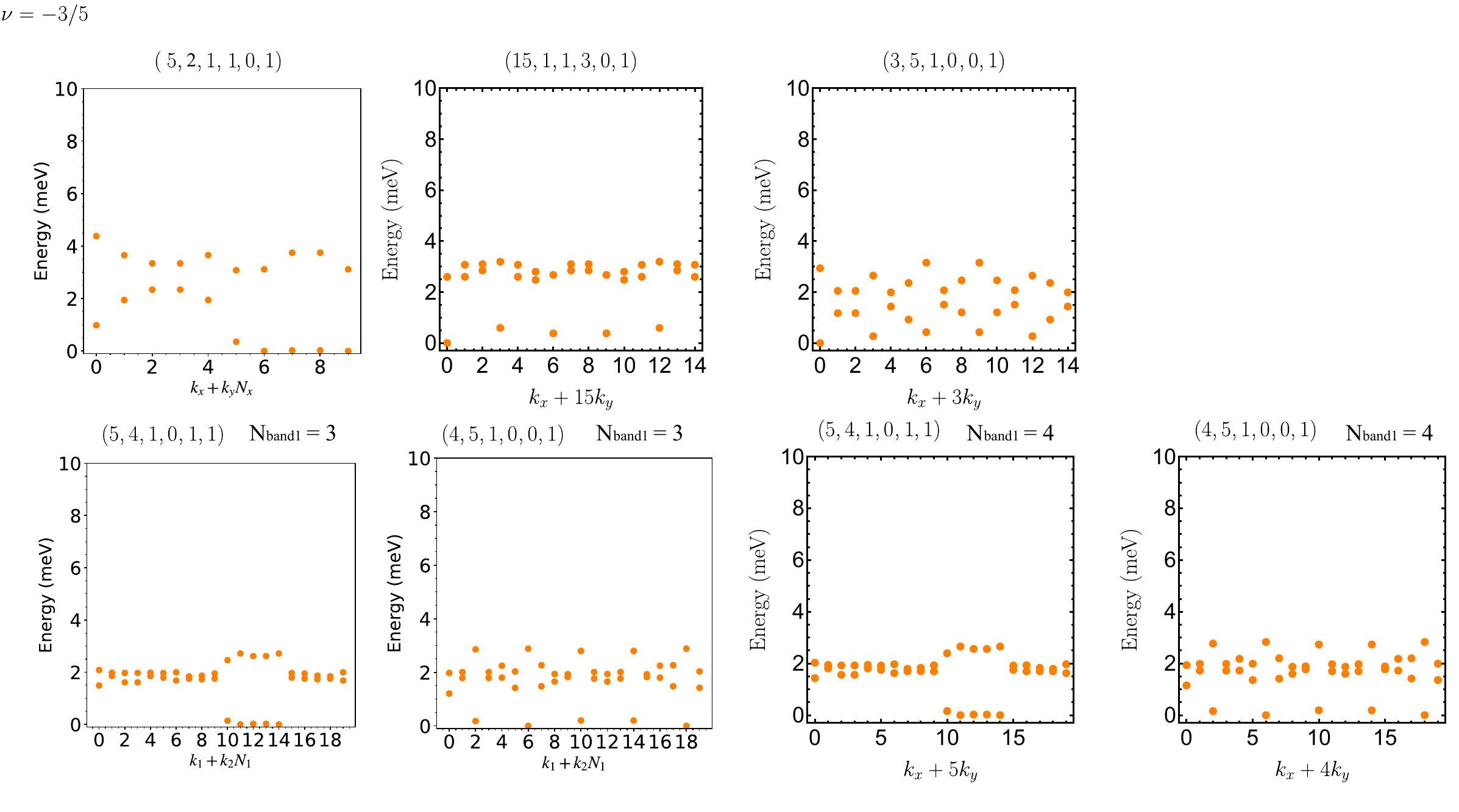}
    \caption{2BPV ED spectra at $\nu=-3/5$ for the FH model with $10/\epsilon=0.9$ and $\theta=3.7^\circ$. The similarity between the spectra in the 20-site system calculated with $N_{\text{band1}}=3$ and 4 indicates convergence with $N_{\text{band1}}$.
    Since we are interested in the FCI gaps, the plots only contain data in the fully spin-polarized case.
    The six integers in the bracket for each plot represent    $(N_x,N_y,\widetilde{n}_{11},\widetilde{n}_{12},\widetilde{n}_{21},\widetilde{n}_{22})$, which specifies the momentum mesh as defined in \eqnref{eq:momentum_mesh}.
    }
    \label{fig:2BPV_finite_size_m3over5}
\end{figure}

For $\nu=-3/5$ in the spin-polarized 2BPV case, we performed spin-polarized ED calculations in the 2BPV case for both $N_x N_y = 15$ sites and $N_x N_y = 20$ sites at $\nu=-3/5$.
We note that the spin-polarized 2BPV calculations can be easily done on $N_x N_y = 15$ sites at $\nu=-3/5$, while the spin-polarized 2BPV calculation on $N_x N_y = 20$ sites at $\nu=-3/5$ has Hilbert-space dimension of $2.8\times 10^8$. Therefore, we restrict the particle number for the 20-site systems with no more than $N_{band1}=3$ or 4 particles, as shown in Fig.~\ref{fig:2BPV_finite_size_m3over5}. The similarity between the spectrum with $N_{\text{band1}}=3$ and 4 indicates convergence with respect to particle number truncation. By extracting the FCI gaps from \cref{fig:2BPV_finite_size_m3over5}, we get the corresponding plot for 2BPV at $\nu=-3/5$ in Fig.~\ref{fig:finite_size_gap_multi}, which again shows that it is hard to get as good convergence for the 2BPV FCI gaps as for the 1BPV case.
We use the FCI gap of 1.3 meV in the largest size available (20 sites) for the filling scaling in Fig.~\ref{fig:FH_gapscaling_1BPV_linear}.

\begin{figure}[H]
    \centering
    \includegraphics[width=0.32\columnwidth]{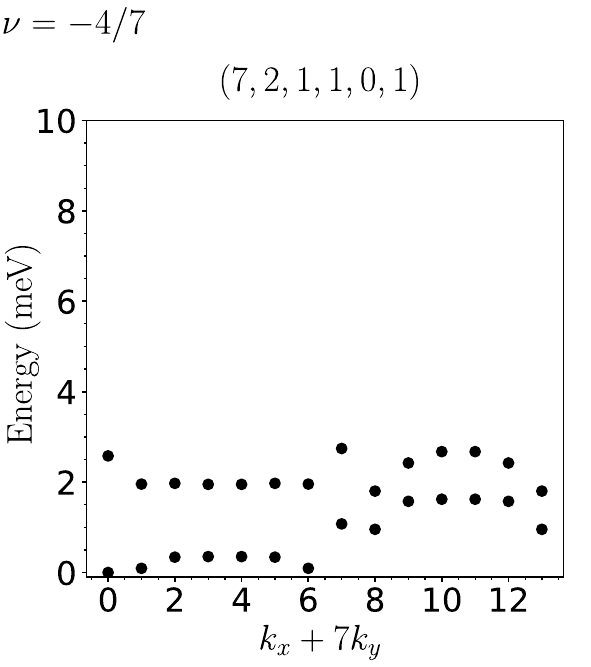}\includegraphics[width=0.293\columnwidth]{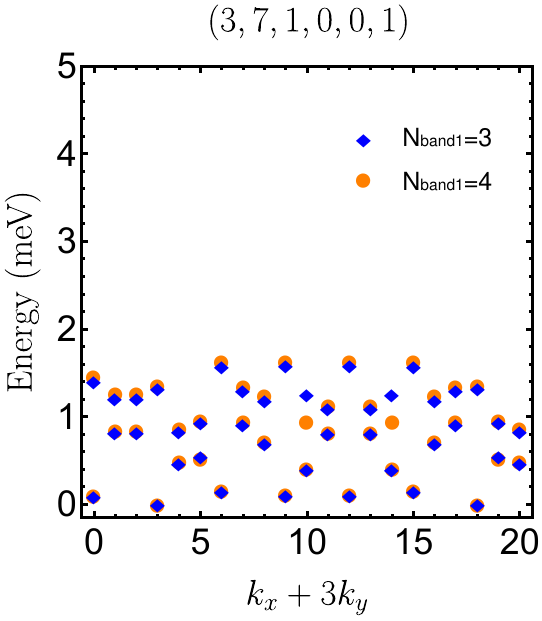}
    \includegraphics[width=0.3\columnwidth]{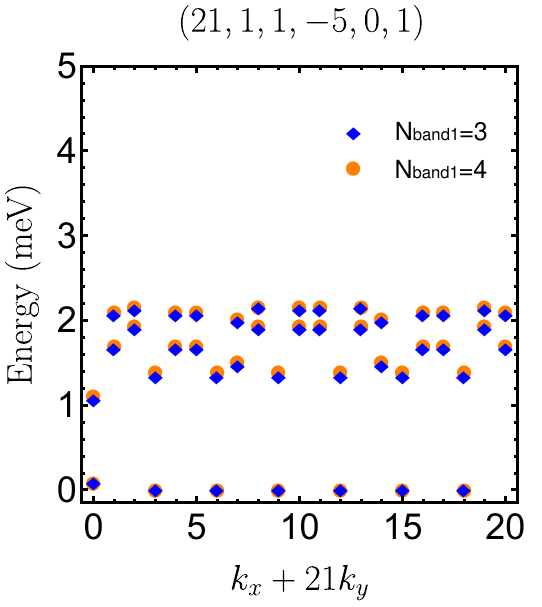}
    \caption{$\theta=3.7^\circ$ 2BPV ED spectrum at $\nu=-4/7$ for the FH model with $10/\epsilon=0.9$.
    The parameters of momentum mesh $(N_x,N_y,\widetilde{n}_{11},\widetilde{n}_{12},\widetilde{n}_{21},\widetilde{n}_{22})$ are shown in the figures.
    Since we are interested in the FCI gaps, we only consider the results in the fully spin-polarized case.
    }
    \label{fig:2BPV_21sites_band1max_m4over7}
\end{figure}

Next we calculate the FCI gap at $\nu=-4/7$ in the spin-polarized 2BPV case.
We performed the spin-polarized ED calculations in the 2BPV case for both $N_x N_y = 14$ sites and $N_x N_y = 21$ sites at $\nu=-4/7$.
We note that the spin-polarized 2BPV calculations can be easily done on $N_x N_y = 14$ sites at $\nu=-4/7$, while the spin-polarized 2BPV calculation on $N_x N_y = 21$ sites at $\nu=-4/7$ has Hilbert-space dimension of $5.3\times 10^8$.
Therefore, to improve the efficiency on $N_x N_y = 21$ sites, we use the band maximum truncation in the remote bands as discussed in \cref{app:band_max}.
In \cref{fig:2BPV_21sites_band1max_m4over7}, we can clearly see that the low-energy spectrum is quite similar between $N_{\text{band1}}=3,4$ on 21 sites, indicating the convergence and the validity of $N_{\text{band1}}=4$ on 21 sites.
Therefore, we use the FCI gap of 1.1 meV in the 21-site system for the filling scaling in Fig.~\ref{fig:FH_gapscaling_1BPV_linear}.

\subsection{Energy spectra in FH and no-fitting models at $10/\epsilon=0.9$}
\label{App_EDcombined}

\begin{figure}
    \centering
    \includegraphics[width=1.0\columnwidth]{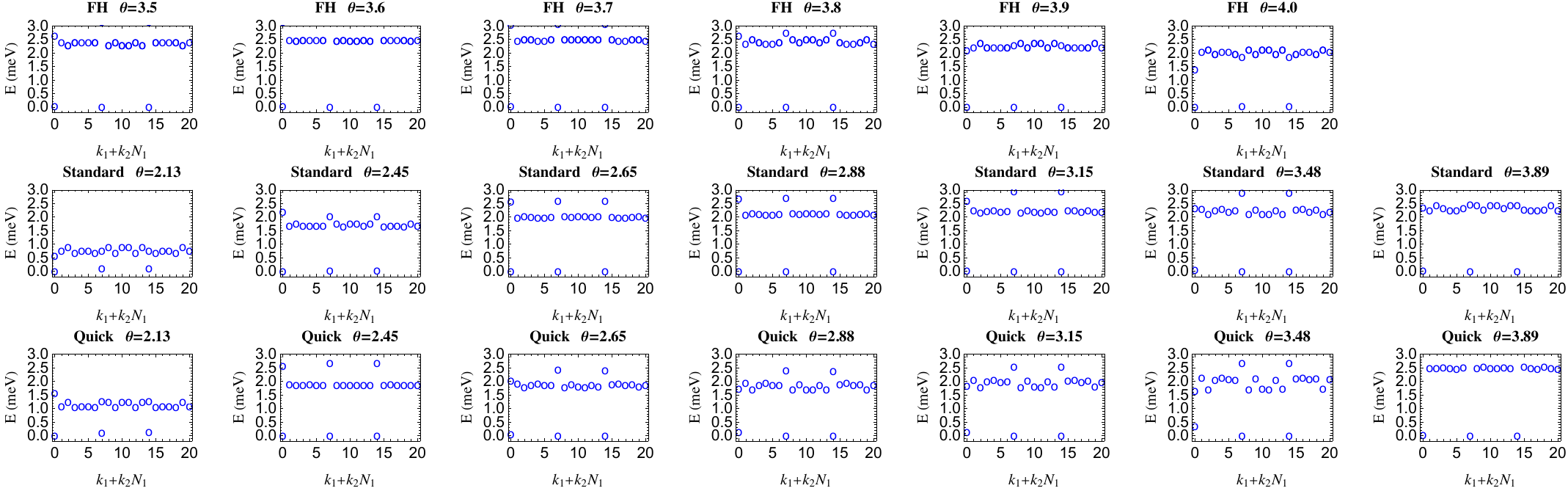}
    \caption{Energy spectrum of FH and no-fitting models in the standard and quick basis for system with $(N_x,N_y,\widetilde{n}_{11},\widetilde{n}_{12},\widetilde{n}_{21},\widetilde{n}_{22})=(21,1,1,-5,0,1)$ at $\nu=-2/3$ and $10/\epsilon=0.9$. The 2BPV calculation is performed in the spin-polarized sector with truncation of bandmax 4.
    }
    \label{fig:ED21_2v3}
\end{figure}

\begin{figure}
    \centering
    \includegraphics[width=1.0\columnwidth]{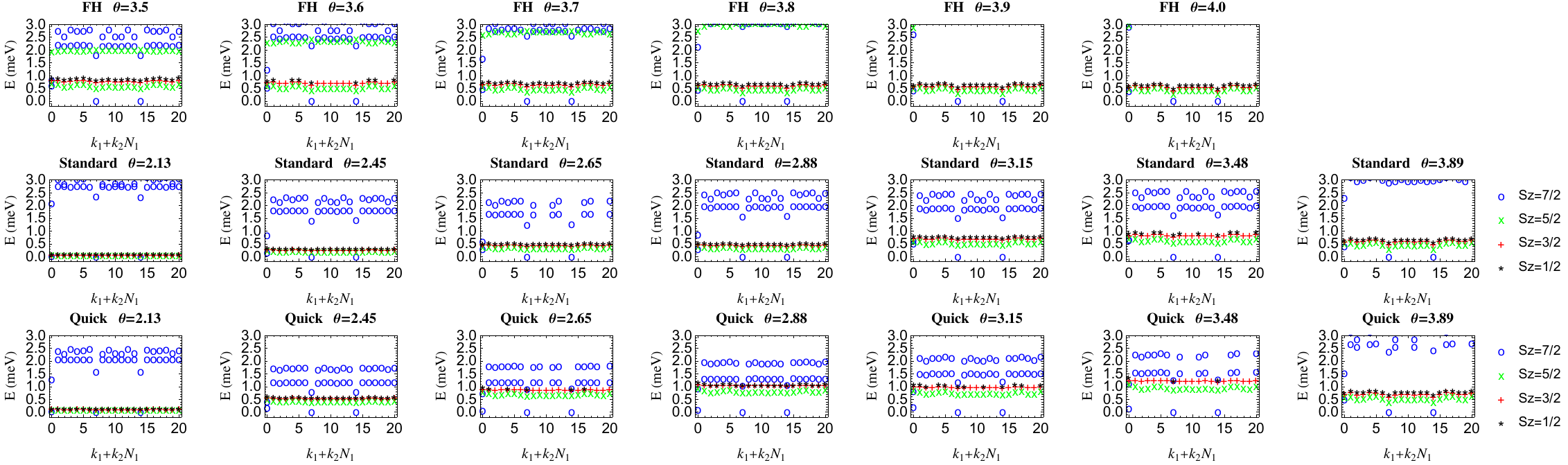}
    \caption{Energy spectrum of FH and no-fitting models in the standard and quick basis for system with $(N_x,N_y,\widetilde{n}_{11},\widetilde{n}_{12},\widetilde{n}_{21},\widetilde{n}_{22})=(21,1,1,-5,0,1)$ at $\nu=-1/3$ and $10/\epsilon=0.9$. The 2BPV calculation is performed without truncation by including all spin sectors.
    }
    \label{fig:ED21_1v3}
\end{figure}

\begin{figure}
    \centering
    \includegraphics[width=1.0\columnwidth]{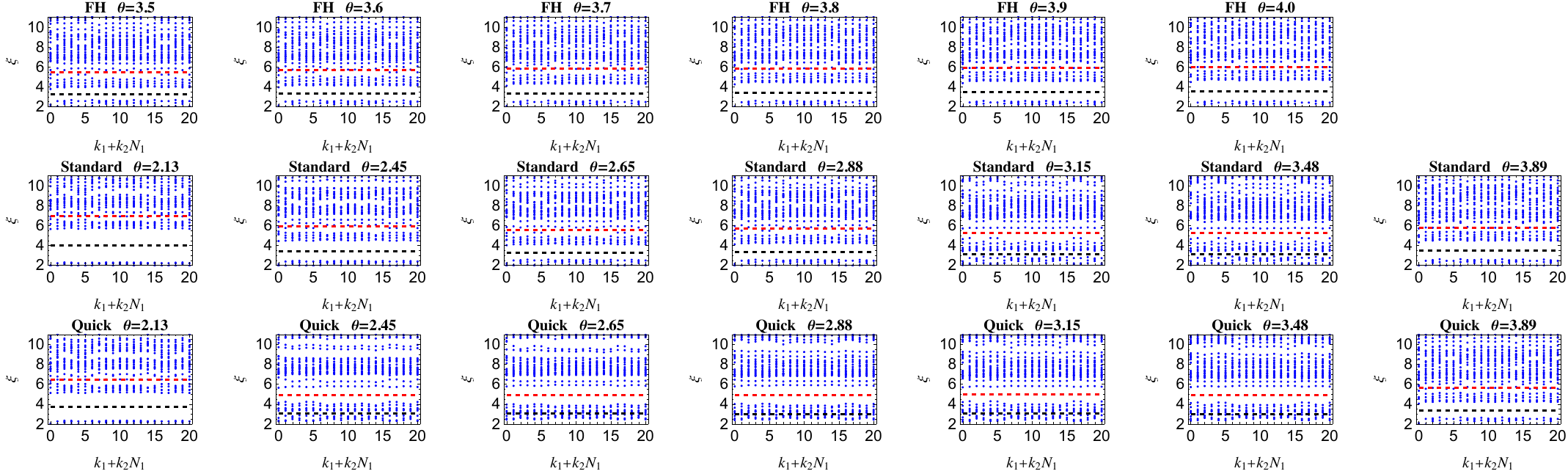}
    \caption{Particle entanglement spectrum (PES) of FH and no-fitting models in the standard and quick basis for system with $(N_x,N_y,\widetilde{n}_{11},\widetilde{n}_{12},\widetilde{n}_{21},\widetilde{n}_{22})=(21,1,1,-5,0,1)$ at $\nu=-1/3$ and $10/\epsilon=0.9$. The PES is computed from the lowest three states in the FCI momenta $k_x+k_yN_x=0,7,14$ in the polarized sector with 2BPV calculation. The red and black dashed lines represent the FCI and CDW counting respectively.
    }
    \label{fig:PES21_1v3}
\end{figure}

\begin{figure}
    \centering
    \includegraphics[width=1.0\columnwidth]{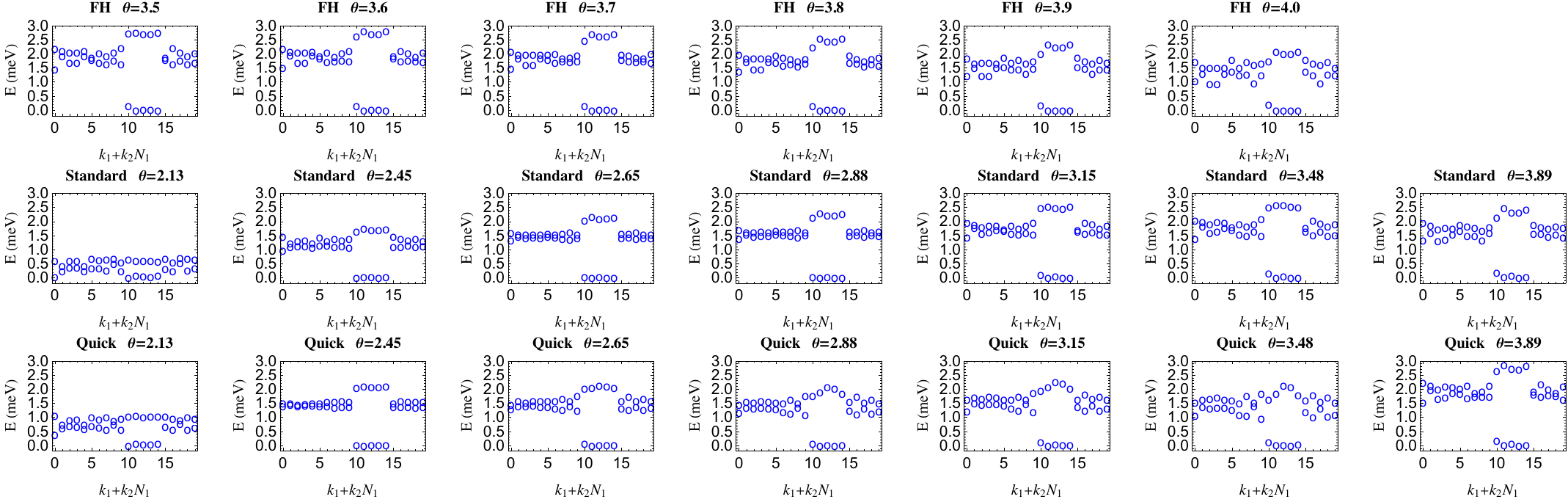}
    \caption{Energy spectrum of FH and no-fitting models in the standard and quick basis for system with $(N_x,N_y,\widetilde{n}_{11},\widetilde{n}_{12},\widetilde{n}_{21},\widetilde{n}_{22})=(5,4,1,0,1,1)$ at $\nu=-3/5$ and $10/\epsilon=0.9$. The 2BPV calculation is performed in the spin-polarized sector with truncation of bandmax 6.
    }
    \label{fig:ED20_3v5}
\end{figure}

\begin{figure}
    \centering
    \includegraphics[width=1.0\columnwidth]{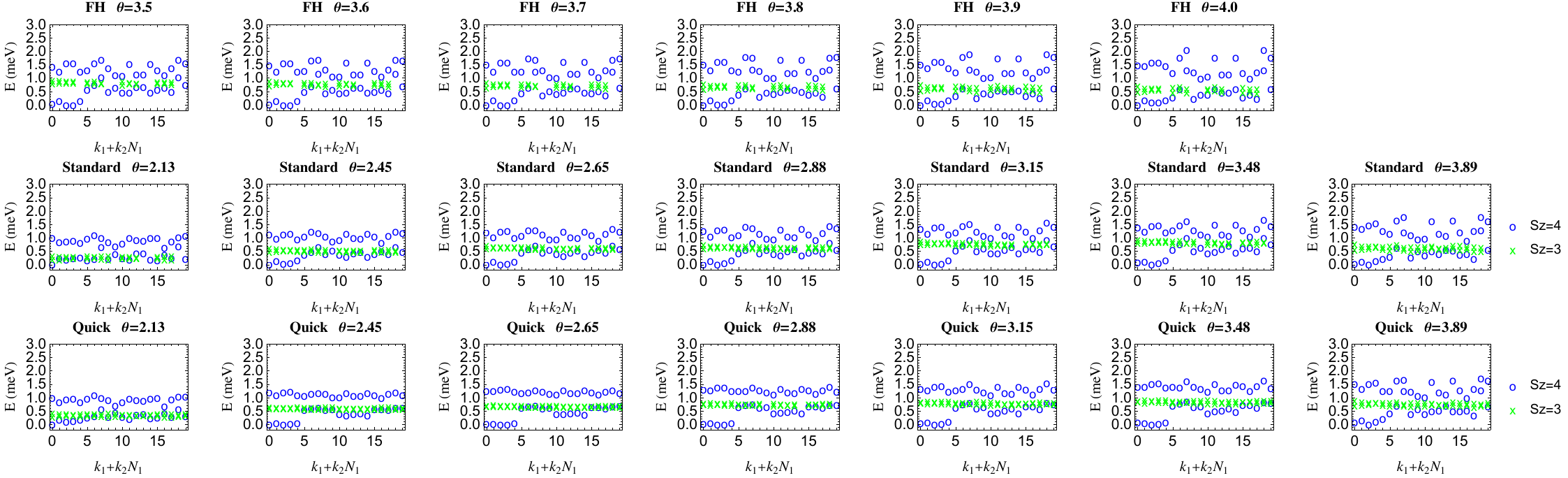}
    \caption{Energy spectrum of FH and no-fitting models in the standard and quick basis for system with $(N_x,N_y,\widetilde{n}_{11},\widetilde{n}_{12},\widetilde{n}_{21},\widetilde{n}_{22})=(5,4,1,0,1,1)$ at $\nu=-2/5$ and $10/\epsilon=0.9$. The 2BPV calculation is performed in the spin sectors with $S_z=3,4$.
    }
    \label{fig:ED20_2v5}
\end{figure}

\begin{figure}
    \centering
    \includegraphics[width=1.0\columnwidth]{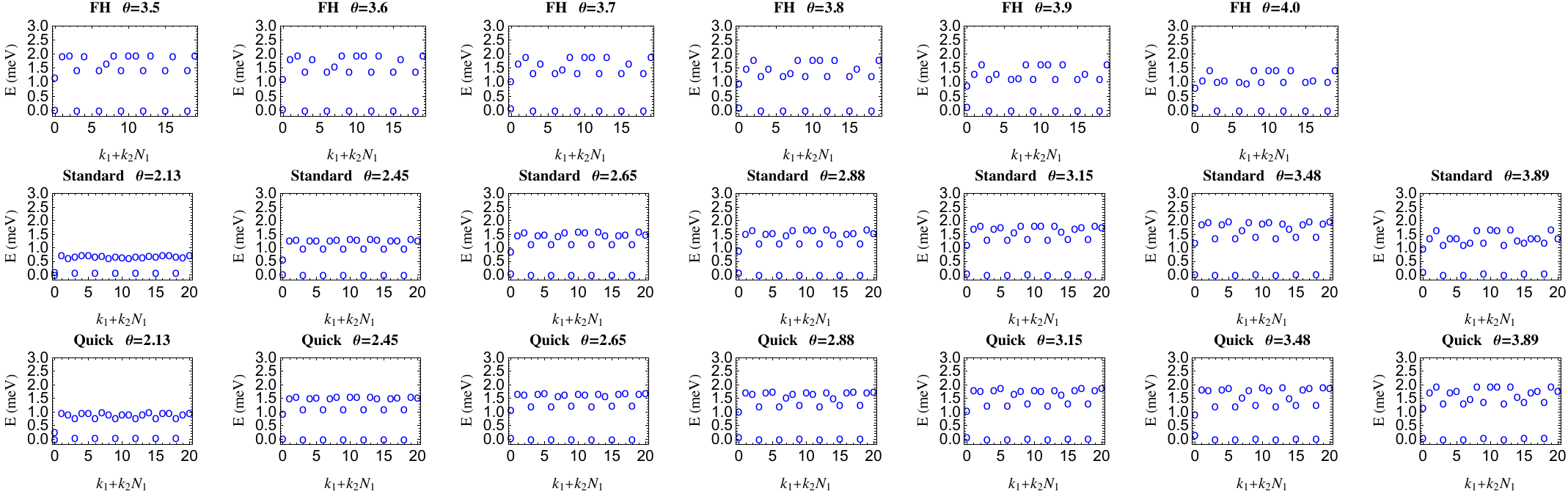}
    \caption{Energy spectrum of FH and no-fitting models in the standard and quick basis for system with $(N_x,N_y,\widetilde{n}_{11},\widetilde{n}_{12},\widetilde{n}_{21},\widetilde{n}_{22})=(21,1,1,-5,0,1)$ at $\nu=-4/7$ and $10/\epsilon=0.9$. The 2BPV calculation is performed in the spin-polarized sector with truncation of bandmax 5.
    }
    \label{fig:ED21_4v7}
\end{figure}

\begin{figure}
    \centering
    \includegraphics[width=1.0\columnwidth]{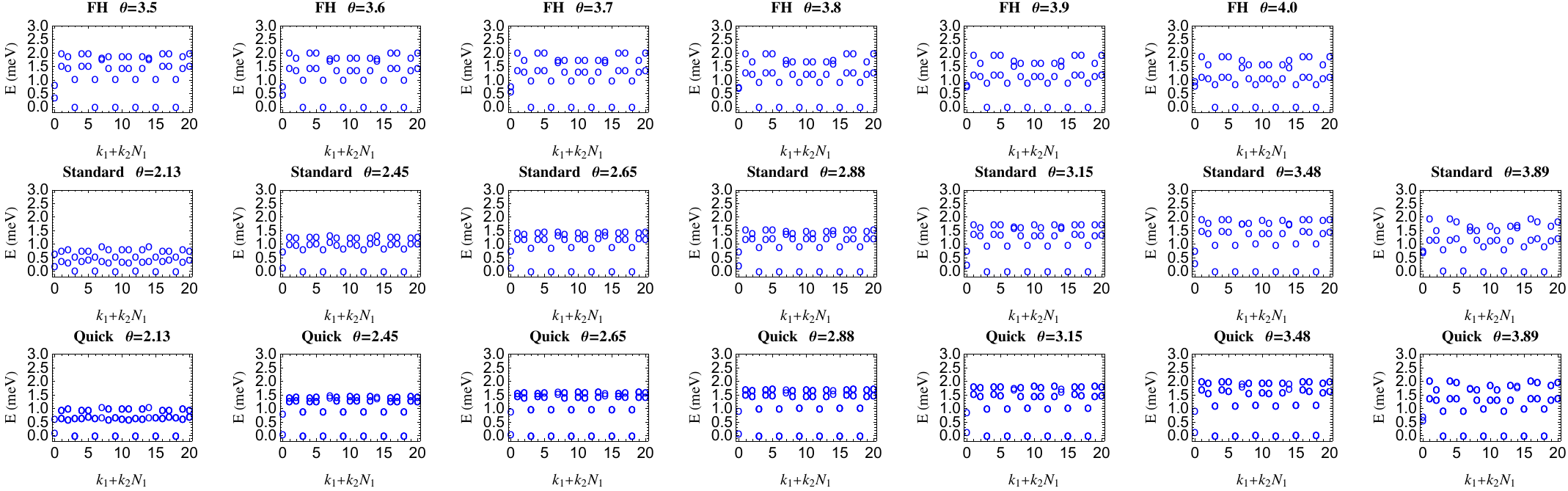}
    \caption{Energy spectrum of FH and no-fitting models in the standard and quick basis for system with $(N_x,N_y,\widetilde{n}_{11},\widetilde{n}_{12},\widetilde{n}_{21},\widetilde{n}_{22})=(21,1,1,-5,0,1)$ at $\nu=-3/7$ and $10/\epsilon=0.9$. 
    }
    \label{fig:ED21_3v7}
\end{figure}

\begin{figure}
    \centering
    \includegraphics[width=0.3\columnwidth]{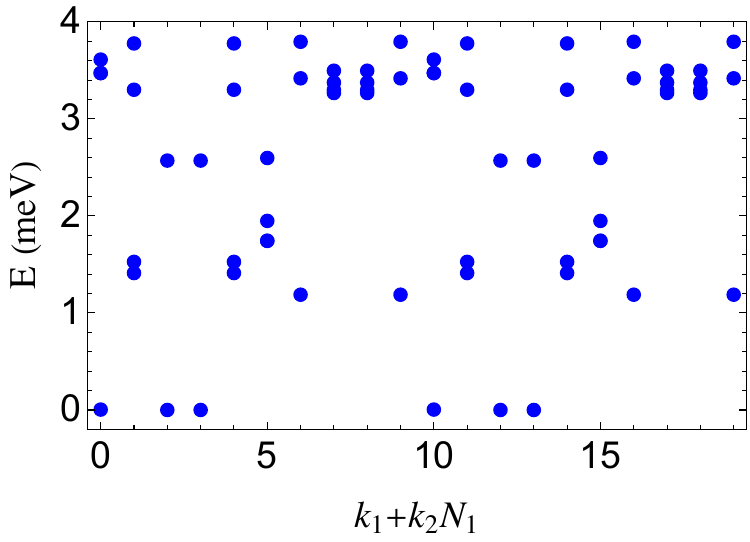}
    \caption{Energy spectrum of ideal lowest Landau level in the momentum mesh with $(N_x,N_y,\widetilde{n}_{11},\widetilde{n}_{12},\widetilde{n}_{21},\widetilde{n}_{22})=(5,4,1,0,1,1)$ at half filling.
    }
    \label{fig:LLL5b4n10}
\end{figure}

\begin{figure}
    \centering
    \includegraphics[width=1.0\columnwidth]{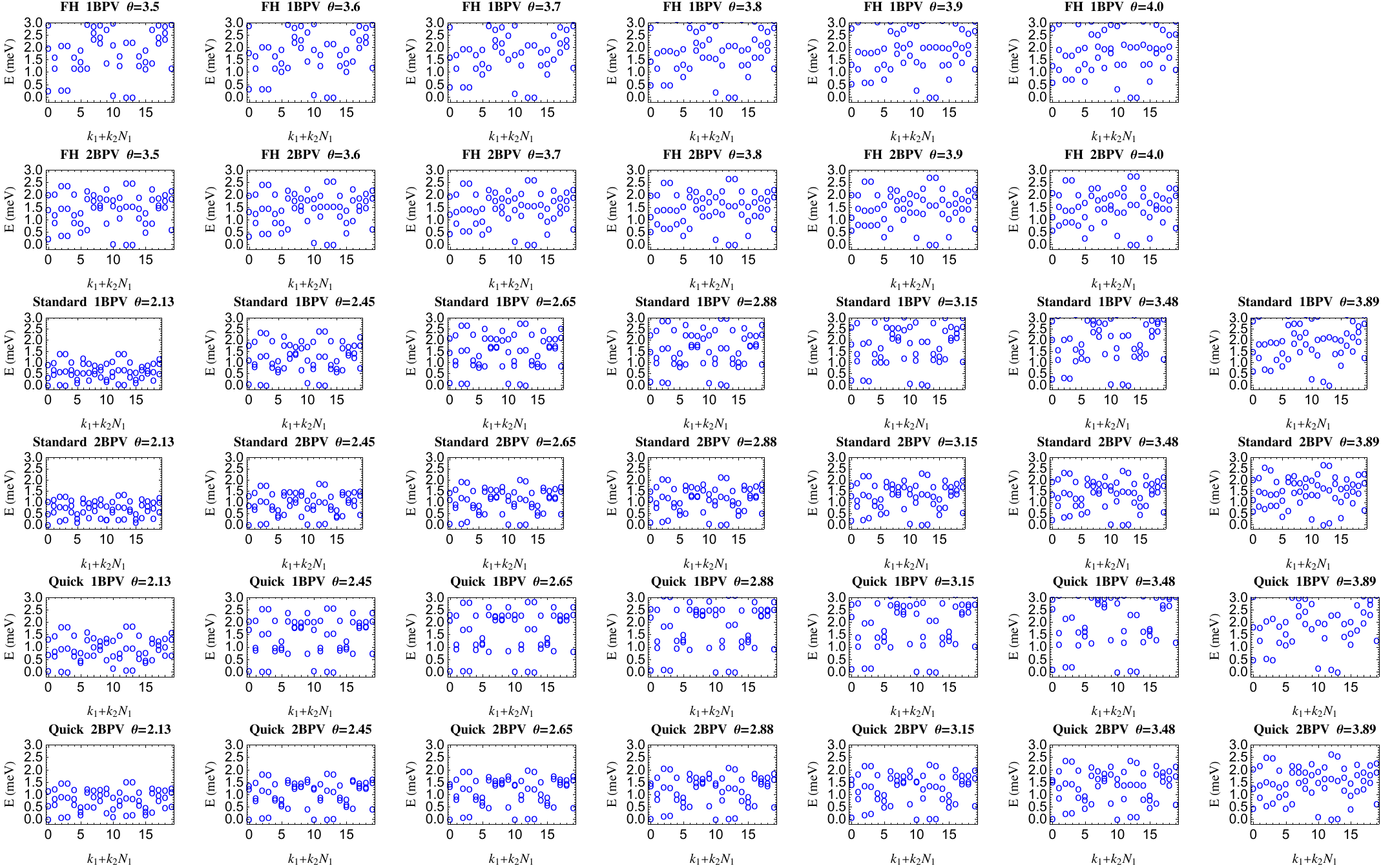}
    \caption{Energy spectrum of 2BPV in FH and no-fitting models in the standard and quick basis for system with $(N_x,N_y,\widetilde{n}_{11},\widetilde{n}_{12},\widetilde{n}_{21},\widetilde{n}_{22})=(5,4,1,0,1,1)$ at $\nu=-1/2$ and $10/\epsilon=0.9$ in the polarized sector. The spectrum in the ideal lowest Landau level at half filling for the same momentum mesh is shown in Fig.~\ref{fig:LLL5b4n10}. The 1BPV energy spectrum is closer to the Landau level spectrum than the 2BPV spectrum.
    }
    \label{ED20_1v2}
\end{figure}

In this section, we present the 2BPV energy spectrum calculated in both the FH and no-fitting models in systems with 20 or 21 sites at $10/\epsilon=0.9$, as shown in Figs.~\ref{fig:ED21_2v3} to \ref{fig:ED21_3v7}. For the 21-site system at $\nu=-1/3$ or $-2/3$, both CDW and FCI have three ground states at the same momenta $\Gamma_M,K_M,K'_M$ with $k_x+k_yN_x=0,7,14$. To distinguish between FCI and CDW at $\nu=-1/3$, we calculate the PES of ground states in 2BPV in Fig.~\ref{fig:PES21_1v3}. When choosing the spin sectors, for computational efficiency, we calculate all spin sectors for $\nu=-1/3$ and spin sectors with $S_z=4,3$ for $\nu=-2/5$. For all  other fillings we only compute the spin-polarized sector. The results are summarized in the phase diagram in Fig.~\ref{fig:phaselargesize}. We also present the 1BPV and 2BPV spectrum at $\nu=-1/2$ in 20-site system in Fig.~\ref{ED20_1v2}. Combining the results obtained from these calculations, several statements can be made:

\begin{itemize}
    \item At fillings $\nu=-2/3,-3/5,-4/7$, the larger system sizes have a stronger tendency towards FCI formation. For example, the $3\times 4$ system at $\nu=-2/3$ in the FH model in Fig.~\ref{fig:FH_2BPV_thirds_new} does not have FCI at twist angle $3.8^\circ - 4.0^\circ$, but in the $21$-site system in Fig.~\ref{fig:ED21_2v3} all twist angles have FCI ground states. Similar behavior is also found for $\nu=-3/5,-4/7$.

    \item At $\nu=-1/3$, the larger system also has more tendency to develop magnetism. For example, at $\nu=-1/3$ at twist angle $3.7^\circ - 4.0^\circ$ the lowest energy state of $9\times 2$ system in Fig.~\ref{fig:FH_2BPV_thirds_new} is not polarized, but for a $21$-site system at $\nu=-1/3$ in Fig.~\ref{fig:ED21_1v3} the lowest energy state is fully polarized for all twist angles.

    \item The spin gap at $\nu=-1/3$ is small, and some states with spin smaller than $S_z^{\text{max}}$ have energies lying in-between the three fully-polarized CDW/FCI ground states, suggesting $\nu=-1/3$ is close to a magnetic transition. The closeness to magnetic transition is consistent with the range of magnetization at $\nu\in [-0.35,-1.1]$ observed experimentally~\cite{cai2023signatures}. The PES of $\nu=-1/3$ in Fig.~\ref{fig:PES21_1v3} shows that near $3.89^\circ$ the ground state is CDW in all three models, which is consistent with experiments~\cite{Xiaodongnew}. However, at smaller twist angles the standard and quick basis give different ground states. For example, at $3.15^\circ,3.48^\circ$ the energy spectrum in Fig.~\ref{fig:ED21_1v3} only has a very small gap in the standard basis, whereas the energy gap in the quick basis remains finite and the ground state becomes FCI as seen from the PES. At $2.45^\circ,2.65^\circ,2.88^\circ$, the PES shows that the ground state in the standard basis is CDW whereas in the quick basis the ground state is FCI.

    \item At filling $\nu=-1/2$, the energy spectrum in the 1BPV calculation in Fig.~\ref{ED20_1v2} in the FH model with $\theta=3.5^\circ-3.7^\circ$ has lowest six states at momenta $k_x+k_yN_x=0,2,3,10,12,13$ which is similar to the spectrum in the lowest Landau level in Fig.~\ref{fig:LLL5b4n10}, therefore the 1BPV ground state is consistent with a CFL. However, when band mixing is considered, the 2BPV spectrum no longer has the lowest six states at those momenta. Therefore the 2BPV ground state is no longer CFL. This weakening of CFL feature with band mixing is similar to the discussion in Fig.~\ref{fig:CFLcor} in the main text.

\end{itemize}

\begin{figure}[H]
    \centering
    \includegraphics[width=\columnwidth]{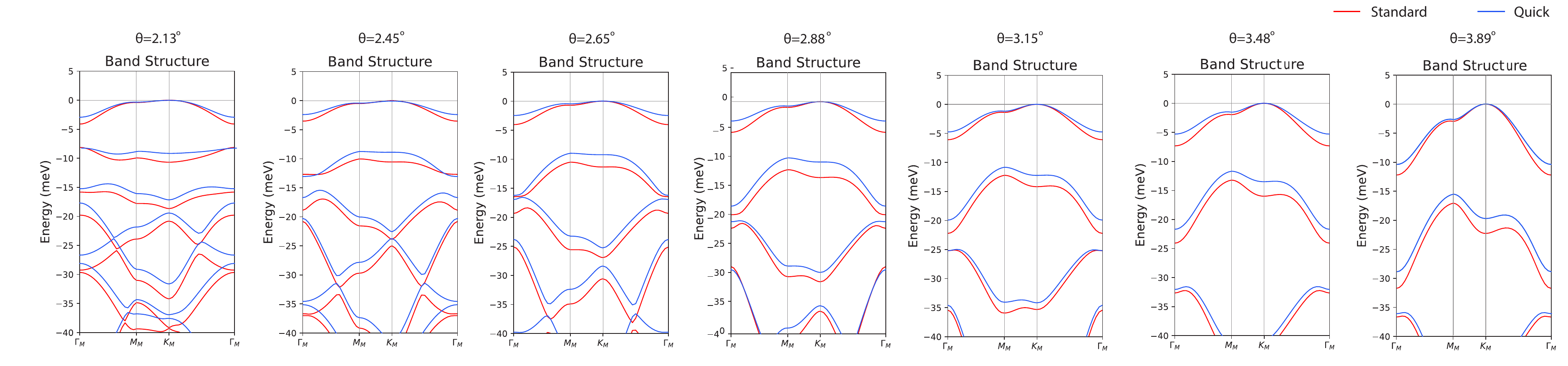}
    \caption{Band structure of the no-fitting model in the standard basis (red) and quick basis (blue) at different twist angles.
    }
    \label{fig:moteband_all}
\end{figure}

\section{Comparison between different DFT basis}
\label{App_quickbasis}

\begin{figure}[H]
    \centering
    \includegraphics[width=\columnwidth]{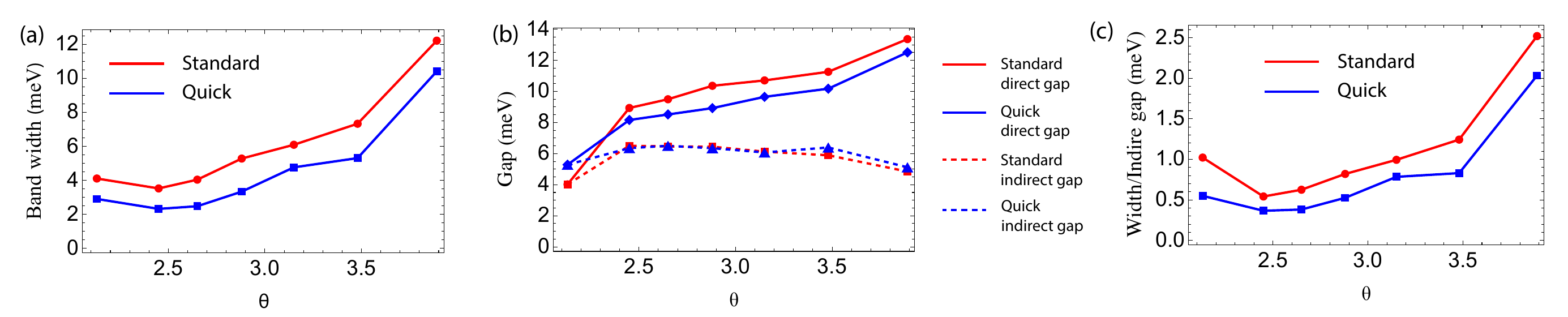}
    \caption{(a) Band width of the top moir\'e band in the standard basis (red) and quick basis (blue). (b) Direct gap (solid lines) and indirect gap (dashed lines) between the top two moir\'e bands in the standard basis (red) and quick basis (blue). (c) The ratio between bandwidth and the indirect gap in the standard basis (red) and quick basis (blue).
    }
    \label{fig:Nofit_gapwidth}
\end{figure}

\begin{figure}[H]
    \centering
    \includegraphics[width=0.8\columnwidth]{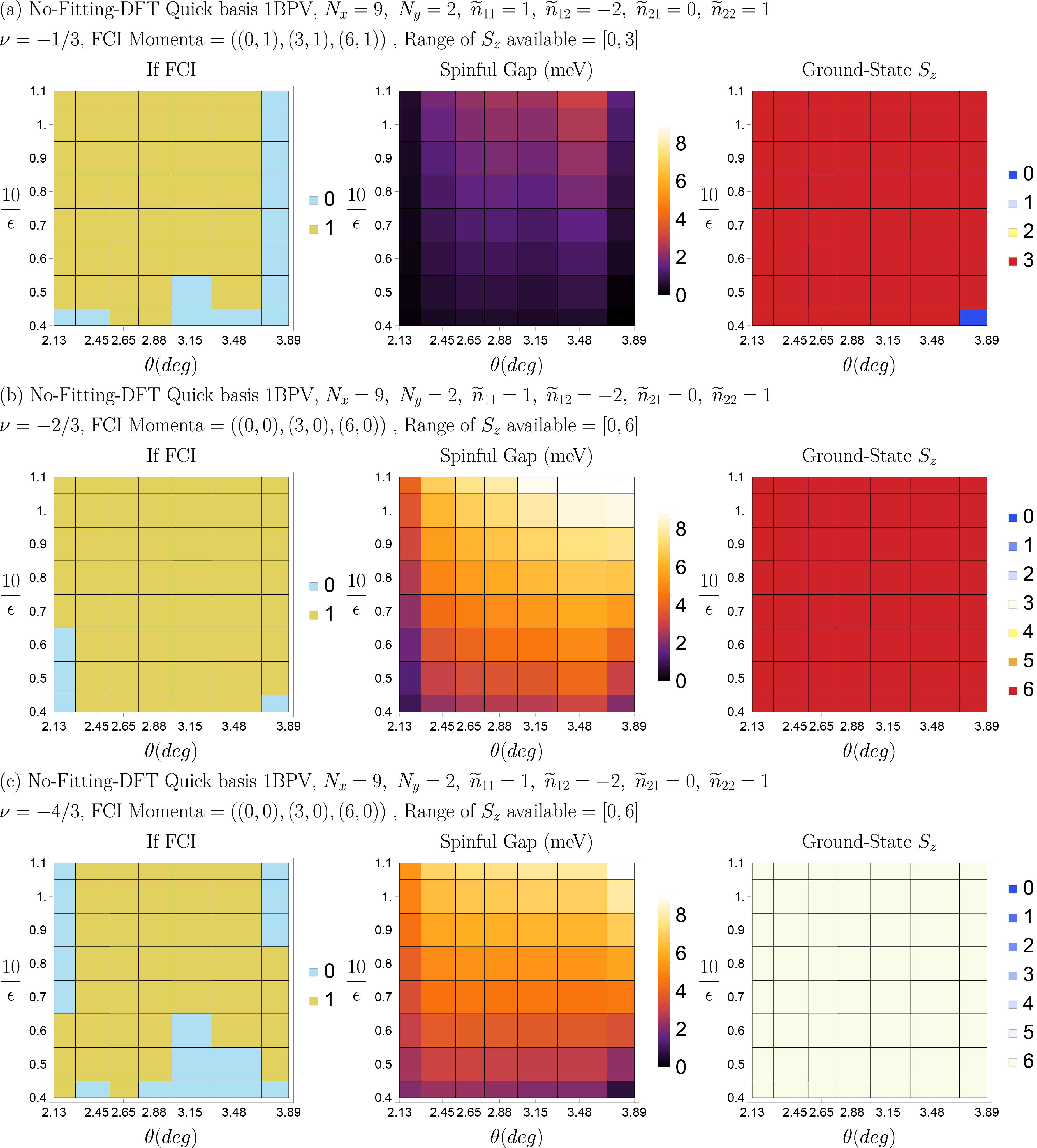}
    \caption{1BPV ED results at $\nu=-1/3,-2/3,-4/3$ for the no-fitting model in the quick basis in $9\times 2$ system.
    The meaning of the labels can be found in the caption of \figref{fig:FH_1BPV_thirds}.
    }
    \label{fig:quickbasis1b}
\end{figure}

\begin{figure}[H]
    \centering
    \includegraphics[width=\columnwidth]{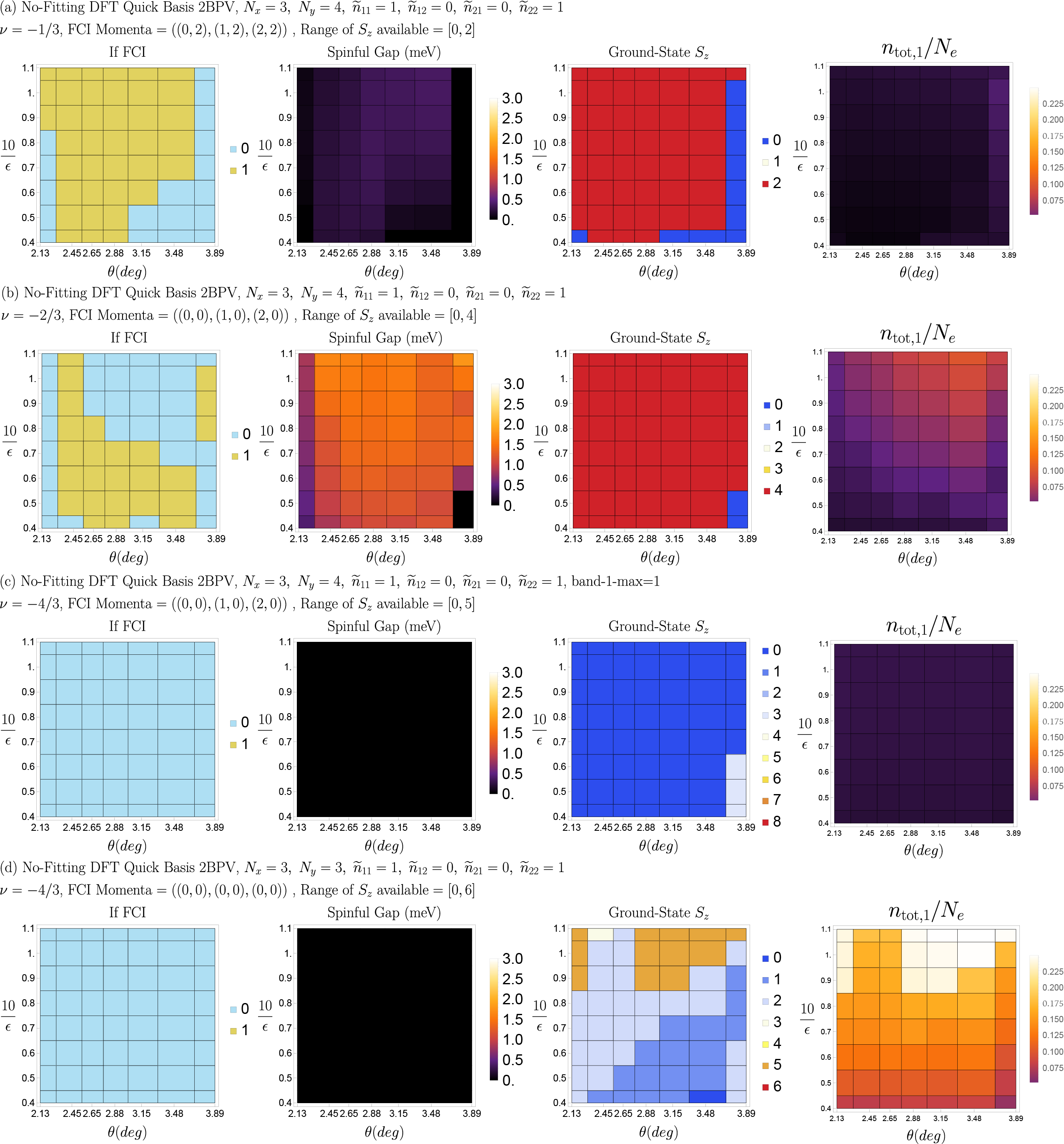}
    \caption{2BPV ED results at $\nu=-1/3,-2/3,-4/3$ for the no-fitting model in the quick basis with the same system sizes as Fig.~\ref{fig:NoFittingDFT_2BPV_thirds} at filling $\nu=-1/3$ (a), $\nu=-2/3$ (b) and $\nu=-4/3$ (c,d). The $3\times 4$ system is used in (a,b,c) and a $3\times 3$ system is used in (d).
    Here $n_{\text{tot},1}$ is the total occupation number in band 1 calculated from the lowest many-body state and $N_e$ is the number of holes.
    The meaning of the labels can be found in the caption of \figref{fig:FH_1BPV_thirds}.
    }
    \label{fig:quickbasis}
\end{figure}

\begin{figure}[H]
    \centering
    \includegraphics[width=0.8\columnwidth]{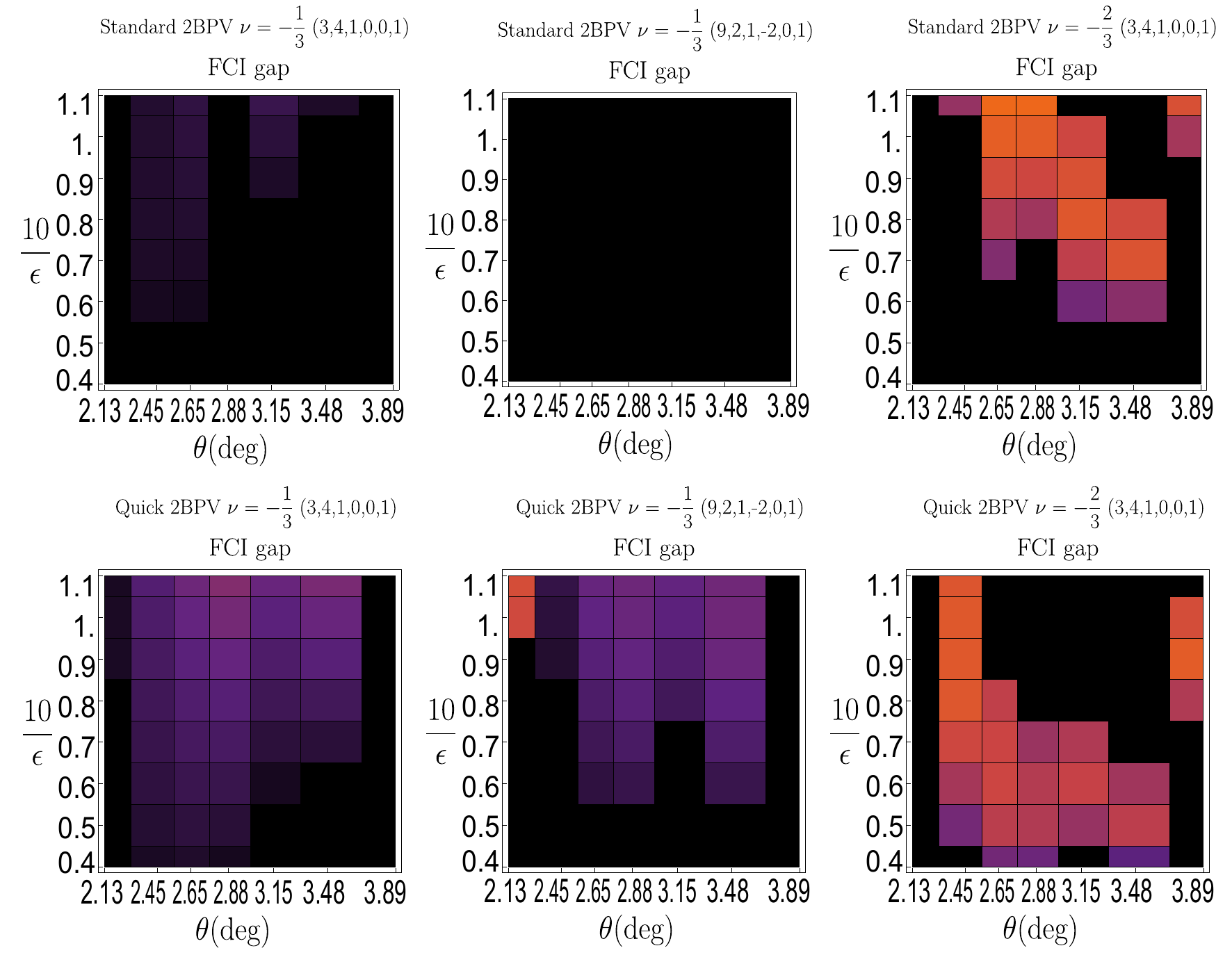}
    \caption{Energy gap above the FCI ground states identified by \propref{prop:FCI} in the polarized sector for the phase diagrams in the no-fitting standard basis and the quick basis. The six numbers in parentheses are $(N_x,N_y,\tilde n_{11},\tilde n_{12},\tilde n_{21},\tilde n_{22})$ respectively, and the system size is $N_x\times N_y$. If the ground state is not an FCI, the gap will be set to zero. All the figures share the same color scheme.
    }
    \label{fig:FCIgap_sq_thirds}
\end{figure}

\begin{figure}[H]
    \centering
    \includegraphics[width=0.8\columnwidth]{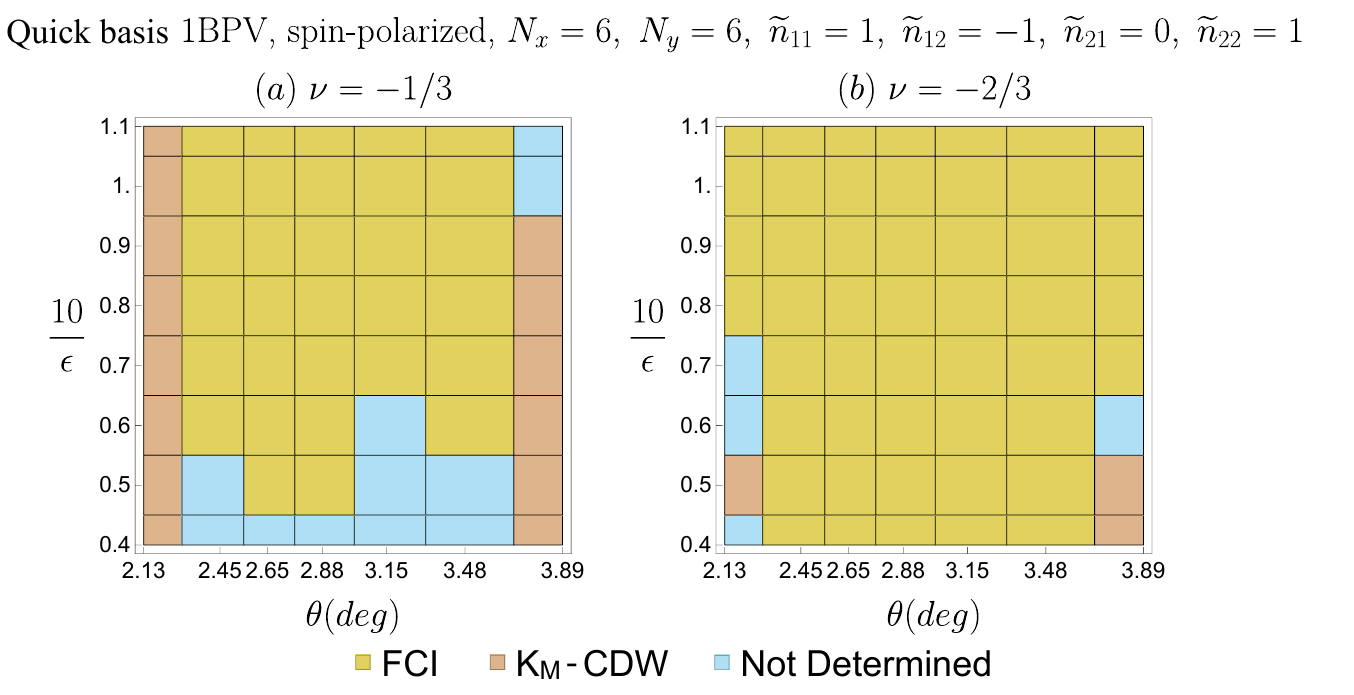}
    \caption{1BPV ED results at $\nu=-1/3,-2/3$ in a $6\times 6$ momentum mesh for the DFT no-fitting model in the quick basis and the fully spin-polarized sector.
    Yellow means the system is in an FCI phase according to \propref{prop:FCI}, brown means the system is in the $K_M$ charge density wave (CDW), and blue means the ground state of the system is neither FCI nor CDW.
    The other labels have the same meaning as in \figref{fig:FH_1BPV_thirds}.
    }
    \label{fig:FCICDW_q_6by6_1BPV}
\end{figure}

\begin{figure}[H]
    \centering
    \includegraphics[width=0.7\columnwidth]{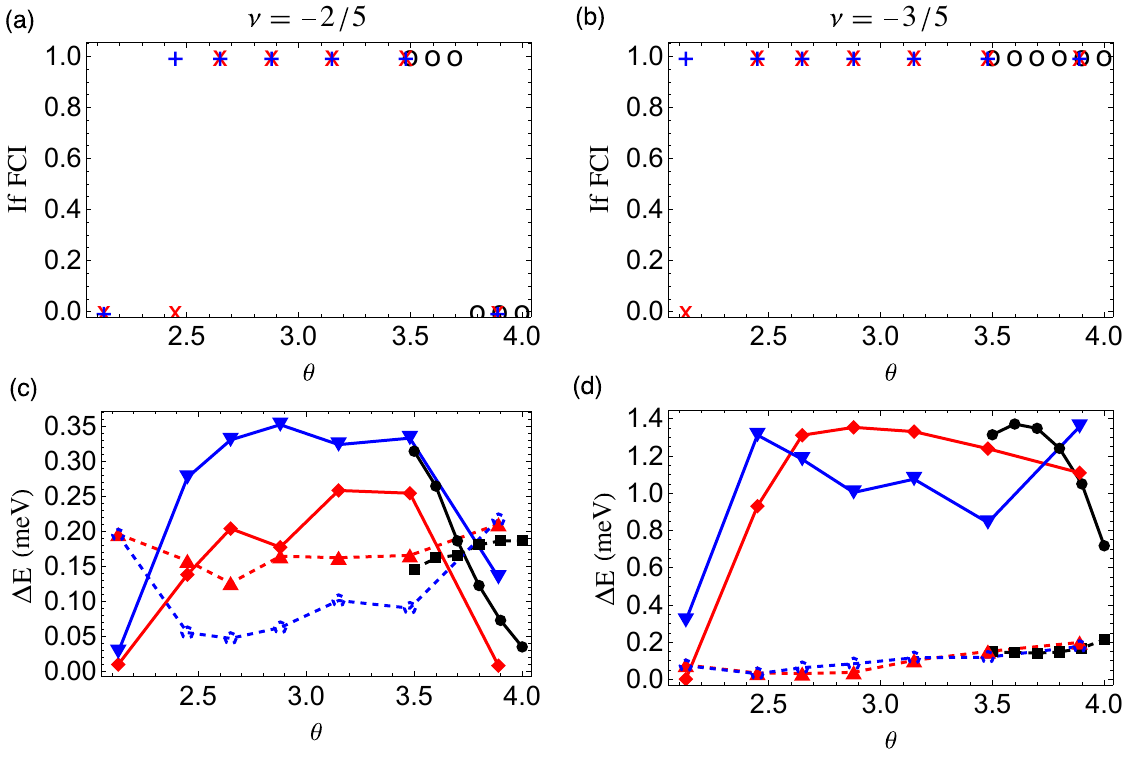}
    \caption{ 2BPV ED results for 20-site system with $(N_x,N_y,\widetilde{n}_{11},\widetilde{n}_{12},\widetilde{n}_{21},\widetilde{n}_{22})=(5,4,1,0,1,1)$ and $10/\epsilon=0.9$ in the polarized sector. The red, blue and black colors represent calculations performed in the standard basis, quick basis and FH model respectively. (a),(c) are calculated for $\nu=-2/5$ with bandmax 8 and (b),(d) are calculated for $\nu=-3/5$ with bandmax 6. The meaning of the plots are identical to the caption of Fig.~\ref{fig:comb2v5}.
    }
    \label{fig:gap20_fifths_sqh}
\end{figure}

\begin{figure}[H]
    \centering
    \includegraphics[width=0.7\columnwidth]{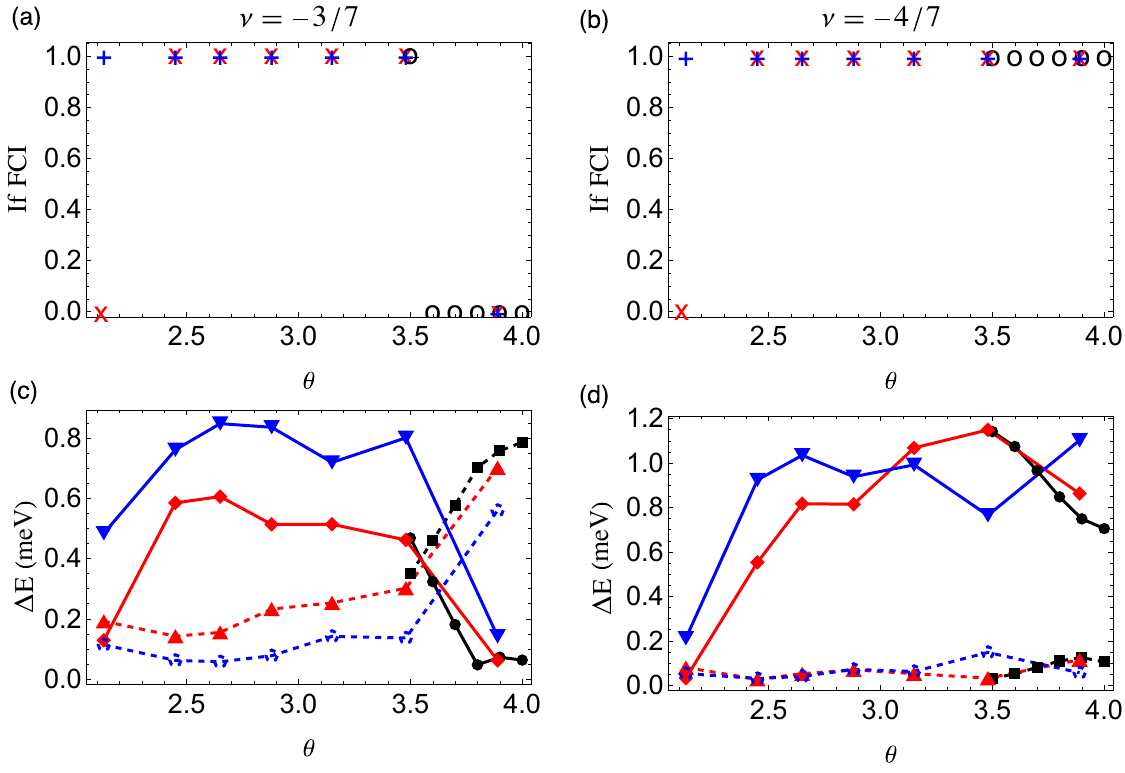}
    \caption{ 2BPV ED results for 21-site system with $(N_x,N_y,\widetilde{n}_{11},\widetilde{n}_{12},\widetilde{n}_{21},\widetilde{n}_{22})=(21,1,1,-5,0,1)$ and $10/\epsilon=0.9$ in the polarized sector. The red, blue and black colors represent calculations performed in the standard basis, quick basis and FH model respectively. (a),(c) are calculated for $\nu=-3/7$ with bandmax 9 and (b),(d) are calculated for $\nu=-4/7$ with bandmax 5. The meaning of the plots are identical to the caption of Fig.~\ref{fig:comb2v5}.
    }
    \label{fig:gap21_sevenths_sqh}
\end{figure}

Depending on the choice of standard basis and the quick basis in the no-fitting model, different spectra and phase diagrams can be obtained. The computations in Sec.~\ref{sec_nofitting} are performed for the standard basis~\cite{DFTnofitting2024}. We now  consider the quick basis which involves a smaller number of orbitals (Mo-s3p2d1 and Te-s3p2d2) as a comparison to the standard basis (Mo-s3p2d2 and Te-s3p2d2f1). The band structures of these models at different twist angles can be found in Fig.~\ref{fig:moteband_all}. The band gap and bandwidth at different twist angles are shown in Fig.~\ref{fig:Nofit_gapwidth}. The 1BPV and 2BPV FCI phase diagrams at fillings $\nu=-1/3,-2/3,-2/5,-3/5,-3/7,-4/7$ that involve calculations in the quick basis can be found in Fig.~\ref{fig:quickbasis1b} to Fig.~\ref{fig:gap21_sevenths_sqh}.

\begin{figure}[H]
    \centering
    \includegraphics[width=\columnwidth]{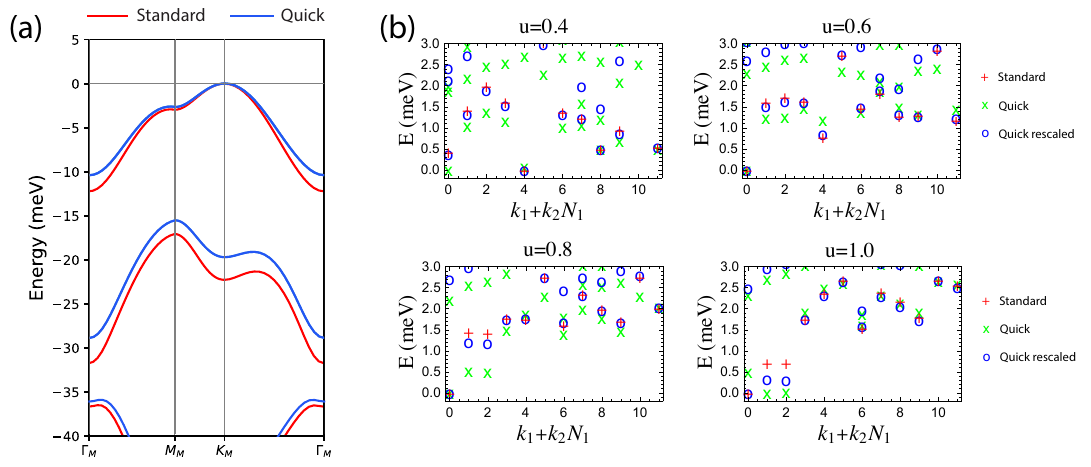}
    \caption{(a) Band structure of the no-fitting model in the standard basis and quick basis, respectively at twist angle $3.89^\circ$. The top band is the highest valence band. (b) Energy spectrum for a $3\times 4$ system at filling -2/3 in the no-fitting model at twist angle $3.89^\circ$ with different interaction strength $u=\frac{10}{\epsilon}$. Red and green labels represent the energy computed in the standard basis and quick basis, respectively. The blue labels are computed from the rescaled quick basis where the energies of all bands are multiplied by $1.18$, so that the first band in the rescaled quick basis has the same bandwidth as the standard basis. This comparison shows that the discrepancy between the standard basis and the quick basis mainly comes from the different bandwidths in the two bases.
    }
    \label{fig:compare_standardquick}
\end{figure}

\begin{figure}[H]
    \centering
    \includegraphics[width=\columnwidth]{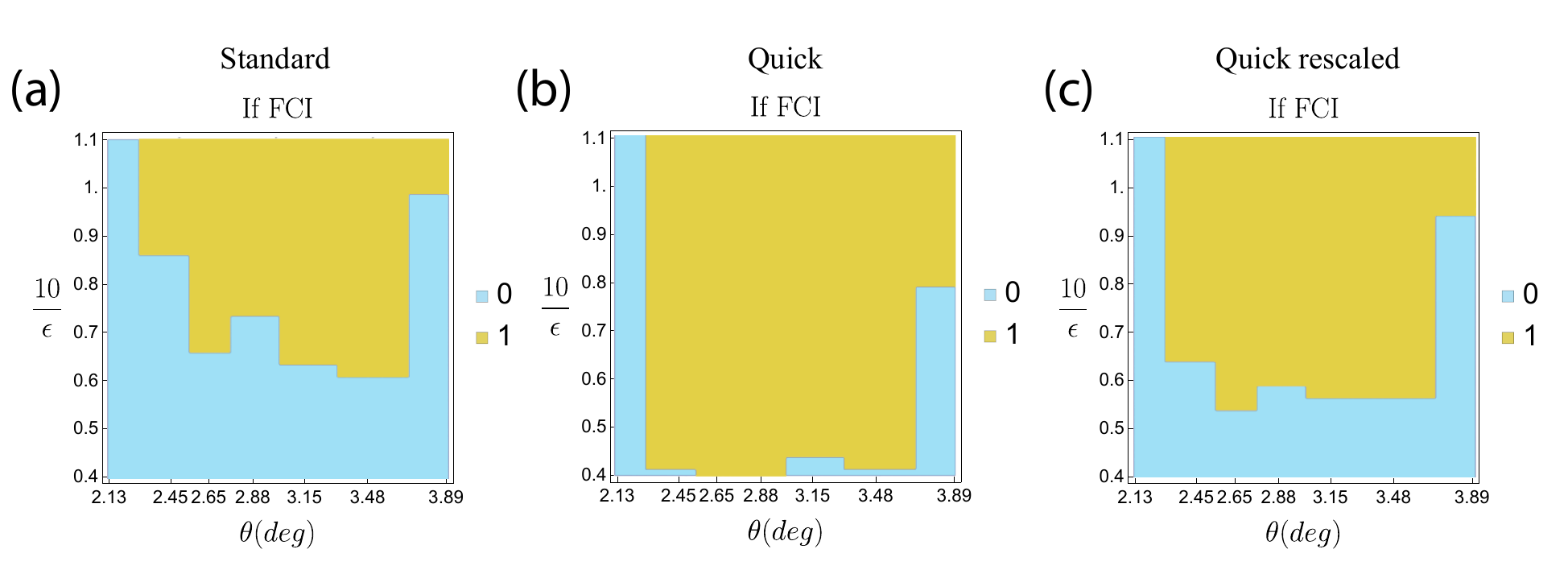}
    \caption{  1BPV FCI phase diagrams in a $3\times 4$ system at $\nu=-2/3$ in the spin-polarized sector for the no-fitting model in the standard basis (a) and quick basis (b). In (c) we plot the FCI phase diagram obtained by scaling the single-particle energy of the quick basis to make the bandwidth at each twist angle match that of the standard basis. 
    }
    \label{fig:SQcompare1BPV}
\end{figure}

If we compare the 2BPV FCI region at $\nu=-2/3$ in the quick basis in Fig.~\ref{fig:quickbasis} with that in the standard basis in Fig.~\ref{fig:NoFittingDFT_2BPV_thirds} in the main text, the FCI region in the quick basis appears to be shifted to locations with smaller interaction. The difference between the phase diagrams obtained from these two bases mostly originates from their difference in bandwidth. As shown in Fig.~\ref{fig:compare_standardquick}(a), the dispersion in the two models are similar, but the width of the first valence band in the model obtained from the standard basis at $3.89^\circ$ is $18\%$ larger than that in the quick basis. To check whether this is the main reason for the difference in phase diagrams of the two models, we perform an ED computation by rescaling the single-particle Hamiltonian obtained from the quick basis by a factor $\lambda=1.18$ while keeping the single-particle wave functions and the interaction unchanged, and compare the ED energy spectrum with those from the standard basis. As shown in Fig.~\ref{fig:compare_standardquick}(b), although the spectrum obtained from the standard basis has sizable difference compared with the quick basis, once we rescale the single-particle Hamiltonian in the quick basis by $\lambda$, the energy spectrum of the rescaled quick basis is very close to that from the standard basis. Since a simultaneous scaling of both the single-particle Hamiltonian and interaction strength leaves the ground state invariant, this also implies the ground state of the model obtained from the standard basis at some interaction strength $\frac{1}{\epsilon}$ will be similar to the ground state obtained from the quick basis at a smaller interaction strength $\frac{1}{\lambda\epsilon}$. To further investigate the effect of the top valence band on the FCI phase diagram, we perform 1BPV calculation for both standard and quick basis in a $3\times 4$ system with $\nu=-2/3$ using a finer mesh of $10/\epsilon$ in Fig.~\ref{fig:SQcompare1BPV}. The FCI phase diagrams in the standard basis (a) and quick basis (b) have sizable difference. However, if we manually scale the single-particle energy of the quick basis at each twist angle to make the bandwidth match that in the standard basis, the resulting phase diagram in Fig.~\ref{fig:SQcompare1BPV}(c) is much closer to the phase diagram in the standard basis in Fig.~\ref{fig:SQcompare1BPV}(a).

\begin{thebibliography}{132}%
\makeatletter
\providecommand \@ifxundefined [1]{%
 \@ifx{#1\undefined}
}%
\providecommand \@ifnum [1]{%
 \ifnum #1\expandafter \@firstoftwo
 \else \expandafter \@secondoftwo
 \fi
}%
\providecommand \@ifx [1]{%
 \ifx #1\expandafter \@firstoftwo
 \else \expandafter \@secondoftwo
 \fi
}%
\providecommand \natexlab [1]{#1}%
\providecommand \enquote  [1]{``#1''}%
\providecommand \bibnamefont  [1]{#1}%
\providecommand \bibfnamefont [1]{#1}%
\providecommand \citenamefont [1]{#1}%
\providecommand \href@noop [0]{\@secondoftwo}%
\providecommand \href [0]{\begingroup \@sanitize@url \@href}%
\providecommand \@href[1]{\@@startlink{#1}\@@href}%
\providecommand \@@href[1]{\endgroup#1\@@endlink}%
\providecommand \@sanitize@url [0]{\catcode `\\12\catcode `\$12\catcode `\&12\catcode `\#12\catcode `\^12\catcode `\_12\catcode `\%12\relax}%
\providecommand \@@startlink[1]{}%
\providecommand \@@endlink[0]{}%
\providecommand \url  [0]{\begingroup\@sanitize@url \@url }%
\providecommand \@url [1]{\endgroup\@href {#1}{\urlprefix }}%
\providecommand \urlprefix  [0]{URL }%
\providecommand \Eprint [0]{\href }%
\providecommand \doibase [0]{https://doi.org/}%
\providecommand \selectlanguage [0]{\@gobble}%
\providecommand \bibinfo  [0]{\@secondoftwo}%
\providecommand \bibfield  [0]{\@secondoftwo}%
\providecommand \translation [1]{[#1]}%
\providecommand \BibitemOpen [0]{}%
\providecommand \bibitemStop [0]{}%
\providecommand \bibitemNoStop [0]{.\EOS\space}%
\providecommand \EOS [0]{\spacefactor3000\relax}%
\providecommand \BibitemShut  [1]{\csname bibitem#1\endcsname}%
\let\auto@bib@innerbib\@empty
\bibitem [{\citenamefont {Neupert}\ \emph {et~al.}(2011)\citenamefont {Neupert}, \citenamefont {Santos}, \citenamefont {Chamon},\ and\ \citenamefont {Mudry}}]{Neupert2011Quantum}%
  \BibitemOpen
  \bibfield  {author} {\bibinfo {author} {\bibfnamefont {T.}~\bibnamefont {Neupert}}, \bibinfo {author} {\bibfnamefont {L.}~\bibnamefont {Santos}}, \bibinfo {author} {\bibfnamefont {C.}~\bibnamefont {Chamon}},\ and\ \bibinfo {author} {\bibfnamefont {C.}~\bibnamefont {Mudry}},\ }\bibfield  {title} {\bibinfo {title} {Fractional quantum hall states at zero magnetic field},\ }\href {https://doi.org/10.1103/PhysRevLett.106.236804} {\bibfield  {journal} {\bibinfo  {journal} {Phys. Rev. Lett.}\ }\textbf {\bibinfo {volume} {106}},\ \bibinfo {pages} {236804} (\bibinfo {year} {2011})}\BibitemShut {NoStop}%
\bibitem [{\citenamefont {Sheng}\ \emph {et~al.}(2011)\citenamefont {Sheng}, \citenamefont {Gu}, \citenamefont {Sun},\ and\ \citenamefont {Sheng}}]{Sheng2011quantum}%
  \BibitemOpen
  \bibfield  {author} {\bibinfo {author} {\bibfnamefont {D.~N.}\ \bibnamefont {Sheng}}, \bibinfo {author} {\bibfnamefont {Z.-C.}\ \bibnamefont {Gu}}, \bibinfo {author} {\bibfnamefont {K.}~\bibnamefont {Sun}},\ and\ \bibinfo {author} {\bibfnamefont {L.}~\bibnamefont {Sheng}},\ }\bibfield  {title} {\bibinfo {title} {Fractional quantum hall effect in the absence of landau levels},\ }\href {https://doi.org/10.1038/ncomms1380} {\bibfield  {journal} {\bibinfo  {journal} {Nat. Commun.}\ }\textbf {\bibinfo {volume} {2}},\ \bibinfo {pages} {389} (\bibinfo {year} {2011})}\BibitemShut {NoStop}%
\bibitem [{\citenamefont {Regnault}\ and\ \citenamefont {Bernevig}(2011)}]{RegnaultBernevig2011Chern}%
  \BibitemOpen
  \bibfield  {author} {\bibinfo {author} {\bibfnamefont {N.}~\bibnamefont {Regnault}}\ and\ \bibinfo {author} {\bibfnamefont {B.~A.}\ \bibnamefont {Bernevig}},\ }\bibfield  {title} {\bibinfo {title} {Fractional chern insulator},\ }\href {https://doi.org/10.1103/PhysRevX.1.021014} {\bibfield  {journal} {\bibinfo  {journal} {Phys. Rev. X}\ }\textbf {\bibinfo {volume} {1}},\ \bibinfo {pages} {021014} (\bibinfo {year} {2011})}\BibitemShut {NoStop}%
\bibitem [{\citenamefont {Tang}\ \emph {et~al.}(2011)\citenamefont {Tang}, \citenamefont {Mei},\ and\ \citenamefont {Wen}}]{Tang2011High-Temperature}%
  \BibitemOpen
  \bibfield  {author} {\bibinfo {author} {\bibfnamefont {E.}~\bibnamefont {Tang}}, \bibinfo {author} {\bibfnamefont {J.-W.}\ \bibnamefont {Mei}},\ and\ \bibinfo {author} {\bibfnamefont {X.-G.}\ \bibnamefont {Wen}},\ }\bibfield  {title} {\bibinfo {title} {High-temperature fractional quantum hall states},\ }\href {https://doi.org/10.1103/PhysRevLett.106.236802} {\bibfield  {journal} {\bibinfo  {journal} {Phys. Rev. Lett.}\ }\textbf {\bibinfo {volume} {106}},\ \bibinfo {pages} {236802} (\bibinfo {year} {2011})}\BibitemShut {NoStop}%
\bibitem [{\citenamefont {Sun}\ \emph {et~al.}(2011)\citenamefont {Sun}, \citenamefont {Gu}, \citenamefont {Katsura},\ and\ \citenamefont {Das~Sarma}}]{Sun2011Flatbands}%
  \BibitemOpen
  \bibfield  {author} {\bibinfo {author} {\bibfnamefont {K.}~\bibnamefont {Sun}}, \bibinfo {author} {\bibfnamefont {Z.}~\bibnamefont {Gu}}, \bibinfo {author} {\bibfnamefont {H.}~\bibnamefont {Katsura}},\ and\ \bibinfo {author} {\bibfnamefont {S.}~\bibnamefont {Das~Sarma}},\ }\bibfield  {title} {\bibinfo {title} {Nearly flatbands with nontrivial topology},\ }\href {https://doi.org/10.1103/PhysRevLett.106.236803} {\bibfield  {journal} {\bibinfo  {journal} {Phys. Rev. Lett.}\ }\textbf {\bibinfo {volume} {106}},\ \bibinfo {pages} {236803} (\bibinfo {year} {2011})}\BibitemShut {NoStop}%
\bibitem [{\citenamefont {Bergholtz}\ and\ \citenamefont {Liu}(2013)}]{BergholtzLiu2013Flat}%
  \BibitemOpen
  \bibfield  {author} {\bibinfo {author} {\bibfnamefont {E.~J.}\ \bibnamefont {Bergholtz}}\ and\ \bibinfo {author} {\bibfnamefont {Z.}~\bibnamefont {Liu}},\ }\bibfield  {title} {\bibinfo {title} {Topological flat band models and fractional chern insulators},\ }\href {https://doi.org/10.1142/S021797921330017X} {\bibfield  {journal} {\bibinfo  {journal} {Int. J. Mod. Phys. B}\ }\textbf {\bibinfo {volume} {27}},\ \bibinfo {pages} {1330017} (\bibinfo {year} {2013})}\BibitemShut {NoStop}%
\bibitem [{\citenamefont {Parameswaran}\ \emph {et~al.}(2013)\citenamefont {Parameswaran}, \citenamefont {Roy},\ and\ \citenamefont {Sondhi}}]{Parameswaran2013quantum}%
  \BibitemOpen
  \bibfield  {author} {\bibinfo {author} {\bibfnamefont {S.~A.}\ \bibnamefont {Parameswaran}}, \bibinfo {author} {\bibfnamefont {R.}~\bibnamefont {Roy}},\ and\ \bibinfo {author} {\bibfnamefont {S.~L.}\ \bibnamefont {Sondhi}},\ }\bibfield  {title} {\bibinfo {title} {Fractional quantum hall physics in topological flat bands},\ }\href {https://doi.org/10.1016/j.crhy.2013.04.003} {\bibfield  {journal} {\bibinfo  {journal} {C. R. Phys.}\ }\textbf {\bibinfo {volume} {14}},\ \bibinfo {pages} {816} (\bibinfo {year} {2013})}\BibitemShut {NoStop}%
\bibitem [{\citenamefont {Cao}\ \emph {et~al.}(2018{\natexlab{a}})\citenamefont {Cao}, \citenamefont {Fatemi}, \citenamefont {Demir}, \citenamefont {Fang}, \citenamefont {Tomarken}, \citenamefont {Luo}, \citenamefont {Sanchez-Yamagishi}, \citenamefont {Watanabe}, \citenamefont {Taniguchi}, \citenamefont {Kaxiras},\ and\ \citenamefont {Jarillo-Herrero}}]{Cao2018ainsulator}%
  \BibitemOpen
  \bibfield  {author} {\bibinfo {author} {\bibfnamefont {Y.}~\bibnamefont {Cao}}, \bibinfo {author} {\bibfnamefont {V.}~\bibnamefont {Fatemi}}, \bibinfo {author} {\bibfnamefont {A.}~\bibnamefont {Demir}}, \bibinfo {author} {\bibfnamefont {S.}~\bibnamefont {Fang}}, \bibinfo {author} {\bibfnamefont {S.~L.}\ \bibnamefont {Tomarken}}, \bibinfo {author} {\bibfnamefont {J.~Y.}\ \bibnamefont {Luo}}, \bibinfo {author} {\bibfnamefont {J.~D.}\ \bibnamefont {Sanchez-Yamagishi}}, \bibinfo {author} {\bibfnamefont {K.}~\bibnamefont {Watanabe}}, \bibinfo {author} {\bibfnamefont {T.}~\bibnamefont {Taniguchi}}, \bibinfo {author} {\bibfnamefont {E.}~\bibnamefont {Kaxiras}},\ and\ \bibinfo {author} {\bibfnamefont {P.}~\bibnamefont {Jarillo-Herrero}},\ }\bibfield  {title} {\bibinfo {title} {Correlated insulator behaviour at half-filling in magic-angle graphene superlattices},\ }\href {https://doi.org/10.1038/nature26154} {\bibfield  {journal} {\bibinfo  {journal} {Nature}\ }\textbf {\bibinfo {volume} {556}},\ \bibinfo {pages}
  {80} (\bibinfo {year} {2018}{\natexlab{a}})}\BibitemShut {NoStop}%
\bibitem [{\citenamefont {Cao}\ \emph {et~al.}(2018{\natexlab{b}})\citenamefont {Cao}, \citenamefont {Fatemi}, \citenamefont {Fang}, \citenamefont {Watanabe}, \citenamefont {Taniguchi}, \citenamefont {Kaxiras},\ and\ \citenamefont {Jarillo-Herrero}}]{Cao2018bsuperconductivity}%
  \BibitemOpen
  \bibfield  {author} {\bibinfo {author} {\bibfnamefont {Y.}~\bibnamefont {Cao}}, \bibinfo {author} {\bibfnamefont {V.}~\bibnamefont {Fatemi}}, \bibinfo {author} {\bibfnamefont {S.}~\bibnamefont {Fang}}, \bibinfo {author} {\bibfnamefont {K.}~\bibnamefont {Watanabe}}, \bibinfo {author} {\bibfnamefont {T.}~\bibnamefont {Taniguchi}}, \bibinfo {author} {\bibfnamefont {E.}~\bibnamefont {Kaxiras}},\ and\ \bibinfo {author} {\bibfnamefont {P.}~\bibnamefont {Jarillo-Herrero}},\ }\bibfield  {title} {\bibinfo {title} {Unconventional superconductivity in magic-angle graphene superlattices},\ }\href {https://doi.org/10.1038/nature26160} {\bibfield  {journal} {\bibinfo  {journal} {Nature}\ }\textbf {\bibinfo {volume} {556}},\ \bibinfo {pages} {43} (\bibinfo {year} {2018}{\natexlab{b}})}\BibitemShut {NoStop}%
\bibitem [{\citenamefont {Abouelkomsan}\ \emph {et~al.}(2020)\citenamefont {Abouelkomsan}, \citenamefont {Liu},\ and\ \citenamefont {Bergholtz}}]{Abouelkomsan2020Duality}%
  \BibitemOpen
  \bibfield  {author} {\bibinfo {author} {\bibfnamefont {A.}~\bibnamefont {Abouelkomsan}}, \bibinfo {author} {\bibfnamefont {Z.}~\bibnamefont {Liu}},\ and\ \bibinfo {author} {\bibfnamefont {E.~J.}\ \bibnamefont {Bergholtz}},\ }\bibfield  {title} {\bibinfo {title} {Particle-hole duality, emergent fermi liquids, and fractional chern insulators in moir{\'e} flatbands},\ }\href {https://doi.org/10.1103/PhysRevLett.124.106803} {\bibfield  {journal} {\bibinfo  {journal} {Phys. Rev. Lett.}\ }\textbf {\bibinfo {volume} {124}},\ \bibinfo {pages} {106803} (\bibinfo {year} {2020})}\BibitemShut {NoStop}%
\bibitem [{\citenamefont {Ledwith}\ \emph {et~al.}(2020)\citenamefont {Ledwith}, \citenamefont {Tarnopolsky}, \citenamefont {Khalaf},\ and\ \citenamefont {Vishwanath}}]{Ledwith2020Chern}%
  \BibitemOpen
  \bibfield  {author} {\bibinfo {author} {\bibfnamefont {P.~J.}\ \bibnamefont {Ledwith}}, \bibinfo {author} {\bibfnamefont {G.}~\bibnamefont {Tarnopolsky}}, \bibinfo {author} {\bibfnamefont {E.}~\bibnamefont {Khalaf}},\ and\ \bibinfo {author} {\bibfnamefont {A.}~\bibnamefont {Vishwanath}},\ }\bibfield  {title} {\bibinfo {title} {Fractional chern insulator states in twisted bilayer graphene: An analytical approach},\ }\href {https://doi.org/10.1103/PhysRevResearch.2.023237} {\bibfield  {journal} {\bibinfo  {journal} {Phys. Rev. Research}\ }\textbf {\bibinfo {volume} {2}},\ \bibinfo {pages} {023237} (\bibinfo {year} {2020})}\BibitemShut {NoStop}%
\bibitem [{\citenamefont {Repellin}\ and\ \citenamefont {Senthil}(2020)}]{RepellinSenthil2020Chern}%
  \BibitemOpen
  \bibfield  {author} {\bibinfo {author} {\bibfnamefont {C.}~\bibnamefont {Repellin}}\ and\ \bibinfo {author} {\bibfnamefont {T.}~\bibnamefont {Senthil}},\ }\bibfield  {title} {\bibinfo {title} {Fractional chern insulators and spin phase transition},\ }\href {https://doi.org/10.1103/PhysRevResearch.2.023238} {\bibfield  {journal} {\bibinfo  {journal} {Phys. Rev. Research}\ }\textbf {\bibinfo {volume} {2}},\ \bibinfo {pages} {023238} (\bibinfo {year} {2020})}\BibitemShut {NoStop}%
\bibitem [{\citenamefont {Parker}\ \emph {et~al.}(2021)\citenamefont {Parker}, \citenamefont {Ledwith}, \citenamefont {Khalaf}, \citenamefont {Soejima}, \citenamefont {Hauschild}, \citenamefont {Xie}, \citenamefont {Pierce}, \citenamefont {Zaletel}, \citenamefont {Yacoby},\ and\ \citenamefont {Vishwanath}}]{Parker2021_arXiv2112tuned}%
  \BibitemOpen
  \bibfield  {author} {\bibinfo {author} {\bibfnamefont {D.}~\bibnamefont {Parker}}, \bibinfo {author} {\bibfnamefont {P.}~\bibnamefont {Ledwith}}, \bibinfo {author} {\bibfnamefont {E.}~\bibnamefont {Khalaf}}, \bibinfo {author} {\bibfnamefont {T.}~\bibnamefont {Soejima}}, \bibinfo {author} {\bibfnamefont {J.}~\bibnamefont {Hauschild}}, \bibinfo {author} {\bibfnamefont {Y.}~\bibnamefont {Xie}}, \bibinfo {author} {\bibfnamefont {A.}~\bibnamefont {Pierce}}, \bibinfo {author} {\bibfnamefont {M.~P.}\ \bibnamefont {Zaletel}}, \bibinfo {author} {\bibfnamefont {A.}~\bibnamefont {Yacoby}},\ and\ \bibinfo {author} {\bibfnamefont {A.}~\bibnamefont {Vishwanath}},\ }\href {https://arxiv.org/abs/2112.13837} {\bibinfo {title} {Field-tuned and zero-field fractional chern insulators in magic angle graphene}} (\bibinfo {year} {2021}),\ \Eprint {https://arxiv.org/abs/2112.13837} {arXiv:2112.13837 [cond-mat.str-el]} \BibitemShut {NoStop}%
\bibitem [{\citenamefont {Wilhelm}\ \emph {et~al.}(2021)\citenamefont {Wilhelm}, \citenamefont {Lang},\ and\ \citenamefont {L{\"a}uchli}}]{Wilhelm2021fractional}%
  \BibitemOpen
  \bibfield  {author} {\bibinfo {author} {\bibfnamefont {P.}~\bibnamefont {Wilhelm}}, \bibinfo {author} {\bibfnamefont {T.~C.}\ \bibnamefont {Lang}},\ and\ \bibinfo {author} {\bibfnamefont {A.~M.}\ \bibnamefont {L{\"a}uchli}},\ }\bibfield  {title} {\bibinfo {title} {Interplay of fractional chern insulator and charge density wave phases in twisted bilayer graphene},\ }\href {https://doi.org/10.1103/PhysRevB.103.125406} {\bibfield  {journal} {\bibinfo  {journal} {Phys. Rev. B}\ }\textbf {\bibinfo {volume} {103}},\ \bibinfo {pages} {125406} (\bibinfo {year} {2021})}\BibitemShut {NoStop}%
\bibitem [{\citenamefont {Sheffer}\ and\ \citenamefont {Stern}(2021)}]{ShefferStern2021magic-angle}%
  \BibitemOpen
  \bibfield  {author} {\bibinfo {author} {\bibfnamefont {Y.}~\bibnamefont {Sheffer}}\ and\ \bibinfo {author} {\bibfnamefont {A.}~\bibnamefont {Stern}},\ }\bibfield  {title} {\bibinfo {title} {Chiral magic-angle twisted bilayer graphene in a magnetic field},\ }\href {https://doi.org/10.1103/PhysRevB.104.L121405} {\bibfield  {journal} {\bibinfo  {journal} {Phys. Rev. B}\ }\textbf {\bibinfo {volume} {104}},\ \bibinfo {pages} {L121405} (\bibinfo {year} {2021})}\BibitemShut {NoStop}%
\bibitem [{\citenamefont {Li}\ \emph {et~al.}(2021)\citenamefont {Li}, \citenamefont {Kumar}, \citenamefont {Sun},\ and\ \citenamefont {Lin}}]{Li2021fractional}%
  \BibitemOpen
  \bibfield  {author} {\bibinfo {author} {\bibfnamefont {H.}~\bibnamefont {Li}}, \bibinfo {author} {\bibfnamefont {U.}~\bibnamefont {Kumar}}, \bibinfo {author} {\bibfnamefont {K.}~\bibnamefont {Sun}},\ and\ \bibinfo {author} {\bibfnamefont {S.-Z.}\ \bibnamefont {Lin}},\ }\bibfield  {title} {\bibinfo {title} {Spontaneous fractional chern insulators in transition metal dichalcogenide moir{\'e} superlattices},\ }\href {https://doi.org/10.1103/PhysRevResearch.3.L032070} {\bibfield  {journal} {\bibinfo  {journal} {Phys. Rev. Research}\ }\textbf {\bibinfo {volume} {3}},\ \bibinfo {pages} {L032070} (\bibinfo {year} {2021})}\BibitemShut {NoStop}%
\bibitem [{\citenamefont {Devakul}\ \emph {et~al.}(2021)\citenamefont {Devakul}, \citenamefont {Cr{\'e}pel}, \citenamefont {Zhang},\ and\ \citenamefont {Fu}}]{Devakul2021NatCommu}%
  \BibitemOpen
  \bibfield  {author} {\bibinfo {author} {\bibfnamefont {T.}~\bibnamefont {Devakul}}, \bibinfo {author} {\bibfnamefont {V.}~\bibnamefont {Cr{\'e}pel}}, \bibinfo {author} {\bibfnamefont {Y.}~\bibnamefont {Zhang}},\ and\ \bibinfo {author} {\bibfnamefont {L.}~\bibnamefont {Fu}},\ }\bibfield  {title} {\bibinfo {title} {Magic in twisted transition metal dichalcogenide bilayers},\ }\href {https://doi.org/10.1038/s41467-021-27042-9} {\bibfield  {journal} {\bibinfo  {journal} {Nature Communications}\ }\textbf {\bibinfo {volume} {12}},\ \bibinfo {pages} {6730} (\bibinfo {year} {2021})}\BibitemShut {NoStop}%
\bibitem [{\citenamefont {Yu}\ \emph {et~al.}(2020)\citenamefont {Yu}, \citenamefont {Chen},\ and\ \citenamefont {Yao}}]{Yu2020GiantBerryHomobilayer}%
  \BibitemOpen
  \bibfield  {author} {\bibinfo {author} {\bibfnamefont {H.}~\bibnamefont {Yu}}, \bibinfo {author} {\bibfnamefont {M.}~\bibnamefont {Chen}},\ and\ \bibinfo {author} {\bibfnamefont {W.}~\bibnamefont {Yao}},\ }\bibfield  {title} {\bibinfo {title} {Giant magnetic field from moiré induced berry phase in homobilayer semiconductors},\ }\href {https://doi.org/10.1093/nsr/nwz117} {\bibfield  {journal} {\bibinfo  {journal} {National Science Review}\ }\textbf {\bibinfo {volume} {7}},\ \bibinfo {pages} {12} (\bibinfo {year} {2020})}\BibitemShut {NoStop}%
\bibitem [{\citenamefont {Pan}\ \emph {et~al.}(2020)\citenamefont {Pan}, \citenamefont {Wu},\ and\ \citenamefont {Das~Sarma}}]{Pan2020WSe2BandTopology}%
  \BibitemOpen
  \bibfield  {author} {\bibinfo {author} {\bibfnamefont {H.}~\bibnamefont {Pan}}, \bibinfo {author} {\bibfnamefont {F.}~\bibnamefont {Wu}},\ and\ \bibinfo {author} {\bibfnamefont {S.}~\bibnamefont {Das~Sarma}},\ }\bibfield  {title} {\bibinfo {title} {Band topology, hubbard model, heisenberg model, and dzyaloshinskii-moriya interaction in twisted bilayer ${\mathrm{wse}}_{2}$},\ }\href {https://doi.org/10.1103/PhysRevResearch.2.033087} {\bibfield  {journal} {\bibinfo  {journal} {Phys. Rev. Res.}\ }\textbf {\bibinfo {volume} {2}},\ \bibinfo {pages} {033087} (\bibinfo {year} {2020})}\BibitemShut {NoStop}%
\bibitem [{\citenamefont {Zhang}\ \emph {et~al.}(2021)\citenamefont {Zhang}, \citenamefont {Liu},\ and\ \citenamefont {Fu}}]{Zhang2021ChargeOrderTMD}%
  \BibitemOpen
  \bibfield  {author} {\bibinfo {author} {\bibfnamefont {Y.}~\bibnamefont {Zhang}}, \bibinfo {author} {\bibfnamefont {T.}~\bibnamefont {Liu}},\ and\ \bibinfo {author} {\bibfnamefont {L.}~\bibnamefont {Fu}},\ }\bibfield  {title} {\bibinfo {title} {Electronic structures, charge transfer, and charge order in twisted transition metal dichalcogenide bilayers},\ }\href {https://doi.org/10.1103/PhysRevB.103.155142} {\bibfield  {journal} {\bibinfo  {journal} {Phys. Rev. B}\ }\textbf {\bibinfo {volume} {103}},\ \bibinfo {pages} {155142} (\bibinfo {year} {2021})}\BibitemShut {NoStop}%
\bibitem [{\citenamefont {Dong}\ \emph {et~al.}(2022)\citenamefont {Dong}, \citenamefont {Wang},\ and\ \citenamefont {Fu}}]{DongWangFu2022electron}%
  \BibitemOpen
  \bibfield  {author} {\bibinfo {author} {\bibfnamefont {J.}~\bibnamefont {Dong}}, \bibinfo {author} {\bibfnamefont {J.}~\bibnamefont {Wang}},\ and\ \bibinfo {author} {\bibfnamefont {L.}~\bibnamefont {Fu}},\ }\href {https://arxiv.org/abs/2208.10516} {\bibinfo {title} {Dirac electron under periodic magnetic field: Platform for fractional chern insulator and generalized wigner crystal}} (\bibinfo {year} {2022}),\ \Eprint {https://arxiv.org/abs/2208.10516} {arXiv:2208.10516 [cond-mat.mes-hall]} \BibitemShut {NoStop}%
\bibitem [{\citenamefont {Spanton}\ \emph {et~al.}(2018)\citenamefont {Spanton}, \citenamefont {Zibrov}, \citenamefont {Zhou}, \citenamefont {Taniguchi}, \citenamefont {Watanabe}, \citenamefont {Zaletel},\ and\ \citenamefont {Young}}]{Spanton2018fractional}%
  \BibitemOpen
  \bibfield  {author} {\bibinfo {author} {\bibfnamefont {E.~M.}\ \bibnamefont {Spanton}}, \bibinfo {author} {\bibfnamefont {A.~A.}\ \bibnamefont {Zibrov}}, \bibinfo {author} {\bibfnamefont {H.}~\bibnamefont {Zhou}}, \bibinfo {author} {\bibfnamefont {T.}~\bibnamefont {Taniguchi}}, \bibinfo {author} {\bibfnamefont {K.}~\bibnamefont {Watanabe}}, \bibinfo {author} {\bibfnamefont {M.~P.}\ \bibnamefont {Zaletel}},\ and\ \bibinfo {author} {\bibfnamefont {A.~F.}\ \bibnamefont {Young}},\ }\bibfield  {title} {\bibinfo {title} {Observation of fractional chern insulators in a van der waals heterostructure},\ }\href {https://doi.org/10.1126/science.aan8458} {\bibfield  {journal} {\bibinfo  {journal} {Science}\ }\textbf {\bibinfo {volume} {360}},\ \bibinfo {pages} {62} (\bibinfo {year} {2018})}\BibitemShut {NoStop}%
\bibitem [{\citenamefont {Xie}\ \emph {et~al.}(2021)\citenamefont {Xie}, \citenamefont {Pierce}, \citenamefont {Park}, \citenamefont {Parker}, \citenamefont {Khalaf}, \citenamefont {Ledwith}, \citenamefont {Cao}, \citenamefont {Lee}, \citenamefont {Chen}, \citenamefont {Forrester} \emph {et~al.}}]{Xie2021Chern}%
  \BibitemOpen
  \bibfield  {author} {\bibinfo {author} {\bibfnamefont {Y.}~\bibnamefont {Xie}}, \bibinfo {author} {\bibfnamefont {A.~T.}\ \bibnamefont {Pierce}}, \bibinfo {author} {\bibfnamefont {J.~M.}\ \bibnamefont {Park}}, \bibinfo {author} {\bibfnamefont {D.~E.}\ \bibnamefont {Parker}}, \bibinfo {author} {\bibfnamefont {E.}~\bibnamefont {Khalaf}}, \bibinfo {author} {\bibfnamefont {P.}~\bibnamefont {Ledwith}}, \bibinfo {author} {\bibfnamefont {Y.}~\bibnamefont {Cao}}, \bibinfo {author} {\bibfnamefont {S.~H.}\ \bibnamefont {Lee}}, \bibinfo {author} {\bibfnamefont {S.}~\bibnamefont {Chen}}, \bibinfo {author} {\bibfnamefont {P.~R.}\ \bibnamefont {Forrester}}, \emph {et~al.},\ }\bibfield  {title} {\bibinfo {title} {Fractional chern insulators in magic-angle twisted bilayer graphene},\ }\href {https://doi.org/10.1038/s41586-021-04002-3} {\bibfield  {journal} {\bibinfo  {journal} {Nature}\ }\textbf {\bibinfo {volume} {600}},\ \bibinfo {pages} {439} (\bibinfo {year} {2021})}\BibitemShut {NoStop}%
\bibitem [{\citenamefont {Cai}\ \emph {et~al.}(2023)\citenamefont {Cai}, \citenamefont {Anderson}, \citenamefont {Wang}, \citenamefont {Zhang}, \citenamefont {Liu}, \citenamefont {Holtzmann}, \citenamefont {Zhang}, \citenamefont {Fan}, \citenamefont {Taniguchi}, \citenamefont {Watanabe}, \citenamefont {Ran}, \citenamefont {Cao}, \citenamefont {Fu}, \citenamefont {Xiao}, \citenamefont {Yao},\ and\ \citenamefont {Xu}}]{cai2023signatures}%
  \BibitemOpen
  \bibfield  {author} {\bibinfo {author} {\bibfnamefont {J.}~\bibnamefont {Cai}}, \bibinfo {author} {\bibfnamefont {E.}~\bibnamefont {Anderson}}, \bibinfo {author} {\bibfnamefont {C.}~\bibnamefont {Wang}}, \bibinfo {author} {\bibfnamefont {X.}~\bibnamefont {Zhang}}, \bibinfo {author} {\bibfnamefont {X.}~\bibnamefont {Liu}}, \bibinfo {author} {\bibfnamefont {W.}~\bibnamefont {Holtzmann}}, \bibinfo {author} {\bibfnamefont {Y.}~\bibnamefont {Zhang}}, \bibinfo {author} {\bibfnamefont {F.}~\bibnamefont {Fan}}, \bibinfo {author} {\bibfnamefont {T.}~\bibnamefont {Taniguchi}}, \bibinfo {author} {\bibfnamefont {K.}~\bibnamefont {Watanabe}}, \bibinfo {author} {\bibfnamefont {Y.}~\bibnamefont {Ran}}, \bibinfo {author} {\bibfnamefont {T.}~\bibnamefont {Cao}}, \bibinfo {author} {\bibfnamefont {L.}~\bibnamefont {Fu}}, \bibinfo {author} {\bibfnamefont {D.}~\bibnamefont {Xiao}}, \bibinfo {author} {\bibfnamefont {W.}~\bibnamefont {Yao}},\ and\ \bibinfo {author} {\bibfnamefont {X.}~\bibnamefont {Xu}},\ }\bibfield  {title}
  {\bibinfo {title} {Signatures of fractional quantum anomalous hall states in twisted mote2},\ }\bibfield  {journal} {\bibinfo  {journal} {Nature}\ }\href {https://doi.org/10.1038/s41586-023-06289-w} {10.1038/s41586-023-06289-w} (\bibinfo {year} {2023})\BibitemShut {NoStop}%
\bibitem [{\citenamefont {Zeng}\ \emph {et~al.}(2023)\citenamefont {Zeng}, \citenamefont {Xia}, \citenamefont {Kang}, \citenamefont {Zhu}, \citenamefont {Kn{\"u}ppel}, \citenamefont {Vaswani}, \citenamefont {Watanabe}, \citenamefont {Taniguchi}, \citenamefont {Mak},\ and\ \citenamefont {Shan}}]{zeng2023integer}%
  \BibitemOpen
  \bibfield  {author} {\bibinfo {author} {\bibfnamefont {Y.}~\bibnamefont {Zeng}}, \bibinfo {author} {\bibfnamefont {Z.}~\bibnamefont {Xia}}, \bibinfo {author} {\bibfnamefont {K.}~\bibnamefont {Kang}}, \bibinfo {author} {\bibfnamefont {J.}~\bibnamefont {Zhu}}, \bibinfo {author} {\bibfnamefont {P.}~\bibnamefont {Kn{\"u}ppel}}, \bibinfo {author} {\bibfnamefont {C.}~\bibnamefont {Vaswani}}, \bibinfo {author} {\bibfnamefont {K.}~\bibnamefont {Watanabe}}, \bibinfo {author} {\bibfnamefont {T.}~\bibnamefont {Taniguchi}}, \bibinfo {author} {\bibfnamefont {K.~F.}\ \bibnamefont {Mak}},\ and\ \bibinfo {author} {\bibfnamefont {J.}~\bibnamefont {Shan}},\ }\bibfield  {title} {\bibinfo {title} {Thermodynamic evidence of fractional chern insulator in moir{\'e} mote2},\ }\bibfield  {journal} {\bibinfo  {journal} {Nature}\ }\href {https://doi.org/10.1038/s41586-023-06452-3} {10.1038/s41586-023-06452-3} (\bibinfo {year} {2023})\BibitemShut {NoStop}%
\bibitem [{\citenamefont {Park}\ \emph {et~al.}(2023)\citenamefont {Park}, \citenamefont {Cai}, \citenamefont {Anderson}, \citenamefont {Zhang}, \citenamefont {Zhu}, \citenamefont {Liu}, \citenamefont {Wang}, \citenamefont {Holtzmann}, \citenamefont {Hu}, \citenamefont {Liu} \emph {et~al.}}]{park2023observation}%
  \BibitemOpen
  \bibfield  {author} {\bibinfo {author} {\bibfnamefont {H.}~\bibnamefont {Park}}, \bibinfo {author} {\bibfnamefont {J.}~\bibnamefont {Cai}}, \bibinfo {author} {\bibfnamefont {E.}~\bibnamefont {Anderson}}, \bibinfo {author} {\bibfnamefont {Y.}~\bibnamefont {Zhang}}, \bibinfo {author} {\bibfnamefont {J.}~\bibnamefont {Zhu}}, \bibinfo {author} {\bibfnamefont {X.}~\bibnamefont {Liu}}, \bibinfo {author} {\bibfnamefont {C.}~\bibnamefont {Wang}}, \bibinfo {author} {\bibfnamefont {W.}~\bibnamefont {Holtzmann}}, \bibinfo {author} {\bibfnamefont {C.}~\bibnamefont {Hu}}, \bibinfo {author} {\bibfnamefont {Z.}~\bibnamefont {Liu}}, \emph {et~al.},\ }\bibfield  {title} {\bibinfo {title} {Observation of fractionally quantized anomalous hall effect},\ }\href {https://doi.org/10.1038/s41586-023-06536-0} {\bibfield  {journal} {\bibinfo  {journal} {Nature}\ }\textbf {\bibinfo {volume} {622}},\ \bibinfo {pages} {74} (\bibinfo {year} {2023})}\BibitemShut {NoStop}%
\bibitem [{\citenamefont {Xu}\ \emph {et~al.}(2023)\citenamefont {Xu}, \citenamefont {Sun}, \citenamefont {Jia}, \citenamefont {Liu}, \citenamefont {Xu}, \citenamefont {Li}, \citenamefont {Gu}, \citenamefont {Watanabe}, \citenamefont {Taniguchi}, \citenamefont {Tong}, \citenamefont {Jia}, \citenamefont {Shi}, \citenamefont {Jiang}, \citenamefont {Zhang}, \citenamefont {Liu},\ and\ \citenamefont {Li}}]{Xu2023FCItMoTe2}%
  \BibitemOpen
  \bibfield  {author} {\bibinfo {author} {\bibfnamefont {F.}~\bibnamefont {Xu}}, \bibinfo {author} {\bibfnamefont {Z.}~\bibnamefont {Sun}}, \bibinfo {author} {\bibfnamefont {T.}~\bibnamefont {Jia}}, \bibinfo {author} {\bibfnamefont {C.}~\bibnamefont {Liu}}, \bibinfo {author} {\bibfnamefont {C.}~\bibnamefont {Xu}}, \bibinfo {author} {\bibfnamefont {C.}~\bibnamefont {Li}}, \bibinfo {author} {\bibfnamefont {Y.}~\bibnamefont {Gu}}, \bibinfo {author} {\bibfnamefont {K.}~\bibnamefont {Watanabe}}, \bibinfo {author} {\bibfnamefont {T.}~\bibnamefont {Taniguchi}}, \bibinfo {author} {\bibfnamefont {B.}~\bibnamefont {Tong}}, \bibinfo {author} {\bibfnamefont {J.}~\bibnamefont {Jia}}, \bibinfo {author} {\bibfnamefont {Z.}~\bibnamefont {Shi}}, \bibinfo {author} {\bibfnamefont {S.}~\bibnamefont {Jiang}}, \bibinfo {author} {\bibfnamefont {Y.}~\bibnamefont {Zhang}}, \bibinfo {author} {\bibfnamefont {X.}~\bibnamefont {Liu}},\ and\ \bibinfo {author} {\bibfnamefont {T.}~\bibnamefont {Li}},\ }\bibfield  {title} {\bibinfo {title}
  {Observation of integer and fractional quantum anomalous hall effects in twisted bilayer ${\mathrm{mote}}_{2}$},\ }\href {https://doi.org/10.1103/PhysRevX.13.031037} {\bibfield  {journal} {\bibinfo  {journal} {Phys. Rev. X}\ }\textbf {\bibinfo {volume} {13}},\ \bibinfo {pages} {031037} (\bibinfo {year} {2023})}\BibitemShut {NoStop}%
\bibitem [{\citenamefont {Ji}\ \emph {et~al.}(2024)\citenamefont {Ji}, \citenamefont {Park}, \citenamefont {Barber}, \citenamefont {Hu}, \citenamefont {Watanabe}, \citenamefont {Taniguchi}, \citenamefont {Chu}, \citenamefont {Xu},\ and\ \citenamefont {Shen}}]{Ji2024LocalProbetMoTe2}%
  \BibitemOpen
  \bibfield  {author} {\bibinfo {author} {\bibfnamefont {Z.}~\bibnamefont {Ji}}, \bibinfo {author} {\bibfnamefont {H.}~\bibnamefont {Park}}, \bibinfo {author} {\bibfnamefont {M.~E.}\ \bibnamefont {Barber}}, \bibinfo {author} {\bibfnamefont {C.}~\bibnamefont {Hu}}, \bibinfo {author} {\bibfnamefont {K.}~\bibnamefont {Watanabe}}, \bibinfo {author} {\bibfnamefont {T.}~\bibnamefont {Taniguchi}}, \bibinfo {author} {\bibfnamefont {J.-H.}\ \bibnamefont {Chu}}, \bibinfo {author} {\bibfnamefont {X.}~\bibnamefont {Xu}},\ and\ \bibinfo {author} {\bibfnamefont {Z.-X.}\ \bibnamefont {Shen}},\ }\bibfield  {title} {\bibinfo {title} {Local probe of bulk and edge states in a fractional chern insulator},\ }\href {https://doi.org/10.1038/s41586-024-08092-7} {\bibfield  {journal} {\bibinfo  {journal} {Nature}\ }\textbf {\bibinfo {volume} {635}},\ \bibinfo {pages} {578} (\bibinfo {year} {2024})}\BibitemShut {NoStop}%
\bibitem [{\citenamefont {Redekop}\ \emph {et~al.}(2024)\citenamefont {Redekop}, \citenamefont {Zhang}, \citenamefont {Park}, \citenamefont {Cai}, \citenamefont {Anderson}, \citenamefont {Sheekey}, \citenamefont {Arp}, \citenamefont {Babikyan}, \citenamefont {Salters}, \citenamefont {Watanabe}, \citenamefont {Taniguchi}, \citenamefont {Huber}, \citenamefont {Xu},\ and\ \citenamefont {Young}}]{Young2024MagtMoTe2}%
  \BibitemOpen
  \bibfield  {author} {\bibinfo {author} {\bibfnamefont {E.}~\bibnamefont {Redekop}}, \bibinfo {author} {\bibfnamefont {C.}~\bibnamefont {Zhang}}, \bibinfo {author} {\bibfnamefont {H.}~\bibnamefont {Park}}, \bibinfo {author} {\bibfnamefont {J.}~\bibnamefont {Cai}}, \bibinfo {author} {\bibfnamefont {E.}~\bibnamefont {Anderson}}, \bibinfo {author} {\bibfnamefont {O.}~\bibnamefont {Sheekey}}, \bibinfo {author} {\bibfnamefont {T.}~\bibnamefont {Arp}}, \bibinfo {author} {\bibfnamefont {G.}~\bibnamefont {Babikyan}}, \bibinfo {author} {\bibfnamefont {S.}~\bibnamefont {Salters}}, \bibinfo {author} {\bibfnamefont {K.}~\bibnamefont {Watanabe}}, \bibinfo {author} {\bibfnamefont {T.}~\bibnamefont {Taniguchi}}, \bibinfo {author} {\bibfnamefont {M.~E.}\ \bibnamefont {Huber}}, \bibinfo {author} {\bibfnamefont {X.}~\bibnamefont {Xu}},\ and\ \bibinfo {author} {\bibfnamefont {A.~F.}\ \bibnamefont {Young}},\ }\bibfield  {title} {\bibinfo {title} {Direct magnetic imaging of fractional chern insulators in twisted mote2},\ }\href
  {https://doi.org/10.1038/s41586-024-08153-x} {\bibfield  {journal} {\bibinfo  {journal} {Nature}\ }\textbf {\bibinfo {volume} {635}},\ \bibinfo {pages} {584} (\bibinfo {year} {2024})}\BibitemShut {NoStop}%
\bibitem [{\citenamefont {Xu}\ \emph {et~al.}(2025{\natexlab{a}})\citenamefont {Xu}, \citenamefont {Chang}, \citenamefont {Xiao}, \citenamefont {Zhang}, \citenamefont {Liu}, \citenamefont {Sun}, \citenamefont {Mao}, \citenamefont {Peshcherenko}, \citenamefont {Li}, \citenamefont {Watanabe}, \citenamefont {Taniguchi}, \citenamefont {Tong}, \citenamefont {Lu}, \citenamefont {Jia}, \citenamefont {Qian}, \citenamefont {Shi}, \citenamefont {Zhang}, \citenamefont {Liu}, \citenamefont {Jiang},\ and\ \citenamefont {Li}}]{Xu2024tMoTe2_3.15}%
  \BibitemOpen
  \bibfield  {author} {\bibinfo {author} {\bibfnamefont {F.}~\bibnamefont {Xu}}, \bibinfo {author} {\bibfnamefont {X.}~\bibnamefont {Chang}}, \bibinfo {author} {\bibfnamefont {J.}~\bibnamefont {Xiao}}, \bibinfo {author} {\bibfnamefont {Y.}~\bibnamefont {Zhang}}, \bibinfo {author} {\bibfnamefont {F.}~\bibnamefont {Liu}}, \bibinfo {author} {\bibfnamefont {Z.}~\bibnamefont {Sun}}, \bibinfo {author} {\bibfnamefont {N.}~\bibnamefont {Mao}}, \bibinfo {author} {\bibfnamefont {N.}~\bibnamefont {Peshcherenko}}, \bibinfo {author} {\bibfnamefont {J.}~\bibnamefont {Li}}, \bibinfo {author} {\bibfnamefont {K.}~\bibnamefont {Watanabe}}, \bibinfo {author} {\bibfnamefont {T.}~\bibnamefont {Taniguchi}}, \bibinfo {author} {\bibfnamefont {B.}~\bibnamefont {Tong}}, \bibinfo {author} {\bibfnamefont {L.}~\bibnamefont {Lu}}, \bibinfo {author} {\bibfnamefont {J.}~\bibnamefont {Jia}}, \bibinfo {author} {\bibfnamefont {D.}~\bibnamefont {Qian}}, \bibinfo {author} {\bibfnamefont {Z.}~\bibnamefont {Shi}}, \bibinfo {author} {\bibfnamefont
  {Y.}~\bibnamefont {Zhang}}, \bibinfo {author} {\bibfnamefont {X.}~\bibnamefont {Liu}}, \bibinfo {author} {\bibfnamefont {S.}~\bibnamefont {Jiang}},\ and\ \bibinfo {author} {\bibfnamefont {T.}~\bibnamefont {Li}},\ }\bibfield  {title} {\bibinfo {title} {Interplay between topology and correlations in the second moir{\'e} band of twisted bilayer mote2},\ }\href {https://doi.org/10.1038/s41567-025-02803-1} {\bibfield  {journal} {\bibinfo  {journal} {Nature Physics}\ }\textbf {\bibinfo {volume} {21}},\ \bibinfo {pages} {542} (\bibinfo {year} {2025}{\natexlab{a}})}\BibitemShut {NoStop}%
\bibitem [{\citenamefont {Kang}\ \emph {et~al.}(2024)\citenamefont {Kang}, \citenamefont {Shen}, \citenamefont {Qiu}, \citenamefont {Zeng}, \citenamefont {Xia}, \citenamefont {Watanabe}, \citenamefont {Taniguchi}, \citenamefont {Shan},\ and\ \citenamefont {Mak}}]{Kang2024_tMoTe2_2.13}%
  \BibitemOpen
  \bibfield  {author} {\bibinfo {author} {\bibfnamefont {K.}~\bibnamefont {Kang}}, \bibinfo {author} {\bibfnamefont {B.}~\bibnamefont {Shen}}, \bibinfo {author} {\bibfnamefont {Y.}~\bibnamefont {Qiu}}, \bibinfo {author} {\bibfnamefont {Y.}~\bibnamefont {Zeng}}, \bibinfo {author} {\bibfnamefont {Z.}~\bibnamefont {Xia}}, \bibinfo {author} {\bibfnamefont {K.}~\bibnamefont {Watanabe}}, \bibinfo {author} {\bibfnamefont {T.}~\bibnamefont {Taniguchi}}, \bibinfo {author} {\bibfnamefont {J.}~\bibnamefont {Shan}},\ and\ \bibinfo {author} {\bibfnamefont {K.~F.}\ \bibnamefont {Mak}},\ }\bibfield  {title} {\bibinfo {title} {Evidence of the fractional quantum spin hall effect in moir{\'e} mote2},\ }\href {https://doi.org/10.1038/s41586-024-07214-5} {\bibfield  {journal} {\bibinfo  {journal} {Nature}\ }\textbf {\bibinfo {volume} {628}},\ \bibinfo {pages} {522} (\bibinfo {year} {2024})}\BibitemShut {NoStop}%
\bibitem [{\citenamefont {Park}\ \emph {et~al.}(2025{\natexlab{a}})\citenamefont {Park}, \citenamefont {Cai}, \citenamefont {Anderson}, \citenamefont {Zhang}, \citenamefont {Liu}, \citenamefont {Holtzmann}, \citenamefont {Li}, \citenamefont {Wang}, \citenamefont {Hu}, \citenamefont {Zhao}, \citenamefont {Taniguchi}, \citenamefont {Watanabe}, \citenamefont {Yang}, \citenamefont {Cobden}, \citenamefont {Chu}, \citenamefont {Regnault}, \citenamefont {Bernevig}, \citenamefont {Fu}, \citenamefont {Cao}, \citenamefont {Xiao},\ and\ \citenamefont {Xu}}]{Park2024tMoTe2_2.6_3.8}%
  \BibitemOpen
  \bibfield  {author} {\bibinfo {author} {\bibfnamefont {H.}~\bibnamefont {Park}}, \bibinfo {author} {\bibfnamefont {J.}~\bibnamefont {Cai}}, \bibinfo {author} {\bibfnamefont {E.}~\bibnamefont {Anderson}}, \bibinfo {author} {\bibfnamefont {X.-W.}\ \bibnamefont {Zhang}}, \bibinfo {author} {\bibfnamefont {X.}~\bibnamefont {Liu}}, \bibinfo {author} {\bibfnamefont {W.}~\bibnamefont {Holtzmann}}, \bibinfo {author} {\bibfnamefont {W.}~\bibnamefont {Li}}, \bibinfo {author} {\bibfnamefont {C.}~\bibnamefont {Wang}}, \bibinfo {author} {\bibfnamefont {C.}~\bibnamefont {Hu}}, \bibinfo {author} {\bibfnamefont {Y.}~\bibnamefont {Zhao}}, \bibinfo {author} {\bibfnamefont {T.}~\bibnamefont {Taniguchi}}, \bibinfo {author} {\bibfnamefont {K.}~\bibnamefont {Watanabe}}, \bibinfo {author} {\bibfnamefont {J.}~\bibnamefont {Yang}}, \bibinfo {author} {\bibfnamefont {D.}~\bibnamefont {Cobden}}, \bibinfo {author} {\bibfnamefont {J.-h.}\ \bibnamefont {Chu}}, \bibinfo {author} {\bibfnamefont {N.}~\bibnamefont {Regnault}}, \bibinfo
  {author} {\bibfnamefont {B.~A.}\ \bibnamefont {Bernevig}}, \bibinfo {author} {\bibfnamefont {L.}~\bibnamefont {Fu}}, \bibinfo {author} {\bibfnamefont {T.}~\bibnamefont {Cao}}, \bibinfo {author} {\bibfnamefont {D.}~\bibnamefont {Xiao}},\ and\ \bibinfo {author} {\bibfnamefont {X.}~\bibnamefont {Xu}},\ }\bibfield  {title} {\bibinfo {title} {Ferromagnetism and topology of the higher flat band in a fractional chern insulator},\ }\href {https://doi.org/10.1038/s41567-025-02804-0} {\bibfield  {journal} {\bibinfo  {journal} {Nature Physics}\ }\textbf {\bibinfo {volume} {21}},\ \bibinfo {pages} {549} (\bibinfo {year} {2025}{\natexlab{a}})}\BibitemShut {NoStop}%
\bibitem [{\citenamefont {Park}\ \emph {et~al.}(2025{\natexlab{b}})\citenamefont {Park}, \citenamefont {Li}, \citenamefont {Hu}, \citenamefont {Beach}, \citenamefont {Gonçalves}, \citenamefont {Mendez-Valderrama}, \citenamefont {Herzog-Arbeitman}, \citenamefont {Taniguchi}, \citenamefont {Watanabe}, \citenamefont {Cobden}, \citenamefont {Fu}, \citenamefont {Bernevig}, \citenamefont {Regnault}, \citenamefont {Chu}, \citenamefont {Xiao},\ and\ \citenamefont {Xu}}]{park2025obsfci}%
  \BibitemOpen
  \bibfield  {author} {\bibinfo {author} {\bibfnamefont {H.}~\bibnamefont {Park}}, \bibinfo {author} {\bibfnamefont {W.}~\bibnamefont {Li}}, \bibinfo {author} {\bibfnamefont {C.}~\bibnamefont {Hu}}, \bibinfo {author} {\bibfnamefont {C.}~\bibnamefont {Beach}}, \bibinfo {author} {\bibfnamefont {M.}~\bibnamefont {Gonçalves}}, \bibinfo {author} {\bibfnamefont {J.~F.}\ \bibnamefont {Mendez-Valderrama}}, \bibinfo {author} {\bibfnamefont {J.}~\bibnamefont {Herzog-Arbeitman}}, \bibinfo {author} {\bibfnamefont {T.}~\bibnamefont {Taniguchi}}, \bibinfo {author} {\bibfnamefont {K.}~\bibnamefont {Watanabe}}, \bibinfo {author} {\bibfnamefont {D.}~\bibnamefont {Cobden}}, \bibinfo {author} {\bibfnamefont {L.}~\bibnamefont {Fu}}, \bibinfo {author} {\bibfnamefont {B.~A.}\ \bibnamefont {Bernevig}}, \bibinfo {author} {\bibfnamefont {N.}~\bibnamefont {Regnault}}, \bibinfo {author} {\bibfnamefont {J.-H.}\ \bibnamefont {Chu}}, \bibinfo {author} {\bibfnamefont {D.}~\bibnamefont {Xiao}},\ and\ \bibinfo {author} {\bibfnamefont
  {X.}~\bibnamefont {Xu}},\ }\href {https://arxiv.org/abs/2503.10989} {\bibinfo {title} {Observation of high-temperature dissipationless fractional chern insulator}} (\bibinfo {year} {2025}{\natexlab{b}}),\ \Eprint {https://arxiv.org/abs/2503.10989} {arXiv:2503.10989 [cond-mat.mes-hall]} \BibitemShut {NoStop}%
\bibitem [{\citenamefont {Xu}\ \emph {et~al.}(2025{\natexlab{b}})\citenamefont {Xu}, \citenamefont {Sun}, \citenamefont {Li}, \citenamefont {Zheng}, \citenamefont {Xu}, \citenamefont {Gao}, \citenamefont {Jia}, \citenamefont {Watanabe}, \citenamefont {Taniguchi}, \citenamefont {Tong}, \citenamefont {Lu}, \citenamefont {Jia}, \citenamefont {Shi}, \citenamefont {Jiang}, \citenamefont {Zhang}, \citenamefont {Zhang}, \citenamefont {Lei}, \citenamefont {Liu},\ and\ \citenamefont {Li}}]{xu_txl2025FCI}%
  \BibitemOpen
  \bibfield  {author} {\bibinfo {author} {\bibfnamefont {F.}~\bibnamefont {Xu}}, \bibinfo {author} {\bibfnamefont {Z.}~\bibnamefont {Sun}}, \bibinfo {author} {\bibfnamefont {J.}~\bibnamefont {Li}}, \bibinfo {author} {\bibfnamefont {C.}~\bibnamefont {Zheng}}, \bibinfo {author} {\bibfnamefont {C.}~\bibnamefont {Xu}}, \bibinfo {author} {\bibfnamefont {J.}~\bibnamefont {Gao}}, \bibinfo {author} {\bibfnamefont {T.}~\bibnamefont {Jia}}, \bibinfo {author} {\bibfnamefont {K.}~\bibnamefont {Watanabe}}, \bibinfo {author} {\bibfnamefont {T.}~\bibnamefont {Taniguchi}}, \bibinfo {author} {\bibfnamefont {B.}~\bibnamefont {Tong}}, \bibinfo {author} {\bibfnamefont {L.}~\bibnamefont {Lu}}, \bibinfo {author} {\bibfnamefont {J.}~\bibnamefont {Jia}}, \bibinfo {author} {\bibfnamefont {Z.}~\bibnamefont {Shi}}, \bibinfo {author} {\bibfnamefont {S.}~\bibnamefont {Jiang}}, \bibinfo {author} {\bibfnamefont {Y.}~\bibnamefont {Zhang}}, \bibinfo {author} {\bibfnamefont {Y.}~\bibnamefont {Zhang}}, \bibinfo {author} {\bibfnamefont
  {S.}~\bibnamefont {Lei}}, \bibinfo {author} {\bibfnamefont {X.}~\bibnamefont {Liu}},\ and\ \bibinfo {author} {\bibfnamefont {T.}~\bibnamefont {Li}},\ }\href {https://arxiv.org/abs/2504.06972} {\bibinfo {title} {Signatures of unconventional superconductivity near reentrant and fractional quantum anomalous hall insulators}} (\bibinfo {year} {2025}{\natexlab{b}}),\ \Eprint {https://arxiv.org/abs/2504.06972} {arXiv:2504.06972 [cond-mat.mes-hall]} \BibitemShut {NoStop}%
\bibitem [{\citenamefont {Sun}\ \emph {et~al.}(2026)\citenamefont {Sun}, \citenamefont {Xu}, \citenamefont {Li}, \citenamefont {Jiang}, \citenamefont {Gao}, \citenamefont {Xu}, \citenamefont {Jia}, \citenamefont {Cheng}, \citenamefont {Zhang}, \citenamefont {Tian}, \citenamefont {Watanabe}, \citenamefont {Taniguchi}, \citenamefont {Jia}, \citenamefont {Jiang}, \citenamefont {Zhang}, \citenamefont {Zhang}, \citenamefont {Lei}, \citenamefont {Liu},\ and\ \citenamefont {Li}}]{sun2026twistangleevolutionvalleypolarizedfractional}%
  \BibitemOpen
  \bibfield  {author} {\bibinfo {author} {\bibfnamefont {Z.}~\bibnamefont {Sun}}, \bibinfo {author} {\bibfnamefont {F.}~\bibnamefont {Xu}}, \bibinfo {author} {\bibfnamefont {J.}~\bibnamefont {Li}}, \bibinfo {author} {\bibfnamefont {Y.}~\bibnamefont {Jiang}}, \bibinfo {author} {\bibfnamefont {J.}~\bibnamefont {Gao}}, \bibinfo {author} {\bibfnamefont {C.}~\bibnamefont {Xu}}, \bibinfo {author} {\bibfnamefont {T.}~\bibnamefont {Jia}}, \bibinfo {author} {\bibfnamefont {K.}~\bibnamefont {Cheng}}, \bibinfo {author} {\bibfnamefont {J.}~\bibnamefont {Zhang}}, \bibinfo {author} {\bibfnamefont {W.}~\bibnamefont {Tian}}, \bibinfo {author} {\bibfnamefont {K.}~\bibnamefont {Watanabe}}, \bibinfo {author} {\bibfnamefont {T.}~\bibnamefont {Taniguchi}}, \bibinfo {author} {\bibfnamefont {J.}~\bibnamefont {Jia}}, \bibinfo {author} {\bibfnamefont {S.}~\bibnamefont {Jiang}}, \bibinfo {author} {\bibfnamefont {Y.}~\bibnamefont {Zhang}}, \bibinfo {author} {\bibfnamefont {Y.}~\bibnamefont {Zhang}}, \bibinfo {author} {\bibfnamefont
  {S.}~\bibnamefont {Lei}}, \bibinfo {author} {\bibfnamefont {X.}~\bibnamefont {Liu}},\ and\ \bibinfo {author} {\bibfnamefont {T.}~\bibnamefont {Li}},\ }\href {https://arxiv.org/abs/2603.16412} {\bibinfo {title} {Twist-angle evolution from valley-polarized fractional topological phases to valley-degenerate superconductivity in twisted bilayer mote2}} (\bibinfo {year} {2026}),\ \Eprint {https://arxiv.org/abs/2603.16412} {arXiv:2603.16412 [cond-mat.mes-hall]} \BibitemShut {NoStop}%
\bibitem [{\citenamefont {Chang}\ \emph {et~al.}(2026)\citenamefont {Chang}, \citenamefont {Liu}, \citenamefont {Xu}, \citenamefont {Xu}, \citenamefont {Xiao}, \citenamefont {Sun}, \citenamefont {Jiao}, \citenamefont {Zhang}, \citenamefont {Wang}, \citenamefont {Shen}, \citenamefont {He}, \citenamefont {Watanabe}, \citenamefont {Taniguchi}, \citenamefont {Zhong}, \citenamefont {Jia}, \citenamefont {Shi}, \citenamefont {Liu}, \citenamefont {Zhang}, \citenamefont {Qian}, \citenamefont {Li},\ and\ \citenamefont {Jiang}}]{Chang2026ecgs}%
  \BibitemOpen
  \bibfield  {author} {\bibinfo {author} {\bibfnamefont {X.}~\bibnamefont {Chang}}, \bibinfo {author} {\bibfnamefont {F.}~\bibnamefont {Liu}}, \bibinfo {author} {\bibfnamefont {F.}~\bibnamefont {Xu}}, \bibinfo {author} {\bibfnamefont {C.}~\bibnamefont {Xu}}, \bibinfo {author} {\bibfnamefont {J.}~\bibnamefont {Xiao}}, \bibinfo {author} {\bibfnamefont {Z.}~\bibnamefont {Sun}}, \bibinfo {author} {\bibfnamefont {P.}~\bibnamefont {Jiao}}, \bibinfo {author} {\bibfnamefont {Y.}~\bibnamefont {Zhang}}, \bibinfo {author} {\bibfnamefont {S.}~\bibnamefont {Wang}}, \bibinfo {author} {\bibfnamefont {B.}~\bibnamefont {Shen}}, \bibinfo {author} {\bibfnamefont {R.}~\bibnamefont {He}}, \bibinfo {author} {\bibfnamefont {K.}~\bibnamefont {Watanabe}}, \bibinfo {author} {\bibfnamefont {T.}~\bibnamefont {Taniguchi}}, \bibinfo {author} {\bibfnamefont {R.}~\bibnamefont {Zhong}}, \bibinfo {author} {\bibfnamefont {J.}~\bibnamefont {Jia}}, \bibinfo {author} {\bibfnamefont {Z.}~\bibnamefont {Shi}}, \bibinfo {author} {\bibfnamefont
  {X.}~\bibnamefont {Liu}}, \bibinfo {author} {\bibfnamefont {Y.}~\bibnamefont {Zhang}}, \bibinfo {author} {\bibfnamefont {D.}~\bibnamefont {Qian}}, \bibinfo {author} {\bibfnamefont {T.}~\bibnamefont {Li}},\ and\ \bibinfo {author} {\bibfnamefont {S.}~\bibnamefont {Jiang}},\ }\bibfield  {title} {\bibinfo {title} {Evidence of competing ground states between fractional chern insulator and antiferromagnetism in moir{\'e} mote2},\ }\href {https://doi.org/10.1038/s41467-026-71479-9} {\bibfield  {journal} {\bibinfo  {journal} {Nature Communications}\ }\textbf {\bibinfo {volume} {17}},\ \bibinfo {pages} {4874} (\bibinfo {year} {2026})}\BibitemShut {NoStop}%
\bibitem [{\citenamefont {Li}\ \emph {et~al.}(2026{\natexlab{a}})\citenamefont {Li}, \citenamefont {Wang~Beach}, \citenamefont {Hu}, \citenamefont {Taniguchi}, \citenamefont {Watanabe}, \citenamefont {Chu}, \citenamefont {Imamo{\u{g}}lu}, \citenamefont {Cao}, \citenamefont {Xiao},\ and\ \citenamefont {Xu}}]{LiXiaodong2026sfc}%
  \BibitemOpen
  \bibfield  {author} {\bibinfo {author} {\bibfnamefont {W.}~\bibnamefont {Li}}, \bibinfo {author} {\bibfnamefont {C.}~\bibnamefont {Wang~Beach}}, \bibinfo {author} {\bibfnamefont {C.}~\bibnamefont {Hu}}, \bibinfo {author} {\bibfnamefont {T.}~\bibnamefont {Taniguchi}}, \bibinfo {author} {\bibfnamefont {K.}~\bibnamefont {Watanabe}}, \bibinfo {author} {\bibfnamefont {J.-H.}\ \bibnamefont {Chu}}, \bibinfo {author} {\bibfnamefont {A.}~\bibnamefont {Imamo{\u{g}}lu}}, \bibinfo {author} {\bibfnamefont {T.}~\bibnamefont {Cao}}, \bibinfo {author} {\bibfnamefont {D.}~\bibnamefont {Xiao}},\ and\ \bibinfo {author} {\bibfnamefont {X.}~\bibnamefont {Xu}},\ }\bibfield  {title} {\bibinfo {title} {Signatures of fractional charges via anyon--trions in twisted mote2},\ }\href {https://doi.org/10.1038/s41586-026-10101-w} {\bibfield  {journal} {\bibinfo  {journal} {Nature}\ }\textbf {\bibinfo {volume} {651}},\ \bibinfo {pages} {48} (\bibinfo {year} {2026}{\natexlab{a}})}\BibitemShut {NoStop}%
\bibitem [{\citenamefont {Wang}\ \emph {et~al.}(2025{\natexlab{a}})\citenamefont {Wang}, \citenamefont {Choe}, \citenamefont {Anderson}, \citenamefont {Li}, \citenamefont {Ingham}, \citenamefont {Arsenault}, \citenamefont {Li}, \citenamefont {Hu}, \citenamefont {Taniguchi}, \citenamefont {Watanabe}, \citenamefont {Roy}, \citenamefont {Basov}, \citenamefont {Xiao}, \citenamefont {Queiroz}, \citenamefont {Hone}, \citenamefont {Xu},\ and\ \citenamefont {Zhu}}]{WangHidden2025}%
  \BibitemOpen
  \bibfield  {author} {\bibinfo {author} {\bibfnamefont {Y.}~\bibnamefont {Wang}}, \bibinfo {author} {\bibfnamefont {J.}~\bibnamefont {Choe}}, \bibinfo {author} {\bibfnamefont {E.}~\bibnamefont {Anderson}}, \bibinfo {author} {\bibfnamefont {W.}~\bibnamefont {Li}}, \bibinfo {author} {\bibfnamefont {J.}~\bibnamefont {Ingham}}, \bibinfo {author} {\bibfnamefont {E.~A.}\ \bibnamefont {Arsenault}}, \bibinfo {author} {\bibfnamefont {Y.}~\bibnamefont {Li}}, \bibinfo {author} {\bibfnamefont {X.}~\bibnamefont {Hu}}, \bibinfo {author} {\bibfnamefont {T.}~\bibnamefont {Taniguchi}}, \bibinfo {author} {\bibfnamefont {K.}~\bibnamefont {Watanabe}}, \bibinfo {author} {\bibfnamefont {X.}~\bibnamefont {Roy}}, \bibinfo {author} {\bibfnamefont {D.}~\bibnamefont {Basov}}, \bibinfo {author} {\bibfnamefont {D.}~\bibnamefont {Xiao}}, \bibinfo {author} {\bibfnamefont {R.}~\bibnamefont {Queiroz}}, \bibinfo {author} {\bibfnamefont {J.~C.}\ \bibnamefont {Hone}}, \bibinfo {author} {\bibfnamefont {X.}~\bibnamefont {Xu}},\ and\ \bibinfo
  {author} {\bibfnamefont {X.-Y.}\ \bibnamefont {Zhu}},\ }\bibfield  {title} {\bibinfo {title} {Hidden states and dynamics of fractional fillings in twisted {MoTe$_2$} bilayers},\ }\href {https://doi.org/10.1038/s41586-025-08954-8} {\bibfield  {journal} {\bibinfo  {journal} {Nature}\ }\textbf {\bibinfo {volume} {641}},\ \bibinfo {pages} {1149} (\bibinfo {year} {2025}{\natexlab{a}})}\BibitemShut {NoStop}%
\bibitem [{\citenamefont {Lu}\ \emph {et~al.}(2024)\citenamefont {Lu}, \citenamefont {Han}, \citenamefont {Yao}, \citenamefont {Reddy}, \citenamefont {Yang}, \citenamefont {Seo}, \citenamefont {Watanabe}, \citenamefont {Taniguchi}, \citenamefont {Fu},\ and\ \citenamefont {Ju}}]{Lu2024PGexp}%
  \BibitemOpen
  \bibfield  {author} {\bibinfo {author} {\bibfnamefont {Z.}~\bibnamefont {Lu}}, \bibinfo {author} {\bibfnamefont {T.}~\bibnamefont {Han}}, \bibinfo {author} {\bibfnamefont {Y.}~\bibnamefont {Yao}}, \bibinfo {author} {\bibfnamefont {A.~P.}\ \bibnamefont {Reddy}}, \bibinfo {author} {\bibfnamefont {J.}~\bibnamefont {Yang}}, \bibinfo {author} {\bibfnamefont {J.}~\bibnamefont {Seo}}, \bibinfo {author} {\bibfnamefont {K.}~\bibnamefont {Watanabe}}, \bibinfo {author} {\bibfnamefont {T.}~\bibnamefont {Taniguchi}}, \bibinfo {author} {\bibfnamefont {L.}~\bibnamefont {Fu}},\ and\ \bibinfo {author} {\bibfnamefont {L.}~\bibnamefont {Ju}},\ }\bibfield  {title} {\bibinfo {title} {Fractional quantum anomalous hall effect in multilayer graphene},\ }\href {https://doi.org/10.1038/s41586-023-07010-7} {\bibfield  {journal} {\bibinfo  {journal} {Nature}\ }\textbf {\bibinfo {volume} {626}},\ \bibinfo {pages} {759} (\bibinfo {year} {2024})}\BibitemShut {NoStop}%
\bibitem [{\citenamefont {Aronson}\ \emph {et~al.}(2025)\citenamefont {Aronson}, \citenamefont {Han}, \citenamefont {Lu}, \citenamefont {Yao}, \citenamefont {Butler}, \citenamefont {Watanabe}, \citenamefont {Taniguchi}, \citenamefont {Ju},\ and\ \citenamefont {Ashoori}}]{Aronson2025DisplacementFCI}%
  \BibitemOpen
  \bibfield  {author} {\bibinfo {author} {\bibfnamefont {S.~H.}\ \bibnamefont {Aronson}}, \bibinfo {author} {\bibfnamefont {T.}~\bibnamefont {Han}}, \bibinfo {author} {\bibfnamefont {Z.}~\bibnamefont {Lu}}, \bibinfo {author} {\bibfnamefont {Y.}~\bibnamefont {Yao}}, \bibinfo {author} {\bibfnamefont {J.~P.}\ \bibnamefont {Butler}}, \bibinfo {author} {\bibfnamefont {K.}~\bibnamefont {Watanabe}}, \bibinfo {author} {\bibfnamefont {T.}~\bibnamefont {Taniguchi}}, \bibinfo {author} {\bibfnamefont {L.}~\bibnamefont {Ju}},\ and\ \bibinfo {author} {\bibfnamefont {R.~C.}\ \bibnamefont {Ashoori}},\ }\bibfield  {title} {\bibinfo {title} {Displacement field-controlled fractional chern insulators and charge density waves in a graphene/hbn moir\'e superlattice},\ }\href {https://doi.org/10.1103/75gl-jzl6} {\bibfield  {journal} {\bibinfo  {journal} {Phys. Rev. X}\ }\textbf {\bibinfo {volume} {15}},\ \bibinfo {pages} {031026} (\bibinfo {year} {2025})}\BibitemShut {NoStop}%
\bibitem [{\citenamefont {Xie}\ \emph {et~al.}(2025{\natexlab{a}})\citenamefont {Xie}, \citenamefont {Huo}, \citenamefont {Lu}, \citenamefont {Feng}, \citenamefont {Zhang}, \citenamefont {Wang}, \citenamefont {Yang}, \citenamefont {Watanabe}, \citenamefont {Taniguchi}, \citenamefont {Liu}, \citenamefont {Song}, \citenamefont {Xie}, \citenamefont {Liu},\ and\ \citenamefont {Lu}}]{Xiaobohexa2025}%
  \BibitemOpen
  \bibfield  {author} {\bibinfo {author} {\bibfnamefont {J.}~\bibnamefont {Xie}}, \bibinfo {author} {\bibfnamefont {Z.}~\bibnamefont {Huo}}, \bibinfo {author} {\bibfnamefont {X.}~\bibnamefont {Lu}}, \bibinfo {author} {\bibfnamefont {Z.}~\bibnamefont {Feng}}, \bibinfo {author} {\bibfnamefont {Z.}~\bibnamefont {Zhang}}, \bibinfo {author} {\bibfnamefont {W.}~\bibnamefont {Wang}}, \bibinfo {author} {\bibfnamefont {Q.}~\bibnamefont {Yang}}, \bibinfo {author} {\bibfnamefont {K.}~\bibnamefont {Watanabe}}, \bibinfo {author} {\bibfnamefont {T.}~\bibnamefont {Taniguchi}}, \bibinfo {author} {\bibfnamefont {K.}~\bibnamefont {Liu}}, \bibinfo {author} {\bibfnamefont {Z.}~\bibnamefont {Song}}, \bibinfo {author} {\bibfnamefont {X.~C.}\ \bibnamefont {Xie}}, \bibinfo {author} {\bibfnamefont {J.}~\bibnamefont {Liu}},\ and\ \bibinfo {author} {\bibfnamefont {X.}~\bibnamefont {Lu}},\ }\bibfield  {title} {\bibinfo {title} {Tunable fractional chern insulators in rhombohedral graphene superlattices},\ }\href
  {https://doi.org/10.1038/s41563-025-02225-7} {\bibfield  {journal} {\bibinfo  {journal} {Nature Materials}\ }\textbf {\bibinfo {volume} {24}},\ \bibinfo {pages} {1042} (\bibinfo {year} {2025}{\natexlab{a}})}\BibitemShut {NoStop}%
\bibitem [{\citenamefont {Choi}\ \emph {et~al.}(2025)\citenamefont {Choi}, \citenamefont {Choi}, \citenamefont {Valentini}, \citenamefont {Patterson}, \citenamefont {Holleis}, \citenamefont {Sheekey}, \citenamefont {Stoyanov}, \citenamefont {Cheng}, \citenamefont {Taniguchi}, \citenamefont {Watanabe},\ and\ \citenamefont {Young}}]{Choi_tetralayer2024}%
  \BibitemOpen
  \bibfield  {author} {\bibinfo {author} {\bibfnamefont {Y.}~\bibnamefont {Choi}}, \bibinfo {author} {\bibfnamefont {Y.}~\bibnamefont {Choi}}, \bibinfo {author} {\bibfnamefont {M.}~\bibnamefont {Valentini}}, \bibinfo {author} {\bibfnamefont {C.~L.}\ \bibnamefont {Patterson}}, \bibinfo {author} {\bibfnamefont {L.~F.~W.}\ \bibnamefont {Holleis}}, \bibinfo {author} {\bibfnamefont {O.~I.}\ \bibnamefont {Sheekey}}, \bibinfo {author} {\bibfnamefont {H.}~\bibnamefont {Stoyanov}}, \bibinfo {author} {\bibfnamefont {X.}~\bibnamefont {Cheng}}, \bibinfo {author} {\bibfnamefont {T.}~\bibnamefont {Taniguchi}}, \bibinfo {author} {\bibfnamefont {K.}~\bibnamefont {Watanabe}},\ and\ \bibinfo {author} {\bibfnamefont {A.~F.}\ \bibnamefont {Young}},\ }\bibfield  {title} {\bibinfo {title} {Superconductivity and quantized anomalous hall effect in rhombohedral graphene},\ }\href {https://doi.org/10.1038/s41586-025-08621-y} {\bibfield  {journal} {\bibinfo  {journal} {Nature}\ }\textbf {\bibinfo {volume} {639}},\ \bibinfo {pages}
  {342} (\bibinfo {year} {2025})}\BibitemShut {NoStop}%
\bibitem [{\citenamefont {Waters}\ \emph {et~al.}(2025)\citenamefont {Waters}, \citenamefont {Okounkova}, \citenamefont {Su}, \citenamefont {Zhou}, \citenamefont {Yao}, \citenamefont {Watanabe}, \citenamefont {Taniguchi}, \citenamefont {Xu}, \citenamefont {Zhang}, \citenamefont {Folk},\ and\ \citenamefont {Yankowitz}}]{Waters2025RPG}%
  \BibitemOpen
  \bibfield  {author} {\bibinfo {author} {\bibfnamefont {D.}~\bibnamefont {Waters}}, \bibinfo {author} {\bibfnamefont {A.}~\bibnamefont {Okounkova}}, \bibinfo {author} {\bibfnamefont {R.}~\bibnamefont {Su}}, \bibinfo {author} {\bibfnamefont {B.}~\bibnamefont {Zhou}}, \bibinfo {author} {\bibfnamefont {J.}~\bibnamefont {Yao}}, \bibinfo {author} {\bibfnamefont {K.}~\bibnamefont {Watanabe}}, \bibinfo {author} {\bibfnamefont {T.}~\bibnamefont {Taniguchi}}, \bibinfo {author} {\bibfnamefont {X.}~\bibnamefont {Xu}}, \bibinfo {author} {\bibfnamefont {Y.-H.}\ \bibnamefont {Zhang}}, \bibinfo {author} {\bibfnamefont {J.}~\bibnamefont {Folk}},\ and\ \bibinfo {author} {\bibfnamefont {M.}~\bibnamefont {Yankowitz}},\ }\bibfield  {title} {\bibinfo {title} {Chern insulators at integer and fractional filling in moir\'e pentalayer graphene},\ }\href {https://doi.org/10.1103/PhysRevX.15.011045} {\bibfield  {journal} {\bibinfo  {journal} {Phys. Rev. X}\ }\textbf {\bibinfo {volume} {15}},\ \bibinfo {pages} {011045} (\bibinfo {year}
  {2025})}\BibitemShut {NoStop}%
\bibitem [{\citenamefont {Lu}\ \emph {et~al.}(2025{\natexlab{a}})\citenamefont {Lu}, \citenamefont {Han}, \citenamefont {Yao}, \citenamefont {Hadjri}, \citenamefont {Yang}, \citenamefont {Seo}, \citenamefont {Shi}, \citenamefont {Ye}, \citenamefont {Watanabe}, \citenamefont {Taniguchi},\ and\ \citenamefont {Ju}}]{LuLong2025}%
  \BibitemOpen
  \bibfield  {author} {\bibinfo {author} {\bibfnamefont {Z.}~\bibnamefont {Lu}}, \bibinfo {author} {\bibfnamefont {T.}~\bibnamefont {Han}}, \bibinfo {author} {\bibfnamefont {Y.}~\bibnamefont {Yao}}, \bibinfo {author} {\bibfnamefont {Z.}~\bibnamefont {Hadjri}}, \bibinfo {author} {\bibfnamefont {J.}~\bibnamefont {Yang}}, \bibinfo {author} {\bibfnamefont {J.}~\bibnamefont {Seo}}, \bibinfo {author} {\bibfnamefont {L.}~\bibnamefont {Shi}}, \bibinfo {author} {\bibfnamefont {S.}~\bibnamefont {Ye}}, \bibinfo {author} {\bibfnamefont {K.}~\bibnamefont {Watanabe}}, \bibinfo {author} {\bibfnamefont {T.}~\bibnamefont {Taniguchi}},\ and\ \bibinfo {author} {\bibfnamefont {L.}~\bibnamefont {Ju}},\ }\bibfield  {title} {\bibinfo {title} {Extended quantum anomalous hall states in graphene/hbn moir{\'e} superlattices},\ }\href {https://doi.org/10.1038/s41586-024-08470-1} {\bibfield  {journal} {\bibinfo  {journal} {Nature}\ }\textbf {\bibinfo {volume} {637}},\ \bibinfo {pages} {1090} (\bibinfo {year}
  {2025}{\natexlab{a}})}\BibitemShut {NoStop}%
\bibitem [{\citenamefont {Huo}\ \emph {et~al.}(2025)\citenamefont {Huo}, \citenamefont {Wang}, \citenamefont {Xie}, \citenamefont {Kwan}, \citenamefont {Herzog-Arbeitman}, \citenamefont {Zhang}, \citenamefont {Yang}, \citenamefont {Wu}, \citenamefont {Watanabe}, \citenamefont {Taniguchi}, \citenamefont {Liu}, \citenamefont {Regnault}, \citenamefont {Bernevig},\ and\ \citenamefont {Lu}}]{Huo2025MoireMatter}%
  \BibitemOpen
  \bibfield  {author} {\bibinfo {author} {\bibfnamefont {Z.}~\bibnamefont {Huo}}, \bibinfo {author} {\bibfnamefont {W.}~\bibnamefont {Wang}}, \bibinfo {author} {\bibfnamefont {J.}~\bibnamefont {Xie}}, \bibinfo {author} {\bibfnamefont {Y.~H.}\ \bibnamefont {Kwan}}, \bibinfo {author} {\bibfnamefont {J.}~\bibnamefont {Herzog-Arbeitman}}, \bibinfo {author} {\bibfnamefont {Z.}~\bibnamefont {Zhang}}, \bibinfo {author} {\bibfnamefont {Q.}~\bibnamefont {Yang}}, \bibinfo {author} {\bibfnamefont {M.}~\bibnamefont {Wu}}, \bibinfo {author} {\bibfnamefont {K.}~\bibnamefont {Watanabe}}, \bibinfo {author} {\bibfnamefont {T.}~\bibnamefont {Taniguchi}}, \bibinfo {author} {\bibfnamefont {K.}~\bibnamefont {Liu}}, \bibinfo {author} {\bibfnamefont {N.}~\bibnamefont {Regnault}}, \bibinfo {author} {\bibfnamefont {B.~A.}\ \bibnamefont {Bernevig}},\ and\ \bibinfo {author} {\bibfnamefont {X.}~\bibnamefont {Lu}},\ }\href {https://arxiv.org/abs/2510.15309} {\bibinfo {title} {Does moire matter? critical moire dependence with quantum
  fluctuations in graphene based integer and fractional chern insulators}} (\bibinfo {year} {2025}),\ \Eprint {https://arxiv.org/abs/2510.15309} {arXiv:2510.15309 [cond-mat.mes-hall]} \BibitemShut {NoStop}%
\bibitem [{\citenamefont {Xie}\ \emph {et~al.}(2025{\natexlab{b}})\citenamefont {Xie}, \citenamefont {Zhang}, \citenamefont {Chen}, \citenamefont {Kwan}, \citenamefont {Huo}, \citenamefont {Herzog-Arbeitman}, \citenamefont {Guo}, \citenamefont {Watanabe}, \citenamefont {Taniguchi}, \citenamefont {Liu}, \citenamefont {Xie}, \citenamefont {Bernevig}, \citenamefont {Song},\ and\ \citenamefont {Lu}}]{Xie2025OrbitalMagnetism}%
  \BibitemOpen
  \bibfield  {author} {\bibinfo {author} {\bibfnamefont {J.}~\bibnamefont {Xie}}, \bibinfo {author} {\bibfnamefont {Z.}~\bibnamefont {Zhang}}, \bibinfo {author} {\bibfnamefont {X.}~\bibnamefont {Chen}}, \bibinfo {author} {\bibfnamefont {Y.~H.}\ \bibnamefont {Kwan}}, \bibinfo {author} {\bibfnamefont {Z.}~\bibnamefont {Huo}}, \bibinfo {author} {\bibfnamefont {J.}~\bibnamefont {Herzog-Arbeitman}}, \bibinfo {author} {\bibfnamefont {L.}~\bibnamefont {Guo}}, \bibinfo {author} {\bibfnamefont {K.}~\bibnamefont {Watanabe}}, \bibinfo {author} {\bibfnamefont {T.}~\bibnamefont {Taniguchi}}, \bibinfo {author} {\bibfnamefont {K.}~\bibnamefont {Liu}}, \bibinfo {author} {\bibfnamefont {X.~C.}\ \bibnamefont {Xie}}, \bibinfo {author} {\bibfnamefont {B.~A.}\ \bibnamefont {Bernevig}}, \bibinfo {author} {\bibfnamefont {Z.-D.}\ \bibnamefont {Song}},\ and\ \bibinfo {author} {\bibfnamefont {X.}~\bibnamefont {Lu}},\ }\href {https://arxiv.org/abs/2506.01485} {\bibinfo {title} {Unconventional orbital magnetism in graphene-based
  fractional chern insulators}} (\bibinfo {year} {2025}{\natexlab{b}}),\ \Eprint {https://arxiv.org/abs/2506.01485} {arXiv:2506.01485 [cond-mat.mes-hall]} \BibitemShut {NoStop}%
\bibitem [{\citenamefont {Li}\ \emph {et~al.}(2025{\natexlab{a}})\citenamefont {Li}, \citenamefont {Zheng}, \citenamefont {Liu}, \citenamefont {Huang}, \citenamefont {Sun}, \citenamefont {Qiao}, \citenamefont {Wei}, \citenamefont {Zhang}, \citenamefont {Xu}, \citenamefont {Watanabe}, \citenamefont {Taniguchi}, \citenamefont {Yang}, \citenamefont {Guan}, \citenamefont {Liu}, \citenamefont {Wang}, \citenamefont {Li}, \citenamefont {Zheng}, \citenamefont {Liu}, \citenamefont {Tong}, \citenamefont {Lu}, \citenamefont {Jia}, \citenamefont {Shi}, \citenamefont {Liu}, \citenamefont {Li}, \citenamefont {Chen}, \citenamefont {Li},\ and\ \citenamefont {Liu}}]{Li2025StackingOrientation}%
  \BibitemOpen
  \bibfield  {author} {\bibinfo {author} {\bibfnamefont {C.}~\bibnamefont {Li}}, \bibinfo {author} {\bibfnamefont {C.}~\bibnamefont {Zheng}}, \bibinfo {author} {\bibfnamefont {K.}~\bibnamefont {Liu}}, \bibinfo {author} {\bibfnamefont {K.}~\bibnamefont {Huang}}, \bibinfo {author} {\bibfnamefont {Z.}~\bibnamefont {Sun}}, \bibinfo {author} {\bibfnamefont {L.}~\bibnamefont {Qiao}}, \bibinfo {author} {\bibfnamefont {Y.}~\bibnamefont {Wei}}, \bibinfo {author} {\bibfnamefont {C.}~\bibnamefont {Zhang}}, \bibinfo {author} {\bibfnamefont {F.}~\bibnamefont {Xu}}, \bibinfo {author} {\bibfnamefont {K.}~\bibnamefont {Watanabe}}, \bibinfo {author} {\bibfnamefont {T.}~\bibnamefont {Taniguchi}}, \bibinfo {author} {\bibfnamefont {H.}~\bibnamefont {Yang}}, \bibinfo {author} {\bibfnamefont {D.}~\bibnamefont {Guan}}, \bibinfo {author} {\bibfnamefont {L.}~\bibnamefont {Liu}}, \bibinfo {author} {\bibfnamefont {S.}~\bibnamefont {Wang}}, \bibinfo {author} {\bibfnamefont {Y.}~\bibnamefont {Li}}, \bibinfo {author} {\bibfnamefont
  {H.}~\bibnamefont {Zheng}}, \bibinfo {author} {\bibfnamefont {C.}~\bibnamefont {Liu}}, \bibinfo {author} {\bibfnamefont {B.}~\bibnamefont {Tong}}, \bibinfo {author} {\bibfnamefont {L.}~\bibnamefont {Lu}}, \bibinfo {author} {\bibfnamefont {J.}~\bibnamefont {Jia}}, \bibinfo {author} {\bibfnamefont {Z.}~\bibnamefont {Shi}}, \bibinfo {author} {\bibfnamefont {J.}~\bibnamefont {Liu}}, \bibinfo {author} {\bibfnamefont {X.}~\bibnamefont {Li}}, \bibinfo {author} {\bibfnamefont {G.}~\bibnamefont {Chen}}, \bibinfo {author} {\bibfnamefont {T.}~\bibnamefont {Li}},\ and\ \bibinfo {author} {\bibfnamefont {X.}~\bibnamefont {Liu}},\ }\href {https://arxiv.org/abs/2505.01767} {\bibinfo {title} {Stacking-orientation and twist-angle control on integer and fractional chern insulators in moir\'e rhombohedral graphene}} (\bibinfo {year} {2025}{\natexlab{a}}),\ \Eprint {https://arxiv.org/abs/2505.01767} {arXiv:2505.01767 [cond-mat.mes-hall]} \BibitemShut {NoStop}%
\bibitem [{\citenamefont {Cr{\'e}pel}\ and\ \citenamefont {Fu}(2023)}]{CrepelFu2023Hall}%
  \BibitemOpen
  \bibfield  {author} {\bibinfo {author} {\bibfnamefont {V.}~\bibnamefont {Cr{\'e}pel}}\ and\ \bibinfo {author} {\bibfnamefont {L.}~\bibnamefont {Fu}},\ }\bibfield  {title} {\bibinfo {title} {Anomalous hall metal and fractional chern insulator in twisted transition metal dichalcogenides},\ }\href {https://doi.org/10.1103/PhysRevB.107.L201109} {\bibfield  {journal} {\bibinfo  {journal} {Phys. Rev. B}\ }\textbf {\bibinfo {volume} {107}},\ \bibinfo {pages} {L201109} (\bibinfo {year} {2023})}\BibitemShut {NoStop}%
\bibitem [{\citenamefont {Wang}\ \emph {et~al.}(2024{\natexlab{a}})\citenamefont {Wang}, \citenamefont {Zhang}, \citenamefont {Liu}, \citenamefont {He}, \citenamefont {Xu}, \citenamefont {Ran}, \citenamefont {Cao},\ and\ \citenamefont {Xiao}}]{wang2023fractional}%
  \BibitemOpen
  \bibfield  {author} {\bibinfo {author} {\bibfnamefont {C.}~\bibnamefont {Wang}}, \bibinfo {author} {\bibfnamefont {X.-W.}\ \bibnamefont {Zhang}}, \bibinfo {author} {\bibfnamefont {X.}~\bibnamefont {Liu}}, \bibinfo {author} {\bibfnamefont {Y.}~\bibnamefont {He}}, \bibinfo {author} {\bibfnamefont {X.}~\bibnamefont {Xu}}, \bibinfo {author} {\bibfnamefont {Y.}~\bibnamefont {Ran}}, \bibinfo {author} {\bibfnamefont {T.}~\bibnamefont {Cao}},\ and\ \bibinfo {author} {\bibfnamefont {D.}~\bibnamefont {Xiao}},\ }\bibfield  {title} {\bibinfo {title} {Fractional chern insulator in twisted bilayer ${\mathrm{mote}}_{2}$},\ }\href {https://doi.org/10.1103/PhysRevLett.132.036501} {\bibfield  {journal} {\bibinfo  {journal} {Phys. Rev. Lett.}\ }\textbf {\bibinfo {volume} {132}},\ \bibinfo {pages} {036501} (\bibinfo {year} {2024}{\natexlab{a}})}\BibitemShut {NoStop}%
\bibitem [{\citenamefont {Qiu}\ \emph {et~al.}(2023)\citenamefont {Qiu}, \citenamefont {Li}, \citenamefont {Luo},\ and\ \citenamefont {Wu}}]{Qiu2023Topological}%
  \BibitemOpen
  \bibfield  {author} {\bibinfo {author} {\bibfnamefont {W.-X.}\ \bibnamefont {Qiu}}, \bibinfo {author} {\bibfnamefont {B.}~\bibnamefont {Li}}, \bibinfo {author} {\bibfnamefont {X.-J.}\ \bibnamefont {Luo}},\ and\ \bibinfo {author} {\bibfnamefont {F.}~\bibnamefont {Wu}},\ }\bibfield  {title} {\bibinfo {title} {Interaction-driven topological phase diagram of twisted bilayer mote$_2$},\ }\href {https://doi.org/10.1103/PhysRevX.13.041026} {\bibfield  {journal} {\bibinfo  {journal} {Phys. Rev. X}\ }\textbf {\bibinfo {volume} {13}},\ \bibinfo {pages} {041026} (\bibinfo {year} {2023})}\BibitemShut {NoStop}%
\bibitem [{\citenamefont {Morales-Dur\'an}\ \emph {et~al.}(2024)\citenamefont {Morales-Dur\'an}, \citenamefont {Wei}, \citenamefont {Shi},\ and\ \citenamefont {MacDonald}}]{MoralesDuran2023bAngles}%
  \BibitemOpen
  \bibfield  {author} {\bibinfo {author} {\bibfnamefont {N.}~\bibnamefont {Morales-Dur\'an}}, \bibinfo {author} {\bibfnamefont {N.}~\bibnamefont {Wei}}, \bibinfo {author} {\bibfnamefont {J.}~\bibnamefont {Shi}},\ and\ \bibinfo {author} {\bibfnamefont {A.~H.}\ \bibnamefont {MacDonald}},\ }\bibfield  {title} {\bibinfo {title} {Magic angles and fractional chern insulators in twisted homobilayer transition metal dichalcogenides},\ }\href {https://doi.org/10.1103/PhysRevLett.132.096602} {\bibfield  {journal} {\bibinfo  {journal} {Phys. Rev. Lett.}\ }\textbf {\bibinfo {volume} {132}},\ \bibinfo {pages} {096602} (\bibinfo {year} {2024})}\BibitemShut {NoStop}%
\bibitem [{\citenamefont {Reddy}\ and\ \citenamefont {Fu}(2023)}]{ReddyFu2023_arXiv2308_10406global}%
  \BibitemOpen
  \bibfield  {author} {\bibinfo {author} {\bibfnamefont {A.~P.}\ \bibnamefont {Reddy}}\ and\ \bibinfo {author} {\bibfnamefont {L.}~\bibnamefont {Fu}},\ }\bibfield  {title} {\bibinfo {title} {Toward a global phase diagram of the fractional quantum anomalous hall effect},\ }\href {https://doi.org/10.1103/PhysRevB.108.245159} {\bibfield  {journal} {\bibinfo  {journal} {Phys. Rev. B}\ }\textbf {\bibinfo {volume} {108}},\ \bibinfo {pages} {245159} (\bibinfo {year} {2023})}\BibitemShut {NoStop}%
\bibitem [{\citenamefont {Abouelkomsan}\ \emph {et~al.}(2023)\citenamefont {Abouelkomsan}, \citenamefont {Yang},\ and\ \citenamefont {Bergholtz}}]{Abouelkomsan2023metric}%
  \BibitemOpen
  \bibfield  {author} {\bibinfo {author} {\bibfnamefont {A.}~\bibnamefont {Abouelkomsan}}, \bibinfo {author} {\bibfnamefont {K.}~\bibnamefont {Yang}},\ and\ \bibinfo {author} {\bibfnamefont {E.~J.}\ \bibnamefont {Bergholtz}},\ }\bibfield  {title} {\bibinfo {title} {Quantum metric induced phases in moir{\'e} materials},\ }\href {https://doi.org/10.1103/PhysRevResearch.5.L012015} {\bibfield  {journal} {\bibinfo  {journal} {Phys. Rev. Research}\ }\textbf {\bibinfo {volume} {5}},\ \bibinfo {pages} {L012015} (\bibinfo {year} {2023})}\BibitemShut {NoStop}%
\bibitem [{\citenamefont {Xu}\ \emph {et~al.}(2024)\citenamefont {Xu}, \citenamefont {Li}, \citenamefont {Xu}, \citenamefont {Bi},\ and\ \citenamefont {Zhang}}]{Xu2024maximallfim}%
  \BibitemOpen
  \bibfield  {author} {\bibinfo {author} {\bibfnamefont {C.}~\bibnamefont {Xu}}, \bibinfo {author} {\bibfnamefont {J.}~\bibnamefont {Li}}, \bibinfo {author} {\bibfnamefont {Y.}~\bibnamefont {Xu}}, \bibinfo {author} {\bibfnamefont {Z.}~\bibnamefont {Bi}},\ and\ \bibinfo {author} {\bibfnamefont {Y.}~\bibnamefont {Zhang}},\ }\bibfield  {title} {\bibinfo {title} {Maximally localized wannier functions, interaction models, and fractional quantum anomalous hall effect in twisted bilayer mote2},\ }\href {https://doi.org/10.1073/pnas.2316749121} {\bibfield  {journal} {\bibinfo  {journal} {Proceedings of the National Academy of Sciences}\ }\textbf {\bibinfo {volume} {121}},\ \bibinfo {pages} {e2316749121} (\bibinfo {year} {2024})}\BibitemShut {NoStop}%
\bibitem [{\citenamefont {Li}\ \emph {et~al.}(2024{\natexlab{a}})\citenamefont {Li}, \citenamefont {Su}, \citenamefont {Kim}, \citenamefont {Kee}, \citenamefont {Sun},\ and\ \citenamefont {Lin}}]{Li2024contrasting}%
  \BibitemOpen
  \bibfield  {author} {\bibinfo {author} {\bibfnamefont {H.}~\bibnamefont {Li}}, \bibinfo {author} {\bibfnamefont {Y.}~\bibnamefont {Su}}, \bibinfo {author} {\bibfnamefont {Y.~B.}\ \bibnamefont {Kim}}, \bibinfo {author} {\bibfnamefont {H.-Y.}\ \bibnamefont {Kee}}, \bibinfo {author} {\bibfnamefont {K.}~\bibnamefont {Sun}},\ and\ \bibinfo {author} {\bibfnamefont {S.-Z.}\ \bibnamefont {Lin}},\ }\bibfield  {title} {\bibinfo {title} {Contrasting twisted bilayer graphene and transition metal dichalcogenides for fractional chern insulators: An emergent gauge picture},\ }\href {https://doi.org/10.1103/PhysRevB.109.245131} {\bibfield  {journal} {\bibinfo  {journal} {Phys. Rev. B}\ }\textbf {\bibinfo {volume} {109}},\ \bibinfo {pages} {245131} (\bibinfo {year} {2024}{\natexlab{a}})}\BibitemShut {NoStop}%
\bibitem [{\citenamefont {Morales-Dur{\'a}n}\ \emph {et~al.}(2023)\citenamefont {Morales-Dur{\'a}n}, \citenamefont {Wang}, \citenamefont {Schleder}, \citenamefont {Angeli}, \citenamefont {Zhu}, \citenamefont {Kaxiras}, \citenamefont {Repellin},\ and\ \citenamefont {Cano}}]{MoralesDuran2023enhanced}%
  \BibitemOpen
  \bibfield  {author} {\bibinfo {author} {\bibfnamefont {N.}~\bibnamefont {Morales-Dur{\'a}n}}, \bibinfo {author} {\bibfnamefont {J.}~\bibnamefont {Wang}}, \bibinfo {author} {\bibfnamefont {G.~R.}\ \bibnamefont {Schleder}}, \bibinfo {author} {\bibfnamefont {M.}~\bibnamefont {Angeli}}, \bibinfo {author} {\bibfnamefont {Z.}~\bibnamefont {Zhu}}, \bibinfo {author} {\bibfnamefont {E.}~\bibnamefont {Kaxiras}}, \bibinfo {author} {\bibfnamefont {C.}~\bibnamefont {Repellin}},\ and\ \bibinfo {author} {\bibfnamefont {J.}~\bibnamefont {Cano}},\ }\bibfield  {title} {\bibinfo {title} {Pressure-enhanced fractional chern insulators along a magic line in twisted wse$_2$},\ }\href {https://doi.org/10.1103/PhysRevResearch.5.L032022} {\bibfield  {journal} {\bibinfo  {journal} {Phys. Rev. Research}\ }\textbf {\bibinfo {volume} {5}},\ \bibinfo {pages} {L032022} (\bibinfo {year} {2023})}\BibitemShut {NoStop}%
\bibitem [{\citenamefont {Song}\ \emph {et~al.}(2024)\citenamefont {Song}, \citenamefont {Zhang},\ and\ \citenamefont {Senthil}}]{SongZhangSenthil2023transitions}%
  \BibitemOpen
  \bibfield  {author} {\bibinfo {author} {\bibfnamefont {X.-Y.}\ \bibnamefont {Song}}, \bibinfo {author} {\bibfnamefont {Y.-H.}\ \bibnamefont {Zhang}},\ and\ \bibinfo {author} {\bibfnamefont {T.}~\bibnamefont {Senthil}},\ }\bibfield  {title} {\bibinfo {title} {Phase transitions out of quantum hall states in moir\'e materials},\ }\href {https://doi.org/10.1103/PhysRevB.109.085143} {\bibfield  {journal} {\bibinfo  {journal} {Phys. Rev. B}\ }\textbf {\bibinfo {volume} {109}},\ \bibinfo {pages} {085143} (\bibinfo {year} {2024})}\BibitemShut {NoStop}%
\bibitem [{\citenamefont {Wang}\ \emph {et~al.}(2023{\natexlab{a}})\citenamefont {Wang}, \citenamefont {Devakul}, \citenamefont {Zaletel},\ and\ \citenamefont {Fu}}]{WangDevakulZaletelFu2023magnetic}%
  \BibitemOpen
  \bibfield  {author} {\bibinfo {author} {\bibfnamefont {T.}~\bibnamefont {Wang}}, \bibinfo {author} {\bibfnamefont {T.}~\bibnamefont {Devakul}}, \bibinfo {author} {\bibfnamefont {M.~P.}\ \bibnamefont {Zaletel}},\ and\ \bibinfo {author} {\bibfnamefont {L.}~\bibnamefont {Fu}},\ }\href {https://arxiv.org/abs/2306.02501} {\bibinfo {title} {Diverse magnetic orders and quantum anomalous hall effect in twisted bilayer mote$_2$ and wse$_2$}} (\bibinfo {year} {2023}{\natexlab{a}}),\ \Eprint {https://arxiv.org/abs/2306.02501} {arXiv:2306.02501 [cond-mat.str-el]} \BibitemShut {NoStop}%
\bibitem [{\citenamefont {Wu}\ \emph {et~al.}(2024{\natexlab{a}})\citenamefont {Wu}, \citenamefont {Shaffer}, \citenamefont {Wu},\ and\ \citenamefont {Santos}}]{Wu2023TRInvariantm}%
  \BibitemOpen
  \bibfield  {author} {\bibinfo {author} {\bibfnamefont {Y.-M.}\ \bibnamefont {Wu}}, \bibinfo {author} {\bibfnamefont {D.}~\bibnamefont {Shaffer}}, \bibinfo {author} {\bibfnamefont {Z.}~\bibnamefont {Wu}},\ and\ \bibinfo {author} {\bibfnamefont {L.~H.}\ \bibnamefont {Santos}},\ }\bibfield  {title} {\bibinfo {title} {Time-reversal invariant topological moir\'e flat band: A platform for the fractional quantum spin hall effect},\ }\href {https://doi.org/10.1103/PhysRevB.109.115111} {\bibfield  {journal} {\bibinfo  {journal} {Phys. Rev. B}\ }\textbf {\bibinfo {volume} {109}},\ \bibinfo {pages} {115111} (\bibinfo {year} {2024}{\natexlab{a}})}\BibitemShut {NoStop}%
\bibitem [{\citenamefont {Kwan}\ \emph {et~al.}(2026)\citenamefont {Kwan}, \citenamefont {Wagner}, \citenamefont {Yu}, \citenamefont {Dagnino}, \citenamefont {Jiang}, \citenamefont {Xu}, \citenamefont {Bernevig}, \citenamefont {Neupert},\ and\ \citenamefont {Regnault}}]{kwan2026FTI}%
  \BibitemOpen
  \bibfield  {author} {\bibinfo {author} {\bibfnamefont {Y.~H.}\ \bibnamefont {Kwan}}, \bibinfo {author} {\bibfnamefont {G.}~\bibnamefont {Wagner}}, \bibinfo {author} {\bibfnamefont {J.}~\bibnamefont {Yu}}, \bibinfo {author} {\bibfnamefont {A.~K.}\ \bibnamefont {Dagnino}}, \bibinfo {author} {\bibfnamefont {Y.}~\bibnamefont {Jiang}}, \bibinfo {author} {\bibfnamefont {X.}~\bibnamefont {Xu}}, \bibinfo {author} {\bibfnamefont {B.~A.}\ \bibnamefont {Bernevig}}, \bibinfo {author} {\bibfnamefont {T.}~\bibnamefont {Neupert}},\ and\ \bibinfo {author} {\bibfnamefont {N.}~\bibnamefont {Regnault}},\ }\bibfield  {title} {\bibinfo {title} {Regarding the existence of abelian fractional topological insulators in twisted mote2 and related systems},\ }\href {https://doi.org/10.1038/s42005-025-02483-6} {\bibfield  {journal} {\bibinfo  {journal} {Communications Physics}\ }\textbf {\bibinfo {volume} {9}},\ \bibinfo {pages} {52} (\bibinfo {year} {2026})}\BibitemShut {NoStop}%
\bibitem [{\citenamefont {Liu}\ \emph {et~al.}(2026)\citenamefont {Liu}, \citenamefont {Wang}, \citenamefont {Chen}, \citenamefont {Zhang}, \citenamefont {Cao},\ and\ \citenamefont {Xiao}}]{liu2025orbitalmagnetizationcorrelatedstates}%
  \BibitemOpen
  \bibfield  {author} {\bibinfo {author} {\bibfnamefont {X.}~\bibnamefont {Liu}}, \bibinfo {author} {\bibfnamefont {C.}~\bibnamefont {Wang}}, \bibinfo {author} {\bibfnamefont {H.}~\bibnamefont {Chen}}, \bibinfo {author} {\bibfnamefont {X.-W.}\ \bibnamefont {Zhang}}, \bibinfo {author} {\bibfnamefont {T.}~\bibnamefont {Cao}},\ and\ \bibinfo {author} {\bibfnamefont {D.}~\bibnamefont {Xiao}},\ }\bibfield  {title} {\bibinfo {title} {Orbital magnetization of correlated states in twisted bilayer transition metal dichalcogenides},\ }\href {https://doi.org/10.1103/k46f-8m8t} {\bibfield  {journal} {\bibinfo  {journal} {Phys. Rev. Lett.}\ }\textbf {\bibinfo {volume} {136}},\ \bibinfo {pages} {166606} (\bibinfo {year} {2026})}\BibitemShut {NoStop}%
\bibitem [{\citenamefont {Gon\ifmmode~\mbox{\c{c}}\else \c{c}\fi{}alves}\ \emph {et~al.}(2026)\citenamefont {Gon\ifmmode~\mbox{\c{c}}\else \c{c}\fi{}alves}, \citenamefont {Mendez-Valderrama}, \citenamefont {Herzog-Arbeitman}, \citenamefont {Yu}, \citenamefont {Xu}, \citenamefont {Xiao}, \citenamefont {Bernevig},\ and\ \citenamefont {Regnault}}]{goncalves2025spinful}%
  \BibitemOpen
  \bibfield  {author} {\bibinfo {author} {\bibfnamefont {M.}~\bibnamefont {Gon\ifmmode~\mbox{\c{c}}\else \c{c}\fi{}alves}}, \bibinfo {author} {\bibfnamefont {J.~F.}\ \bibnamefont {Mendez-Valderrama}}, \bibinfo {author} {\bibfnamefont {J.}~\bibnamefont {Herzog-Arbeitman}}, \bibinfo {author} {\bibfnamefont {J.}~\bibnamefont {Yu}}, \bibinfo {author} {\bibfnamefont {X.}~\bibnamefont {Xu}}, \bibinfo {author} {\bibfnamefont {D.}~\bibnamefont {Xiao}}, \bibinfo {author} {\bibfnamefont {B.~A.}\ \bibnamefont {Bernevig}},\ and\ \bibinfo {author} {\bibfnamefont {N.}~\bibnamefont {Regnault}},\ }\bibfield  {title} {\bibinfo {title} {Spinless and spinful charge excitations in moir\'e fractional chern insulators},\ }\href {https://doi.org/10.1103/gmkp-bxf2} {\bibfield  {journal} {\bibinfo  {journal} {Phys. Rev. Lett.}\ }\textbf {\bibinfo {volume} {136}},\ \bibinfo {pages} {196503} (\bibinfo {year} {2026})}\BibitemShut {NoStop}%
\bibitem [{\citenamefont {Lu}\ \emph {et~al.}(2025{\natexlab{b}})\citenamefont {Lu}, \citenamefont {Wu},\ and\ \citenamefont {Santos}}]{LuWuSantos2025ele}%
  \BibitemOpen
  \bibfield  {author} {\bibinfo {author} {\bibfnamefont {T.}~\bibnamefont {Lu}}, \bibinfo {author} {\bibfnamefont {Y.-M.}\ \bibnamefont {Wu}},\ and\ \bibinfo {author} {\bibfnamefont {L.~H.}\ \bibnamefont {Santos}},\ }\bibfield  {title} {\bibinfo {title} {Electromagnetic response and emergent topological orders in transition metal dichalcogenide ${\mathrm{mote}}_{2}$ bilayers},\ }\href {https://doi.org/10.1103/3crp-gb3d} {\bibfield  {journal} {\bibinfo  {journal} {Phys. Rev. B}\ }\textbf {\bibinfo {volume} {112}},\ \bibinfo {pages} {085138} (\bibinfo {year} {2025}{\natexlab{b}})}\BibitemShut {NoStop}%
\bibitem [{\citenamefont {Zaklama}\ \emph {et~al.}(2025)\citenamefont {Zaklama}, \citenamefont {Luo},\ and\ \citenamefont {Fu}}]{Zaklama2025StructureFactor}%
  \BibitemOpen
  \bibfield  {author} {\bibinfo {author} {\bibfnamefont {T.}~\bibnamefont {Zaklama}}, \bibinfo {author} {\bibfnamefont {D.}~\bibnamefont {Luo}},\ and\ \bibinfo {author} {\bibfnamefont {L.}~\bibnamefont {Fu}},\ }\bibfield  {title} {\bibinfo {title} {Structure factor and topological bound of twisted bilayer semiconductors at fractional fillings},\ }\href {https://doi.org/10.1103/wlvf-tdq1} {\bibfield  {journal} {\bibinfo  {journal} {Phys. Rev. B}\ }\textbf {\bibinfo {volume} {112}},\ \bibinfo {pages} {L041115} (\bibinfo {year} {2025})}\BibitemShut {NoStop}%
\bibitem [{\citenamefont {Qiu}\ and\ \citenamefont {Wu}(2025)}]{QiuWu2025MagnonsDomainWalls}%
  \BibitemOpen
  \bibfield  {author} {\bibinfo {author} {\bibfnamefont {W.-X.}\ \bibnamefont {Qiu}}\ and\ \bibinfo {author} {\bibfnamefont {F.}~\bibnamefont {Wu}},\ }\bibfield  {title} {\bibinfo {title} {Topological magnons and domain walls in twisted bilayer ${\mathrm{mote}}_{2}$},\ }\href {https://doi.org/10.1103/sl5k-c825} {\bibfield  {journal} {\bibinfo  {journal} {Phys. Rev. B}\ }\textbf {\bibinfo {volume} {112}},\ \bibinfo {pages} {085132} (\bibinfo {year} {2025})}\BibitemShut {NoStop}%
\bibitem [{\citenamefont {Liu}\ \emph {et~al.}(2025)\citenamefont {Liu}, \citenamefont {Li}, \citenamefont {Shi},\ and\ \citenamefont {Wu}}]{Liu2025FCIQuasiparticles}%
  \BibitemOpen
  \bibfield  {author} {\bibinfo {author} {\bibfnamefont {Z.}~\bibnamefont {Liu}}, \bibinfo {author} {\bibfnamefont {B.}~\bibnamefont {Li}}, \bibinfo {author} {\bibfnamefont {Y.}~\bibnamefont {Shi}},\ and\ \bibinfo {author} {\bibfnamefont {F.}~\bibnamefont {Wu}},\ }\bibfield  {title} {\bibinfo {title} {Characterization of fractional chern insulator quasiparticles in twisted homobilayer ${\mathrm{mote}}_{2}$},\ }\href {https://doi.org/10.1103/nddl-729x} {\bibfield  {journal} {\bibinfo  {journal} {Phys. Rev. B}\ }\textbf {\bibinfo {volume} {112}},\ \bibinfo {pages} {245104} (\bibinfo {year} {2025})}\BibitemShut {NoStop}%
\bibitem [{\citenamefont {Wu}\ \emph {et~al.}(2024{\natexlab{b}})\citenamefont {Wu}, \citenamefont {Sarkar}, \citenamefont {Wan}, \citenamefont {Sun},\ and\ \citenamefont {Lin}}]{Wu2024MetricInversionFCI}%
  \BibitemOpen
  \bibfield  {author} {\bibinfo {author} {\bibfnamefont {A.-K.}\ \bibnamefont {Wu}}, \bibinfo {author} {\bibfnamefont {S.}~\bibnamefont {Sarkar}}, \bibinfo {author} {\bibfnamefont {X.}~\bibnamefont {Wan}}, \bibinfo {author} {\bibfnamefont {K.}~\bibnamefont {Sun}},\ and\ \bibinfo {author} {\bibfnamefont {S.-Z.}\ \bibnamefont {Lin}},\ }\bibfield  {title} {\bibinfo {title} {Quantum-metric-induced quantum hall conductance inversion and reentrant transition in fractional chern insulators},\ }\href {https://doi.org/10.1103/PhysRevResearch.6.L032063} {\bibfield  {journal} {\bibinfo  {journal} {Phys. Rev. Res.}\ }\textbf {\bibinfo {volume} {6}},\ \bibinfo {pages} {L032063} (\bibinfo {year} {2024}{\natexlab{b}})}\BibitemShut {NoStop}%
\bibitem [{\citenamefont {Luo}\ \emph {et~al.}(2025)\citenamefont {Luo}, \citenamefont {Zaklama},\ and\ \citenamefont {Fu}}]{Luo2025solvingfra}%
  \BibitemOpen
  \bibfield  {author} {\bibinfo {author} {\bibfnamefont {D.}~\bibnamefont {Luo}}, \bibinfo {author} {\bibfnamefont {T.}~\bibnamefont {Zaklama}},\ and\ \bibinfo {author} {\bibfnamefont {L.}~\bibnamefont {Fu}},\ }\href {https://arxiv.org/abs/2503.13585} {\bibinfo {title} {Solving fractional electron states in twisted mote$_2$ with deep neural network}} (\bibinfo {year} {2025}),\ \Eprint {https://arxiv.org/abs/2503.13585} {arXiv:2503.13585 [cond-mat.str-el]} \BibitemShut {NoStop}%
\bibitem [{\citenamefont {Hart}\ \emph {et~al.}(2026)\citenamefont {Hart}, \citenamefont {Azam}, \citenamefont {Li}, \citenamefont {Li}, \citenamefont {Bi}, \citenamefont {Pan},\ and\ \citenamefont {Yu}}]{Hart2026representability}%
  \BibitemOpen
  \bibfield  {author} {\bibinfo {author} {\bibfnamefont {J.~B.}\ \bibnamefont {Hart}}, \bibinfo {author} {\bibfnamefont {A.~A.}\ \bibnamefont {Azam}}, \bibinfo {author} {\bibfnamefont {T.}~\bibnamefont {Li}}, \bibinfo {author} {\bibfnamefont {Y.}~\bibnamefont {Li}}, \bibinfo {author} {\bibfnamefont {Y.}~\bibnamefont {Bi}}, \bibinfo {author} {\bibfnamefont {H.}~\bibnamefont {Pan}},\ and\ \bibinfo {author} {\bibfnamefont {J.}~\bibnamefont {Yu}},\ }\href {https://arxiv.org/abs/2605.20326} {\bibinfo {title} {Representability-aware neural networks for reduced density matrices: Application to fractional chern insulators}} (\bibinfo {year} {2026}),\ \Eprint {https://arxiv.org/abs/2605.20326} {arXiv:2605.20326 [cond-mat.str-el]} \BibitemShut {NoStop}%
\bibitem [{\citenamefont {Chen}\ \emph {et~al.}(2026)\citenamefont {Chen}, \citenamefont {Li}, \citenamefont {Wang},\ and\ \citenamefont {Li}}]{CHEN20261034}%
  \BibitemOpen
  \bibfield  {author} {\bibinfo {author} {\bibfnamefont {J.}~\bibnamefont {Chen}}, \bibinfo {author} {\bibfnamefont {Q.}~\bibnamefont {Li}}, \bibinfo {author} {\bibfnamefont {X.}~\bibnamefont {Wang}},\ and\ \bibinfo {author} {\bibfnamefont {W.}~\bibnamefont {Li}},\ }\bibfield  {title} {\bibinfo {title} {Fractional chern insulator and quantum anomalous hall crystal in twisted mote2},\ }\href {https://doi.org/https://doi.org/10.1016/j.scib.2026.01.014} {\bibfield  {journal} {\bibinfo  {journal} {Science Bulletin}\ }\textbf {\bibinfo {volume} {71}},\ \bibinfo {pages} {1034} (\bibinfo {year} {2026})}\BibitemShut {NoStop}%
\bibitem [{\citenamefont {Wang}\ \emph {et~al.}(2026)\citenamefont {Wang}, \citenamefont {Minarik}, \citenamefont {Li}, \citenamefont {Kwan}, \citenamefont {Yuan}, \citenamefont {Anderson}, \citenamefont {Hu}, \citenamefont {Ingham}, \citenamefont {Choe}, \citenamefont {Taniguchi}, \citenamefont {Watanabe}, \citenamefont {Roy}, \citenamefont {Chu}, \citenamefont {Queiroz}, \citenamefont {Hone}, \citenamefont {Regnault}, \citenamefont {Xu},\ and\ \citenamefont {Zhu}}]{Wang2026fti}%
  \BibitemOpen
  \bibfield  {author} {\bibinfo {author} {\bibfnamefont {Y.}~\bibnamefont {Wang}}, \bibinfo {author} {\bibfnamefont {G.~E.}\ \bibnamefont {Minarik}}, \bibinfo {author} {\bibfnamefont {W.}~\bibnamefont {Li}}, \bibinfo {author} {\bibfnamefont {Y.}~\bibnamefont {Kwan}}, \bibinfo {author} {\bibfnamefont {S.}~\bibnamefont {Yuan}}, \bibinfo {author} {\bibfnamefont {E.}~\bibnamefont {Anderson}}, \bibinfo {author} {\bibfnamefont {C.}~\bibnamefont {Hu}}, \bibinfo {author} {\bibfnamefont {J.}~\bibnamefont {Ingham}}, \bibinfo {author} {\bibfnamefont {J.}~\bibnamefont {Choe}}, \bibinfo {author} {\bibfnamefont {T.}~\bibnamefont {Taniguchi}}, \bibinfo {author} {\bibfnamefont {K.}~\bibnamefont {Watanabe}}, \bibinfo {author} {\bibfnamefont {X.}~\bibnamefont {Roy}}, \bibinfo {author} {\bibfnamefont {J.-H.}\ \bibnamefont {Chu}}, \bibinfo {author} {\bibfnamefont {R.}~\bibnamefont {Queiroz}}, \bibinfo {author} {\bibfnamefont {J.~C.}\ \bibnamefont {Hone}}, \bibinfo {author} {\bibfnamefont {N.}~\bibnamefont {Regnault}}, \bibinfo
  {author} {\bibfnamefont {X.}~\bibnamefont {Xu}},\ and\ \bibinfo {author} {\bibfnamefont {X.}~\bibnamefont {Zhu}},\ }\bibfield  {title} {\bibinfo {title} {Candidate for a fractional topological insulator in twisted ${\mathrm{mote}}_{2}$},\ }\href {https://doi.org/10.1103/bvrb-z4hj} {\bibfield  {journal} {\bibinfo  {journal} {Phys. Rev. X}\ }\textbf {\bibinfo {volume} {16}},\ \bibinfo {pages} {031009} (\bibinfo {year} {2026})}\BibitemShut {NoStop}%
\bibitem [{\citenamefont {Yu}\ \emph {et~al.}(2024)\citenamefont {Yu}, \citenamefont {Herzog-Arbeitman}, \citenamefont {Wang}, \citenamefont {Vafek}, \citenamefont {Bernevig},\ and\ \citenamefont {Regnault}}]{Yu2024mote}%
  \BibitemOpen
  \bibfield  {author} {\bibinfo {author} {\bibfnamefont {J.}~\bibnamefont {Yu}}, \bibinfo {author} {\bibfnamefont {J.}~\bibnamefont {Herzog-Arbeitman}}, \bibinfo {author} {\bibfnamefont {M.}~\bibnamefont {Wang}}, \bibinfo {author} {\bibfnamefont {O.}~\bibnamefont {Vafek}}, \bibinfo {author} {\bibfnamefont {B.~A.}\ \bibnamefont {Bernevig}},\ and\ \bibinfo {author} {\bibfnamefont {N.}~\bibnamefont {Regnault}},\ }\bibfield  {title} {\bibinfo {title} {Fractional chern insulators versus nonmagnetic states in twisted bilayer ${\mathrm{mote}}_{2}$},\ }\href {https://doi.org/10.1103/PhysRevB.109.045147} {\bibfield  {journal} {\bibinfo  {journal} {Phys. Rev. B}\ }\textbf {\bibinfo {volume} {109}},\ \bibinfo {pages} {045147} (\bibinfo {year} {2024})}\BibitemShut {NoStop}%
\bibitem [{\citenamefont {Abouelkomsan}\ \emph {et~al.}(2024)\citenamefont {Abouelkomsan}, \citenamefont {Reddy}, \citenamefont {Fu},\ and\ \citenamefont {Bergholtz}}]{Fu2023BandMixingFCItMoTe2}%
  \BibitemOpen
  \bibfield  {author} {\bibinfo {author} {\bibfnamefont {A.}~\bibnamefont {Abouelkomsan}}, \bibinfo {author} {\bibfnamefont {A.~P.}\ \bibnamefont {Reddy}}, \bibinfo {author} {\bibfnamefont {L.}~\bibnamefont {Fu}},\ and\ \bibinfo {author} {\bibfnamefont {E.~J.}\ \bibnamefont {Bergholtz}},\ }\bibfield  {title} {\bibinfo {title} {Band mixing in the quantum anomalous hall regime of twisted semiconductor bilayers},\ }\href {https://doi.org/10.1103/PhysRevB.109.L121107} {\bibfield  {journal} {\bibinfo  {journal} {Phys. Rev. B}\ }\textbf {\bibinfo {volume} {109}},\ \bibinfo {pages} {L121107} (\bibinfo {year} {2024})}\BibitemShut {NoStop}%
\bibitem [{\citenamefont {He}\ \emph {et~al.}(2025)\citenamefont {He}, \citenamefont {Simon},\ and\ \citenamefont {Parameswaran}}]{He2025fractionalcherninsulatorscompeting}%
  \BibitemOpen
  \bibfield  {author} {\bibinfo {author} {\bibfnamefont {Y.}~\bibnamefont {He}}, \bibinfo {author} {\bibfnamefont {S.~H.}\ \bibnamefont {Simon}},\ and\ \bibinfo {author} {\bibfnamefont {S.~A.}\ \bibnamefont {Parameswaran}},\ }\href {https://arxiv.org/abs/2505.06354} {\bibinfo {title} {Fractional chern insulators and competing states in a twisted mote$_2$ lattice model}} (\bibinfo {year} {2025}),\ \Eprint {https://arxiv.org/abs/2505.06354} {arXiv:2505.06354 [cond-mat.str-el]} \BibitemShut {NoStop}%
\bibitem [{\citenamefont {Hou}\ and\ \citenamefont {Nevidomskyy}(2025)}]{hou2025stabilizingfractionalchernstates}%
  \BibitemOpen
  \bibfield  {author} {\bibinfo {author} {\bibfnamefont {R.}~\bibnamefont {Hou}}\ and\ \bibinfo {author} {\bibfnamefont {A.~H.}\ \bibnamefont {Nevidomskyy}},\ }\href {https://arxiv.org/abs/2511.16641} {\bibinfo {title} {Stabilizing fractional chern states in twisted mote2: Multi-band correlations via non-perturbative renormalization group}} (\bibinfo {year} {2025}),\ \Eprint {https://arxiv.org/abs/2511.16641} {arXiv:2511.16641 [cond-mat.str-el]} \BibitemShut {NoStop}%
\bibitem [{\citenamefont {Wang}\ \emph {et~al.}(2023{\natexlab{b}})\citenamefont {Wang}, \citenamefont {Wang}, \citenamefont {Kim}, \citenamefont {Louie}, \citenamefont {Fu},\ and\ \citenamefont {Zaletel}}]{Wang2023HigherIntegerMoTe2}%
  \BibitemOpen
  \bibfield  {author} {\bibinfo {author} {\bibfnamefont {T.}~\bibnamefont {Wang}}, \bibinfo {author} {\bibfnamefont {M.}~\bibnamefont {Wang}}, \bibinfo {author} {\bibfnamefont {W.}~\bibnamefont {Kim}}, \bibinfo {author} {\bibfnamefont {S.~G.}\ \bibnamefont {Louie}}, \bibinfo {author} {\bibfnamefont {L.}~\bibnamefont {Fu}},\ and\ \bibinfo {author} {\bibfnamefont {M.~P.}\ \bibnamefont {Zaletel}},\ }\href {https://arxiv.org/abs/2312.12531} {\bibinfo {title} {Topology, magnetism and charge order in twisted mote2 at higher integer hole fillings}} (\bibinfo {year} {2023}{\natexlab{b}}),\ \Eprint {https://arxiv.org/abs/2312.12531} {arXiv:2312.12531 [cond-mat.str-el]} \BibitemShut {NoStop}%
\bibitem [{\citenamefont {Kwan}\ \emph {et~al.}(2024)\citenamefont {Kwan}, \citenamefont {Wagner}, \citenamefont {Yu}, \citenamefont {Dagnino}, \citenamefont {Jiang}, \citenamefont {Xu}, \citenamefont {Bernevig}, \citenamefont {Neupert},\ and\ \citenamefont {Regnault}}]{kwan2024abelianfractionaltopologicalinsulators}%
  \BibitemOpen
  \bibfield  {author} {\bibinfo {author} {\bibfnamefont {Y.~H.}\ \bibnamefont {Kwan}}, \bibinfo {author} {\bibfnamefont {G.}~\bibnamefont {Wagner}}, \bibinfo {author} {\bibfnamefont {J.}~\bibnamefont {Yu}}, \bibinfo {author} {\bibfnamefont {A.~K.}\ \bibnamefont {Dagnino}}, \bibinfo {author} {\bibfnamefont {Y.}~\bibnamefont {Jiang}}, \bibinfo {author} {\bibfnamefont {X.}~\bibnamefont {Xu}}, \bibinfo {author} {\bibfnamefont {B.~A.}\ \bibnamefont {Bernevig}}, \bibinfo {author} {\bibfnamefont {T.}~\bibnamefont {Neupert}},\ and\ \bibinfo {author} {\bibfnamefont {N.}~\bibnamefont {Regnault}},\ }\href {https://arxiv.org/abs/2407.02560} {\bibinfo {title} {When could abelian fractional topological insulators exist in twisted mote$_2$ (and other systems)}} (\bibinfo {year} {2024}),\ \Eprint {https://arxiv.org/abs/2407.02560} {arXiv:2407.02560 [cond-mat.str-el]} \BibitemShut {NoStop}%
\bibitem [{\citenamefont {Sheng}\ \emph {et~al.}(2024)\citenamefont {Sheng}, \citenamefont {Reddy}, \citenamefont {Abouelkomsan}, \citenamefont {Bergholtz},\ and\ \citenamefont {Fu}}]{Sheng2024QAHCrystal}%
  \BibitemOpen
  \bibfield  {author} {\bibinfo {author} {\bibfnamefont {D.~N.}\ \bibnamefont {Sheng}}, \bibinfo {author} {\bibfnamefont {A.~P.}\ \bibnamefont {Reddy}}, \bibinfo {author} {\bibfnamefont {A.}~\bibnamefont {Abouelkomsan}}, \bibinfo {author} {\bibfnamefont {E.~J.}\ \bibnamefont {Bergholtz}},\ and\ \bibinfo {author} {\bibfnamefont {L.}~\bibnamefont {Fu}},\ }\bibfield  {title} {\bibinfo {title} {Quantum anomalous hall crystal at fractional filling of moir\'e superlattices},\ }\href {https://doi.org/10.1103/PhysRevLett.133.066601} {\bibfield  {journal} {\bibinfo  {journal} {Phys. Rev. Lett.}\ }\textbf {\bibinfo {volume} {133}},\ \bibinfo {pages} {066601} (\bibinfo {year} {2024})}\BibitemShut {NoStop}%
\bibitem [{\citenamefont {Tuo}\ \emph {et~al.}(2025)\citenamefont {Tuo}, \citenamefont {Li},\ and\ \citenamefont {Yao}}]{Tuo2025fqah}%
  \BibitemOpen
  \bibfield  {author} {\bibinfo {author} {\bibfnamefont {C.}~\bibnamefont {Tuo}}, \bibinfo {author} {\bibfnamefont {M.-R.}\ \bibnamefont {Li}},\ and\ \bibinfo {author} {\bibfnamefont {H.}~\bibnamefont {Yao}},\ }\href {https://arxiv.org/abs/2512.23608} {\bibinfo {title} {Fractional quantum anomalous hall and anyon density-wave halo in a minimal interacting lattice model of twisted bilayer mote$_2$}} (\bibinfo {year} {2025}),\ \Eprint {https://arxiv.org/abs/2512.23608} {arXiv:2512.23608 [cond-mat.str-el]} \BibitemShut {NoStop}%
\bibitem [{\citenamefont {Shen}\ \emph {et~al.}(2024)\citenamefont {Shen}, \citenamefont {Wang}, \citenamefont {Guo}, \citenamefont {Xu}, \citenamefont {Duan},\ and\ \citenamefont {Xu}}]{Shen2024ExchangeFCI}%
  \BibitemOpen
  \bibfield  {author} {\bibinfo {author} {\bibfnamefont {X.}~\bibnamefont {Shen}}, \bibinfo {author} {\bibfnamefont {C.}~\bibnamefont {Wang}}, \bibinfo {author} {\bibfnamefont {R.}~\bibnamefont {Guo}}, \bibinfo {author} {\bibfnamefont {Z.}~\bibnamefont {Xu}}, \bibinfo {author} {\bibfnamefont {W.}~\bibnamefont {Duan}},\ and\ \bibinfo {author} {\bibfnamefont {Y.}~\bibnamefont {Xu}},\ }\href {https://arxiv.org/abs/2405.12294} {\bibinfo {title} {Stabilizing fractional chern insulators via exchange interaction in moir\'e systems}} (\bibinfo {year} {2024}),\ \Eprint {https://arxiv.org/abs/2405.12294} {arXiv:2405.12294 [cond-mat.str-el]} \BibitemShut {NoStop}%
\bibitem [{\citenamefont {Song}\ and\ \citenamefont {Senthil}(2024)}]{SongSenthil2024AnyonHalo}%
  \BibitemOpen
  \bibfield  {author} {\bibinfo {author} {\bibfnamefont {X.-Y.}\ \bibnamefont {Song}}\ and\ \bibinfo {author} {\bibfnamefont {T.}~\bibnamefont {Senthil}},\ }\bibfield  {title} {\bibinfo {title} {Density wave halo around anyons in fractional quantum anomalous {Hall} states},\ }\href {https://doi.org/10.1103/PhysRevB.110.085120} {\bibfield  {journal} {\bibinfo  {journal} {Phys. Rev. B}\ }\textbf {\bibinfo {volume} {110}},\ \bibinfo {pages} {085120} (\bibinfo {year} {2024})}\BibitemShut {NoStop}%
\bibitem [{\citenamefont {Li}\ and\ \citenamefont {Wu}(2025)}]{LiWu2025VariationalMapping}%
  \BibitemOpen
  \bibfield  {author} {\bibinfo {author} {\bibfnamefont {B.}~\bibnamefont {Li}}\ and\ \bibinfo {author} {\bibfnamefont {F.}~\bibnamefont {Wu}},\ }\bibfield  {title} {\bibinfo {title} {Variational mapping of {Chern} bands to {Landau} levels: Application to fractional {Chern} insulators in twisted {MoTe$_2$}},\ }\href {https://doi.org/10.1103/PhysRevB.111.125122} {\bibfield  {journal} {\bibinfo  {journal} {Phys. Rev. B}\ }\textbf {\bibinfo {volume} {111}},\ \bibinfo {pages} {125122} (\bibinfo {year} {2025})}\BibitemShut {NoStop}%
\bibitem [{\citenamefont {Shen}\ \emph {et~al.}(2026)\citenamefont {Shen}, \citenamefont {Wang}, \citenamefont {Hu}, \citenamefont {Guo}, \citenamefont {Yao}, \citenamefont {Wang}, \citenamefont {Duan},\ and\ \citenamefont {Xu}}]{Shen2026Magnetorotons}%
  \BibitemOpen
  \bibfield  {author} {\bibinfo {author} {\bibfnamefont {X.}~\bibnamefont {Shen}}, \bibinfo {author} {\bibfnamefont {C.}~\bibnamefont {Wang}}, \bibinfo {author} {\bibfnamefont {X.}~\bibnamefont {Hu}}, \bibinfo {author} {\bibfnamefont {R.}~\bibnamefont {Guo}}, \bibinfo {author} {\bibfnamefont {H.}~\bibnamefont {Yao}}, \bibinfo {author} {\bibfnamefont {C.}~\bibnamefont {Wang}}, \bibinfo {author} {\bibfnamefont {W.}~\bibnamefont {Duan}},\ and\ \bibinfo {author} {\bibfnamefont {Y.}~\bibnamefont {Xu}},\ }\bibfield  {title} {\bibinfo {title} {Magnetorotons in moir{\'e} fractional {Chern} insulators},\ }\href {https://doi.org/10.1103/t4gb-jrxg} {\bibfield  {journal} {\bibinfo  {journal} {Phys. Rev. B}\ }\textbf {\bibinfo {volume} {113}},\ \bibinfo {pages} {L081403} (\bibinfo {year} {2026})}\BibitemShut {NoStop}%
\bibitem [{\citenamefont {Cr{\'e}pel}\ and\ \citenamefont {Millis}(2024)}]{CrepelMillis2024TMDTightBinding}%
  \BibitemOpen
  \bibfield  {author} {\bibinfo {author} {\bibfnamefont {V.}~\bibnamefont {Cr{\'e}pel}}\ and\ \bibinfo {author} {\bibfnamefont {A.}~\bibnamefont {Millis}},\ }\bibfield  {title} {\bibinfo {title} {Bridging the small and large in twisted transition metal dichalcogenide homobilayers: A tight binding model capturing orbital interference and topology across a wide range of twist angles},\ }\href {https://doi.org/10.1103/PhysRevResearch.6.033127} {\bibfield  {journal} {\bibinfo  {journal} {Phys. Rev. Research}\ }\textbf {\bibinfo {volume} {6}},\ \bibinfo {pages} {033127} (\bibinfo {year} {2024})}\BibitemShut {NoStop}%
\bibitem [{\citenamefont {Zeng}\ \emph {et~al.}(2024)\citenamefont {Zeng}, \citenamefont {Guerci}, \citenamefont {Cr{\'e}pel}, \citenamefont {Millis},\ and\ \citenamefont {Cano}}]{Zeng2024SublatticeTopology}%
  \BibitemOpen
  \bibfield  {author} {\bibinfo {author} {\bibfnamefont {Y.}~\bibnamefont {Zeng}}, \bibinfo {author} {\bibfnamefont {D.}~\bibnamefont {Guerci}}, \bibinfo {author} {\bibfnamefont {V.}~\bibnamefont {Cr{\'e}pel}}, \bibinfo {author} {\bibfnamefont {A.~J.}\ \bibnamefont {Millis}},\ and\ \bibinfo {author} {\bibfnamefont {J.}~\bibnamefont {Cano}},\ }\bibfield  {title} {\bibinfo {title} {Sublattice structure and topology in spontaneously crystallized electronic states},\ }\href {https://doi.org/10.1103/PhysRevLett.132.236601} {\bibfield  {journal} {\bibinfo  {journal} {Phys. Rev. Lett.}\ }\textbf {\bibinfo {volume} {132}},\ \bibinfo {pages} {236601} (\bibinfo {year} {2024})}\BibitemShut {NoStop}%
\bibitem [{\citenamefont {Li}\ \emph {et~al.}(2025{\natexlab{b}})\citenamefont {Li}, \citenamefont {Chen}, \citenamefont {Li}, \citenamefont {Chen}, \citenamefont {Wu}, \citenamefont {Chen},\ and\ \citenamefont {Ren}}]{Li2025DeepLearningTopologicalInsulators}%
  \BibitemOpen
  \bibfield  {author} {\bibinfo {author} {\bibfnamefont {X.}~\bibnamefont {Li}}, \bibinfo {author} {\bibfnamefont {Y.}~\bibnamefont {Chen}}, \bibinfo {author} {\bibfnamefont {B.}~\bibnamefont {Li}}, \bibinfo {author} {\bibfnamefont {H.}~\bibnamefont {Chen}}, \bibinfo {author} {\bibfnamefont {F.}~\bibnamefont {Wu}}, \bibinfo {author} {\bibfnamefont {J.}~\bibnamefont {Chen}},\ and\ \bibinfo {author} {\bibfnamefont {W.}~\bibnamefont {Ren}},\ }\href {https://arxiv.org/abs/2503.11756} {\bibinfo {title} {Deep learning sheds light on integer and fractional topological insulators}} (\bibinfo {year} {2025}{\natexlab{b}}),\ \Eprint {https://arxiv.org/abs/2503.11756} {arXiv:2503.11756 [cond-mat.str-el]} \BibitemShut {NoStop}%
\bibitem [{\citenamefont {Liu}\ \emph {et~al.}(2024)\citenamefont {Liu}, \citenamefont {He}, \citenamefont {Wang}, \citenamefont {Zhang}, \citenamefont {Cao},\ and\ \citenamefont {Xiao}}]{LiuWangZhangCaoXiao2023antiferromagnetic}%
  \BibitemOpen
  \bibfield  {author} {\bibinfo {author} {\bibfnamefont {X.}~\bibnamefont {Liu}}, \bibinfo {author} {\bibfnamefont {Y.}~\bibnamefont {He}}, \bibinfo {author} {\bibfnamefont {C.}~\bibnamefont {Wang}}, \bibinfo {author} {\bibfnamefont {X.-W.}\ \bibnamefont {Zhang}}, \bibinfo {author} {\bibfnamefont {T.}~\bibnamefont {Cao}},\ and\ \bibinfo {author} {\bibfnamefont {D.}~\bibnamefont {Xiao}},\ }\bibfield  {title} {\bibinfo {title} {Gate-tunable antiferromagnetic chern insulator in twisted bilayer transition metal dichalcogenides},\ }\href {https://doi.org/10.1103/PhysRevLett.132.146401} {\bibfield  {journal} {\bibinfo  {journal} {Phys. Rev. Lett.}\ }\textbf {\bibinfo {volume} {132}},\ \bibinfo {pages} {146401} (\bibinfo {year} {2024})}\BibitemShut {NoStop}%
\bibitem [{\citenamefont {Li}\ \emph {et~al.}(2024{\natexlab{b}})\citenamefont {Li}, \citenamefont {Qiu},\ and\ \citenamefont {Wu}}]{Fengcheng2023tMoTe2HFnum1}%
  \BibitemOpen
  \bibfield  {author} {\bibinfo {author} {\bibfnamefont {B.}~\bibnamefont {Li}}, \bibinfo {author} {\bibfnamefont {W.-X.}\ \bibnamefont {Qiu}},\ and\ \bibinfo {author} {\bibfnamefont {F.}~\bibnamefont {Wu}},\ }\bibfield  {title} {\bibinfo {title} {Electrically tuned topology and magnetism in twisted bilayer ${\mathrm{mote}}_{2}$ at ${\ensuremath{\nu}}_{h}=1$},\ }\href {https://doi.org/10.1103/PhysRevB.109.L041106} {\bibfield  {journal} {\bibinfo  {journal} {Phys. Rev. B}\ }\textbf {\bibinfo {volume} {109}},\ \bibinfo {pages} {L041106} (\bibinfo {year} {2024}{\natexlab{b}})}\BibitemShut {NoStop}%
\bibitem [{\citenamefont {Sharma}\ \emph {et~al.}(2024)\citenamefont {Sharma}, \citenamefont {Peng},\ and\ \citenamefont {Sheng}}]{Sharma_2024qp}%
  \BibitemOpen
  \bibfield  {author} {\bibinfo {author} {\bibfnamefont {P.}~\bibnamefont {Sharma}}, \bibinfo {author} {\bibfnamefont {Y.}~\bibnamefont {Peng}},\ and\ \bibinfo {author} {\bibfnamefont {D.~N.}\ \bibnamefont {Sheng}},\ }\bibfield  {title} {\bibinfo {title} {Topological quantum phase transitions driven by a displacement field in twisted ${\mathrm{mote}}_{2}$ bilayers},\ }\href {https://doi.org/10.1103/PhysRevB.110.125142} {\bibfield  {journal} {\bibinfo  {journal} {Phys. Rev. B}\ }\textbf {\bibinfo {volume} {110}},\ \bibinfo {pages} {125142} (\bibinfo {year} {2024})}\BibitemShut {NoStop}%
\bibitem [{\citenamefont {Wang}\ \emph {et~al.}(2024{\natexlab{b}})\citenamefont {Wang}, \citenamefont {Wang},\ and\ \citenamefont {Vafek}}]{WangVafek2024MagneticFieldMoTe2}%
  \BibitemOpen
  \bibfield  {author} {\bibinfo {author} {\bibfnamefont {M.}~\bibnamefont {Wang}}, \bibinfo {author} {\bibfnamefont {X.}~\bibnamefont {Wang}},\ and\ \bibinfo {author} {\bibfnamefont {O.}~\bibnamefont {Vafek}},\ }\bibfield  {title} {\bibinfo {title} {Phase diagram of twisted bilayer ${\mathrm{mote}}_{2}$ in a magnetic field with an account for the electron-electron interaction},\ }\href {https://doi.org/10.1103/PhysRevB.110.L201107} {\bibfield  {journal} {\bibinfo  {journal} {Phys. Rev. B}\ }\textbf {\bibinfo {volume} {110}},\ \bibinfo {pages} {L201107} (\bibinfo {year} {2024}{\natexlab{b}})}\BibitemShut {NoStop}%
\bibitem [{\citenamefont {Shi}\ and\ \citenamefont {Liu}(2026)}]{ShiLiu2026}%
  \BibitemOpen
  \bibfield  {author} {\bibinfo {author} {\bibfnamefont {Y.}~\bibnamefont {Shi}}\ and\ \bibinfo {author} {\bibfnamefont {Z.}~\bibnamefont {Liu}},\ }\bibfield  {title} {\bibinfo {title} {Intervalley band crossing and transition of fractional chern insulators in floquet twisted bilayer mote2},\ }\href {https://doi.org/10.1088/0256-307X/43/5/050716} {\bibfield  {journal} {\bibinfo  {journal} {Chinese Physics Letters}\ }\textbf {\bibinfo {volume} {43}},\ \bibinfo {pages} {050716} (\bibinfo {year} {2026})}\BibitemShut {NoStop}%
\bibitem [{\citenamefont {Dong}\ \emph {et~al.}(2023)\citenamefont {Dong}, \citenamefont {Wang}, \citenamefont {Ledwith}, \citenamefont {Vishwanath},\ and\ \citenamefont {Parker}}]{Dong2023CFLtMoTe2}%
  \BibitemOpen
  \bibfield  {author} {\bibinfo {author} {\bibfnamefont {J.}~\bibnamefont {Dong}}, \bibinfo {author} {\bibfnamefont {J.}~\bibnamefont {Wang}}, \bibinfo {author} {\bibfnamefont {P.~J.}\ \bibnamefont {Ledwith}}, \bibinfo {author} {\bibfnamefont {A.}~\bibnamefont {Vishwanath}},\ and\ \bibinfo {author} {\bibfnamefont {D.~E.}\ \bibnamefont {Parker}},\ }\bibfield  {title} {\bibinfo {title} {Composite fermi liquid at zero magnetic field in twisted ${\mathrm{mote}}_{2}$},\ }\href {https://doi.org/10.1103/PhysRevLett.131.136502} {\bibfield  {journal} {\bibinfo  {journal} {Phys. Rev. Lett.}\ }\textbf {\bibinfo {volume} {131}},\ \bibinfo {pages} {136502} (\bibinfo {year} {2023})}\BibitemShut {NoStop}%
\bibitem [{\citenamefont {Goldman}\ \emph {et~al.}(2023)\citenamefont {Goldman}, \citenamefont {Reddy}, \citenamefont {Paul},\ and\ \citenamefont {Fu}}]{Goldman2023Composite}%
  \BibitemOpen
  \bibfield  {author} {\bibinfo {author} {\bibfnamefont {H.}~\bibnamefont {Goldman}}, \bibinfo {author} {\bibfnamefont {A.~P.}\ \bibnamefont {Reddy}}, \bibinfo {author} {\bibfnamefont {N.}~\bibnamefont {Paul}},\ and\ \bibinfo {author} {\bibfnamefont {L.}~\bibnamefont {Fu}},\ }\bibfield  {title} {\bibinfo {title} {Zero-field composite fermi liquid in twisted semiconductor bilayers},\ }\href {https://doi.org/10.1103/PhysRevLett.131.136501} {\bibfield  {journal} {\bibinfo  {journal} {Phys. Rev. Lett.}\ }\textbf {\bibinfo {volume} {131}},\ \bibinfo {pages} {136501} (\bibinfo {year} {2023})}\BibitemShut {NoStop}%
\bibitem [{\citenamefont {Wang}\ \emph {et~al.}(2025{\natexlab{b}})\citenamefont {Wang}, \citenamefont {Zhang}, \citenamefont {Liu}, \citenamefont {Wang}, \citenamefont {Cao},\ and\ \citenamefont {Xiao}}]{Wang2025higherLL}%
  \BibitemOpen
  \bibfield  {author} {\bibinfo {author} {\bibfnamefont {C.}~\bibnamefont {Wang}}, \bibinfo {author} {\bibfnamefont {X.-W.}\ \bibnamefont {Zhang}}, \bibinfo {author} {\bibfnamefont {X.}~\bibnamefont {Liu}}, \bibinfo {author} {\bibfnamefont {J.}~\bibnamefont {Wang}}, \bibinfo {author} {\bibfnamefont {T.}~\bibnamefont {Cao}},\ and\ \bibinfo {author} {\bibfnamefont {D.}~\bibnamefont {Xiao}},\ }\bibfield  {title} {\bibinfo {title} {Higher landau-level analogs and signatures of non-abelian states in twisted bilayer ${\mathrm{mote}}_{2}$},\ }\href {https://doi.org/10.1103/PhysRevLett.134.076503} {\bibfield  {journal} {\bibinfo  {journal} {Phys. Rev. Lett.}\ }\textbf {\bibinfo {volume} {134}},\ \bibinfo {pages} {076503} (\bibinfo {year} {2025}{\natexlab{b}})}\BibitemShut {NoStop}%
\bibitem [{\citenamefont {Xu}\ \emph {et~al.}(2025{\natexlab{c}})\citenamefont {Xu}, \citenamefont {Mao}, \citenamefont {Zeng},\ and\ \citenamefont {Zhang}}]{Xu2025nonabelian}%
  \BibitemOpen
  \bibfield  {author} {\bibinfo {author} {\bibfnamefont {C.}~\bibnamefont {Xu}}, \bibinfo {author} {\bibfnamefont {N.}~\bibnamefont {Mao}}, \bibinfo {author} {\bibfnamefont {T.}~\bibnamefont {Zeng}},\ and\ \bibinfo {author} {\bibfnamefont {Y.}~\bibnamefont {Zhang}},\ }\bibfield  {title} {\bibinfo {title} {Multiple chern bands in twisted ${\mathrm{mote}}_{2}$ and possible non-abelian states},\ }\href {https://doi.org/10.1103/PhysRevLett.134.066601} {\bibfield  {journal} {\bibinfo  {journal} {Phys. Rev. Lett.}\ }\textbf {\bibinfo {volume} {134}},\ \bibinfo {pages} {066601} (\bibinfo {year} {2025}{\natexlab{c}})}\BibitemShut {NoStop}%
\bibitem [{\citenamefont {Ahn}\ \emph {et~al.}(2024)\citenamefont {Ahn}, \citenamefont {Lee}, \citenamefont {Yananose}, \citenamefont {Kim},\ and\ \citenamefont {Cho}}]{Ahn2024nonabelian}%
  \BibitemOpen
  \bibfield  {author} {\bibinfo {author} {\bibfnamefont {C.-E.}\ \bibnamefont {Ahn}}, \bibinfo {author} {\bibfnamefont {W.}~\bibnamefont {Lee}}, \bibinfo {author} {\bibfnamefont {K.}~\bibnamefont {Yananose}}, \bibinfo {author} {\bibfnamefont {Y.}~\bibnamefont {Kim}},\ and\ \bibinfo {author} {\bibfnamefont {G.~Y.}\ \bibnamefont {Cho}},\ }\bibfield  {title} {\bibinfo {title} {Non-abelian fractional quantum anomalous hall states and first landau level physics of the second moir\'e band of twisted bilayer ${\mathrm{mote}}_{2}$},\ }\href {https://doi.org/10.1103/PhysRevB.110.L161109} {\bibfield  {journal} {\bibinfo  {journal} {Phys. Rev. B}\ }\textbf {\bibinfo {volume} {110}},\ \bibinfo {pages} {L161109} (\bibinfo {year} {2024})}\BibitemShut {NoStop}%
\bibitem [{\citenamefont {Chen}\ \emph {et~al.}(2025)\citenamefont {Chen}, \citenamefont {Luo}, \citenamefont {Zhu},\ and\ \citenamefont {Sheng}}]{Chen2025nonabelian}%
  \BibitemOpen
  \bibfield  {author} {\bibinfo {author} {\bibfnamefont {F.}~\bibnamefont {Chen}}, \bibinfo {author} {\bibfnamefont {W.-W.}\ \bibnamefont {Luo}}, \bibinfo {author} {\bibfnamefont {W.}~\bibnamefont {Zhu}},\ and\ \bibinfo {author} {\bibfnamefont {D.~N.}\ \bibnamefont {Sheng}},\ }\bibfield  {title} {\bibinfo {title} {Robust non-abelian even-denominator fractional chern insulator in twisted bilayer mote2},\ }\href {https://doi.org/10.1038/s41467-025-57326-3} {\bibfield  {journal} {\bibinfo  {journal} {Nature Communications}\ }\textbf {\bibinfo {volume} {16}},\ \bibinfo {pages} {2115} (\bibinfo {year} {2025})}\BibitemShut {NoStop}%
\bibitem [{\citenamefont {Reddy}\ \emph {et~al.}(2024)\citenamefont {Reddy}, \citenamefont {Paul}, \citenamefont {Abouelkomsan},\ and\ \citenamefont {Fu}}]{Reddy2024NonAbelianMinibands}%
  \BibitemOpen
  \bibfield  {author} {\bibinfo {author} {\bibfnamefont {A.~P.}\ \bibnamefont {Reddy}}, \bibinfo {author} {\bibfnamefont {N.}~\bibnamefont {Paul}}, \bibinfo {author} {\bibfnamefont {A.}~\bibnamefont {Abouelkomsan}},\ and\ \bibinfo {author} {\bibfnamefont {L.}~\bibnamefont {Fu}},\ }\bibfield  {title} {\bibinfo {title} {Non-abelian fractionalization in topological minibands},\ }\href {https://doi.org/10.1103/PhysRevLett.133.166503} {\bibfield  {journal} {\bibinfo  {journal} {Phys. Rev. Lett.}\ }\textbf {\bibinfo {volume} {133}},\ \bibinfo {pages} {166503} (\bibinfo {year} {2024})}\BibitemShut {NoStop}%
\bibitem [{\citenamefont {Reddy}\ \emph {et~al.}(2026)\citenamefont {Reddy}, \citenamefont {Sheng}, \citenamefont {Abouelkomsan}, \citenamefont {Bergholtz},\ and\ \citenamefont {Fu}}]{Reddy2026nonabelian}%
  \BibitemOpen
  \bibfield  {author} {\bibinfo {author} {\bibfnamefont {A.~P.}\ \bibnamefont {Reddy}}, \bibinfo {author} {\bibfnamefont {D.~N.}\ \bibnamefont {Sheng}}, \bibinfo {author} {\bibfnamefont {A.}~\bibnamefont {Abouelkomsan}}, \bibinfo {author} {\bibfnamefont {E.~J.}\ \bibnamefont {Bergholtz}},\ and\ \bibinfo {author} {\bibfnamefont {L.}~\bibnamefont {Fu}},\ }\bibfield  {title} {\bibinfo {title} {Anti-topological crystal and non-abelian liquid in twisted semiconductor bilayers},\ }\href {https://doi.org/10.1038/s41467-026-70916-z} {\bibfield  {journal} {\bibinfo  {journal} {Nature Communications}\ }\textbf {\bibinfo {volume} {17}},\ \bibinfo {pages} {3814} (\bibinfo {year} {2026})}\BibitemShut {NoStop}%
\bibitem [{\citenamefont {Li}\ \emph {et~al.}(2026{\natexlab{b}})\citenamefont {Li}, \citenamefont {Ouyang},\ and\ \citenamefont {Wu}}]{LiWuAbelian2026}%
  \BibitemOpen
  \bibfield  {author} {\bibinfo {author} {\bibfnamefont {B.}~\bibnamefont {Li}}, \bibinfo {author} {\bibfnamefont {Y.}~\bibnamefont {Ouyang}},\ and\ \bibinfo {author} {\bibfnamefont {F.}~\bibnamefont {Wu}},\ }\bibfield  {title} {\bibinfo {title} {Abelian and non-abelian fractionalized states in twisted ${\mathrm{mote}}_{2}$: A generalized landau-level theory},\ }\href {https://doi.org/10.1103/dvry-pfnb} {\bibfield  {journal} {\bibinfo  {journal} {Phys. Rev. B}\ }\textbf {\bibinfo {volume} {113}},\ \bibinfo {pages} {195129} (\bibinfo {year} {2026}{\natexlab{b}})}\BibitemShut {NoStop}%
\bibitem [{Xia()}]{Xiaodongnew}%
  \BibitemOpen
  \bibinfo {note} {Seminar talk by Xiaodong Xu at Moire 2.0 conference, Princeton University, November 2025}\BibitemShut {NoStop}%
\bibitem [{\citenamefont {Pan}\ \emph {et~al.}(2026)\citenamefont {Pan}, \citenamefont {Yang}, \citenamefont {Wang}, \citenamefont {Cai}, \citenamefont {Wang}, \citenamefont {Zhao}, \citenamefont {Watanabe}, \citenamefont {Taniguchi}, \citenamefont {Zhang}, \citenamefont {Liu}, \citenamefont {Yang},\ and\ \citenamefont {Gao}}]{Pan1v3FCI2026}%
  \BibitemOpen
  \bibfield  {author} {\bibinfo {author} {\bibfnamefont {H.}~\bibnamefont {Pan}}, \bibinfo {author} {\bibfnamefont {S.}~\bibnamefont {Yang}}, \bibinfo {author} {\bibfnamefont {Y.}~\bibnamefont {Wang}}, \bibinfo {author} {\bibfnamefont {X.}~\bibnamefont {Cai}}, \bibinfo {author} {\bibfnamefont {W.}~\bibnamefont {Wang}}, \bibinfo {author} {\bibfnamefont {Y.}~\bibnamefont {Zhao}}, \bibinfo {author} {\bibfnamefont {K.}~\bibnamefont {Watanabe}}, \bibinfo {author} {\bibfnamefont {T.}~\bibnamefont {Taniguchi}}, \bibinfo {author} {\bibfnamefont {L.}~\bibnamefont {Zhang}}, \bibinfo {author} {\bibfnamefont {Y.}~\bibnamefont {Liu}}, \bibinfo {author} {\bibfnamefont {B.}~\bibnamefont {Yang}},\ and\ \bibinfo {author} {\bibfnamefont {W.}~\bibnamefont {Gao}},\ }\bibfield  {title} {\bibinfo {title} {Optical signatures of $\ensuremath{-}\frac{1}{3}$ fractional quantum anomalous hall state in twisted ${\mathrm{mote}}_{2}$},\ }\href {https://doi.org/10.1103/f4dj-7sts} {\bibfield  {journal} {\bibinfo  {journal} {Phys. Rev.
  Lett.}\ }\textbf {\bibinfo {volume} {136}},\ \bibinfo {pages} {056601} (\bibinfo {year} {2026})}\BibitemShut {NoStop}%
\bibitem [{\citenamefont {Kang}\ \emph {et~al.}(2025)\citenamefont {Kang}, \citenamefont {Qiu}, \citenamefont {Shen}, \citenamefont {Lee}, \citenamefont {Xia}, \citenamefont {Zeng}, \citenamefont {Watanabe}, \citenamefont {Taniguchi}, \citenamefont {Shan},\ and\ \citenamefont {Mak}}]{KangTRBFQSH2025}%
  \BibitemOpen
  \bibfield  {author} {\bibinfo {author} {\bibfnamefont {K.}~\bibnamefont {Kang}}, \bibinfo {author} {\bibfnamefont {Y.}~\bibnamefont {Qiu}}, \bibinfo {author} {\bibfnamefont {B.}~\bibnamefont {Shen}}, \bibinfo {author} {\bibfnamefont {K.}~\bibnamefont {Lee}}, \bibinfo {author} {\bibfnamefont {Z.}~\bibnamefont {Xia}}, \bibinfo {author} {\bibfnamefont {Y.}~\bibnamefont {Zeng}}, \bibinfo {author} {\bibfnamefont {K.}~\bibnamefont {Watanabe}}, \bibinfo {author} {\bibfnamefont {T.}~\bibnamefont {Taniguchi}}, \bibinfo {author} {\bibfnamefont {J.}~\bibnamefont {Shan}},\ and\ \bibinfo {author} {\bibfnamefont {K.~F.}\ \bibnamefont {Mak}},\ }\href {https://arxiv.org/abs/2501.02525} {\bibinfo {title} {Time-reversal symmetry breaking fractional quantum spin hall insulator in moir\'e mote2}} (\bibinfo {year} {2025}),\ \Eprint {https://arxiv.org/abs/2501.02525} {arXiv:2501.02525 [cond-mat.mes-hall]} \BibitemShut {NoStop}%
\bibitem [{\citenamefont {Anderson}\ \emph {et~al.}(2024)\citenamefont {Anderson}, \citenamefont {Cai}, \citenamefont {Reddy}, \citenamefont {Park}, \citenamefont {Holtzmann}, \citenamefont {Davis}, \citenamefont {Taniguchi}, \citenamefont {Watanabe}, \citenamefont {Smolenski}, \citenamefont {Imamo{\u{g}}lu}, \citenamefont {Cao}, \citenamefont {Xiao}, \citenamefont {Fu}, \citenamefont {Yao},\ and\ \citenamefont {Xu}}]{Anderson2024TrionCFL}%
  \BibitemOpen
  \bibfield  {author} {\bibinfo {author} {\bibfnamefont {E.}~\bibnamefont {Anderson}}, \bibinfo {author} {\bibfnamefont {J.}~\bibnamefont {Cai}}, \bibinfo {author} {\bibfnamefont {A.~P.}\ \bibnamefont {Reddy}}, \bibinfo {author} {\bibfnamefont {H.}~\bibnamefont {Park}}, \bibinfo {author} {\bibfnamefont {W.}~\bibnamefont {Holtzmann}}, \bibinfo {author} {\bibfnamefont {K.}~\bibnamefont {Davis}}, \bibinfo {author} {\bibfnamefont {T.}~\bibnamefont {Taniguchi}}, \bibinfo {author} {\bibfnamefont {K.}~\bibnamefont {Watanabe}}, \bibinfo {author} {\bibfnamefont {T.}~\bibnamefont {Smolenski}}, \bibinfo {author} {\bibfnamefont {A.}~\bibnamefont {Imamo{\u{g}}lu}}, \bibinfo {author} {\bibfnamefont {T.}~\bibnamefont {Cao}}, \bibinfo {author} {\bibfnamefont {D.}~\bibnamefont {Xiao}}, \bibinfo {author} {\bibfnamefont {L.}~\bibnamefont {Fu}}, \bibinfo {author} {\bibfnamefont {W.}~\bibnamefont {Yao}},\ and\ \bibinfo {author} {\bibfnamefont {X.}~\bibnamefont {Xu}},\ }\bibfield  {title} {\bibinfo {title} {Trion sensing of a
  zero-field composite fermi liquid},\ }\href {https://doi.org/10.1038/s41586-024-08134-0} {\bibfield  {journal} {\bibinfo  {journal} {Nature}\ }\textbf {\bibinfo {volume} {635}},\ \bibinfo {pages} {590} (\bibinfo {year} {2024})}\BibitemShut {NoStop}%
\bibitem [{\citenamefont {Wu}\ \emph {et~al.}(2019)\citenamefont {Wu}, \citenamefont {Lovorn}, \citenamefont {Tutuc}, \citenamefont {Martin},\ and\ \citenamefont {MacDonald}}]{Wu2019TIintTMD}%
  \BibitemOpen
  \bibfield  {author} {\bibinfo {author} {\bibfnamefont {F.}~\bibnamefont {Wu}}, \bibinfo {author} {\bibfnamefont {T.}~\bibnamefont {Lovorn}}, \bibinfo {author} {\bibfnamefont {E.}~\bibnamefont {Tutuc}}, \bibinfo {author} {\bibfnamefont {I.}~\bibnamefont {Martin}},\ and\ \bibinfo {author} {\bibfnamefont {A.~H.}\ \bibnamefont {MacDonald}},\ }\bibfield  {title} {\bibinfo {title} {Topological insulators in twisted transition metal dichalcogenide homobilayers},\ }\href {https://doi.org/10.1103/PhysRevLett.122.086402} {\bibfield  {journal} {\bibinfo  {journal} {Phys. Rev. Lett.}\ }\textbf {\bibinfo {volume} {122}},\ \bibinfo {pages} {086402} (\bibinfo {year} {2019})}\BibitemShut {NoStop}%
\bibitem [{\citenamefont {Jia}\ \emph {et~al.}(2024)\citenamefont {Jia}, \citenamefont {Yu}, \citenamefont {Liu}, \citenamefont {Herzog-Arbeitman}, \citenamefont {Qi}, \citenamefont {Pi}, \citenamefont {Regnault}, \citenamefont {Weng}, \citenamefont {Bernevig},\ and\ \citenamefont {Wu}}]{MFCII}%
  \BibitemOpen
  \bibfield  {author} {\bibinfo {author} {\bibfnamefont {Y.}~\bibnamefont {Jia}}, \bibinfo {author} {\bibfnamefont {J.}~\bibnamefont {Yu}}, \bibinfo {author} {\bibfnamefont {J.}~\bibnamefont {Liu}}, \bibinfo {author} {\bibfnamefont {J.}~\bibnamefont {Herzog-Arbeitman}}, \bibinfo {author} {\bibfnamefont {Z.}~\bibnamefont {Qi}}, \bibinfo {author} {\bibfnamefont {H.}~\bibnamefont {Pi}}, \bibinfo {author} {\bibfnamefont {N.}~\bibnamefont {Regnault}}, \bibinfo {author} {\bibfnamefont {H.}~\bibnamefont {Weng}}, \bibinfo {author} {\bibfnamefont {B.~A.}\ \bibnamefont {Bernevig}},\ and\ \bibinfo {author} {\bibfnamefont {Q.}~\bibnamefont {Wu}},\ }\bibfield  {title} {\bibinfo {title} {Moir\'e fractional chern insulators. i. first-principles calculations and continuum models of twisted bilayer ${\mathrm{mote}}_{2}$},\ }\href {https://doi.org/10.1103/PhysRevB.109.205121} {\bibfield  {journal} {\bibinfo  {journal} {Phys. Rev. B}\ }\textbf {\bibinfo {volume} {109}},\ \bibinfo {pages} {205121} (\bibinfo {year}
  {2024})}\BibitemShut {NoStop}%
\bibitem [{\citenamefont {Mao}\ \emph {et~al.}(2024)\citenamefont {Mao}, \citenamefont {Xu}, \citenamefont {Li}, \citenamefont {Bao}, \citenamefont {Liu}, \citenamefont {Xu}, \citenamefont {Felser}, \citenamefont {Fu},\ and\ \citenamefont {Zhang}}]{Mao2024translearn}%
  \BibitemOpen
  \bibfield  {author} {\bibinfo {author} {\bibfnamefont {N.}~\bibnamefont {Mao}}, \bibinfo {author} {\bibfnamefont {C.}~\bibnamefont {Xu}}, \bibinfo {author} {\bibfnamefont {J.}~\bibnamefont {Li}}, \bibinfo {author} {\bibfnamefont {T.}~\bibnamefont {Bao}}, \bibinfo {author} {\bibfnamefont {P.}~\bibnamefont {Liu}}, \bibinfo {author} {\bibfnamefont {Y.}~\bibnamefont {Xu}}, \bibinfo {author} {\bibfnamefont {C.}~\bibnamefont {Felser}}, \bibinfo {author} {\bibfnamefont {L.}~\bibnamefont {Fu}},\ and\ \bibinfo {author} {\bibfnamefont {Y.}~\bibnamefont {Zhang}},\ }\bibfield  {title} {\bibinfo {title} {Transfer learning relaxation, electronic structure and continuum model for twisted bilayer mote2},\ }\href {https://doi.org/10.1038/s42005-024-01754-y} {\bibfield  {journal} {\bibinfo  {journal} {Communications Physics}\ }\textbf {\bibinfo {volume} {7}},\ \bibinfo {pages} {262} (\bibinfo {year} {2024})}\BibitemShut {NoStop}%
\bibitem [{\citenamefont {Zhang}\ \emph {et~al.}(2024{\natexlab{a}})\citenamefont {Zhang}, \citenamefont {Wang}, \citenamefont {Liu}, \citenamefont {Fan}, \citenamefont {Cao},\ and\ \citenamefont {Xiao}}]{Zhang2024pol}%
  \BibitemOpen
  \bibfield  {author} {\bibinfo {author} {\bibfnamefont {X.-W.}\ \bibnamefont {Zhang}}, \bibinfo {author} {\bibfnamefont {C.}~\bibnamefont {Wang}}, \bibinfo {author} {\bibfnamefont {X.}~\bibnamefont {Liu}}, \bibinfo {author} {\bibfnamefont {Y.}~\bibnamefont {Fan}}, \bibinfo {author} {\bibfnamefont {T.}~\bibnamefont {Cao}},\ and\ \bibinfo {author} {\bibfnamefont {D.}~\bibnamefont {Xiao}},\ }\bibfield  {title} {\bibinfo {title} {Polarization-driven band topology evolution in twisted mote2 and wse2},\ }\href {https://doi.org/10.1038/s41467-024-48511-x} {\bibfield  {journal} {\bibinfo  {journal} {Nature Communications}\ }\textbf {\bibinfo {volume} {15}},\ \bibinfo {pages} {4223} (\bibinfo {year} {2024}{\natexlab{a}})}\BibitemShut {NoStop}%
\bibitem [{\citenamefont {Zhang}\ \emph {et~al.}(2024{\natexlab{b}})\citenamefont {Zhang}, \citenamefont {Pi}, \citenamefont {Liu}, \citenamefont {Miao}, \citenamefont {Qi}, \citenamefont {Regnault}, \citenamefont {Weng}, \citenamefont {Dai}, \citenamefont {Bernevig}, \citenamefont {Wu},\ and\ \citenamefont {Yu}}]{DFTnofitting2024}%
  \BibitemOpen
  \bibfield  {author} {\bibinfo {author} {\bibfnamefont {Y.}~\bibnamefont {Zhang}}, \bibinfo {author} {\bibfnamefont {H.}~\bibnamefont {Pi}}, \bibinfo {author} {\bibfnamefont {J.}~\bibnamefont {Liu}}, \bibinfo {author} {\bibfnamefont {W.}~\bibnamefont {Miao}}, \bibinfo {author} {\bibfnamefont {Z.}~\bibnamefont {Qi}}, \bibinfo {author} {\bibfnamefont {N.}~\bibnamefont {Regnault}}, \bibinfo {author} {\bibfnamefont {H.}~\bibnamefont {Weng}}, \bibinfo {author} {\bibfnamefont {X.}~\bibnamefont {Dai}}, \bibinfo {author} {\bibfnamefont {B.~A.}\ \bibnamefont {Bernevig}}, \bibinfo {author} {\bibfnamefont {Q.}~\bibnamefont {Wu}},\ and\ \bibinfo {author} {\bibfnamefont {J.}~\bibnamefont {Yu}},\ }\href {https://arxiv.org/abs/2411.08108} {\bibinfo {title} {Universal moir\'e-model-building method without fitting: Application to twisted mote$_2$ and wse$_2$}} (\bibinfo {year} {2024}{\natexlab{b}}),\ \Eprint {https://arxiv.org/abs/2411.08108} {arXiv:2411.08108 [cond-mat.mes-hall]} \BibitemShut {NoStop}%
\bibitem [{\citenamefont {Zhang}\ \emph {et~al.}(2025)\citenamefont {Zhang}, \citenamefont {Yang}, \citenamefont {Wang}, \citenamefont {Liu}, \citenamefont {Cao},\ and\ \citenamefont {Xiao}}]{Zhangtwisttransferable2025}%
  \BibitemOpen
  \bibfield  {author} {\bibinfo {author} {\bibfnamefont {X.-W.}\ \bibnamefont {Zhang}}, \bibinfo {author} {\bibfnamefont {K.}~\bibnamefont {Yang}}, \bibinfo {author} {\bibfnamefont {C.}~\bibnamefont {Wang}}, \bibinfo {author} {\bibfnamefont {X.}~\bibnamefont {Liu}}, \bibinfo {author} {\bibfnamefont {T.}~\bibnamefont {Cao}},\ and\ \bibinfo {author} {\bibfnamefont {D.}~\bibnamefont {Xiao}},\ }\bibfield  {title} {\bibinfo {title} {Twist-angle transferable continuum model and second flat chern band in twisted mote2 and wse2},\ }\href {https://doi.org/10.1038/s41535-025-00828-6} {\bibfield  {journal} {\bibinfo  {journal} {npj Quantum Materials}\ }\textbf {\bibinfo {volume} {10}},\ \bibinfo {pages} {110} (\bibinfo {year} {2025})}\BibitemShut {NoStop}%
\bibitem [{\citenamefont {Zhang}\ \emph {et~al.}(2017)\citenamefont {Zhang}, \citenamefont {Jain},\ and\ \citenamefont {Eisenstein}}]{PhysRevB.95.195105}%
  \BibitemOpen
  \bibfield  {author} {\bibinfo {author} {\bibfnamefont {Y.}~\bibnamefont {Zhang}}, \bibinfo {author} {\bibfnamefont {J.~K.}\ \bibnamefont {Jain}},\ and\ \bibinfo {author} {\bibfnamefont {J.~P.}\ \bibnamefont {Eisenstein}},\ }\bibfield  {title} {\bibinfo {title} {Tunnel transport and interlayer excitons in bilayer fractional quantum hall systems},\ }\href {https://doi.org/10.1103/PhysRevB.95.195105} {\bibfield  {journal} {\bibinfo  {journal} {Phys. Rev. B}\ }\textbf {\bibinfo {volume} {95}},\ \bibinfo {pages} {195105} (\bibinfo {year} {2017})}\BibitemShut {NoStop}%
\bibitem [{\citenamefont {Zhao}\ \emph {et~al.}(2023)\citenamefont {Zhao}, \citenamefont {Balram},\ and\ \citenamefont {Jain}}]{PhysRevLett.130.186302}%
  \BibitemOpen
  \bibfield  {author} {\bibinfo {author} {\bibfnamefont {T.}~\bibnamefont {Zhao}}, \bibinfo {author} {\bibfnamefont {A.~C.}\ \bibnamefont {Balram}},\ and\ \bibinfo {author} {\bibfnamefont {J.~K.}\ \bibnamefont {Jain}},\ }\bibfield  {title} {\bibinfo {title} {Composite fermion pairing induced by landau level mixing},\ }\href {https://doi.org/10.1103/PhysRevLett.130.186302} {\bibfield  {journal} {\bibinfo  {journal} {Phys. Rev. Lett.}\ }\textbf {\bibinfo {volume} {130}},\ \bibinfo {pages} {186302} (\bibinfo {year} {2023})}\BibitemShut {NoStop}%
\bibitem [{\citenamefont {Reddy}\ \emph {et~al.}(2023)\citenamefont {Reddy}, \citenamefont {Alsallom}, \citenamefont {Zhang}, \citenamefont {Devakul},\ and\ \citenamefont {Fu}}]{reddy2023fractional}%
  \BibitemOpen
  \bibfield  {author} {\bibinfo {author} {\bibfnamefont {A.~P.}\ \bibnamefont {Reddy}}, \bibinfo {author} {\bibfnamefont {F.}~\bibnamefont {Alsallom}}, \bibinfo {author} {\bibfnamefont {Y.}~\bibnamefont {Zhang}}, \bibinfo {author} {\bibfnamefont {T.}~\bibnamefont {Devakul}},\ and\ \bibinfo {author} {\bibfnamefont {L.}~\bibnamefont {Fu}},\ }\bibfield  {title} {\bibinfo {title} {Fractional quantum anomalous hall states in twisted bilayer ${\mathrm{mote}}_{2}$ and ${\mathrm{wse}}_{2}$},\ }\href {https://doi.org/10.1103/PhysRevB.108.085117} {\bibfield  {journal} {\bibinfo  {journal} {Phys. Rev. B}\ }\textbf {\bibinfo {volume} {108}},\ \bibinfo {pages} {085117} (\bibinfo {year} {2023})}\BibitemShut {NoStop}%
\bibitem [{\citenamefont {Shi}\ \emph {et~al.}(2024)\citenamefont {Shi}, \citenamefont {Morales-Dur\'an}, \citenamefont {Khalaf},\ and\ \citenamefont {MacDonald}}]{Shi2024adiabatic}%
  \BibitemOpen
  \bibfield  {author} {\bibinfo {author} {\bibfnamefont {J.}~\bibnamefont {Shi}}, \bibinfo {author} {\bibfnamefont {N.}~\bibnamefont {Morales-Dur\'an}}, \bibinfo {author} {\bibfnamefont {E.}~\bibnamefont {Khalaf}},\ and\ \bibinfo {author} {\bibfnamefont {A.~H.}\ \bibnamefont {MacDonald}},\ }\bibfield  {title} {\bibinfo {title} {Adiabatic approximation and aharonov-casher bands in twisted homobilayer transition metal dichalcogenides},\ }\href {https://doi.org/10.1103/PhysRevB.110.035130} {\bibfield  {journal} {\bibinfo  {journal} {Phys. Rev. B}\ }\textbf {\bibinfo {volume} {110}},\ \bibinfo {pages} {035130} (\bibinfo {year} {2024})}\BibitemShut {NoStop}%
\bibitem [{\citenamefont {Laturia}\ \emph {et~al.}(2018)\citenamefont {Laturia}, \citenamefont {Van~de Put},\ and\ \citenamefont {Vandenberghe}}]{laturia_dielectric_2018}%
  \BibitemOpen
  \bibfield  {author} {\bibinfo {author} {\bibfnamefont {A.}~\bibnamefont {Laturia}}, \bibinfo {author} {\bibfnamefont {M.~L.}\ \bibnamefont {Van~de Put}},\ and\ \bibinfo {author} {\bibfnamefont {W.~G.}\ \bibnamefont {Vandenberghe}},\ }\bibfield  {title} {\bibinfo {title} {Dielectric properties of hexagonal boron nitride and transition metal dichalcogenides: from monolayer to bulk},\ }\href {https://doi.org/10.1038/s41699-018-0050-x} {\bibfield  {journal} {\bibinfo  {journal} {npj 2D Materials and Applications}\ }\textbf {\bibinfo {volume} {2}},\ \bibinfo {pages} {6} (\bibinfo {year} {2018})}\BibitemShut {NoStop}%
\bibitem [{\citenamefont {{Lieb}}\ \emph {et~al.}(1961)\citenamefont {{Lieb}}, \citenamefont {{Schultz}},\ and\ \citenamefont {{Mattis}}}]{1961AnPhy..16..407L}%
  \BibitemOpen
  \bibfield  {author} {\bibinfo {author} {\bibfnamefont {E.}~\bibnamefont {{Lieb}}}, \bibinfo {author} {\bibfnamefont {T.}~\bibnamefont {{Schultz}}},\ and\ \bibinfo {author} {\bibfnamefont {D.}~\bibnamefont {{Mattis}}},\ }\bibfield  {title} {\bibinfo {title} {{Two soluble models of an antiferromagnetic chain}},\ }\href {https://doi.org/10.1016/0003-4916(61)90115-4} {\bibfield  {journal} {\bibinfo  {journal} {Annals of Physics}\ }\textbf {\bibinfo {volume} {16}},\ \bibinfo {pages} {407} (\bibinfo {year} {1961})}\BibitemShut {NoStop}%
\bibitem [{\citenamefont {Bernevig}\ and\ \citenamefont {Regnault}(2012)}]{BernevigPhysRevB.85.075128}%
  \BibitemOpen
  \bibfield  {author} {\bibinfo {author} {\bibfnamefont {B.~A.}\ \bibnamefont {Bernevig}}\ and\ \bibinfo {author} {\bibfnamefont {N.}~\bibnamefont {Regnault}},\ }\bibfield  {title} {\bibinfo {title} {Emergent many-body translational symmetries of abelian and non-abelian fractionally filled topological insulators},\ }\href {https://doi.org/10.1103/PhysRevB.85.075128} {\bibfield  {journal} {\bibinfo  {journal} {Phys. Rev. B}\ }\textbf {\bibinfo {volume} {85}},\ \bibinfo {pages} {075128} (\bibinfo {year} {2012})}\BibitemShut {NoStop}%
\bibitem [{\citenamefont {Wu}\ \emph {et~al.}(2014)\citenamefont {Wu}, \citenamefont {Regnault},\ and\ \citenamefont {Bernevig}}]{2014PhRvB..89o5113W}%
  \BibitemOpen
  \bibfield  {author} {\bibinfo {author} {\bibfnamefont {Y.-L.}\ \bibnamefont {Wu}}, \bibinfo {author} {\bibfnamefont {N.}~\bibnamefont {Regnault}},\ and\ \bibinfo {author} {\bibfnamefont {B.~A.}\ \bibnamefont {Bernevig}},\ }\bibfield  {title} {\bibinfo {title} {Haldane statistics for fractional chern insulators with an arbitrary chern number},\ }\href {https://doi.org/10.1103/PhysRevB.89.155113} {\bibfield  {journal} {\bibinfo  {journal} {Phys. Rev. B}\ }\textbf {\bibinfo {volume} {89}},\ \bibinfo {pages} {155113} (\bibinfo {year} {2014})}\BibitemShut {NoStop}%
\bibitem [{\citenamefont {Sterdyniak}\ \emph {et~al.}(2011)\citenamefont {Sterdyniak}, \citenamefont {Regnault},\ and\ \citenamefont {Bernevig}}]{SterdyniakPhysRevLett.106.100405}%
  \BibitemOpen
  \bibfield  {author} {\bibinfo {author} {\bibfnamefont {A.}~\bibnamefont {Sterdyniak}}, \bibinfo {author} {\bibfnamefont {N.}~\bibnamefont {Regnault}},\ and\ \bibinfo {author} {\bibfnamefont {B.~A.}\ \bibnamefont {Bernevig}},\ }\bibfield  {title} {\bibinfo {title} {Extracting excitations from model state entanglement},\ }\href {https://doi.org/10.1103/PhysRevLett.106.100405} {\bibfield  {journal} {\bibinfo  {journal} {Phys. Rev. Lett.}\ }\textbf {\bibinfo {volume} {106}},\ \bibinfo {pages} {100405} (\bibinfo {year} {2011})}\BibitemShut {NoStop}%
\bibitem [{\citenamefont {Yu}\ \emph {et~al.}(2025)\citenamefont {Yu}, \citenamefont {Herzog-Arbeitman}, \citenamefont {Kwan}, \citenamefont {Regnault},\ and\ \citenamefont {Bernevig}}]{MFCIIV}%
  \BibitemOpen
  \bibfield  {author} {\bibinfo {author} {\bibfnamefont {J.}~\bibnamefont {Yu}}, \bibinfo {author} {\bibfnamefont {J.}~\bibnamefont {Herzog-Arbeitman}}, \bibinfo {author} {\bibfnamefont {Y.~H.}\ \bibnamefont {Kwan}}, \bibinfo {author} {\bibfnamefont {N.}~\bibnamefont {Regnault}},\ and\ \bibinfo {author} {\bibfnamefont {B.~A.}\ \bibnamefont {Bernevig}},\ }\bibfield  {title} {\bibinfo {title} {Moir\'e fractional chern insulators. iv. fluctuation-driven collapse in multiband exact diagonalization calculations on rhombohedral graphene},\ }\href {https://doi.org/10.1103/PhysRevB.112.075110} {\bibfield  {journal} {\bibinfo  {journal} {Phys. Rev. B}\ }\textbf {\bibinfo {volume} {112}},\ \bibinfo {pages} {075110} (\bibinfo {year} {2025})}\BibitemShut {NoStop}%
\bibitem [{\citenamefont {Li}\ \emph {et~al.}(2025{\natexlab{c}})\citenamefont {Li}, \citenamefont {Bernevig},\ and\ \citenamefont {Regnault}}]{Li2025multibanditer}%
  \BibitemOpen
  \bibfield  {author} {\bibinfo {author} {\bibfnamefont {H.}~\bibnamefont {Li}}, \bibinfo {author} {\bibfnamefont {B.~A.}\ \bibnamefont {Bernevig}},\ and\ \bibinfo {author} {\bibfnamefont {N.}~\bibnamefont {Regnault}},\ }\bibfield  {title} {\bibinfo {title} {Multiband exact diagonalization and an iteration approach to search for fractional chern insulators in rhombohedral multilayer graphene},\ }\href {https://doi.org/10.1103/zlqg-sj86} {\bibfield  {journal} {\bibinfo  {journal} {Phys. Rev. B}\ }\textbf {\bibinfo {volume} {112}},\ \bibinfo {pages} {075130} (\bibinfo {year} {2025}{\natexlab{c}})}\BibitemShut {NoStop}%
\bibitem [{\citenamefont {Liu}\ \emph {et~al.}(2013)\citenamefont {Liu}, \citenamefont {Repellin}, \citenamefont {Bernevig},\ and\ \citenamefont {Regnault}}]{PhysRevB.87.205136}%
  \BibitemOpen
  \bibfield  {author} {\bibinfo {author} {\bibfnamefont {T.}~\bibnamefont {Liu}}, \bibinfo {author} {\bibfnamefont {C.}~\bibnamefont {Repellin}}, \bibinfo {author} {\bibfnamefont {B.~A.}\ \bibnamefont {Bernevig}},\ and\ \bibinfo {author} {\bibfnamefont {N.}~\bibnamefont {Regnault}},\ }\bibfield  {title} {\bibinfo {title} {Fractional chern insulators beyond laughlin states},\ }\href {https://doi.org/10.1103/PhysRevB.87.205136} {\bibfield  {journal} {\bibinfo  {journal} {Phys. Rev. B}\ }\textbf {\bibinfo {volume} {87}},\ \bibinfo {pages} {205136} (\bibinfo {year} {2013})}\BibitemShut {NoStop}%
\bibitem [{\citenamefont {Lu}\ and\ \citenamefont {Santos}(2024)}]{Lu2024FCImoteCFL}%
  \BibitemOpen
  \bibfield  {author} {\bibinfo {author} {\bibfnamefont {T.}~\bibnamefont {Lu}}\ and\ \bibinfo {author} {\bibfnamefont {L.~H.}\ \bibnamefont {Santos}},\ }\bibfield  {title} {\bibinfo {title} {Fractional chern insulators in twisted bilayer ${\mathrm{mote}}_{2}$: A composite fermion perspective},\ }\href {https://doi.org/10.1103/PhysRevLett.133.186602} {\bibfield  {journal} {\bibinfo  {journal} {Phys. Rev. Lett.}\ }\textbf {\bibinfo {volume} {133}},\ \bibinfo {pages} {186602} (\bibinfo {year} {2024})}\BibitemShut {NoStop}%
\bibitem [{\citenamefont {Murthy}\ \emph {et~al.}(1998)\citenamefont {Murthy}, \citenamefont {Park}, \citenamefont {Shankar},\ and\ \citenamefont {Jain}}]{Murthy1998scaling}%
  \BibitemOpen
  \bibfield  {author} {\bibinfo {author} {\bibfnamefont {G.}~\bibnamefont {Murthy}}, \bibinfo {author} {\bibfnamefont {K.}~\bibnamefont {Park}}, \bibinfo {author} {\bibfnamefont {R.}~\bibnamefont {Shankar}},\ and\ \bibinfo {author} {\bibfnamefont {J.~K.}\ \bibnamefont {Jain}},\ }\bibfield  {title} {\bibinfo {title} {Scaling relations for gaps in fractional quantum hall states},\ }\href {https://doi.org/10.1103/PhysRevB.58.15363} {\bibfield  {journal} {\bibinfo  {journal} {Phys. Rev. B}\ }\textbf {\bibinfo {volume} {58}},\ \bibinfo {pages} {15363} (\bibinfo {year} {1998})}\BibitemShut {NoStop}%
\bibitem [{\citenamefont {Zhao}\ \emph {et~al.}(2022)\citenamefont {Zhao}, \citenamefont {Kudo}, \citenamefont {Faugno}, \citenamefont {Balram},\ and\ \citenamefont {Jain}}]{ZhaoJain2022scaling}%
  \BibitemOpen
  \bibfield  {author} {\bibinfo {author} {\bibfnamefont {T.}~\bibnamefont {Zhao}}, \bibinfo {author} {\bibfnamefont {K.}~\bibnamefont {Kudo}}, \bibinfo {author} {\bibfnamefont {W.~N.}\ \bibnamefont {Faugno}}, \bibinfo {author} {\bibfnamefont {A.~C.}\ \bibnamefont {Balram}},\ and\ \bibinfo {author} {\bibfnamefont {J.~K.}\ \bibnamefont {Jain}},\ }\bibfield  {title} {\bibinfo {title} {Revisiting excitation gaps in the fractional quantum hall effect},\ }\href {https://doi.org/10.1103/PhysRevB.105.205147} {\bibfield  {journal} {\bibinfo  {journal} {Phys. Rev. B}\ }\textbf {\bibinfo {volume} {105}},\ \bibinfo {pages} {205147} (\bibinfo {year} {2022})}\BibitemShut {NoStop}%
\bibitem [{\citenamefont {Jain}(1989)}]{JainCFL1989}%
  \BibitemOpen
  \bibfield  {author} {\bibinfo {author} {\bibfnamefont {J.~K.}\ \bibnamefont {Jain}},\ }\bibfield  {title} {\bibinfo {title} {Composite-fermion approach for the fractional quantum hall effect},\ }\href {https://doi.org/10.1103/PhysRevLett.63.199} {\bibfield  {journal} {\bibinfo  {journal} {Phys. Rev. Lett.}\ }\textbf {\bibinfo {volume} {63}},\ \bibinfo {pages} {199} (\bibinfo {year} {1989})}\BibitemShut {NoStop}%
\bibitem [{\citenamefont {Lopez}\ and\ \citenamefont {Fradkin}(1991)}]{Lopez1991}%
  \BibitemOpen
  \bibfield  {author} {\bibinfo {author} {\bibfnamefont {A.}~\bibnamefont {Lopez}}\ and\ \bibinfo {author} {\bibfnamefont {E.}~\bibnamefont {Fradkin}},\ }\bibfield  {title} {\bibinfo {title} {Fractional quantum hall effect and chern-simons gauge theories},\ }\href {https://doi.org/10.1103/PhysRevB.44.5246} {\bibfield  {journal} {\bibinfo  {journal} {Phys. Rev. B}\ }\textbf {\bibinfo {volume} {44}},\ \bibinfo {pages} {5246} (\bibinfo {year} {1991})}\BibitemShut {NoStop}%
\bibitem [{\citenamefont {Halperin}\ \emph {et~al.}(1993)\citenamefont {Halperin}, \citenamefont {Lee},\ and\ \citenamefont {Read}}]{Halperin1993}%
  \BibitemOpen
  \bibfield  {author} {\bibinfo {author} {\bibfnamefont {B.~I.}\ \bibnamefont {Halperin}}, \bibinfo {author} {\bibfnamefont {P.~A.}\ \bibnamefont {Lee}},\ and\ \bibinfo {author} {\bibfnamefont {N.}~\bibnamefont {Read}},\ }\bibfield  {title} {\bibinfo {title} {Theory of the half-filled landau level},\ }\href {https://doi.org/10.1103/PhysRevB.47.7312} {\bibfield  {journal} {\bibinfo  {journal} {Phys. Rev. B}\ }\textbf {\bibinfo {volume} {47}},\ \bibinfo {pages} {7312} (\bibinfo {year} {1993})}\BibitemShut {NoStop}%
\bibitem [{\citenamefont {Rezayi}\ and\ \citenamefont {Read}(1994)}]{Rezayi1994}%
  \BibitemOpen
  \bibfield  {author} {\bibinfo {author} {\bibfnamefont {E.}~\bibnamefont {Rezayi}}\ and\ \bibinfo {author} {\bibfnamefont {N.}~\bibnamefont {Read}},\ }\bibfield  {title} {\bibinfo {title} {Fermi-liquid-like state in a half-filled landau level},\ }\href {https://doi.org/10.1103/PhysRevLett.72.900} {\bibfield  {journal} {\bibinfo  {journal} {Phys. Rev. Lett.}\ }\textbf {\bibinfo {volume} {72}},\ \bibinfo {pages} {900} (\bibinfo {year} {1994})}\BibitemShut {NoStop}%
\bibitem [{\citenamefont {Geraedts}\ \emph {et~al.}(2016)\citenamefont {Geraedts}, \citenamefont {Zaletel}, \citenamefont {Mong}, \citenamefont {Metlitski}, \citenamefont {Vishwanath},\ and\ \citenamefont {Motrunich}}]{Geraedts2016}%
  \BibitemOpen
  \bibfield  {author} {\bibinfo {author} {\bibfnamefont {S.~D.}\ \bibnamefont {Geraedts}}, \bibinfo {author} {\bibfnamefont {M.~P.}\ \bibnamefont {Zaletel}}, \bibinfo {author} {\bibfnamefont {R.~S.~K.}\ \bibnamefont {Mong}}, \bibinfo {author} {\bibfnamefont {M.~A.}\ \bibnamefont {Metlitski}}, \bibinfo {author} {\bibfnamefont {A.}~\bibnamefont {Vishwanath}},\ and\ \bibinfo {author} {\bibfnamefont {O.~I.}\ \bibnamefont {Motrunich}},\ }\bibfield  {title} {\bibinfo {title} {The half-filled landau level: The case for dirac composite fermions},\ }\href {https://doi.org/10.1126/science.aad4302} {\bibfield  {journal} {\bibinfo  {journal} {Science}\ }\textbf {\bibinfo {volume} {352}},\ \bibinfo {pages} {197} (\bibinfo {year} {2016})},\ \Eprint {https://arxiv.org/abs/https://www.science.org/doi/pdf/10.1126/science.aad4302} {https://www.science.org/doi/pdf/10.1126/science.aad4302} \BibitemShut {NoStop}%
\bibitem [{\citenamefont {Herzog-Arbeitman}\ \emph {et~al.}(2024)\citenamefont {Herzog-Arbeitman}, \citenamefont {Wang}, \citenamefont {Liu}, \citenamefont {Tam}, \citenamefont {Qi}, \citenamefont {Jia}, \citenamefont {Efetov}, \citenamefont {Vafek}, \citenamefont {Regnault}, \citenamefont {Weng}, \citenamefont {Wu}, \citenamefont {Bernevig},\ and\ \citenamefont {Yu}}]{MFCI2}%
  \BibitemOpen
  \bibfield  {author} {\bibinfo {author} {\bibfnamefont {J.}~\bibnamefont {Herzog-Arbeitman}}, \bibinfo {author} {\bibfnamefont {Y.}~\bibnamefont {Wang}}, \bibinfo {author} {\bibfnamefont {J.}~\bibnamefont {Liu}}, \bibinfo {author} {\bibfnamefont {P.~M.}\ \bibnamefont {Tam}}, \bibinfo {author} {\bibfnamefont {Z.}~\bibnamefont {Qi}}, \bibinfo {author} {\bibfnamefont {Y.}~\bibnamefont {Jia}}, \bibinfo {author} {\bibfnamefont {D.~K.}\ \bibnamefont {Efetov}}, \bibinfo {author} {\bibfnamefont {O.}~\bibnamefont {Vafek}}, \bibinfo {author} {\bibfnamefont {N.}~\bibnamefont {Regnault}}, \bibinfo {author} {\bibfnamefont {H.}~\bibnamefont {Weng}}, \bibinfo {author} {\bibfnamefont {Q.}~\bibnamefont {Wu}}, \bibinfo {author} {\bibfnamefont {B.~A.}\ \bibnamefont {Bernevig}},\ and\ \bibinfo {author} {\bibfnamefont {J.}~\bibnamefont {Yu}},\ }\bibfield  {title} {\bibinfo {title} {Moir\'e fractional chern insulators. ii. first-principles calculations and continuum models of rhombohedral graphene superlattices},\ }\href
  {https://doi.org/10.1103/PhysRevB.109.205122} {\bibfield  {journal} {\bibinfo  {journal} {Phys. Rev. B}\ }\textbf {\bibinfo {volume} {109}},\ \bibinfo {pages} {205122} (\bibinfo {year} {2024})}\BibitemShut {NoStop}%
\bibitem [{\citenamefont {Kwan}\ \emph {et~al.}(2025)\citenamefont {Kwan}, \citenamefont {Yu}, \citenamefont {Herzog-Arbeitman}, \citenamefont {Efetov}, \citenamefont {Regnault},\ and\ \citenamefont {Bernevig}}]{MFCI3}%
  \BibitemOpen
  \bibfield  {author} {\bibinfo {author} {\bibfnamefont {Y.~H.}\ \bibnamefont {Kwan}}, \bibinfo {author} {\bibfnamefont {J.}~\bibnamefont {Yu}}, \bibinfo {author} {\bibfnamefont {J.}~\bibnamefont {Herzog-Arbeitman}}, \bibinfo {author} {\bibfnamefont {D.~K.}\ \bibnamefont {Efetov}}, \bibinfo {author} {\bibfnamefont {N.}~\bibnamefont {Regnault}},\ and\ \bibinfo {author} {\bibfnamefont {B.~A.}\ \bibnamefont {Bernevig}},\ }\bibfield  {title} {\bibinfo {title} {Moir\'e fractional chern insulators. iii. hartree-fock phase diagram, magic angle regime for chern insulator states, role of moir\'e potential, and goldstone gaps in rhombohedral graphene superlattices},\ }\href {https://doi.org/10.1103/PhysRevB.112.075109} {\bibfield  {journal} {\bibinfo  {journal} {Phys. Rev. B}\ }\textbf {\bibinfo {volume} {112}},\ \bibinfo {pages} {075109} (\bibinfo {year} {2025})}\BibitemShut {NoStop}%
\end{thebibliography}

%

\end{document}